\documentclass[twocolumn,numberedappendix,twocolappendix,appendixfloats]{openjournal}
\usepackage[dvipsnames]{xcolor}
\usepackage{textgreek}
\usepackage[utf8]{inputenc}
\usepackage[english]{babel}
\usepackage{overpic}

\usepackage{hyperref}
\hypersetup{
    unicode, 
    colorlinks=true,
    linkcolor=linkcolor,
    citecolor=linkcolor,
    filecolor=linkcolor,
    urlcolor=linkcolor,
}
\usepackage{color,colortbl}
\definecolor{linkcolor}{rgb}{0.0,0.3,0.5}
\usepackage{tensind}
\tensordelimiter{?}
\DeclareGraphicsExtensions{.bmp,.png,.jpg,.pdf}
\usepackage{verbatim}
\usepackage[normalem]{ulem}
\usepackage{orcidlink}
\usepackage{soul}
\usepackage{graphicx}	% Including figure files
\usepackage{amsmath}	% Advanced maths commands
\usepackage{cleveref}   % For references to sections and equations
\usepackage{subcaption} % For figures with multiple captions
\usepackage{physics}
\usepackage{listings}
\usepackage{booktabs}
\usepackage{pifont}

\graphicspath{ {./figs/} }

\newcommand{\ramses}{{\sc ramses}}
\newcommand{\ramsesrt}{{\sc ramses-rt}}
\newcommand{\ramsesrtz}{{\sc ramses-rtz}}

\newcommand{\prism}{{\sc prism}}
\newcommand{\calima}{{\sc calima}}
\newcommand{\hmol}{H$_2$}

\newcommand{\cmark}{\textcolor{ForestGreen}{\ding{51}}} % green check
\newcommand{\xmark}{\textcolor{BrickRed}{\ding{55}}}    % red cross
\newcommand{\smark}{\textcolor{YellowOrange}{\ding{81}}}      % neutral / intermediate

\begin{document}
\title{\textbf{{\sc calima}}: On-the-fly dust and PAH evolution for radiation-hydrodynamics galaxy formation simulations\vspace{-15mm}}

\author{
    Francisco Rodr\'iguez Montero$^{1,*}$\orcidlink{0000-0001-6535-1766},
    Yohan Dubois$^{2}$\orcidlink{0000-0003-0225-6387},
    Harley Katz$^{1,3}$\orcidlink{0000-0003-1561-3814},\\
    Adrianne Slyz$^{4}$,
    Julien Devriendt$^{4}$\orcidlink{0000-0002-8140-0422}
}

\affiliation{$^{1}$Kavli Institute for Cosmological Physics, University of Chicago, Chicago IL 60637, USA}
\affiliation{$^{2}$Institut d'Astrophysique de Paris, Sorbonne Université, CNRS, UMR 7095, 98 bis bd Arago, 75014 Paris, France}
\affiliation{$^{3}$Department of Astronomy \& Astrophysics, University of Chicago, 5640 S Ellis Avenue, Chicago, IL 60637, USA}
\affiliation{$^{4}$Sub-department of Astrophysics, University of Oxford, Keble Road, Oxford OX1 3RH, United Kingdom}

\thanks{$^*$E-mail: \href{mailto:currodri@uchicago.edu}{currodri@uchicago.edu}}

\begin{abstract}
    Dust grains and polycyclic aromatic hydrocarbons (PAHs) actively contribute to the thermodynamics, chemistry, and radiative state of the interstellar medium (ISM), yet most ISM models and galaxy simulations either exclude them altogether or adopt simplified treatments. We present {\sc calima}, a new module for dust and PAH formation and evolution in radiation-hydrodynamics simulations for {\sc ramses}, designed to self-consistently couple dust physics to radiative transfer and non-equilibrium thermochemistry in a multiphase ISM.  The model employs a two-size, two-composition dust framework with log-normal grain populations, explicitly evolving stellar dust injection, turbulence-informed gas-phase accretion, shattering, coagulation, thermal and non-thermal sputtering, and shock destruction, while PAHs are separate components with their own evolution. The evolving dust populations and radiation field determine local, wavelength-dependent opacities, photoelectric heating efficiencies, grain-assisted recombination, dust–gas collisional heating/cooling, and H$_2$ formation on both grains and PAHs. Updated treatments of thermal sputtering and collisional cooling that include finite grain sizes and modern ion–solid physics reduce sputtering rates at high temperatures and extend the regime where dust significantly cools hot gas. One-zone ISM tests show that dust and PAH evolution modifies classical thermal phase diagrams and C-bearing chemistry, while isolated disc galaxy simulations reveal environment-dependent variations in dust-to-metal ratio, small-to-large grain ratio, PAH fraction, and interstellar radiation field intensity that drive non-trivial structure in infrared emission, UV transparency, and H$_2$ formation. {\sc calima} provides a physically motivated framework to interpret dust- and PAH-based observables and to assess dust-mediated feedback in galaxy formation across cosmic time.
\end{abstract}

% Write your keywords here

\maketitle

%%%%%%%%%%%%%%%%%%%%%%%%%%%%%%%%%%%%%%%%%%%%%%%%%%

%%%%%%%%%%%%%%%%% BODY OF PAPER %%%%%%%%%%%%%%%%%%

\section{Introduction}\label{sec:intro}
Interstellar dust grains and polycyclic aromatic hydrocarbons (PAHs) are not passive tracers of galaxy evolution. Rather they are active agents that regulate the thermodynamics, chemistry, and radiative state of the interstellar medium (ISM). Dust locks up a substantial fraction of refractory elements\footnote{Refractory elements are those that condense into solids at relatively high temperatures compared to volatile elements. These usually include Fe, Mg, Si, Ca, Al, and Ni.}, modifies the charge balance and heating/cooling processes, mediates the formation of molecules, and sets the opacity that links stellar radiation to the gas \citep{Draine2011PhysicsMedium,Klessen2016PhysicalMedium}. PAHs and very small grains in particular dominate the photoelectric heating of neutral gas, providing a direct feedback loop between young stars and the thermal state of the ISM \citep[e.g.][]{Weingartner2001PhotoelectricHeating,Verstraete2021TheMedium}. Consequently, any attempt to interpret galaxy evolution through dust- and PAH-based observables must treat these components as dynamically evolving, thermodynamically active constituents rather than static by-products.

Dust grains remove heavy elements from the gas phase via depletion, thereby altering both the cooling channels and the chemical composition of the gas-phase ISM. Depletion is strong function of the local ISM properties \citep[][]{Jenkins2009AMedium,DeCia2016Dust-depletionGalaxy,DeVis2017Herschel-ATLAS:Relations}, and cosmic evolution \citep[][]{Wiseman2017EvolutionGRB-DLAs,Konstantopoulou2022DustISM,Konstantopoulou2024DustComposition}. These trends imply that most refractory atoms spend a significant fraction of their lifetime bound in dust grains and that the partition between gas-phase and solid-phase metals evolves as gas cycles between ISM phases.
Simultaneously, grains play a fundamental role in the free charge budget of the ISM. Through photo-emission, collisional charging and grain-assisted recombination, dust efficiently removes electrons and ions from the gas, especially in dense, shielded regions where grains tend to carry a net negative charge \citep[e.g.][]{Draine1987CollisionalGrains,Ibanez-Mejia2019DustMedium,Glatzle2022RadiativeRegions}. This modifies ion–molecule chemistry \citep{Langer1976TimeReactions.,Weingartner2001ElectronIonHydrocarbons,Gong2019AISM,Rimola2021InteractionImplications} and alters ambipolar diffusion \citep{Mouschovias1991MagneticMasses,Kunz2010TheResults,Khesali2012ThermalDiffusion}, influencing the fragmentation properties of magnetised clouds. Moreover, the charge balance is highly sensitive to the local grain-size distribution and charging state \citep{Masson2015AmbipolarCase,Guillet2020DustCollapse}.

One of the cornerstone results of modern ISM theory is the central role of photoelectric heating (PEH) by small grains and PAHs in the neutral, diffuse ISM \citep[e.g.][]{Bakes1994TheHydrocarbons,Weingartner2001PhotoelectricHeating,Wolfire2003NeutralGalaxy,Glover2007SimulatingConditions,Kim2013THREE-DIMENSIONALRATES}. Because the ionization potentials of many grains and PAHs lie below the Lyman limit, far-ultraviolet (FUV) photons efficiently eject valence electrons from the surface of PAHs, and the kinetic energy of these electrons is transferred to the gas. Theoretical studies consistently find that very small grains and PAHs have the largest photoelectric yields and efficiencies. The overall heating rate strongly depends on the abundance and charge distribution of the small-grain end of the size spectrum \citep{Bakes1994TheHydrocarbons,Weingartner2001PhotoelectricHeating,Berne2022ContributionObservations}. In solar-metallicity neutral gas, PEH may dominate the heating budget over a wide range of densities and radiation field strengths, with the thermal balance set primarily by the competition between this heating and cooling through CII and OI fine-structure lines \citep[see e.g.][]{Wolfire2003NeutralGalaxy,Bialy2019ThermalGas,Kim2023PhotochemistrySimulations,Katz2022PRISM:Galaxies}. Variations in the abundance or charging of PAHs and small grains therefore directly affect the position and stability of the cold and warm neutral medium (CNM/WNM) equilibrium.

Dust also provides a major cooling mechanism for star-forming gas. By absorbing UV/optical photons and re-radiating in the infrared, dust both removes energy from the radiation field and increases shielding of dense gas from young stars \citep[see the reviews by][]{Li2003InDust,Draine2003InterstellarGrains}. At sufficiently high densities, gas–dust collisions couple the gas temperature to the dust internal temperature \citep{Dwek1981TheCondensates,Hollenbach1989MOLECULECLOUDS}, allowing thermal energy to be efficiently radiated away in the IR and enabling gas to cool to low temperatures conducive to fragmentation. The necessity of dust cooling for low-mass fragmentation was established theoretically by \citet{Schneider2002FirstMetals,Schneider2006FragmentationDust} and \citet{Omukai2005ThermalEnvironments}, who showed that gas–dust collisional cooling at high densities enables a second fragmentation phase inaccessible to molecular or fine-structure line cooling. Subsequent work \citep{Schneider2010MetalsClouds,Omukai2010Low-metallicitySymmetry} demonstrated that dust cooling is the only mechanism capable of producing sub-solar mass fragments at metallicities well below the fine-structure critical metallicity threshold, thereby enabling the Pop III–Pop II transition.

Chemically, dust grains act as the dominant catalytic surfaces in the ISM. They provide a thermal bath for adsorbed atoms and molecules, enabling surface reactions that are otherwise inefficient in the gas phase. Of particular importance is molecular hydrogen: under typical ISM conditions, \hmol~formation proceeds mainly on grain surfaces, with rates determined by the total grain surface area, surface properties, and dust temperature \citep{Cazaux2002MolecularMedium,Cazaux2004Surfaces,Bron2014SurfaceFluctuations}. Dust grains also serve as the seeds for the first solid phases of many molecules in the cold and dense ISM. The build-up of icy mantles on grain surfaces is a critical step in the formation of complex organic molecules and the onset of abiogenesis \citep[e.g.][]{Tielens1982ModelMantles,Allamandola1988PhotochemicalAnalogs,Schutte1995TheIces}.

Dust may play an important dynamical role through radiation pressure, both by (single-)scattering of UV/optical photons and through the scattering, emission, and re-emission (usually termed multi-scattering) of infrared (IR) photons. Radiation pressure on dust plays markedly different roles across scales. On molecular cloud scales, it couples efficiently to dense, optically thick gas and acts as a robust feedback mechanism that contributes to cloud disruption and the regulation of star formation \citep[e.g.][]{Krumholz2010SurvivalGalaxies,Murray2010TheGalaxies}. Even so, its importance on galactic scales is far less certain: early models invoking efficient infrared multi-scattering suggested radiation pressure could drive large-scale outflows \citep{Murray2005,Thompson2005RadiationFueling}, but subsequent multidimensional radiation-hydrodynamic simulations have shown that instabilities, porosity, and photon leakage substantially reduce momentum coupling, rendering radiation pressure generally subdominant to supernova feedback except in extreme starbursts \citep{Krumholz2013NumericalWinds,Hopkins2014,Rosdahl2015GalaxiesGalaxies}. These conclusions remain sensitive to poorly constrained dust properties, including dust-to-gas ratio, grain size distribution, and opacity, which are expected to vary strongly with environment and redshift. Thus, self-consistent evolution of dust is a necessary step towards understanding the role of radiation pressure on galaxy formation and the ISM. This is particularly relevant nowadays, given the revived interest in radiation-driven ejection of dust to explain the `apparent' attenuation-free galaxies observed in the early Universe \citep[e.g.][]{Ferrara2022OnJWST,Pallottini2023StochasticImplications,Ferrara2024}.

Self-consistent evolution of dust and PAHs is not only necessary for their feedback on the ISM and galaxy formation, but also for a direct prediction of their observables across cosmic time. PAH emission bands in the 3–20~\micron~range predominantly originate in photo-dissociation regions (PDRs), at the surfaces of molecular clouds, and within or around H~{\small II} regions \citep[see the review by][]{Tielens2008InterstellarMolecules}. As their presence correlates with the far-UV (FUV) field, observations with ISO and Spitzer established that PAH emission scales with star formation rate \citep[e.g.][]{Genzel1997WhatGalaxies,Peeters2004PAHsFormation,Calzetti2007TheIndicators}. However, the overall strength, spectral shape, and relative mid-infrared band fluxes of PAH emission depend sensitively on local physical conditions, including gas density \citep{Joblin1996SpatialHydrocarbons,Verstraete2001TheModel,Sandstrom2022PHANGS-JWSTGalaxies,Chastenet2023PHANGS-JWSTMetallicity}, radiation field intensity and hardness \citep{Boulanger1998TheSpectra,Peeters2002ThePAHs,Galliano2008VariationsGalaxies,Egorov2023PHANGS-JWSTMUSE,Egorov2025PolycyclicGalaxies}, and metallicity \citep{Engelbracht2005MetallicityGalaxies,Madden2005ISMGalaxies,Sandstrom2012TheEnvironment,Chastenet2023PHANGS-JWSTMetallicity}. In particular, the PAH-to-dust mass ratio exhibits a much steeper dependence on metallicity than either the dust-to-gas or dust-to-metal ratios, with PAH emission strongly suppressed in low-metallicity environments \citep[e.g.][]{Whitcomb2024TheSpectroscopy,Shivaei2024AMIRI}. The combined effects of PAH destruction in hard radiation fields, this strong metallicity dependence, and the contribution of evolved stellar populations to PAH excitation \citep[e.g.][]{Helou2001EvidenceHydrocarbons,Calzetti2007TheIndicators,Bendo2010TheM81,Chastenet2023PHANGS-JWSTMetallicity} render PAH emission a complex and intrinsically environment-dependent tracer of star formation.

Early dust evolution models extended classical chemical evolution frameworks, evolving dust in one-zone or multi-zone schemes alongside gas, stars, and metals \citep[e.g.][]{Dwek1998TheGalaxy,Zhukovska2008EvolutionNeighbourhood,Asano2013WhatGalaxies,Hirashita2015Two-sizeGalaxies}. These models included stellar dust injection by SNe and AGB stars, destruction in SN shocks, and grain growth in the ISM, and were computationally inexpensive yet highly effective at exploring dust-to-gas relations, dust mass build-up, and extinction in idealized galaxies \citep[see the compilation by][]{Parente2025ModelingSimulations}. Coupled to semi-analytic galaxy formation models, they reproduced global dust scaling relations and the cosmic dust budget \citep[e.g.][]{Lisenfeld1998DusttoGasGalaxies,Calura2007TheTypes,Popping2017The9,Vijayan2019DetailedFormation}, but could not capture the spatially resolved coupling between dust, radiation, and ISM thermodynamics. Hydrodynamical simulations of galaxy formation have long tracked metal abundances, motivating early dust treatments based on simple scalings between metals and dust calibrated on the Milky Way (MW) or nearby galaxies. As resolved observations of dust abundances have pointed out, a fixed DTM overpredicts dust masses in low-metallicity environments and underpredicts them in metal-rich systems where ISM grain growth dominates. This leads to systematic biases in dust masses, attenuation, and IR luminosities in cosmological simulations \citep[see e.g.][for a comparison between these simplified approaches and live dust evolution models]{McKinnon2017SimulatingFailures,Byun2025HowSimulation}.

Live dust models address these issues by evolving dust as a separate component with explicit sources and sinks, so that local dust-to-metal ratios emerge from the physical conditions and the thermodynamic evolution of the gas. The increased availability of large-scale computational resources has driven a rapid expansion in the development of live dust models for cosmological galaxy formation simulations \citep[e.g.][]{Bekki2013CoevolutionHydrogen,McKinnon2016DustGalaxies,Aoyama2017GalaxyDestruction,Hou2017EvolutionSimulation,Gjergo2018DustSimulations,Li2019The6,Aoyama2020GalaxyDistribution,Granato2021DustFormation,Parente2022DustVolumes,Romano2022TheGalaxy,Choban2022TheFIRE,Dubois2024GalaxiesSimulations}. Sub-grid prescriptions track different size bins or species, including stellar seeding, accretion from the gas phase, shattering, coagulation, and destruction in SN shocks and hot gas. This new generation of simulations has achieved important successes: matching the cosmic dust density evolution to at least cosmic noon \citep[e.g.][]{McKinnon2017SimulatingFailures,Aoyama2018CosmologicalDestruction,Li2019The6,Ragone-Figueroa2024IntertwinedSimulations,Trayford2025ModellingMedium}, recovering observed relations between metallicity and dust-to-gas/dust-to-metal ratios \citep[e.g.][]{Zhukovska2014DustSupernovae,Hou2019DustSimulation,Dubois2024GalaxiesSimulations}, capturing the transition from stellar-source–dominated to ISM growth–dominated dust production at higher metallicities \citep[e.g.][]{Hou2019DustSimulation,Granato2021DustFormation,Choban2024AGalaxies,Narayanan2025TheMonsters}, and reproducing environment-dependent grain processing \citep[e.g.][]{Gjergo2018DustSimulations}, radial dust profiles \citep[e.g.][]{Granato2021DustFormation,Parente2022DustVolumes}, spatial dust inhomogeneities across ISM phases \citep[e.g.][]{Zhukovska2016MODELINGISM,Hu2019ThermalMedium,Matsumoto2024ObservationalSimulations}, and extinction-curve trends driven by evolving grain size distributions \citep[e.g.][]{Aoyama2020GalaxyDistribution,Romano2022DustGalaxy,Dubois2024GalaxiesSimulations}.

Despite this progress, the coupling of live dust to radiation transfer and thermochemistry remains incomplete in most galaxy and ISM-scale simulations. Many models evolve dust mass (and sometimes size distributions) but still employ static or metallicity-scaled opacities in radiative transfer, so changes in small-grain and PAH abundances, coagulation, or shattering do not feed back into UV/optical attenuation or radiation pressure. Conversely, simulations that prioritize detailed multiphase ISM structure and non-equilibrium chemistry typically rely on empirical dust prescriptions: metal depletions are not consistent with the adopted cooling curves, photoelectric heating efficiencies are fixed or tied only to bulk dust abundance, and \hmol~formation rates depend on Milky Way–calibrated efficiencies rather than on the properties of an evolving grain population. Cosmological live-dust simulations often lack on-the-fly radiative transfer, using approximate UV backgrounds and line-of-sight treatments that ignore detailed dust properties, while PAHs are almost always implemented via fixed templates or post-processing rather than as a live component that modifies local heating and opacities.

We present a new implementation of dust and PAH evolution in the \ramses\ hydrodynamical galaxy formation code \citep{Teyssier2002}, in its version adapted for on-the-fly radiative transfer \citep[\ramsesrt,][]{Rosdahl2013Ramses-rt:Context,Rosdahl2015ARAMSES-RT} and its direct coupling to the non-equilibrium chemistry of primordial, metal, and molecular species \citep[\ramsesrtz,][]{Katz2022RAMSES-RTZ:Hydrodynamics}. Section~\ref{sec:twosize_model} describes the two-size dust evolution model, including its coupling to gas heating, cooling, and radiative transfer. Section~\ref{sec:pahs} extends this framework to include the chemistry, evolution, and impact of PAHs in the ISM. In Section~\ref{subsec:eq_tests} we assess the impact of individual dust and PAH processes on the equilibrium structure of the local ISM, and in Section~\ref{subsec:isolated_galaxy} we apply the full model to an isolated galaxy simulation. Finally, Section~\ref{sec:discussion} discusses how this implementation relates to previous dust evolution models in \ramses\ and outlines its limitations, while Section~\ref{sec:conclusions} summarises our main results and future prospects.

\begin{figure*}
\centering
\begin{overpic}[width=\textwidth]{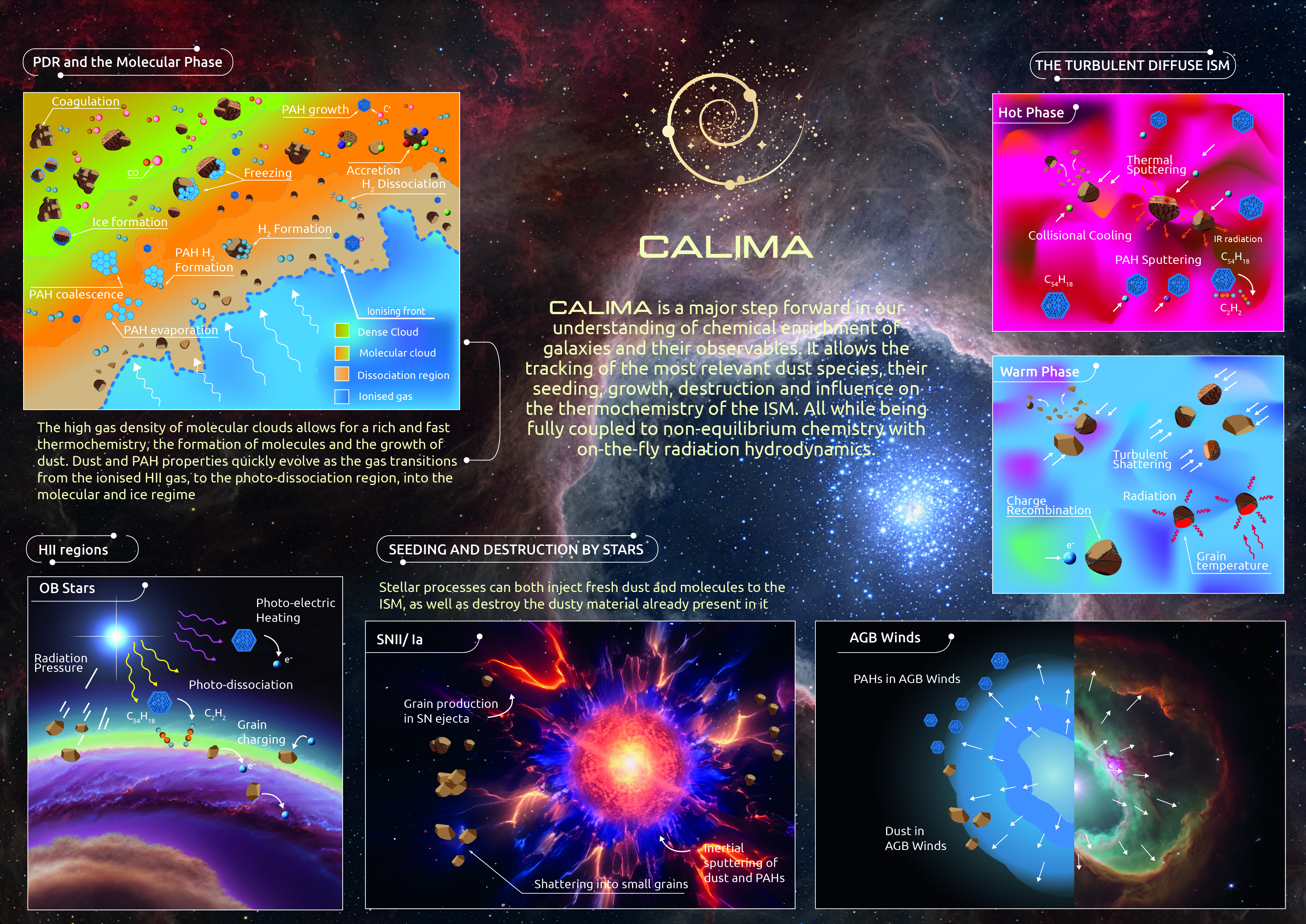}
  % Coordinates are percentages of the figure size
  \put(04,63){\hyperref[subsec:coagulation]{\phantom{coagul}}}
  \put(22,62){\hyperref[subsec:pah_accretion]{\phantom{accret}}}
  \put(19,57){\hyperref[subsec:pah_freezing]{\phantom{accr}}}
  \put(27,57.5){\hyperref[subsec:accretion]{\phantom{accrt}}}
  \put(07,53){\hyperref[subsec:accretion]{\phantom{ice form}}}
  \put(20,53){\hyperref[subsec:h2_formation_dust]{\phantom{h2 form}}}
  \put(13,51){\hyperref[subsec:h2_formation_pah]{\phantom{h2 fo}}}
  \put(02,48){\hyperref[subsec:pah_clustering]{\phantom{pah cluste}}}
  \put(09.5,45){\hyperref[subsec:pah_evaporation]{\phantom{pah cluste}}}
  \put(03,20){\hyperref[subsec:optical properties]{\phantom{pah c}}}
  \put(19,24){\hyperref[subsec:photoelectric_heating]{\phantom{photoele}}}
  \put(15,18){\hyperref[subsec:pah_dissociation]{\phantom{photodisso}}}
  \put(18,14){\hyperref[subsec:charging]{\phantom{charg}}}
  \put(31,16){\hyperref[subsec:grain_seeding]{\phantom{grain seed}}}
  \put(65,19){\hyperref[subsec:pah_seeding]{\phantom{grain seedi}}}
  \put(68,06){\hyperref[subsec:grain_seeding]{\phantom{grain s}}}
  \put(54,04){\hyperref[subsec:sn_shocks]{\phantom{sn shoc}}}
  \put(41,03){\hyperref[subsec:sn_shocks]{\phantom{sn shock destru}}}
  \put(86,58){\hyperref[subsec:sputtering]{\phantom{sputte}}}
  \put(78.5,52){\hyperref[subsec:collisional_cooling]{\phantom{collisionalc}}}
  \put(85.5,50.5){\hyperref[subsec:pah_thermal_sputtering]{\phantom{collisiona}}}
  \put(85,35){\hyperref[subsec:shattering]{\phantom{shatte}}}
  \put(77,32){\hyperref[subsec:dust_photoelectric_heating]{\phantom{recombi}}}
  \put(91.5,27.5){\hyperref[subsec:optical properties]{\phantom{tempera}}}
\end{overpic}
\caption{Overview of the \calima~model. 
Click on a process name to jump to its detailed description.}
\label{fig:general_diagram}
\end{figure*}
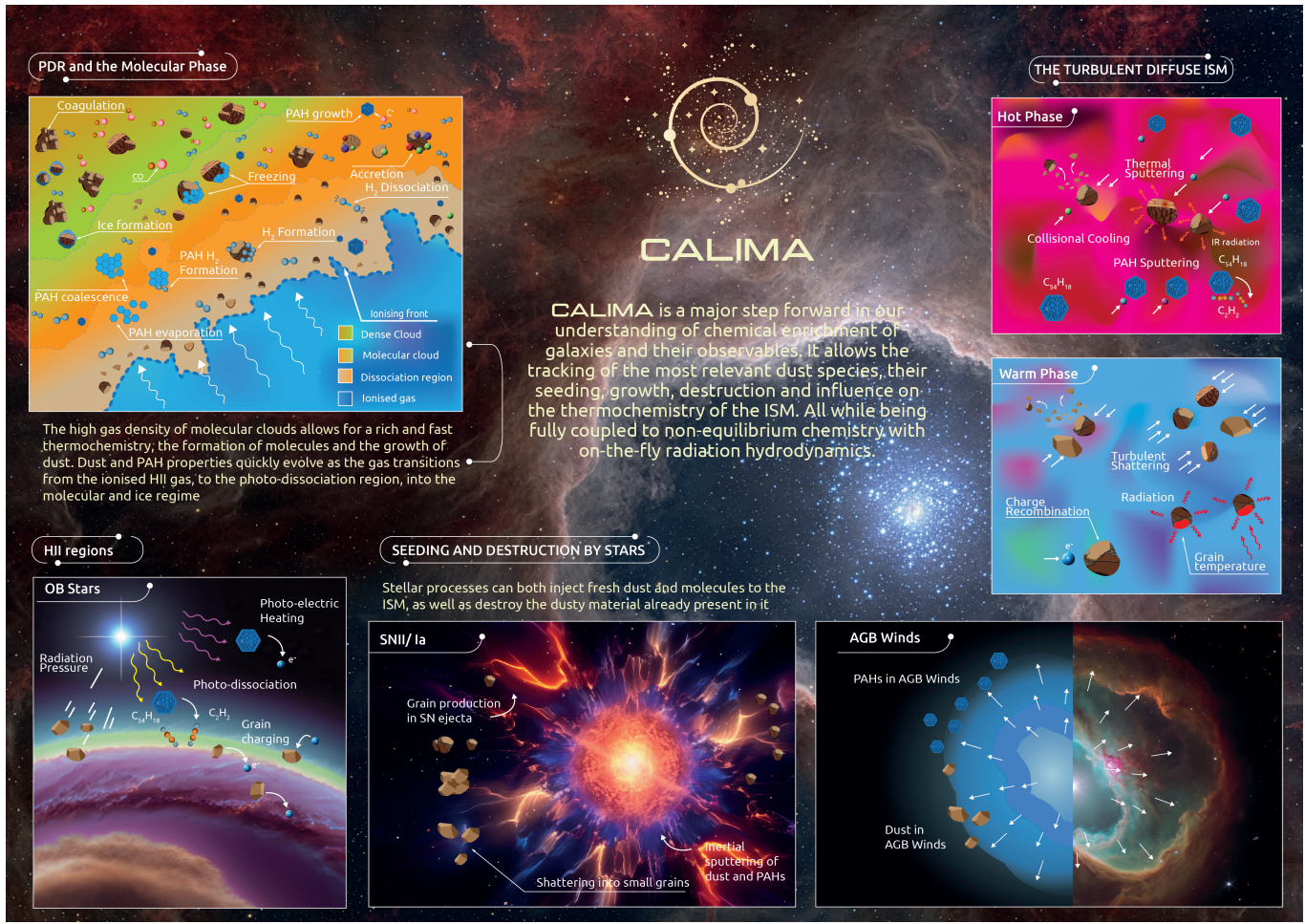

\section{The two-size dust model} \label{sec:twosize_model}
The dust size distribution is the result of the complex interplay between dust formation and destruction. Because interstellar extinction spans a wide range of wavelengths, grains with radii $a$ satisfying $2\pi a/ \lambda \sim 1$ tend to contribute most efficiently to the extinction at wavelength $\lambda$. The grain-size distribution (GSD) derived from this premise extends over many orders of magnitude and is usually explained by two fundamental processes: collisional fragmentation and turbulent grain growth. If one considers an initial distribution dominated by large dust grains (possibly seeded by AGB winds or SNe Type~II ejecta, see Section~\ref{subsec:grain_seeding} for further details) the large relative velocities of grains can easily result in collisions that break apart these grains into fragments of smaller sizes. This process of shattering was originally discussed by \citet{Hellyer1970TheAsteroids} in the context of asteroid collisions, and later introduced within the framework of interstellar dust grains by \cite{Dorschner1982InterstellarCollisions}. For a simplified derivation, see Chapter 5.5.2 of \cite{Krugel2008AnDust}, but the fundamental result is the prediction of a power-law grain distribution such that the grain number density $n$ with radius $a$ scales according to $n(a) \propto a^{-3.5}$, in agreement with the canonical Mathis-Rumpl-Nordsieck (MRN) distribution \citep{Mathis1977TheGrains.} for the MW diffuse ISM. While the collisional fragmentation model predicts that an initial large dust grain distribution evolves towards a power law distribution, based on the conservation of dust mass in high velocity collisions and without evaporation, (see also Section~\ref{subsec:shattering}), turbulent grain growth is needed to explain the growth of dust mass after it is injected into the ISM. 

There is well established observational evidence for supersonic ISM turbulence, which according to theory and simulations gives rise to an approximately log-normal\footnote{There is strong evidence that the probability density function (PDF) deviates from log-normal at high Mach numbers \citep[e.g.][]{Federrath2012OnClouds,Hopkins2013ATurbulence,Federrath2015TheTurbulence,Hennebelle2024InefficientModel}.} distribution of gas densities \citep[see reviews by][]{Low2004ControlTurbulence,Elmegreen2004InterstellarProcesses,McKee2007,Hennebelle2012TurbulentClouds}. In Section~\ref{subsec:accretion} we carefully examine how the growth of dust grains by gas-phase accretion of metals depends on gas density. Given the large variance of the gas density field caused by supersonic turbulence, dust accretion occurs on different timescales within a turbulent ISM. Therefore, the assumption that gas and dust grains are highly coupled implies that dust grain size correlates with gas density distribution. Even if the initial dust grain size distribution follows a power law, the memory of the initial conditions will be quickly erased by the underlying gas density field \citep{Mattsson2020GalacticTurbulence}. 

These two scenarios explain the origin of two widely used GSDs: power-laws and log-normals \citep[e.g.][]{Compiegne2010TheDustEM,Jones2017TheSolids,Zubko2004InterstellarConstraints,Hensley2021ObservationalEra}. However, these models make different assumptions about grain type and GSD spatial variation. Nevertheless, both models agree on the following aspects of the diffuse ISM grain properties \citep{Krugel2008AnDust}:
\begin{enumerate}%[itemsep=5pt, topsep=10pt, partopsep=0pt, parsep=0pt]
    \item The sizes of very small carbonaceous grains (VSGs) need to peak in the range between $3 \leq a \leq 30$\AA, so that they can be stochastically heated to the high temperature necessary to explain mid-IR features (see Section~\ref{sec:pahs}).
    \item Small (but larger than VSGs) grains in the range from $\sim 5$~nm to $\sim 30$~nm contribute to the mid-IR continuum. Whether or not silicate grains are also found at these sizes determines whether there are $9.7$~and $18$\,$\mu$m emission features due to O-Si-O bending.
    \item The sizes of large grains need to peak at $\sim 0.1$~$\mu$m and have a sharp cut-off at larger $a$. This constraint arises from multiple observational probes: the near-IR and optical extinction slope \citep[e.g.][]{Cardelli1989TheExtinction,Fitzpatrick1999CorrectingExtinction,Weingartner2001DustCloud}, scattering of diffuse galactic light \citep[e.g.][]{Brandt2011THELIGHT}, and scattering from X-ray sources \citep[e.g.][]{Predehl1995X-rayingHalos.,Draine2003TheDust}
\end{enumerate}
\begin{table*}
    \centering
    \begin{tabular}{c|c|c|c|c|c|c}
         & $s$ [g\,cm$^{-3}$] & $N_{\rm C}$ & $a_0$ [\micron] & $a_{\rm min}$ [\micron] & $a_{\rm max}$ [\micron] & $\sigma$\\
         \hline
         \textbf{small PAHs} & 2.0 & 54 & $5\times 10^{-4}$ & $3\times 10^{-4}$ & $1\times 10^{-3}$ & 0.3\\
         \textbf{large PAHs} & 2.0 & 418 & $1\times 10^{-3}$ & $3\times 10^{-4}$ & $9\times 10^{-3}$ & 0.4 \\
         \textbf{small C} & 2.2 & - & $1\times 10^{-2}$ & $1\times 10^{-3}$ & $1\times 10^{-1}$ & 0.7\\
         \textbf{large C} & 2.2 & - & $1\times 10^{-1}$ & $5\times 10^{-3}$ & $1.0$ & 0.8\\
         \textbf{small Sil} & 3.3 & - & $5\times 10^{-3}$ & $5\times 10^{-4}$ & $1\times 10^{-1}$ & 0.75\\
         \textbf{large Sil} & 3.3 & - & $1\times 10^{-1}$ & $5\times 10^{-3}$ & $1.0$ & 0.75\\
    \end{tabular}
    \caption{Fiducial parameters for the full \calima~model. All grain/molecules assume a modified log-normal GSD given by Eq.~\ref{eq:lognormal_grain_size_distribution}. From left to right columns: material density, number of carbon atoms in the PAH molecule, grain/molecule centroid radius, minimum and maximum size cut-offs, and width of the log-normal distribution.}
    \label{tab:dusty_size_parameters}
\end{table*}
Therefore, temporarily ignoring the source of aromatic features attributed to VSGs, the GSD can be roughly described by a population of small grains with $a \leq 30$\,nm and a population of large grains with $a > 30$\,nm. This is considered to be the minimal two-size approximation for the evolution of the GSD in the ISM \citep{Hirashita2015Two-sizeGalaxies}, and captures the main features of the full GSD \citep{Asano2013WhatGalaxies}. Similar to the two-size approximation adopted by other galaxy formation simulations \citep{Aoyama2017GalaxyDestruction,Gjergo2018DustSimulations,Granato2021DustFormation}, we model small and large grains with central bin sizes given in Table~\ref{tab:dusty_size_parameters}. PAHs are also separated in small ($a_0=5$\AA) and large ($a_0=10$\AA) bins (see Section~\ref{sec:pahs} for further details with respect to this discretisation). We assume that the underlying GSD within each bin is a modified log-normal:
\begin{align}\label{eq:lognormal_grain_size_distribution}
    n(a) \dd a= \frac{C}{a^4}\exp \left(\frac{[\log{(a/a_{\rm 0})}]^2}{2\sigma^2} \right) \dd a,
\end{align}
where the grain number density $n(a)$ has been defined as the number of grains per unit volume with radius in the range $a \rightarrow a + \dd a$. This log-normal is centred at $a_0$, and with standard deviation given by $\sigma$. The normalisation constant $C$ is obtained for each cell and each grain type $j$:
\begin{align}
    C_j = \frac{m_j n_{{\rm d},j}}{\int_{a_{{\rm min},j}}^{a_{{\rm max},j}}\frac{4}{3}\pi \frac{s_j}{a}\exp \left(\frac{[\log{(a/a_{\rm 0})}]^2}{2\sigma^2} \right) \dd a},
\end{align}
which arises from relating the local dust density to the integral of the underlying distribution. Here $m_j=\frac{4}{3} \pi s_j a_0^3$ is the mass of a single dust grain, $n_{{\rm d},j}$ is the dust number density, and $s_j$ is the grain material density.

\begin{figure}
    \centering
	\includegraphics[width=\columnwidth]{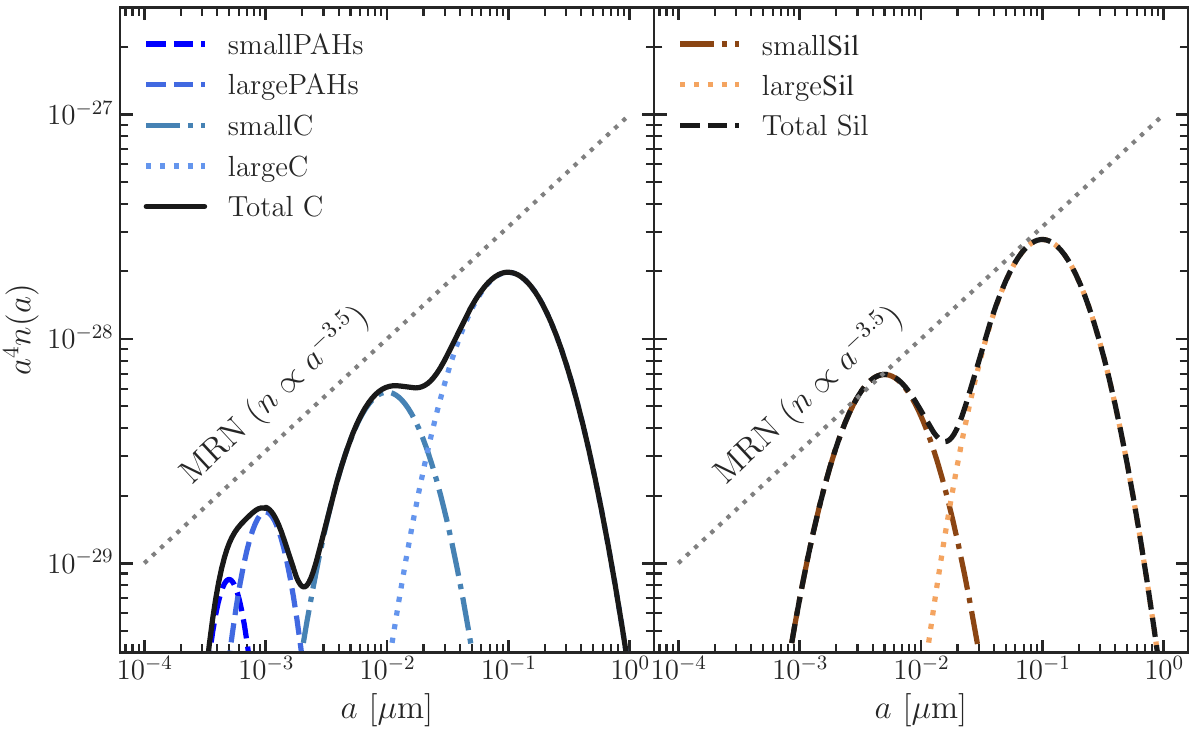}
    \caption{Underlying grain size distribution assumed within each size bin in the \calima~model for PAHs, carbonaceous grains and silicate grains. This has been computed for a gas density of $1$\,H cm$^{-3}$ and using the DTG for each individual component (PAH, carbonaceous, silicate) of the BARE-GR-S model in \citet{Zubko2004InterstellarConstraints}.}
    \label{fig:lognormal_distribution_pahs}
\end{figure}
Fig.~\ref{fig:lognormal_distribution_pahs} shows the decomposition of the underlying GSD for the full \calima~model, separating carbonaceous grains in the left panel, from silicate grains in the right panel. Abundance for each grain type is inferred using the BARE-GR-S model in \citet{Zubko2004InterstellarConstraints}. We compare our results to the slope of the MRN distribution: our choice of underlying GSD and MW abundances assumption agrees with an MRN extending from large grains to the smallest PAHs.

In addition to dividing the dust grain population into large and small size bins, we separate the dust composition into amorphous carbonaceous grains and silicate grains with a fixed stoichiometry for a given simulation. We briefly review our choice of silicate properties. 

Silicate compounds are built around the silica tetrahedra SiO$_{4}^{4-}$. Due to their overall negative charge, they pair by ionic bonds with interstellar cations, which due to cosmic abundance are limited to Mg$^{2+}$ and Fe$^{2+}$ \citep[e.g.][]{Asplund2009TheSun}. Olivine and pyroxene are typically considered the relevant silicate compositions for astrophysical dust grains \citep[see the reviews by ][]{Dorschner1995DustGalaxy,Draine2003InterstellarGrains,Henning2010CosmicSilicates}. Olivine crystals are characterised by a orthorhombic network of silica tetrahedrae joined by dications, while in pyroxene the crystalline structure can be either orthorhombic or monoclinic. The general composition of olivine is Mg$_{2x}$Fe$_{2-2x}$SiO$_4$, and of pyroxene is Mg$_{x}$Fe$_{1-x}$SiO$_3$, with $x$ ranging from 0 to 1 depending on the Mg-Fe inclusion ratio. As Mg and Fe are interchangeable on the crystalline lattice, both olivine and pyroxene can be considered as a solid solution of magnesium and iron. 

Since we fix the silicate composition in any given \calima~ simulation, our choice will influence the exact depletion of refractory elements from the gas phase. 
Different studies have tried to put constraints on the (Fe+Mg)/Si ratio of astrophysical silicates. \cite{Sofia2001InterstellarRevisited} finds that interstellar depletion results in a 2:1 ratio, suggesting a (Mg,Fe)SiO$_4$ composition, with X-ray spectroscopy finding evidence of a mix of silicates with olivine- and pyroxene-type\footnote{The olivine and pyroxene terms are specific to crystalline silicate structures, but we use them across this paper as a way to refer to the \textit{effective} stoichiometry of a given amorphous silicate.} composition \citep{Costantini2001ChandraProperties}. However, these compositions appear to be in disagreement with the depletion factors determined by \cite{Jenkins2009AMedium}, which suggest a Si:O ratio close to $6.6 \pm 2.5$. It remains difficult to account for the depletion of oxygen in galactic dust models, even if it also forms oxides \citep[e.g.][]{Choban2022TheFIRE}. 
Given the modelling difficulty involved, we leave the exploration of different silicate compositions for future work and instead assume a composition of olivine with equal amounts of Mg and Fe inclusions (MgFeSiO$_4$) as commonly done to represent the interstellar silicate absorption profiles \citep[i.e. the 9.7 and 18 \micron~features,][]{Draine1984OpticalGrains,Li2001InfraredMedium,Kemper2004TheMedium,Chiar2005PixieMedium}, although we note that there is tentative evidence of Mg-rich silicates in starburst systems \citep{Min2007TheGrains}.

This minimal model (two grain size bins, and two chemical species, i.e.~carbonaceous and silicate grains) is capable of capturing the primary features of the Galactic GSD, while remaining computationally efficient in terms of memory usage and simplifying the number of inter-grain reaction rates.

\subsection{Grain seeding}\label{subsec:grain_seeding}
The amount of silicates $m_{\rm key}$, traced by its key element (here Si), released by a single star is given by the mass of the least abundant element entering its composition:
\begin{align}
m_{\rm key}=\min_{\ell=\rm{Mg,Fe,Si,O}} \left ( \frac{M_\ell}{A_\ell N_\ell} \right ) \, ,
\end{align}
with $M_\ell$ the total mass of the $\ell$-th element released by the star, $A_\ell$ its atomic weight and $N_\ell$ the number of $\ell$-th element atoms entering the composition of the silicate. The key element condensates within the ejecta into dust of mass $D_{{\rm ej,sil},\ell}^k$ with a given efficiency $\delta_{\rm Sil}$ according to the formula:
\begin{align}
D_{{\rm ej,sil},\ell}^k=\delta_{\rm Sil}^k m_{\rm key} A_\ell N_\ell\
\end{align}
which guarantees a constant mass ratio with values $A_\ell N_\ell/\sum(A_\ell N_\ell)=0.14,0.33,0.16$, 0.37 for $\ell=\rm Mg$, Fe, Si, and O respectively for the case of $x=1$ olivine\footnote{$A_\ell N_\ell/\sum(A_\ell N_\ell)=0.11,0.24,0.24,0.41$ for $\ell=\rm Mg$, Fe, Si, and O respectively for the case of $x=1$ pyroxene.}. 
The superscript $k$ denotes the origin of the stellar mass release with $k=$ AGB, SNII or SNIa.
The condensation efficiency for SNII following~\citet{Dwek1998TheGalaxy} is equal to $\delta_{\rm Sil}^{\rm SNII}=0.8$. Contrary to \citet{Dwek1998TheGalaxy}, we do not consider dust formation in SN Ia, as models suggest that formation efficiency is drastically reduced in SN Ia~\citep[e.g.][]{Nozawa2011FormationSupernovae}, and observations have yet to find newly formed dust around them \citep{Gomez2012DustHerschel}. For AGB stars, the condensation efficiency depends on the ratio of C/O: if C/O~$\ge1$ all the oxygen is associated to carbon in CO molecules, and, hence, silicates do not condensate anymore due to the unavailability of O, while with C/O~$<1$ oxygen is still available to condense into silicates.
We adopt $\delta_{\rm Sil}^{\rm AGB, C/O<1}=0.8$ and $\delta_{\rm Sil}^{\rm AGB, C/O\ge1}=0$ as in~\citet{Dwek1998TheGalaxy}.
For the same reasons, accretion of elements onto silicates from the ISM will be limited by the least abundant of the elements in the ISM entering the composition of the grain, replacing the total mass $M_\ell$ released by the star by the gas mass of the given element $\ell$ in the cell.

Similarly to silicates, the amount of carbonaceous dust and its condensation efficiency $\delta_{\rm C}^{k}$ vary with the nature of the stellar ejecta. We assume $\delta_{\rm C}^{\rm SNII}=0.5$ and $\delta_{\rm C}^{\rm SNIa}=0$, while for AGB stars this efficiency depends on the ratio of C/O in the ejecta. If the ratio C/O~$<1$, then all the C elements form CO molecules that are then not available for dust $\delta_{\rm C}^{\rm AGB, C/O<1}=0$\footnote{For consistency, we should then release CO in this situation, but contrary to dust grains, \ramsesrtz~includes direct formation of CO from gas-phase C and O, so we assume that the direct formation is predicted by the chemistry solver.}, while for C/O~$\ge1$ the efficiency is 1 with a mass released into C dust:
\begin{align}
D_{\rm ej, C}^{\rm AGB}=\delta_{\rm C}^{\rm AGB, C/O\ge 1} \left (M^{\rm AGB}_{\rm C}-\frac{A_{\rm C}}{A_{\rm O}} M^{\rm AGB}_{\rm O} \right)\, .
\end{align}

The dust produced in stellar ejecta is divided into a population of small and large grains, i.e.~$\dot D_{{\rm ej},i,{\rm S}}^{k}=f_{{\rm ej},i,{\rm S}}\dot D_{{\rm ej},i}$ and $\dot D_{{\rm ej},i,{\rm L}}^k=(1-f_{{\rm ej},i,{\rm S}})\dot D_{{\rm ej},i}$, where $f_{{\rm ej},i,{\rm S}}$ is the fraction of small grains released for a given chemical type $i$ (carbonaceous or silicate grains).
Observational studies of dust in young SN remnants (SNRs) indicate that SN-condensed dust is frequently dominated by relatively large grains. Far-IR spectral modelling of the Crab Nebula requires size distributions with significant mass in grains $\gtrsim 0.1$\micron~\citep[e.g.][]{Blair1997Overview,Temim2013THENEBULA}, and analyses of Galactic SNRs show trends in extinction properties consistent with size redistribution towards large grains \citep[][]{Zhao2025ObservationalPhase}. Additionally, modelling of SN dust survival indicates that larger grains have much higher survival fractions through reverse shock processing \citep[e.g.][]{Nozawa2007EvolutionMedium,Slavin2020TheGrains}.
We assume that all the dust released in the ejecta condenses into the large grain population only, thus, $f_{{\rm ej},i,{\rm S}}=0$ (see appendix B in \citealt{Dubois2024GalaxiesSimulations} for the predictions of extinction curves with different values of $f_{{\rm ej},i,{\rm S}}$). These efficiencies can easily be used with the metal yields of any stellar model. Finally, it is important to note that the predicted and observed dust condensation efficiencies are very uncertain and can vary by up to an order of magnitude~\citep[see the review by][]{Schneider2024TheSources}. 

\subsection{Grain charging}\label{subsec:charging}
Dust grains constantly interact with ions and molecules in ISM, generally causing the grain to acquire a charge\footnote{Throughout this work it is assume that grain charge $Q_{\rm g}$ is in units of the elementary charge $e$.} $Q_{\rm g}$. In the diffuse ISM, where grains are subject to more intense radiation fields, they may be stripped of electrons by photons, resulting in an overall positive charge, whereas in dense clouds, dust grains can capture free electrons rendering them negatively charged. This has a non-negligible effect on the interaction of grains with each other and the gas, since accretion of gas phase metals is affected by electrostatic forces between the grain and ion \citep[e.g.][]{Weingartner1999InterstellarGrains,Zhukovska2018IronDepletions}, photo-electric yields are highly dependent on the ionisation potential of the grain \citep[][]{Bakes1994TheHydrocarbons,Weingartner2001PhotoelectricHeating}, charged dust dynamics is affected by electrostatic drags and acceleration by magnetic field fluctuations \citep[e.g.][]{Yan2004DustTurbulence}, and magnetic resistivities on collapsing protostellar cores depends on the charge acquired by grains under low ionisation fractions and high densities. Therefore, dust charging is a very important process for the thermo-dynamical interaction of dust grains with the ISM. Despite this, it has been restricted to just a few implementations in photo-ionisation/spectral synthesis codes \citep[{\sc Meudon PDR}, {\sc cloudy}, and {\sc mocassin},][respectively]{LePetit2006ACode,Ferland2017TheCloudy,Ercolano20083DRegions}, Monte-Carlo radiative-transfer \citep[{\sc CRASH},][]{Glatzle2022RadiativeRegions}, and post-processing dust modelling tools for SN ejecta \citep[{\sc paperboats},][]{Kirchschlager2019DustDensities}. 

Introducing explicit grain-charging in radiation hydrodynamics (RHD) simulations is non-trivial, since dust grains have a large reservoir of valence electrons as well as sufficiently low electron affinity to allow capture of free electrons. This results in  grain charge distributions (GCDs) that extend over a very broad range of positive and negative charges, established from the equilibrium of collisional and photo-emission charging \citep[see][]{Weingartner2001PhotoelectricHeating}. Given the computational and memory cost of obtaining the equilibrium GCD at every cooling time-step and in every cell of our \ramses\ simulation, we instead make use of tabulated values of grain mean charge and the width of the GSD for the local conditions. The underlying assumptions from these tables are: (i) an averaged interstellar radiation field (ISRF) of photo-emission efficiency, and (ii) charging timescales much shorter than the cooling and hydrodynamical time-steps \citep[see][showing that this is a good assumption for grains larger than PAHs]{Ibanez-Mejia2019DustMedium}.

\calima~uses pre-computed tables of the GSDs using the theoretical framework of \citet{Weingartner2001PhotoelectricHeating}. This allows us to predict the evolution of the GCD for the individual grain sizes and compositions used in \calima. We provide further details of the algorithm to compute the equilibrium GCD in Appendix~\ref{ap:dust_charging}. In Fig.~\ref{fig:dust_charge_silicate_0.005micron} we represent the GCD mean charge $\langle Q_{\rm g} \rangle$ and width $\sigma_{Q_{\rm g}}$ of small silicate grains for a range of values of the charging parameter\footnote{$G_0$ is the radiation energy density in the Habing band (i.e.~6-13.6\,eV) normalised to the \citet{Habing1968TheA} estimate of $u_{\rm Hab}=5.33\times 10^{-14}\, \rm erg\, \rm cm^{-3}$} $\gamma=G_0\sqrt{T}/n_{\rm e}$ ($T$ is the gas temperature and $n_{\rm e}$ is the free electron number density). Previous works have argued that the GCD can be solely parametrised by $\gamma$ \citep[e.g.][]{Weingartner2001PhotoelectricHeating,Ibanez-Mejia2019DustMedium} when considering an average ISRF like the one from \citet{Mathis1983InterstellarClouds}. In order to test this assumption, we have computed the GCD for different combinations of $n_{\rm e}$ and $T$ that result in the same value of $\gamma$, represented in Fig.~\ref{fig:dust_charge_silicate_0.005micron} by different marker size (varying $n_{\rm e}$) and different colours (varying $T$). We find that there is a non-trivial second dependence on $T$, which suggest that fitting functions, like the ones from \citet{Ibanez-Mejia2019DustMedium}, may fail to fully capture the variation in grain charge with temperature (see the orange curves in Fig.~\ref{fig:dust_charge_silicate_0.005micron} that result from using their fitting parameters). To help in the physical understanding of this temperature dependence, we have overplotted running medians of $\langle Q_{\rm g} \rangle$ and $\sigma_{Q_{\rm g}}$ for low temperature $T<10^4$\,K (green, solid lines) and high temperature $T>10^4$\,K (red, dot-dashed lines). While for low temperature the collisional charging is not sufficient to overcome the photo-detachment and photo-emission caused by the \citet{Mathis1983InterstellarClouds} ISRF, sufficiently high gas temperatures allow for the collisional charging rate to exceed any photo-emitting rate, which causes these small grains to acquire a negative charge as $\gamma$ (and, effectively, $n_{\rm e}$) decreases. The grain can keep decreasing its charge until the minimum allowed charge (see the definition of $Q_{\rm min}$ in Appendix~\ref{ap:dust_charging}) is reached, which for small silicate grains is $Q_{\rm min}=-22$ (indicated by a horizontal dashed line in the top panel of Fig.~\ref{fig:dust_charge_silicate_0.005micron}). In practice, we have tabulated the mean charge and the width of the GCD varying $\gamma$ and $T$, which allows quick computation of the GCD properties for each grain in \calima~considering the properties of the local radiation field.

\begin{figure}
    \centering
	\includegraphics[width=\columnwidth]{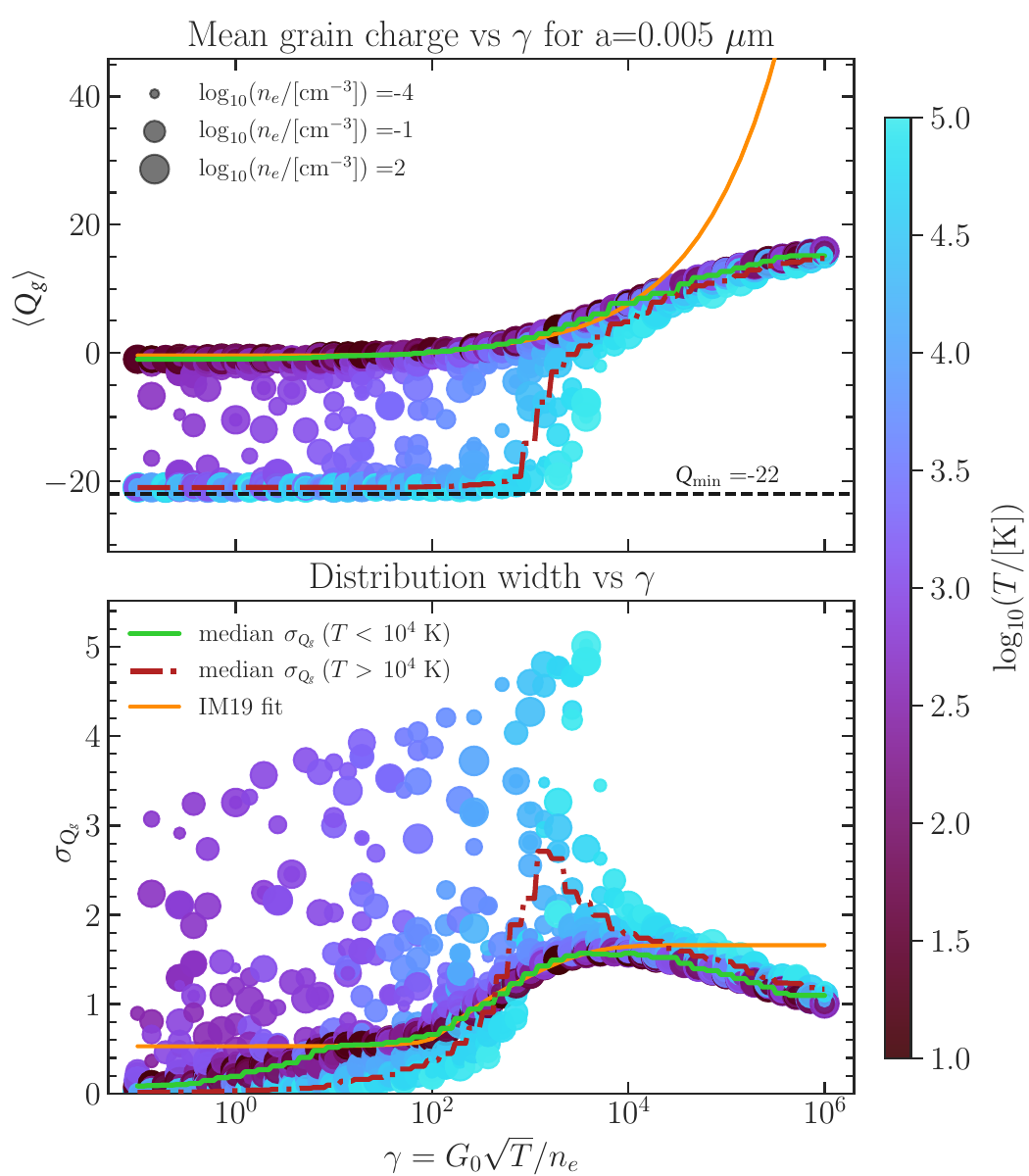}
    \caption{Equilibrium grain charge of the small silicate grains ($a=0.005$\,\micron) under the \citet{Mathis1983InterstellarClouds} ISRF. Each point describes the charge probability distribution by the mean $\langle Q_{\rm g} \rangle$ (top panel) and width $\sigma_{Q_{\rm g}}$ (bottom panel) at a given value of the charging parameter $\gamma = G_0 \sqrt{T}/n_{\rm e}$. For a fixed $\gamma$, we also vary $T$ between $10$\,K and $10^5$\,K and $n_{\rm e}$ between $10^{-4}$ and $10^2$\,cm$^{-3}$, represented by the colourmap and the size of the point, respectively. The result from the equilibrium modelling by \citet{Ibanez-Mejia2019DustMedium} is overplotted as solid orange lines, compare to the median of our distribution for neutral ($T<10^4$\,K) and ionised gas ($T>10^4$\,K).}
    \label{fig:dust_charge_silicate_0.005micron}
\end{figure}

\subsection{Dust photo-electric heating and recombination cooling}\label{subsec:dust_photoelectric_heating}
Photo-electric heating (PEH) is the dominant source of heating in the diffuse ISM as the gas metallicity approaches solar values \citep[e.g.][]{Wolfire2008ChemicalFormation,Bialy2019ThermalGas,Kim2023PhotochemistrySimulations,Katz2022PRISM:Galaxies}. PEH is also the dominant process that couples the sub-ionising ($E< 13.6$\,eV) far-UV radiation field to the diffuse gas, therefore controlling the ionisation state of PDRs and the phase structure of the ISM. Originally proposed by \citet{Spitzer1948TheI.}, it has been the subject of extensive theoretical modelling \citep[e.g.][]{Bakes1994TheHydrocarbons,Weingartner2001ElectronIonHydrocarbons}. Essentially, absorption of a far-UV photon results in the release of an electron, which loses energy as it travels through the grain material. If the electron reaches the grain surface with sufficient energy to overcome the work function $W$ of the material, it may escape the grain and become a free electron that contributes to the thermal state of the gas. 

\begin{figure}
    \centering
	\includegraphics[width=\columnwidth]{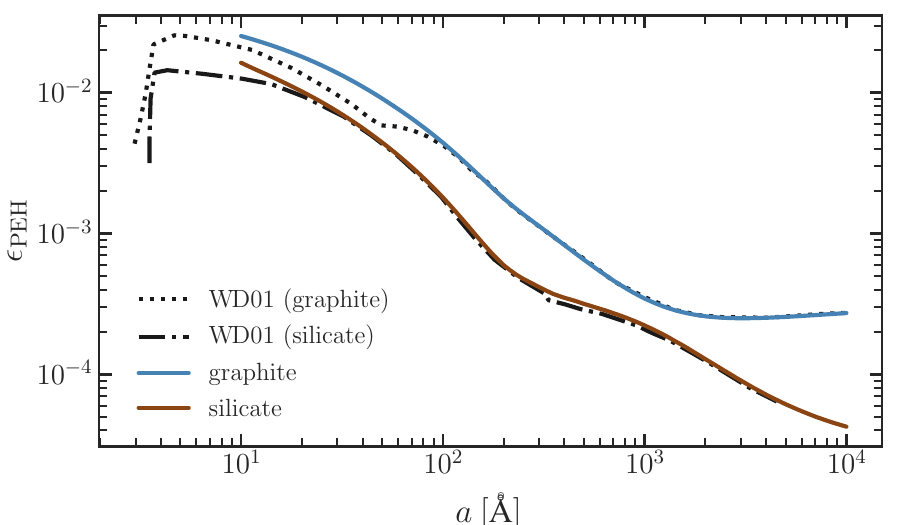}
    \caption{Photo-electric net heating efficiency for graphite (blue line) and silicate (brown line) grains embedded in the \citet{Mathis1983InterstellarClouds} ISRF with the WNM phase conditions. Our modelling is in excellent agreement with the results of \citet{Weingartner2001PhotoelectricHeating} down to $a\lesssim 100$\,\AA~for graphite grains due to the different choice of optical properties in their work compared to \calima.}
    \label{fig:dust_peh_efficiency_WNM}
\end{figure}

The implementation of PEH in thermo-chemical models of the ISM relies on fitting functions to the total gas heating per H nucleus and per Habing band flux $G_0$, assuming a particular composition mix of dust grains and a GSD \citep[see e.g.][]{Bakes1994TheHydrocarbons,Weingartner2001PhotoelectricHeating}. However, this classical approach would be inconsistent with the multi-size, multi-composition capabilities of \calima. Instead we make use of our modelling of dust charging (Section~\ref{subsec:charging} and Appendix~\ref{ap:dust_charging}) to revisit the properties of PEH for individual grain sizes and compositions\footnote{We cover the modelling of PEH for PAHs in Section~\ref{subsec:photoelectric_heating}}. The dust charging model can be easily extended to consider the per-grain heating \citep{Weingartner2001PhotoelectricHeating}:
\begin{align}
    \Gamma_{\rm PEH} (a) = \sum_{Q_{\rm g}} f(Q_{\rm g},a) \left[\Gamma_{\rm PEH,v}(Q_{\rm g},a) + \Gamma_{\rm pd}(Q_{\rm g},a)\right],
\end{align}
where $f$ is the probability of grain $a$ to have charge $Q_{\rm g}$, and $\Gamma_{\rm PEH,v}$ and $\Gamma_{\rm pd}$ are the PEH contribution from valence electrons and photo-detachment of electrons in the lowest unoccupied molecular orbit, respectively. The sum is over the GCD computed with the method in Appendix~\ref{ap:dust_charging}. These heating rates are computed from Eqs.~39 and 40 in \citet{Weingartner2001PhotoelectricHeating}. Recombination of electrons and ions onto the grain result in a loss $\Lambda_{\rm rec}(a)$ of thermal energy from the gas phase, which we compute following Eq.~3.6 in \citet{Draine1987CollisionalGrains}. We benchmark our implementation by computing the total efficiency for conversion of absorbed radiation into net gas heating:
\begin{align}\label{eq:peh_efficiency_dust}
    \epsilon_{\rm PEH} (a) = \frac{\Gamma_{\rm PEH}(a)-\Lambda_{\rm rec}(a)}{P_{\rm abs}(a)},
\end{align}
where $P_{\rm abs}(a)$ is the integrated radiation absorbed power by a grain of radius $a$.
We compute this efficiency for graphite and silicate grains within the \citet{Mathis1983InterstellarClouds} ISRF and the definition of WNM phase in \citet{Weingartner2001PhotoelectricHeating}. Our results are plotted in Fig.~\ref{fig:dust_peh_efficiency_WNM} (blue for graphite, brown for silicate) in comparison to the results of \citet{Weingartner2001PhotoelectricHeating} in their Figure~12 (carbonaceous, dotted line) and Figure~13 (silicates, dot-dashed line). Besides deviations at $a\lesssim 100$\,\AA~caused by different choices of graphite optical properties (see Appendix~\ref{ap:dust_charging} for further details) in \citet{Weingartner2001PhotoelectricHeating}, our results seamlessly match theirs. Given the computational and memory imprint of obtaining the full GCD, we continue our approach in Section~\ref{subsec:charging} by pre-computing $\Gamma_{\rm PEH} (a)$ and $\Lambda_{\rm rec}(a)$ for each grain size and composition followed in \calima~for a range of $\gamma$ and $T$.

\subsection{Gas-phase accretion}\label{subsec:accretion}
The primary mode of dust mass growth in the ISM is by accretion of refractory elements from the gas phase. As in \citet{Dubois2024GalaxiesSimulations}, we model accretion of metals onto grains by a modified version of the collisional argument laid out in \citet{Dwek1998TheGalaxy}. The mass density accretion rate onto dust is given by:
\begin{align}
    \dot{\rho}_{{\rm acc},j} = \left(1-\frac{\rho_{d,j}}{\rho_{{\rm Z},j}} \right) \frac{\rho_{d,j}}{\tau_{{\rm acc},j}},
\end{align}
where the index $j$ runs over each dust bin (small, large) and chemical species (silicate, carbonaceous), $\rho_{d,j}$ is the mass density of dust in bin $j$, and $\rho_{{\rm Z},j}$ is the mass density of metals (combining gas and dust in bin $j$). The accretion timescale is written as:
\begin{align}\label{eq:accretion_timescale}
    (\tau_{{\rm acc},j})^{-1} = \alpha \frac{\pi \overline{a}_j^2}{f_X m_{{\rm gr},j}}\rho_X v_X,
\end{align}
where $m_{{\rm gr},j}$ is the mass of a single grain of type $j$, $\alpha$ is the sticking coefficient, and $\pi \overline{a}_j^2$ is the effective grain cross-section, which considers the effect of Coulomb focusing \citep{Weingartner1999InterstellarGrains} with the factor $F_{\rm C}$ as $\pi \overline{a}_j^2\equiv F_{\rm C} \pi a_j^2$. The quantities $\rho_X$, $v_X$ and $f_X$ denote the gas-phase mass density, gas thermal velocity and fractional contribution of the \textit{limiting} element $X$ to the grain $j$ composition. The concept of \textit{limiting} element requires some care. In our current implementation for dust composition in \calima, the stoichiometry of a given dust species is kept fixed across the simulation. This does not complicate the treatment of accretion for amorphous carbonaceous grains, as for these $X=C$. However, when determining the accretion rate of the different elements that compose a silicate grain, we ensure that the increase in dust mass corresponds to a decrease in metal mass in order to keep the stoichiometry of a given grain fixed. This means that at a given time, the abundance of a particular element $X$ contributing to the silicate composition may be lower and/or the element $X$ may experience a lower interaction rate with the grain, hence \textit{limiting} the final growth of mass by accretion. 
More specifically, we make use of the following `frequentist argument': considering the average Maxwellian thermal velocity\footnote{Reaction rates use the Maxwellian average because reactions depend on the full velocity distribution — especially the high-energy tail — not on the most probable particle speed.} given by $v_X = (8k_{\rm B}T/(\pi m_X))^{1/2}$, we compute the \textit{limiting} element parameters for the accretion time-scale using:
\begin{align}\label{eq:limiting_element}
   \frac{\rho_X v_X}{f_X} = \left(\frac{8 k_{\rm B}T}{\pi}\right)^{1/2}\min \left[\frac{\rho_\ell}{f_\ell m_\ell^{1/2}} \right]_{\ell=\,\rm Mg,Fe,Si,O}, 
\end{align}
where $\ell$ runs over the gas elements that contribute to the silicate composition and $m_\ell$ is the atomic mass of element $\ell$.

\begin{figure}
    \centering
	\includegraphics[width=\columnwidth]{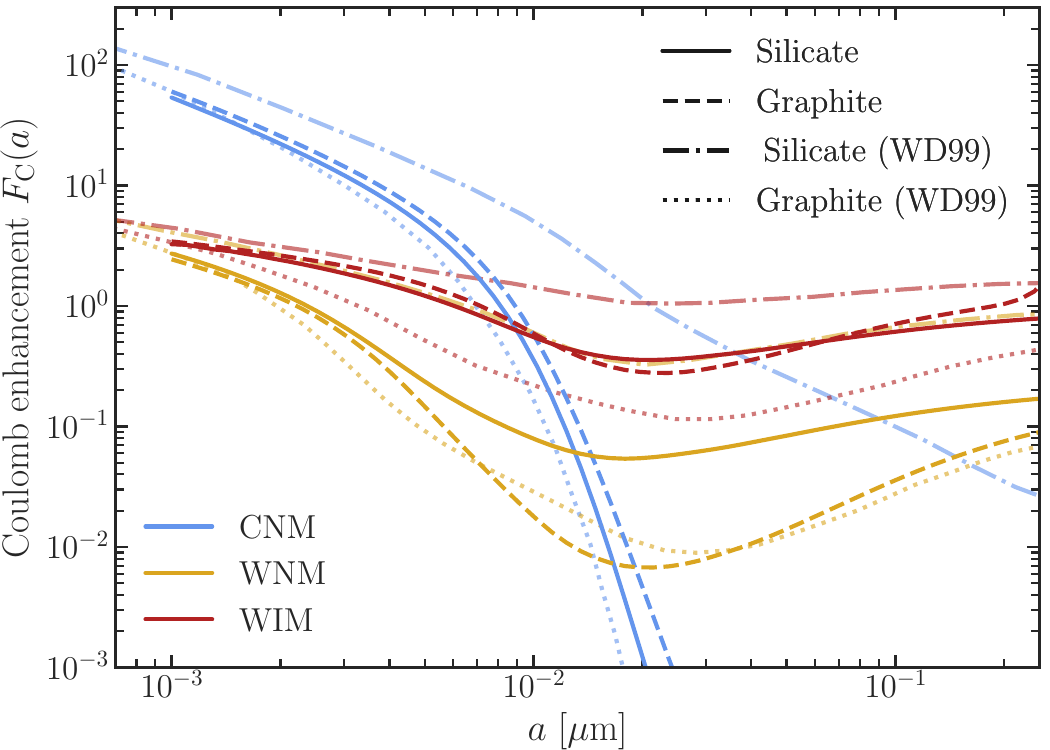}
    \caption{Coulomb enhancement factor $F_{\rm C}$ as function of grain size for different ISM phases (CNM, blue; WNM, yellow; WIM, red) assuming a colliding ion with charge $Z_k=1$. Solid-dashed and dotted lines correspond to the results of \citet{Weingartner1999InterstellarGrains} for silicate and graphite, while solid and dashed correspond to our computed $F_{\rm C}$ for, respectively, silicate and graphite grains. Our results agree reasonably well with \citet{Weingartner1999InterstellarGrains}, except large deviations for silicate grains (see text for discussion).}
    \label{fig:dust_coulomb_enhancement}
\end{figure}

The effective grain cross-section includes the contribution from electrostatic interactions between ionised gas and dust grains and requires knowing a dust grain's charge acquired through many ISM processes. We use the results from Section~\ref{subsec:charging} to compute on-the-fly GCDs. We obtain the local Coulomb focusing factor $F_{\rm C}$ \citep{Weingartner1999InterstellarGrains}:
\begin{align}
    F_{\rm C}(a,T,Q_k) = \sum_{Q_{\rm g}} f(Q_{\rm g},a) B(Q_{{\rm g}},Q_k,a,T),
\end{align}
where the sum goes over each charge state present in the distribution of the grain charge $Q_{\rm g}$, $Q_k$ is the charge of the ion $k$ and where \citep{Draine1987CollisionalGrains}:
\begin{align}
    B(Q_{{\rm g},i},Q_k,a,T) = \begin{cases}
        \exp \left(- \frac{Q_{{\rm g}}Q_k e^2}{k_{\rm B}T a}\right) ,& \text{if } Q_{{\rm g}} Q_k > 0, \\
        \left(1- \frac{Q_{{\rm g}}Q_k e^2}{k_{\rm B}Ta}\right) ,& \text{if } Q_{{\rm g}}Q_k <0,\\
        1+ \left(\frac{\pi Q_k^2e^2}{2k_{\rm B}Ta}\right) ,& \text{if } Q_{{\rm g}} = 0.
    \end{cases}
\end{align}
In practice, we compute these accretion timescales for the individual metal ion populations as predicted by the non-equilibrium solver in \ramsesrtz, hence depleting individual ions. To conserve charge and to not be forced to include this charge influx to the computation of equilibrium dust charge distributions, we assume that any given accreted ion instantaneously recombines. This assumes that charging timescales are much faster than accretion timescales, which is a fair assumption for dust grains with $a\gtrsim 50$~\AA\ \citep[see Figure~5 in][]{Ibanez-Mejia2019DustMedium}.
In Fig.~\ref{fig:dust_coulomb_enhancement} we show the value of $F_{\rm C}$ for ISM phases traditionally considered in dust charging models (see Table 1 in \citealp{Weingartner1999InterstellarGrains}). As an example, we use an ion with $Q_k=1$ (e.g.~C$^+$ or Mg$^+$), such that we can compare with the values of the Coulomb correction factor presented in Figure~1 of \citet{Weingartner1999InterstellarGrains}, for the CNM, the WNM and the WIM (as defined in their work) ISM phases. We note that in \citet{Weingartner1999InterstellarGrains} the dust optical properties followed \citet{Draine1984OpticalGrains} and \citet{Laor1993SpectroscopicNuclei}, without the modifications to the small graphitic grains and the suppression of the absorption feature in  silicates introduced in \citet{Weingartner1999InterstellarGrains} and \citet{Li2001InfraredMedium}. This makes the comparison with their work on dust charging and photo-electric heating particularly challenging, as well as with our choice of optical properties. Despite this, our results seem to suggest similar trends as the original work in \citet{Weingartner1999InterstellarGrains}: small grains are either neutral or negatively charge in the cold phases of the ISM, resulting in positive attraction with positive ions and therefore $F_{\rm C}>1$, while big grains are easily stripped of many electrons in the ISRF, resulting in accretion being slowed down in the WNM and WIM, and completely halted in the CNM. As expected from different choice of optical properties for silicate grains, we find for this composition the largest discrepancy with \citet{Weingartner1999InterstellarGrains}. However, the size of this discrepancy in the CNM is concerning, because our results otherwise agree well with the charging model of \citet{Weingartner2001PhotoelectricHeating} (see Fig.~\ref{fig:average_potential_comp} for the good agreement for large grains). This difference is unlikely to be driven by the adopted optical properties, which typically differ by no more than 20\%. Instead, it appears to arise from their treatment of grain charging in the CNM: the large charge contrast between silicate and graphite grains that we obtain is not present in the CNM solutions of \citet{Weingartner2001PhotoelectricHeating} (see their figure 10). We therefore caution against using the plotted values of $F_{\rm C}$ from \citet{Weingartner1999InterstellarGrains} without considering these limitations.

Efficient accretion occurs in the dense, cold phases of the ISM, which are barely reproduced by current cosmological galaxy simulations. In \citet{Dubois2024GalaxiesSimulations} they found that simulations with different maximum spatial resolution (from $\Delta x=72$~pc to $9$~pc resolution) led to a significant increase (up to a factor of 2) in the final DTM. This lack of numerical convergence for the DTM can  be attributed to the lack of very cold and dense gas where accretion of metals from the gas phase is expected to be more efficient (see Eq.~\ref{eq:accretion_timescale}). To alleviate this problem, in \citet{Dubois2024GalaxiesSimulations}, they employed a sub-grid model for the unresolved gas density similar to the one used in turbulence-driven star formation models \citep[see e.g.][]{Kimm2017Feedback-regulatedReionisation}. Sub-grid accretion is locally estimated by the gas density of the cell is actually the mean gas density $\overline{n}$ of a log-normal PDF of gas densities shaped by turbulence:
\begin{align}\label{eq:lognormal_density_pdf}
    p(s) = \frac{1}{\sqrt{2 \pi}\sigma_s}\exp \left(-\frac{(s-s_0)^2}{2\sigma_s^2} \right),
\end{align}
where $s=\ln(n/\overline{n})$, $s_0=-\sigma_s^2/2$ and $\sigma_s=\ln(1+b^2\mathcal{M}^2)$, with $b$ the compression ratio (we take $b=0.4$ for a mixture of compressible and solenoidal turbulence which is the same value later used in the for the star formation efficiency, see section~\ref{sec:sims}) and $\mathcal{M}$ the turbulent Mach number. The effective accretion timescale $t_{\rm acc,eff}$ is obtained by integrating the mass accretion rate over the log-normal PDF for a given set of mean density, temperature and metallicity ($\bar n, \bar T, \bar Z$), hence:
\begin{equation}
\frac{t_{\rm acc}(\bar n, \bar T, \bar Z)}{t_{\rm acc,eff}}=\int_{-\infty}^{s_{\rm max}} \frac{t_{\rm acc}(\bar n, \bar T, \bar Z)}{t_{\rm acc}(n, T, Z)} \frac{n}{\bar n} p(s)ds\, ,
\end{equation}
up to a maximum density $s_{\rm max}=\ln(n_{\rm max}/\bar n)$, where the maximum density is $n_{\rm max}=10^4 \, \rm H\, cm ^{-3}$, i.e.~the density where grains start to be significantly coated with a water ice mantle  \citep[$n\ge 10^3 \,\rm H\, cm^{-3}$,][]{Cuppen2007SimulationGrains,Hollenbach2008WaterClouds,Lamberts2014TheExothermicity,Elyajouri2024PDRs4AllBar} which suppresses the accretion of refractory material onto the grain surface. This unresolved density distribution should also give rise to an unresolved temperature and gas metal density distribution. However, there is no simple correspondence between these three random fields. Instead, we assume that the average metal fraction and temperature are those given by the local value of these quantities (i.e.~$T=\overline{T}$ and $n_X/n_{\rm H}=\overline{n}_X/\overline{n}_{\rm H}$). This leads to:
\begin{eqnarray}
\frac{t_{\rm acc}(\bar n,\bar T,\bar Z)}{t_{\rm acc,eff}}&=&\int_{-\infty}^{s_{\rm max}} \exp(2s) p(s)ds\, \nonumber \\
&=&\frac{e^{\sigma_s^2}}{2}{\rm Erfc}\left(\frac{3\sigma_{s}^2/2-s_{\rm max}}{\sqrt{2}\sigma_s} \right)  \, ,
\end{eqnarray}
where $\rm Erfc$ is the complementary error function.

Many different expressions have been suggested in the literature for the sticking coefficient $\alpha(T)$ \citep[see the discussion in][]{Zhukovska2016MODELINGISM}, but much of our knowledge of the sticking probability of gas phase atoms onto grain surfaces is based on the physical-chemistry modelling and experiments that have studied the sticking of H atoms to form molecular hydrogen either by the Langmuir-Hinshelwood or the Eley-Rideal mechanisms \citep[see e.g.][but also Section~\ref{subsec:h2_formation_dust}]{Chaabouni2012StickingConditions,LeBourlot2012SurfaceMechanisms,Bron2014SurfaceFluctuations}. It remains unclear how heavier atoms (that are also found in different ionisation states in the ISM) stick to surface grains, which has resulted in one-zone models simplifying the problem by assuming $\alpha=1$ \citep[e.g.][either at all $T$ or only for a range of low temperatures $\lesssim 300$\,K]{Asano2013WhatGalaxies,McKinnon2016DustGalaxies,Vogelsberger2019DustCooling,Li2019The6,Lewis2022DUSTiERRAMSES-CUDATON,Choban2022TheFIRE,Hu2023Co-evolutionMedium}. \citet{Zhukovska2016MODELINGISM} have shown that using a fixed value of $\alpha=1$ tends to over predict the depletion of gas-phase metals in the Galactic ISM, and instead argues that a lower value of $\alpha=0.3$ is more appropriate to match current constraints of metal depletion \citep[e.g.][]{Jenkins2009AMedium}. We identify the choice of $\alpha(T)$ as a significant source of uncertainty in our model \citep[as it has been for previous implementations of dust evolution in one-zone models and hydrodynamical simulations][]{Hirashita2011EffectsGrowth,Grassi2014KROME-aSimulations,Zhukovska2016MODELINGISM,Aoyama2017GalaxyDestruction,Gjergo2018DustSimulations} and leave the sticking coefficient as a free parameter. Sticking coefficients values we explored are detailed in appendix D of \citet{Dubois2024GalaxiesSimulations}.

\subsection{Thermal sputtering}\label{subsec:sputtering}
Dust particles in high temperature coronal gas are subject to the destructive effect of collisions with fast moving ions. If the kinetic energy of the collision is sufficiently large, the energy of a dust grain may be able to overcome the dust grain binding energy. This process of sputtering is capable of removing a large fraction of the dust grain mass in high temperature gas and is hence of high importance for the hot phase DTM. We follow a derivation similar to the one outlined in \citet{Kirchschlager2019DustDensities}, which is similar to the original approach in \citet{Tielens1994TheShocks} but considers the finite size $a$ of the target dust grain and the effect of grain charge in the interaction energy. Our implementation is detailed in Appendix~\ref{ap:dust_sputtering}, where we have modified previous theoretical derivations by using more updated atomic experimental data and by considering more appropriate theoretical modelling for these sub-relativistic ion-solid interactions.

\begin{figure}
    \centering
	\includegraphics[width=\columnwidth]{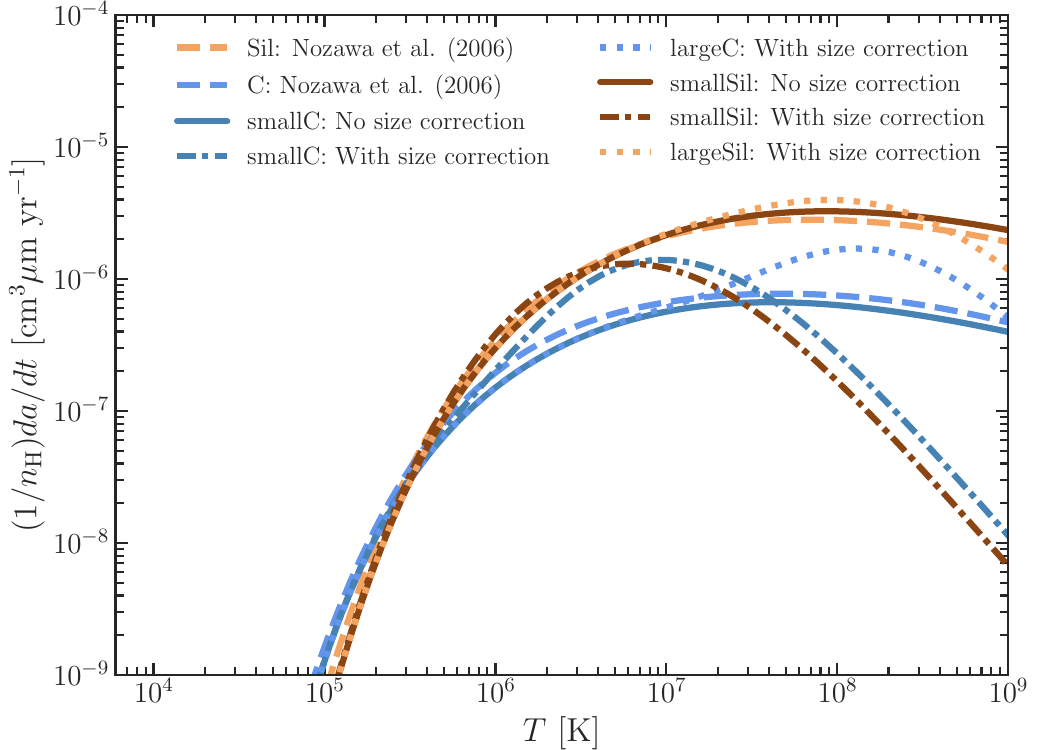}
    \caption{Erosion rate (i.e.~$(\dd a/\dd t)/n_{\rm H}$) by thermal sputtering given by Eq.~\ref{eq:definition_sputtering_rate} for each dust species. These have been computed for a gas composition of $Z=10^{-4}Z_{\odot}$. We compare the results of \citet{Nozawa2006DustUniverse} to our updated model that considers the finite size of the grain for the effective energy deposited by ions. This results in small grains being less sputtered for $T\sim 10^7$\,K.}
    \label{fig:sputtering_rate}
\end{figure}

In order to compare to the rates obtained by \citet{Nozawa2006DustUniverse}, we have computed thermal sputtering for a gas with $10^{-4}Z_{\odot}$\footnote{Across this work we assume the \citet{Asplund2009TheSun} solar metallicity value of $Z_\odot=0.01345$.} (see Fig.~\ref{fig:sputtering_rate}). Our vanilla rates (solid brown and blue lines) closely follow the shape and absolute values obtained by \citet{Nozawa2006DustUniverse} (dashed orange and blue lines), confirming the higher sputtering yield predicted for silicate grains compared to carbonaceous. When the finite size correction is applied (dotted lines), our large and small grain rates deviate from each other at temperatures above $\sim 10^6$\,K, resulting in rates that clearly peak at different temperatures. The most remarkable result of these updated erosion rates is their decrease by $\gtrsim 2$\,dex and $\gtrsim 1$\,dex for small and large grains above $5\times 10^7$\,K. This could have a significant effect on the sputtering timescale of dust grains in SN and shock heated X-ray gas within galaxy clusters \citep{Vogelsberger2019DustCooling}. We pre-compute these rates for individual ions for each grain species and size, allowing on-the-fly interpolation to determine the combined sputtering rates of the ions followed in \ramsesrtz.

\subsection{Gas collisional cooling/heating}\label{subsec:collisional_cooling}
So far, we have not explicitly detailed the direct collisional transfer of energy between dust grains and gas atoms. In Section~\ref{subsec:sputtering}, we introduced how high velocity collisions that take place at high temperature cause sputtering of atoms from dust grains. This fragmentation reaction takes place via the transfer of thermal energy of the gas to the internal energy of the grain, which effectively means an additional source of cooling for gas enriched with dust. Collisional heating may be an important source of dust heating (gas cooling) in high density shocks, which could explain the infrared-to-X-ray (IRX) relation observed for SNRs in the Galaxy and the LMC \citep[e.g.][]{Graham1987IRASRemnants,Seok2015DustCloud}. Furthermore, in highly obscured dense regions, collisional heating significantly impacts the equilibrium temperature of dust grains \citep{Hollenbach1989MOLECULECLOUDS,Glover2007StarDensities,Dopcke2011THECLOUDS}. Dust-gas collisional cooling has been implemented in various ISM chemistry models \citep[e.g.][]{Wolfire2003NeutralGalaxy,Grassi2014KROME-aSimulations,Krumholz2014,Bialy2019ThermalGas,Kim2023PhotochemistrySimulations,Katz2022PRISM:Galaxies} and hydrodynamical simulations with explicit dust modelling \citep[e.g.][]{Vogelsberger2019DustCooling,Granato2021DustFormation,Dubois2024GalaxiesSimulations}. Models of dust cooling are primarily based on the seminal work by \citet{Hollenbach1979MOLECULEPROCESSES} and \citet{Dwek1981TheCondensates}, and indeed the high-temperature dust-electron collisional cooling in our original dust model for \ramses~\citep{Dubois2024GalaxiesSimulations} is based on a fitting function provided in Appendix~A of \citet{Dwek1981TheCondensates}. However, we would like to emphasise that these models have been applied, within the ISM and galaxy formation community, with little scrutiny as to the extent of their applicability. In Appendix~\ref{ap:dust_collisional_cooling} we review these aspects and propose a more appropriate modelling of dust-grain collisional cooling.

\begin{figure}
    \centering
	\includegraphics[width=\columnwidth]{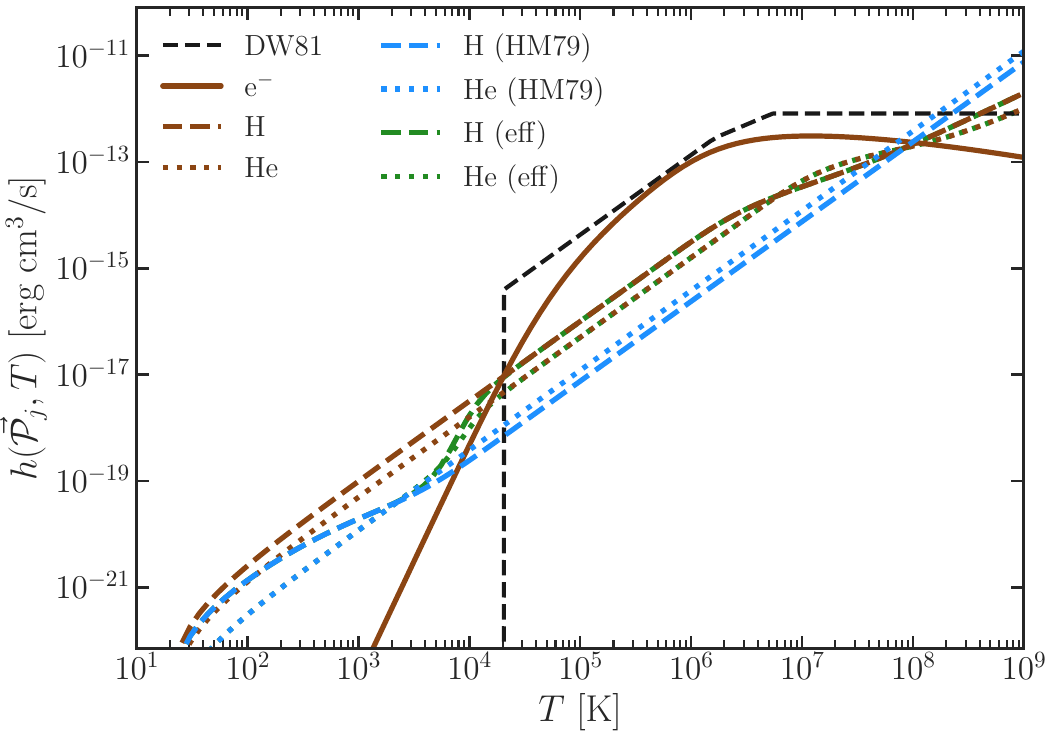}
    \caption{Collisional cooling efficiency $h$ as defined by Eq.~\ref{eq:collisional_efficiency} for small silicate ($a=0.005$\,\micron) at different gas temperatures, and computed for electrons (solid brown line), H (dashed brown line), and He (dotted brown line). We compare our results to the model at low temperature from \citet{Hollenbach1979MOLECULEPROCESSES} for H (dashed blue line) and He (dotted blue line), and at high temperature cooling from electrons by \citet{Dwek1981TheCondensates} (dashed black line). To transition between the neutral and ionised gas regimes, we apply a logistic smoothing at $T\sim 10^4$\,K represented by the green lines (dashed for H and dotted for He).}
    \label{fig:collisional_cooling_smallSil}
\end{figure}

For the low temperature regime ($T\lesssim 10^4$\,K), we use the \textit{soft cube} model from \citet{Burke1974DustPhysics}, which is implemented for dust collisional cooling in shocks by \citet{Hollenbach1979MOLECULEPROCESSES}. The prediction of this model for our small silicate grains ($a=0.005$\,\micron) is shown in Fig.~\ref{fig:collisional_cooling_smallSil} for H (dashed blue line) and He (dotted blue line) collisional partners. The quantity $h$ (Eq.~\ref{eq:collisional_efficiency}) in Fig.~\ref{fig:collisional_cooling_smallSil} encapsulates the efficiency of energy transfer in individual grain-atom collisions. As we argue in Appendix~\ref{ap:dust_collisional_cooling}, the \textit{soft cube} model is only applicable for neutral atoms and low energies for which the interaction is limited to inelastic collisions with the grain surface. For hot ionised gas ($T> 10^4$\,K), we instead need to consider the energy deposited by atoms that penetrate the grain material. This is the regime of interactions developed by \citet{Dwek1981TheCondensates} and that we have updated for \calima~in Section~\ref{subsec:sputtering}. Therefore, for ionised gas we simultaneously predict the gas cooling due to ion-grain collisions and the consequent the sputtering of grain material. We show the results of this computation in Fig.~\ref{fig:collisional_cooling_smallSil} for small silicate grains considering electrons (solid brown), H (dashed brown) and He (dotted brown). We compare our electron cooling efficiency to the fitting equation from Appendix~A in \citet{Dwek1981TheCondensates} (dashed black line), which is the main cooling agent in previous works \citep[e.g.][]{Vogelsberger2019DustCooling,Granato2021DustFormation,Dubois2024GalaxiesSimulations}. Our treatment of finite grain size results in deviations from their modelling at high temperature, while the correct treatment of low energy collisions allows us to extend the efficiency to lower temperature than the limit imposed in \citet{Dwek1981TheCondensates}.

Inversely to the \textit{soft cube} model, the hard-sphere regime of ion collisions is not applicable for sufficiently low energy collisions that result in minimal grain penetration. Therefore, we adopt a combination of the \textit{soft cube} model and the the high-energy collisional cooling by means of a smoothly transitioning from one to the other using a logistic factor at $T\sim 10^4$\,K in logarithmic space with width of 10. We show how this appears in practice for H and He collisions with the dashed and dotted green lines in Fig.~\ref{fig:collisional_cooling_smallSil}.

\subsection{Non-thermal sputtering in SN shocks}\label{subsec:sn_shocks}

SNe (type II and Ia) destroy dust already present in the ISM, where the dust destruction is produced by inertial (non-thermal) sputtering and grain-grain collisions~\citep{Kirchschlager2020SilicateA,Vasiliev2025DestructionSupernova,Vasiliev2025DustExplosions}. As opposed to thermal sputtering, non-thermal sputtering is independent of temperature, and the energy of the collision is instead related to the kinetic energy imparted by a travelling shock wave in the ISM (i.e.~a SN shock).
The amount of SN-destroyed dust $\Delta M_{{\rm SN},j}$ is a fraction of the mass of gas $M_{\rm 100}$ shocked to velocities above $100 \, \rm km \, s^{-1}$ defined as 
\begin{align}
\Delta M_{{\rm SN},j}=\varepsilon_{\rm SN}(a_j) {\rm min}\left ( \frac{M_{\rm 100}}{M_{\rm gas}},1\right)M_{{\rm d},j}\, ,
\end{align}
with a size-dependent destruction efficiency given by ~\citet{Aoyama2020GalaxyDistribution} (their equation 11):
\begin{align}\label{eq:SN_destruction_efficiency}
\varepsilon_{\rm SN}(a_j)=1-\exp\left(-\frac{\delta_{{\rm SN},j}}{a_{0.1,j}}\right)\, ,
\end{align}
and $M_{\rm gas}$ is the gas mass in the current cell, $M_{{\rm d},j}$ the local dust mass in bin $j$, $a_{0.1,j}$ the dust grain size in units of $0.1 \,\rm \mu m$, and $\delta_{{\rm SN},i}$ the destruction efficiency.
Small grains are more efficiently decelerated by drag forces, trapping them near the shock region where, opposite to large grains, thermal sputtering can quickly destroy them~\citep{Nozawa2006DustUniverse}.
The dependence of the destruction efficiency Eq.~\ref{eq:SN_destruction_efficiency} on grain size $a$ reflects this.
The mass of shocked gas is determined by the Sedov solution in a medium of homogeneous gas density \citep[i.e.~$M_{\rm 100}=6800 E_{\rm SN, 51}\, \rm M_\odot$][]{McKee1989DustMedium}, where $E_{\rm SN, 51}$ is the energy of the SN explosion in units of $10^{51}\, \rm erg$.
Multiple SNe occur in one time step $\Delta t$ (each individual SN engulfing the gas swept up by the previous explosion), following \citep[see][]{Hou2017EvolutionSimulation}:
\begin{equation}
\Delta M_{{\rm SNe},j}=\left [ 1-\left ( 1-\frac{\Delta M_{{\rm SN},j}}{M_{{\rm d},j}}\right)^{N_{\rm SN}} \right ] M_{{\rm d},j}\, ,
\end{equation}
with $N_{\rm SN}$ the number of SNe.
We emphasise that in this model the SN blast wave does not  destroy the dust newly formed in stellar ejecta. It only destroys pre-existing dust: the dust condensation efficiency (see Section~\ref{subsec:grain_seeding}) implicitly takes this into account.

We use a different value of the dust destruction efficiency for carbonaceous ($0.10$) and silicate ($0.15$) grains, following results from hydrodynamical simulations of dust sputtering in a turbulent multiphase ISM~\citep{Hu2019ThermalMedium}, where approximately 50 percent more silicate grains are destroyed than carbonaceous grains. Grain-grain collisions in shocks could also lead to efficient shattering of large grains into small grains \citep[e.g.][]{Jones1996GrainDistribution,Nozawa2006DustUniverse,SerraDiaz-Cano2008CarbonaceousGraphite,Bocchio2014AWaves,Slavin2015DestructionWaves,Kirchschlager2021SupernovaTurbulence}. It has been argued that this shattering in SN shocks could allow for the formation of large PAH clusters ($<15$\,\AA) from up to $\sim 10$~\% of the carbon dust in shocks with $v_{\rm s}\sim 100\,\rm km\,s^{-1}$ \citep{Jones1996GrainDistribution}. Nevertheless, there is still little evidence for the formation of very small grains in SN shocks \citep[see e.g.][for a discussion of the problem]{Tielens2008InterstellarMolecules}, and it is not clear how much of these new small grains will survive in the hot post-shock gas. \calima~therefore allows for different small grain production efficiencies in SN shocks, leaving them as free parameters of the model.

\subsection{Turbulent shattering of large grains}\label{subsec:shattering}

As particles suspended in gas, dust grains experience collisions and mixing in the turbulent ISM (see e.g.~\cite{McKee2007,Elmegreen2004InterstellarProcesses} for reviews about the presence of turbulence in the ISM). Close-range interactions can lead to elastic collisions or can compromise the physical integrity of dust grains, by deformations, coagulation or shattering. In this section we begin by detailing how large energy collisions can cause grains to shatter into smaller fragments, strongly altering their size distribution and, consequently, the extinction curve. Small grains account for the majority of the cross-sectional area of the dust distribution, and thus they are of great importance to modelling the dust-radiation coupling as well as the chemistry on the surface of grains (e.g.~\hmol~formation). Hence the top-down processing of the dust distribution from large grain to their transformation into small grains, ultimately including PAHs, cannot be neglected in a dust model for galaxy formation \citep[e.g.][]{Seok2014FormationGalaxies,Narayanan2023ASimulations}.

The statistical properties of magnetised, compressible gas turbulence are fundamental for determining the motion of dust grains. The dynamics of grains is primarily driven by hydrodynamic drag and Lorentz force acceleration. Efficient dust grain acceleration (against damping mechanisms) occurs via gyroresonance when the Doppler-shifted frequency of the fast magnetosonic wave in the frame of reference of the grain is a multiple of the gyrofrequency of the grain in the local magnetic field. In the original work of \citet{Yan2004DustTurbulence} \citep[see][for the theoretical MHD turbulence framework]{Lazarian2001GrainGas,Yan2002GrainMechanism}, large grains ($a\geq 0.2$\,$\mu$m) can decouple from the turbulent flow in the WNM (due to a higher ratio of stopping time to flow characteristic time, i.e.~Stokes' number) and be accelerated to $\sim\,1 \rm \, km\,\rm s^{-1}$, (i.e. above the shattering threshold) by the Lorentz force \citep{Jones1996GrainDistribution}. However, the computations in \citet{Yan2004DustTurbulence} are limited to grains with $a\geq 0.01$\,$\mu$m, and to specific thermodynamical conditions rather than the full multiphase conditions that arise in realistic galaxy formation simulations. Full modelling of dust dynamics in MHD turbulence remains a difficult problem \citep[e.g.][]{Fromang2009GlobalSettling,Zhu2015DUSTDIFFUSION,Lee2017TheClouds,Commercon2023DynamicsImplementations}, and translating the results of simulations where the large grain decoupling is resolved ($\ll 0.01$\,pc) to galactic scales simulations without magnetic fields is not a trivial task. We leave further developments of the dust and MHD coupling within the \ramses~code \citep[see e.g.][for current efforts]{Lebreuilly2019SmallMethods,Moseley2022Acceleration,Commercon2023DynamicsImplementations,Moseley2025DustGrains} to future work.

Instead, we follow the purely HD formulation of turbulent-induced collision velocities for dust grains by \citet{Ormel2007AstrophysicsNote} and the fragmentation modelling by \citet{Kobayashi2010FragmentationCascades} with the material properties of \citet{Jones1996GrainDistribution} (see Appendix~\ref{ap:dust_turbulence} for further details).

\subsection{Turbulent coagulation of small grains}\label{subsec:coagulation}

When two grains collide at sufficiently low velocities, they can fuse together, coagulating into a larger grain. In \calima~we consider both the coagulation of small grains to form large grains, and the coagulation of small and large grains. The process of dust grain coagulation has been invoked to explain the lack of small grains contributing to the $60$\,$\mu$m emission at ISM gas densities $\gtrsim 10^3$\,cm$^{-3}$ \citep[e.g.][]{Stepnik2003EvolutionFilament,Roy2012ChangesCloud} and the flattening of the extinction curve in the near-IR and mid-IR \citep[e.g.][]{Moore2005TheAV,Foster2013EvidencePerseus,Li2023TheSurveys}. Coagulation has been extensively explored in the study of proto-stellar core collapse \citep{Ormel2009DustDistribution,Guillet2020DustCollapse,Marchand2021FastDerivation,Kawasaki2022ImplementationCore,Lebreuilly2022ProtostellarFragmentation} with the unanimous conclusion that dust grains grow by coagulation in high density gas. Previous models including coagulation \citep[e.g.][]{Chokshi1993DustCoagulation,Hirashita2009ShatteringTurbulence,Aoyama2017GalaxyDestruction,Yan2004DustTurbulence} assume that coagulation between grains $i$ and $j$ only takes place below a threshold velocity
\begin{align}\label{eq:coagulation_velocity}
    v_{\rm coa}^{ij} = 21.4 \left[\frac{a_i^3+a_j^3}{(a_i+a_j)^3}\right]^{1/2}\frac{\gamma^{5/6}}{E^{1/3}R_{i,j}^{5/6}s^{1/2}},
\end{align}
above which bouncing (i.e.~elastic collision) will be the end result of the collision. Again, $a_i$ and $a_j$ are the radii of target and projectile grains. This expression depends on a series of material properties of the dust grains \citep[see Table 3 of][for the values of $\gamma$, the grain material surface energy per unit area, $R_{ij}=a_i a_j/(a_i+a_j)$, the reduced radius of the grains, and $E$, which is related to the Poisson's ratios and Young's modulus of the grain material]{Chokshi1993DustCoagulation}. Experimental studies with micron-size silica spheres \citep{Poppe1997ExperimentsGrowth} have shown that there is a sharp transition velocity from coagulation to grain bouncing, but the estimate from Eq.~\ref{eq:coagulation_velocity} is 4 times lower than what is expected for grains with icy mantels. Consistent with our maximum density $n_{\rm H} = 10^4$\,cm$^{-3}$ for accretion of refractory materials limited by the formation of icy mantles (see Section~\ref{subsec:accretion}), we consider that above a $n_{\rm H}\geq 10^4$\,cm$^{-3}$ water, CO and CH$_4$ ices freeze on the surfaces of dust grains \citep{Hollenbach2008WaterClouds,Silsbee2021IceSize}. Therefore, we increase the velocity threshold for coagulation by a factor of 4 \citep{Guillet2020DustCollapse} compared the one predicted by Eq.~\ref{eq:coagulation_velocity}. Similar to the shattering timescales, we compute coagulation timescales using the models for the relative collision velocities from Appendix~\ref{ap:dust_turbulence}. Additionally we also apply the un-resolved turbulence enhancement of the gas density\footnote{In the vanilla dust model for \ramses~that we have presented in \citet{Dubois2024GalaxiesSimulations} we do not consider the un-resolved density PDF. Instead, we assume that half the gas mass above $n_{\rm H}>10\,\rm cm^{-3}$ has an actual density of $10^3\,\rm cm^{-3}$.} as done for accretion (see Section~\ref{subsec:accretion}). 

Instead of abruptly shutting off coagulation when $\Delta V_{ij}>v_{\rm coa}$, we apply a sigmoidal smoothing that returns a maximum $t_{\rm coa}$ of $10^5$\,Myr (i.e.~one order of magnitude larger than the Hubble time) in order to effectively stop coagulation. Previous works \citep[e.g.][]{Asano2014EvolutionGalaxies} have shown that the extinction curve of the MW cannot be explained by a sharp cutoff in coagulation.

\subsection{Molecular hydrogen formation on dust surfaces}\label{subsec:h2_formation_dust}
Molecular hydrogen is the most abundant molecule in the Universe \citep[e.g.][]{Fraser2002TheUniverse,Bolatto2013}. It strongly influences the thermal state of star-forming gas as well as the formation of other molecules in the ISM. While primordial H$_2$ formation plays a dominant role in the dust-free formation of the first stars \citep[e.g.][]{Abel2000TheClouds,Abel2002TheUniverse}, H$_2$ is believed to be primarily formed via catalytic reactions on dust grain surfaces. The consistent, on-the-fly modelling of H$_2$ formation on dust grains is then one of the strengths of \calima. We follow the modelling of molecular hydrogen formation developed by \citet{Cazaux2002MolecularMedium,Cazaux2004Surfaces} through the Langmuir-Hinshelwood and the Eley-Rideal processes. In practice, we compute the contribution of each grain size bin to the total local formation rate of molecular hydrogen, considering the on-the-fly dust temperature $T_{\rm d}$. We use the physical parameters for olivine silicate surfaces and amorphous graphite given by \citet{Cazaux2004HSUB2/SUBSurfaces}, which are themselves derived from the experimental results from laboratory irradiation of grain surfaces performed in \citet{Pirronello1996LaboratoryInterest,Pirronello1997EfficiencySilicates,Pirronello1999MeasurementsGrains}. We emphasise that the self-consistency in dust-mediated \hmol~formation is a singular feature of \calima, compared to efficiencies usually fixed to an empirically derived value for the MW \citet[e.g.][]{Wolfire2008ChemicalFormation,Gnedin2009ModelingSimulations}.

\subsection{Dust optical properties}\label{subsec:optical properties}
We use the optical properties for graphite and ``UV smoothed'' astronomical silicates by \citet{Draine1984OpticalGrains} \citep[see also][]{Laor1993SpectroscopicNuclei,Weingartner2000DustSMC}\footnote{We use the publicly available optical properties from Bruce T. Draine's personal website (\url{https://www.astro.princeton.edu/~draine/dust/dust.diel.html}).} for our carbonaceous and silicate grains. The total local dust optical depth is given by:
\begin{align}
    \tau (\lambda) = L \sum_i C_{{\rm ext},i}(\lambda)n_{{\rm d},i},
\end{align}
where $L$ is the path length for light propagating through a medium of dust density $n_{{\rm d},i}$, $C_{{\rm ext},i}$ is the wavelength dependent extinction cross-section, and the sum is over dust components (large, small, silicate, carbonaceous grains, PAHs, etc.). 

Based on these optical properties, we can derive the dust temperature $T_{\rm d}$ in the equilibrium approximation using the local radiation properties \footnote{This is only performed for regular dust grains, while PAH photo-heating is described in Section~\ref{subsec:pah_dissociation} within the description of photo-dissociation}. Ignoring the small amount of energy that goes into luminescence or photo-electrons for regular grains \citep[e.g.][]{Weingartner2001PhotoelectricHeating}, the heating rate of a grain by absorption of radiation is:
\begin{align}
    \frac{\dd E_{\rm abs}}{\dd t} = \int \frac{u_{\nu}\dd \nu}{h \nu} c h \nu C_{\rm abs}(\nu),
\end{align}
where $\frac{u_{\nu}\dd \nu}{h \nu}$ is the number density of photons with frequencies in the range $[\nu, \nu + \dd \nu]$ for the radiation field energy density $u_{\nu}$. In practice, we compute a spectrum-averaged\footnote{This is based on the stellar-polling approximation in \ramsesrt, which recomputes the average spectrum in each radiation bin at every coarse timestep by performing a weighted average over all stellar sources in the simulation volume \citep[e.g.][]{Rosdahl2018}. In the case of including radiation from other bright sources with very different spectrum to the dominant stellar emission (e.g. the active galactic nuclei radiation from the accretion disc of a supermassive black hole) this approximation may fail in capturing the effective radiation field near the secondary source, and would require the use of photon tracer algorithms \citep[e.g.][]{Trebitsch2021TheProtoclusters}.} absorption cross-section for each radiation bin $j$ in \ramsesrtz:
\begin{align}
    \overline{C}_{{\rm abs},j} &= \frac{\int_{\lambda_{\rm min}^j}^{\lambda_{\rm max}^j}\lambda C_{\rm abs}(\lambda)u_{\lambda}\dd \lambda}{\int_{\lambda_{\rm min}^j}^{\lambda_{\rm max}^j}\lambda u_{\lambda}\dd \lambda},
\end{align}
where the integral is over the minimum $\lambda_{\rm min}^j$ and maximum $\lambda_{\rm max}^j$ wavelengths covered by radiation bin $j$. The energy loss due to IR emission is given by:
\begin{align}
    \frac{\dd E_{\rm em}}{\dd t} = \int \dd \nu 4\pi B_{\nu}(T_{\rm d}) C_{\rm abs}(\nu),
\end{align}
with $B_{\nu}$ being Planck's blackbody function. To simplify this calculation at runtime, we can compute a Planck-averaged emission efficiency defined as:
\begin{align}
    \overline{C}_{{\rm abs},T_{\rm d}} = \frac{\int \dd \nu B_{\nu}(T_{\rm d}) C_{\rm abs}(\nu)}{\int \dd \nu B_{\nu}(T_{\rm d})},
\end{align}
which allows us to write the IR emission rate as:
\begin{align}
    \frac{\dd E_{\rm em}}{\dd t} = 4  \overline{C}_{{\rm abs},T_{\rm d}} \sigma_{\rm SB} T_{\rm d}^4,
\end{align}
with $\sigma_{\rm SB}$ the Stefan-Boltzmann constant. With these simplifications, we can determine the steady state temperature for grain bin $i$ \citep{Draine2011PhysicsMedium} based on the balance of absorbed energy for all radiation bins (below the Lyman limit of $13.6$\,eV) and collisional heating sources (Eq.~\ref{eq:collisional_cooling_rate} and~\ref{eq:collisional_heating_rate_integrated}, see Section~\ref{subsec:collisional_cooling}):
\begin{align}
    4 \overline{C}_{{\rm abs},T_{{\rm d},i}} \sigma_{\rm SB} T_{{\rm d},i}^4 = \sum_j \overline{C}_{{\rm abs},i,j} u_j c + \sum_k H_{{\rm coll},k} (\vec{P}_j,T,n_k)
\end{align}
where $k$ loops over all gas species contributing to the collisional heating and $\vec{P}_j$ is the vector of grain material properties (see Eq.~\ref{eq:collisional_heating_rate_integrated} for the expression of $H_{\rm coll}$) and:
\begin{align}
    u_i = \int_{\lambda_{\rm min}^i}^{\lambda_{\rm max}^i}\lambda u_{\lambda}\dd \lambda.
\end{align}
We set a floor for the dust temperature given by the current CMB temperature ($T_{\rm CMB}(z)$) at redshift $z$.

\section{Polycyclic aromatic hydrocarbons in galaxy formation} \label{sec:pahs}

In this section we detail our extension of the two-size grain approximation to account for the properties of very small particles containing C atoms: PAHs. By computing the ratio of flux in IR emission features (attributed to PAHs) to the far-IR flux, one can obtain the fraction of UV flux that is absorbed by PAHs and, by adopting standard dust parameters \citep{Allamandola1989InterstellarImplications}, predict the fraction of C locked in different PAHs. We again follow the modified log-normal distribution in Section~\ref{sec:twosize_model} for each PAH bin. The original small carbonaceous (small C) grains centroid of $5\times 10^{-3}$\,\micron~\citep[as in][]{Dubois2024GalaxiesSimulations} has been displaced to $1\times 10^{-2}$\,\micron~to account for the higher discretisation of very small grains allowed by our two PAH bins. Our choice of PAH centroids is based on the abundance constraints of the MW SED \citep{Allamandola1989InterstellarImplications,Draine2007Era} in resemblance to Table~2 in \citet{Tielens2008InterstellarMolecules}. We separate traditional PAHs (composed here of $N_{\rm C}= 54$ carbon atoms per PAH corresponding to circumcoronene) that are photolytically stable in the ISM \citep{Guhathakurta1989TemperatureGrains} from large PAH clusters (here $a= 10$\,\AA, i.e. $N_{\rm C}= 418$), which contribute to the emission plateau underlying the IR emission features \citet{Tielens2008InterstellarMolecules}. The effective radius $a$ of a PAH molecule is given by Equation~8 in \citet{Draine2021ExcitationIntensity}:
\begin{align}\label{eq:pah_radius}
    a = 10 \left(\frac{N_{\rm C}}{418}\right)^{1/3}\, \text{\AA}.
\end{align}

\subsection{PAH optical properties}\label{subsec:pah_optical_properties}
For flexibility, we use the optical properties for PAHs derived by \citet{Li2001InfraredMedium} and publicly available\footnote{Again, from B. T. Draine's personal website (\url{https://www.astro.princeton.edu/~draine/dust/dust.diel.html}).}, allowing us to obtain GSD-averaged cross-sections for any choice of PAH centroid $a_0$ (see Eq.~\ref{eq:lognormal_grain_size_distribution}). The intrinsic spectral properties of PAHs vary primarily because of their ionisation state \citep[the 3.3\,\micron~is reduced in ionised PAHs, and the features at the 11-15\,\micron~are considerably boosted][]{BauschlicherJr.2008ThePAHs} and the size of the molecule \citep[contribution to the mid-IR plateau increases with increasing size, and the peak position of the major bands becomes more stable, see][]{Hudgins1999TheSize,BauschlicherJr.2008ThePAHs}. The optical properties in \citet{Li2001InfraredMedium} allow us to capture these changes in the strength of the different features and the scale of the mid-IR continuum with different charge states and sizes.

\subsection{PAH seeding}\label{subsec:pah_seeding}
There is tentative evidence for the formation of PAHs in carbon-rich AGB stars \citep{Allamandola1989InterstellarImplications,Sloan2007TheAliphatics,Smolders2010WhenStars,Matsuura2014SpitzerMetallicities,Sloan2017Carbon-richNebulae}. However, observations are not uniform in their estimated production efficiencies, since the detections are only in post-AGB phases or in binary systems where there is enough UV flux to excite the PAH molecule. Despite this, these estimates appear to suggest that AGB stars do not deplete more than a few percent of their carbon in PAHs. Theoretical modelling of PAH formation in winds was pioneered in \citet{Frenklach1989FormationEnvelopes} and \citet{Cherchneff1992PolycyclicEnvelopes}, with predictions suggesting that it is only efficient in the narrow temperature window between 900 and 1100\,K. However, the predicted efficiency was confined to $10^{-5}-10^{-4}$ of their carbon production, orders of magnitude lower than the observe abundances. Even when considering more complex and realistic wind models \citep{Cau2002FormationDimers}, these estimates reach at most $3.3\times10^{-3}$. Wolf-Rayet binaries have also been observed to show the distinctive mid-IR aromatic features attributed to PAHs \citep[e.g.][]{Marchenko2017SearchStars,Taniguchi2025TheJWST}. However, their PAH formation efficiencies are even more uncertain than for AGB stars, since yields are highly dependent on companion presence, wind mixing, and local physical conditions. We leave an exploration of PAH seeding mechanisms for future work, and just assume a condensation efficiency of $3.3\times10^{-3}$ for C-rich AGB winds and equally shared in mass fraction between small and large PAHs.

\subsection{PAH growth}\label{subsec:pah_accretion}
The formation of PAHs directly from the gas phase probes the molecular origins of these interstellar particles. The theoretical foundations of bottom-up PAH formation originate in combustion chemistry \citep[see the mechanistic review by][]{Altarawneh2024FormationReview}, where the formation of soot in hydrocarbon flames was extensively studied from the 1970s onward \citep[e.g.][]{Frenklach1989FormationEnvelopes,Frenklach2002ReactionFlames}. These studies showed that aromatic rings could form efficiently from small hydrocarbons (e.g.~acetylene) through radical-driven reaction sequences activated at high-temperature ($T\sim 10^3\,\rm K$), moderate-density conditions, as was expected for carbon-rich AGB stellar winds (see Section~\ref{subsec:pah_seeding}).

As these models matured \citep[e.g.~see the review by][]{Reizer2022FormationMini-review}, several distinct bottom-up mechanisms were identified, each predicting efficiencies that depended differently on the local temperature, UV field, and ionisation state. The most widely explored pathway is the hydrogen abstraction-acetylene addition (HACA) mechanism, in which the PAH growth happens via continuous abstraction of peripheral hydrogen atoms followed by accretion of C$_2$H$_2$ \citep[e.g.][]{Frenklach1989FormationEnvelopes,Cherchneff1992PolycyclicEnvelopes,Richter2002FormationFlames}. However, the effectiveness of the HACA mechanisms is limited to high gas temperatures ($T\sim 10^3\,\rm K$) and densities ($n_{\rm H}\sim 10^9-10^{11}\,\rm cm^{-3}$), which propelled a search for alternative growth mechanisms that could be more appropriate for ISM gas conditions. 
One such alternative is the hydrogen abstraction-vinylacetylene addition (HAVA) mechanism, which replaces acetylene with C$_4$H$_4$, enabling faster ring growth per reaction cycle \citep[][]{Zhang2010FormationStudy,Parker2012LowMedium,Liu2019ComputationalAddition,Zhao2018Low-temperatureAtmosphere}. The HAVA$^*$ further extends this pathway by reducing the reaction barrier thanks to vibrationally or electronically excited radicals \citep[][]{Shukla2012AMechanisms,Zhao2018Low-temperatureAtmosphere}. In parallel, laboratory cross-beam experiments demonstrated that radical-radical reactions between PAH compounds, summarised under the phenyl addition cyclisation framework \citep[][]{Shukla2008RoleHydrocarbons,Shukla2010AHydrocarbons,Zhang2025KineticPropyne}, can drive efficient PAH growth through largely barrier-less pathways. These developments significantly broadened the parameter space for bottom-up PAH formation, suggesting that PAH growth could occur not only in hot circumstellar envelopes, but also in cooler, UV-irradiated or shock-processed regions of the ISM. Recent JWST observations, including results from the PDRs4All program (ID 1288), reveal tight spatial correlations between PAHs, C$_2$H, and excited \hmol~in PDRs and molecular interfaces, providing strong empirical support for radical-mediated growth channels \citep[e.g.][]{Goicoechea2025PDRs4AllHydrocarbons,Khan2025PDRs4AllJWST}.

Observations of the Taurus Molecular Cloud (TMC-1) have been of particular relevance to the understanding of radical-mediated PAH growth mechanisms \citep[e.g.][]{Burkhardt2021DiscoveryTMC-1,Cernicharo2021PureIndene,McGuire2021DetectionFiltering,Wenzel2024DetectionHydrocarbon,Wenzel2025DiscoveryTMC-1,Cabezas2025DiscoverySpace}. Cyclopentadiene \citep[e.g.][]{Cernicharo2021DiscoveryCycles,McCarthy2021InterstellarCyanocyclopentadiene,KelvinLee2021TMC-1} and benzene have been abundantly detected in TMC-1 \citep[e.g.][]{McGuire2018DetectionMedium,Loru2023Detection12-diethynylbenzene}, molecules, with respectively $N_{\rm C}=5$ and 6, that are considered building blocks in the formation of larger aromatic molecules. However, astrochemical models still predict too-low abundances for many of the PAH precursors \citep[e.g.][]{McGuire2021DetectionFiltering,Cernicharo2022DiscoverySurvey,Mallo2025Ion-moleculeReaction}, and fundamental intermediate products in the formation sequence to cyclopentadiene and benzene have not been observed yet \citep[e.g.~1,3-butadiene][]{Agundez2025A13-butadiene}. Furthermore, these routes to the formation of benzene have been recently challenged by laboratory experiments \citep{Kocheril2025TerminationC6H5+}, leaving the issue of bottom-up PAH formation an unsolved problem at the time of writing \citep[e.g.][]{Loison2025EvidenceConditions}. Given these large uncertainties, we do not explore a direct coupling of astrochemical models of PAH precursors with \calima~in this work and leave this exploration for future work.

We instead opt by following the simpler approach of direct C accretion onto pre-existing PAHs. In previous works that have predicted the PAH mass fraction via aromatisation of carbonaceous grains \citep{Rau2019ModellingGalaxies,Hirashita2020Self-consistentEvolution,Narayanan2023ASimulations}, this takes place following the same framework as was considered for regular dust grains (see e.g.~Section~\ref{subsec:accretion}). Experiments have shown that C$^+$ accretion can be fairly efficient \citep{Canosa1995ReactionTemperature}, but if not balanced by an efficient process of C desorption from the PAHs could lead to the almost complete depletion of interstellar C within a few Myr \citep[e.g.][]{Snow2008IonMedium,Omont2021Intermediate-sizeHydrocarbons}. In this work we have instead used the results from \citet{Omont2025CarbonHydrocarbons} with respect to the photo-stability of [C-PAH] complexes: small anion and neutral PAHs have a 20\% probability of accreting C$^+$ without releasing acetylene \citep{Marciniak2021PhotodissociationConditions} if they are hydrogenated. The hydrogenation state is computed based on Section~\ref{subsec:h2_formation_pah} and we also apply the sub-grid density enhancement from Section~\ref{subsec:accretion}.

\subsection{PAH sputtering}\label{subsec:pah_thermal_sputtering}
In the same manner as for classical dust grains, PAHs immersed in a hot gas experience frequent collisions with energetic ions. These collisions may impart sufficiently large energies to cause PAH molecular dissociation. The survival of PAHs in hot supernova gas remains a debated topic, with some observations presenting tentative evidence of PAH emission co-spatial with known SNRs \citep{Reach2005AGalaxy,Tappe2006ShockN132D}. Extra-galactic observations find that PAH emission follows ionised gas in starburst winds \citep{McCormick2013DustyGalaxiesb,Bolatto2024JWSTWind}. Therefore, understanding the processing of PAHs in hot gas is a fundamental step in modelling the PAH life-cycle within a realistic galactic environment. We follow the modelling of thermal sputtering of PAHs presented in \citet{Micelotta2010PolycyclicGas}, with a series of modifications and adaptations to our PAH model (see Appendix~\ref{ap:pah_sputtering}). Our resulting sputtering rates are separated by electrons, H, He, C and O ions (dominant sputtering species), such that interpolated rates can be computed on-the-fly by \calima~considering the effects of local ionisation states and abundances.
We compute the sputtering timescale as:
\begin{align}\label{eq:pah_sputtering_timescale}
    \tau_{\rm spu}(N_{\rm C}) = \frac{N_{\rm C}}{2R_{\rm elec} + \sum_k (2R_{\rm e}^k+R_{\rm n}^k)},
\end{align}
where electron excitations (either by electron collision $R_{\rm elec}$ or by nuclear collision $R_{\rm e}^k$) are multiplied by 2 as each dissociation event leads to the ejection of two carbon atoms. The sum is over the $k$~ions. 

We also include non-thermal sputtering in SN shocks for PAHs, as we did for regular grains (Section~\ref{subsec:sn_shocks}). Again, we do not track the properties of individual shocks in \calima, but instead determine the amount of PAHs destroyed based on the amount of gas mass shocked above $100\,\rm km\,s^{-1}$~\citep{McKee1989DustMedium}. We use the results of \citet{Micelotta2010PolycyclicShocks} and assume that small PAHs are fully destroyed in all gas above $100\,\rm km\,s^{-1}$, but only $50$\% of the large PAHs are destroyed. This lost mass is returned to the local carbon gas-phase density.

\subsection{Photo-electric heating by PAHs}\label{subsec:photoelectric_heating} 
In Section~\ref{subsec:dust_photoelectric_heating} we introduced the physical process of photo-electric heating by classical dust grains. Within our PAH modelling, we can now extend this physical process into the molecular regime, by first considering a fundamental difference between large dust grains and very small grains. Since the absorption depth of far-UV radiation ($\sim 100 \rm \,\text{\AA}$) can be significantly smaller than the mean free path of emitted electrons, the free electron yield quickly decreases for large grains. Additionally, large grains easily acquire a positive charge (see Section~\ref{subsec:accretion}), making it even harder for an electron to escape the electric potential of the grain. This means that in practice $\gtrsim 50$\% of the PEH for an MRN-like GSD originates in grains with $N_{\rm C}< 1500$ \citep{Bakes1994TheHydrocarbons}, and grains larger than $a>100 \rm \,\text{\AA}$ have a secondary contribution. Therefore, for a long time, PAHs have been recognised as fundamental contributors to the very small dust grain PEH \citep{Verstraete1990IonizationGas}. Recently, \citet{Berne2022ContributionObservations} developed an updated model of the PAH contribution to the PEH using new molecular parameters from laboratory experiments and quantum mechanical calculations. They show that with only the contribution of small PAHs (i.e.~circumcoronene, $N_{\rm C}\sim 54$) one can reconcile the observed ratio of gas emission in the IR (by cooling lines like [CII] and [OI]) and the total IR emission also including the dust emission \citep{Okada2013ProbingRegions}. We follow a modified version of the model in \citet{Berne2022ContributionObservations} for the contribution to PEH by small and large PAHs (see Appendix~\ref{ap:pah_photoelectric_heating}). The benefit of this approach is that solving for the PAH charge distribution is not as computational and memory expensive as for regular dust grains, allowing the full computation of the photo-emission, attachment and recombination network on-the-fly in \calima. Therefore, self-consistent PEH modelling of PAHs renders \calima~as a unique tool for understanding the role of PAHs in coupling the sub-ionising UV radiation field with the thermal state of the gas in the ISM.

\subsection{PAH clustering}\label{subsec:pah_clustering}
The high polarisability of PAH molecules results in large van der Waals forces, providing a strong bonding force between aromatic planes of PAHs \citep{Zacharia2004InterlayerHydrocarbons} and the formation of PAH clusters.
We follow the dimerisation (or clustering) of small PAHs to form large PAH clusters in light of kinetic theory and the binding energy of monomers forming large carbon clusters \citep[e.g.][]{Rapacioli2005StackedMolecules,Rapacioli2006FormationMedium,Tielens2021MolecularAstrophysics,Khabazipur2023DevelopmentModel}. At low temperatures ($\sim 10$\,K)\footnote{We note that this implies that for higher CMB temperatures at high redshift, efficient clustering may be halted by the overall hotter gas temperature.}, the collision energy is significantly smaller than the interaction energy, which implies that long-lived compounds easily form in the interaction, allowing for the PAH monomers to re-arrange into the most energetically favoured orientations \citep[in the case of circumcoronene monomers stacking generally yields the most stable cluster;][]{Rapacioli2005StackedMolecules}. However, if the collision imparts too much energy to the system, monomer evaporation may occur before this re-orientation can take place. We extend the above rates by considering a variable sticking efficiency $C_{\rm eff}$ fitted to numerical results in Figure~10 of \citet{Totton2012ATemperatures}:
\begin{align}
    C_{\rm eff}(T) = \frac{1}{1+A[\log_{10}(T)]^B},
\end{align}
where the fitting parameters are $A=9.928\times 10^{-7}$ and $B=1.38$. In this case, the timescale for coalescence is given by:
\begin{align}\label{eq:coalescence_timescale}
    \tau_{\rm clus} = (\sigma_{\rm smallPAH,col}\Delta V C_{\rm eff} n_{\rm smallPAH})^{-1},
\end{align}
where $\sigma_{\rm smallPAH,col}=4\pi a_{\rm smallPAH}^2$ is the cross-section for collision between two small PAHs, $\Delta V$ is the collision velocity given by Brownian motion (Eq.~\ref{eq:brownian_relative_velocity}), and $n_{\rm smallPAH}$ is the local number density of small PAHs.

\subsection{PAH evaporation}\label{subsec:pah_evaporation}
We also consider PAH cluster evaporation \citep{Rapacioli2006FormationMedium}. Similar to the photo-dissociation of small PAH molecules due to excitation by UV photons, large PAH clusters may lose their monomers (in our case, circumcoronene molecules) if their internal energy exceeds the monomer binding energy. Therefore, determining the dissociation rate requires the calculation of the full energy probability distribution for a PAH cluster exposed to a given photon energy $E$. This is achieved by using the Arrhenius expression derived from the Rice-Ramsperger-Kassel-Marcus (RRKM) theory \citep[][]{Tielens2005TheMedium}. While solving these equations for thermal collisions (see Section~\ref{subsec:pah_thermal_sputtering}) can be simplified by assuming that the probability of dissociation changes very little after each IR photon emission, for the case of the absorption of UV photons by large PAH clusters one needs to fully compute the energy probability distribution in an iterative fashion. Indeed, the larger heat capacity of these systems \citep{Bakes2001TheoreticalI.,Tielens2005TheMedium} allows them to maintain a significant fraction of their internal energy between UV excitations (or equivalently, the cooling rate is slower than the absorption rate). This means that for these large molecular clusters the dissociation rates deviate from the linear scaling with the radiation intensity in the Habing band $G_0$, instead needing more photons to dissociate \citep[e.g.~$3-4$~photons for the dissociation of C$_{96}$H$_{24}$;][]{Montillaud2013EvolutionStates}. This multi-photon tracking of the heating/cooling of large PAHs, either via Monte-Carlo methods \citep[e.g.][]{Lange2021StabilityDiscs,Lange2023TurbulentPAHs} or statistical iterative methods \citep{Purcell1976TemperatureGrains.,Bakes2001TheoreticalI.,Andrews2016HydrogenationH2formation,Montillaud2014AbsoluteClusters}, is beyond the scope of our current modelling. We instead refer to the work of \citet{Rapacioli2006FormationMedium} and \citet{Montillaud2014AbsoluteClusters} on the effective dissociation rate of large PAH clusters. Considering that the large PAHs used in \calima~($N_{\rm C}\sim 400$) are formed from the clustering of our small PAHs (circumcoronene, C$_{54}$H$_{18}$), we use the results by \citet{Montillaud2014AbsoluteClusters} for the evaporation timescales of circumcoronene clusters with a number of monomers ranging from 4 to 12. They find that for these large monomer clusters the evaporation time scales more steeply than $1/G_0$, until the radiation field becomes very intense ($G_0\sim 10^6$), at which point the time between two UV photon absorptions becomes shorter than the cooling timescale. We thus use a scaling of the evaporation time given by the maximum of two functions fitted to their results:
\begin{align}\label{eq:pah_cluster_evaporation}
    \tau_{\rm evap} = \max \left[\frac{0.19306}{G_0},10^{-3.169 x + 13.564} \right],
\end{align}
where $x\equiv \log_{10}(G_0)$.

\subsection{PAH freezing}\label{subsec:pah_freezing}
Similar to PAH or grain coalescence (Sections~\ref{subsec:pah_clustering} and~\ref{subsec:coagulation} respectively), we also consider the adsorption of small and large PAHs on dust grains (also called freeze-out or sequestration of PAHs). For this, we follow exactly the same procedure as in Section~\ref{subsec:coagulation} for the coagulation of small grains, but computing the relative velocity ($\Delta V$, Eq.~\ref{eq:OC07_relative_velocity}) between the small and large PAHs and the small and large carbonaceous grains\footnote{We do not consider freezing of PAHs on silicate grains, as this would alter their chemical composition.}. With a similar argument as for the threshold velocity for coagulation (Section~\ref{subsec:coagulation}), we impose the requirement that the collision energy does not exceed the van der Waals bond energy \citep[see also][]{Lange2023TurbulentPAHs} attaching the PAH molecule/cluster to the grain surface \citep[$\sim 1$\,eV;][]{Rapacioli2006FormationMedium}:
\begin{align}
    \tau_{\rm free} = (\sigma_{\alpha\beta}\Delta V_{\alpha\beta} \psi_{\rm vdW}(E) n_{{\rm d},\beta})^{-1},
\end{align}
where the indices $\alpha$ and $\beta$ go over the combination of PAH sizes ($\beta$) and carbonaceous grain sizes ($\alpha$), respectively, $\sigma_{\alpha\beta}=\pi (a_{\alpha}+a_{\beta})^2$ is the collision geometrical cross-section, and $E$ is the kinetic energy of the collision in eV, used to compute the efficiency for the adsorption via van der Waals forces:
\begin{align}\label{eq:freezing_efficiency}
    \psi_{\rm vdW}(E) = \frac{1}{1+\exp [(E-1)/0.1]}.
\end{align}

\subsection{PAH photo-dissociation}\label{subsec:pah_dissociation}
Due to the temperature spikes experienced by a small PAH, there is a non vanishing probability that it can eject H, H$_2$ or even C$_2$H$_2$ when it experiences an intense UV radiation field in the diffuse ISM and in PDRs \citep{Jochims1994SizeImplications,Allain1996PhotodestructionGroup,LePage2001HydrogenationModel}. There is observational evidence of this destruction effect around young stellar clusters and in H~{\small II} regions \citep{Galametz2016TheN11,Relano2016Dust33}, which suggests that PAH fractions (or VSG fractions) can be used to trace the UV radiation field intensity \citep{Pilleri2012EvaporatingRegions,Bouteraon2019NanoComponents}. When a molecule is photo-excited, it divides its energy among vibrational degrees of freedom via intra-molecular vibrational redistribution. An assumption of the uni-molecular dissociation reaction \citep[e.g.][]{LePage2001HydrogenationModel,Tielens2005TheMedium} is that there exists a transition state (${\rm C}_{54}{\rm H}_{n})^*$) that, if reached, will lead to bond dissociation (e.g.~H-loss). In the reverse process, H atoms can be re-attached to the molecule \citep{Rauls2008CatalyzedConditions,Goumans2011HydrogenTunnelling,Jones2014TheDust}. The balance of these two reactions determines the hydrogen covering fraction of the molecule, termed the hydrogenation state of the PAH. Similar to Sections~\ref{subsec:pah_thermal_sputtering} and~\ref{subsec:pah_clustering}, the acetylene (C$_2$H$_2$) dissociation rate can be treated using a statistical approach based on RRKM theory \citep[e.g.][]{Tielens2005TheMedium,Montillaud2013EvolutionStates,Andrews2016HydrogenationH2formation}. However, at the photon energies common of PDRs and the diffuse ISM, photo-dissociation by loss of H and H$_2$ is faster by $\sim 2$~dex compared to the loss of acetylene \citep{Allain1996PhotodestructionGroup,LePage2001HydrogenationModel,Montillaud2013EvolutionStates,Andrews2016HydrogenationH2formation,Castellanos2018PhotoinducedH2-loss}. Experiments have shown that for small PAHs, H-loss dominates over C-loss \citep{Jochims1994SizeImplications}, and for larger PAHs stripping of hydrogen atoms takes place before the C-loss starts \citep{Zhen2014LaboratoryChemistry,Zhen2015LABORATORYFRAGMENTATION,Castellanos2018PhotoinducedDehydrogenation}. Therefore, a prior requirement for the fragmentation of aromatic rings in a small PAH is its full de-hydrogenation. This requires tracking the hydrogen coverage of our PAH bins, adding further computational cost to our model. We present here an alternative approach. One can define the dissociation parameter:
\begin{align}
    \psi = \frac{p_{\rm d} k_{\rm UV}}{k_{\rm a}n_{\rm H}},
\end{align}
where $p_{\rm d}$ is the probability for H dissociation, $k_{\rm UV}$ is the photon absorption rate, $k_{\rm a}$ is the H attachment rate and $n_{\rm H}$ the hydrogen number density. This parameter effectively captures the hydrogenation balance of a PAH. Considering an attachment rate of $k_{\rm a}=2\times 10^{-8} \sqrt{N_{\rm C}/50}\rm\,cm^3\,s^{-1}$ and an average UV absorption rate  $k_{\rm UV}\simeq 2.3 G_0$\,s$^{-1}$, one arrives at the relation $\psi \propto p_{\rm d}G_0/n_{\rm H}$  \citep{Tielens2005TheMedium,Berne2011FormationSpace}. Numerical computation of the hydrogenation of small and intermediate size PAHs has shown that the transition from fully de-hydrogenated to normally hydrogenated is very sharp in the $G_0$-$n_{\rm H}$ plane \citep{Montillaud2013EvolutionStates,Andrews2016HydrogenationH2formation}. Since the relevant quantity for C-loss is the fraction of small PAHs that are de-hydrogenated ($\chi_{\rm dH}$), we use fitting functions for the limiting region between de-hydrogenated and normally hydrogenated circumcoronene provided in figure 10 of \citet{Montillaud2013EvolutionStates}, allowing for a $1$\,dex dispersion, where PAHs transition from being 10\% to 90\% totally de-hydrogenated. We then assume that upon photon absorption, the energy is either spent on IR emission or on C-loss. Therefore, the probability for dissociation of a PAH upon absorption of a photon with energy $E$ is given by (see also Section~\ref{subsec:pah_thermal_sputtering}):
\begin{align}
    P_{\rm diss}^{{\rm C}_2{\rm H}_2} (E) = \frac{k_{\rm diss}^{{\rm C}_2{\rm H}_2}(E)}{k_{\rm diss}^{{\rm C}_2{\rm H}_2}(E) + k_{\rm IR}/(n_{\rm max}-1)},
\end{align}
where we compute the dissociation rate as in the thermal sputtering calculation of Section~\ref{subsec:pah_thermal_sputtering} \citep[see also][]{Murga2020ImpactPDRs}. From this, one can calculate the dissociation rate for a \citet{Draine1979DestructionDust.} radiation field in the Habing band as:
\begin{align}
    R_{\rm diss}^{{\rm C}_2{\rm H}_2}(G_0) = \int_{6\rm eV}^{13.6\rm eV} \chi_{\rm dH}\sigma_{\rm PAH}(E) P_{\rm diss}^{{\rm C}_2{\rm H}_2} (E) \frac{F(E)}{E}\dd E.
\end{align}
The numerical value of this rate for diffuse ISM conditions ($n_{\rm H}=0.1$\,cm$^{-3}$ and $G_0=1$) is $4.68\times 10^{-17}$\,s$^{-1}$.

\subsection{Molecular hydrogen formation mediated by PAHs}\label{subsec:h2_formation_pah}
In Section~\ref{subsec:h2_formation_dust} we based our modelling of molecular hydrogen formation on the assumption that the attachment surface for hydrogen on dust grains can be separated into physisorbed and chemisorbed sites. The former are characterised by weak van-der-Waals forces of a few tens of meV, which implies that when the dust temperature increases beyond $\sim 20$\,K H atoms quickly evaporate from the grain surface. This causes a drop in the H$_2$ formation efficiency, since beyond this point the total formation rate relies solely on the recombination via the Eley-Rideal mechanism on chemisorbed sites. This appears to be in disagreement with observations of PDRs \citep{Habart2003SomeRegions,Allers2005H_2Bar}, where estimated formation rates close to $\sim 3\times 10^{-17}\rm\,cm^3\,s^{-1}$ are well above the predictions for dust at high temperature. This points to additional physical mechanisms that can efficiently form H$_2$ in these regimes in which traditional dust grain formation mechanisms appear to fail.

Recently, more accurate modelling of PAH photo-physics \citep[e.g.][]{Montillaud2013EvolutionStates,Boschman2015H2formationHydrogen,Andrews2016HydrogenationH2formation,Barrera2023FormationStudy} and accelerated progress in experimental studies of PAH-hydrogen reactivity \citep[e.g.][]{Thrower2012ExperimentalFormation,Boschman2012HydrogenationFormation,Mennella2012TheFormation,Castellanos2018PhotoinducedH2-loss} enable to re-evaluate the pathways to H$_2$ formation in the ISM \citep[see the review by][]{Wakelam2017H2Observations}. Observed correlations between H$_2$ and PAH emission \citep{Habart2003HSUB2/SUBCloud,Habart2003SomeRegions} appear to support the scenario where PAHs can efficiently form H$_2$ in a highly excited state within PDRs. We therefore include molecular hydrogen formation catalysed by PAHs in \calima.

H$_2$ formation on PAHs is usually considered to take place via two possible routes: 
\begin{itemize}
    \item A hydrogen atom forms a covalent bond with a PAH molecule, and then it is abstracted by a second colliding hydrogen atom.
    \item UV photon absorption by a super-hydrogenated PAH molecule leads to relaxation by the loss of H$_2$.
\end{itemize}
Molecular hydrogen abstraction has been observed in desorption experiments on coronene deuterium bombardment \citep{Thrower2012ExperimentalFormation}, as well as predicted by theoretical modelling \citep{Rauls2008CatalyzedConditions}. The abstraction (as well as the hydrogenation) is subject to energy barriers in neutral coronene \citep{Boschman2012HydrogenationFormation} of $10$\,meV, while \citet{Bauschlicher2001TheHydrogen} predicts no barriers for H$_2$ abstraction from coronene and circumcoronene anions. Given the experimentally determined abstraction cross-section of $0.06$\,\AA$^2$ per coronene C atom, we assume for all abstraction of super-hydrogenated anion and neutral PAHs (both small and large) the Eley-Rideal rate:
\begin{align}
    k_{\rm ER} \simeq 8.7 \times 10^{-13} \sqrt{\frac{T}{100{\rm K}}}\left(\frac{n_{\rm H}}{1{\rm cm^{-3}}}\right)\,\rm cm^3\,s^{-1}.
\end{align}
We now need to determine what fraction of the PAH population is in a super-hydrogenated state. Similar to the approximations made in Section~\ref{subsec:pah_dissociation}, it is safe to assume that the transition from normally hydrogenated to super-hydrogenated is very sharp in $G_0$-$n_{\rm H}$ space \citep[see Figure 9 in][]{Andrews2016HydrogenationH2formation}, and that this transition is approximately independent of size for PAHs with $N_{\rm C}\geq 54$. We obtain a transition curve given by:
\begin{align}
    \log_{10}G_0 = 0.995 \log_{10} \left(\frac{n_{\rm H}}{1{\rm cm^{-3}}}\right) - 1.945,
\end{align}
with the width of the transition being $\Delta \simeq 0.05$ in log-log space as extracted from figure 9 in \citet{Andrews2016HydrogenationH2formation}. H$_2$ abstraction is only allowed for anion and neutral PAHs.

\begin{figure}
    \centering
	\includegraphics[width=\columnwidth]{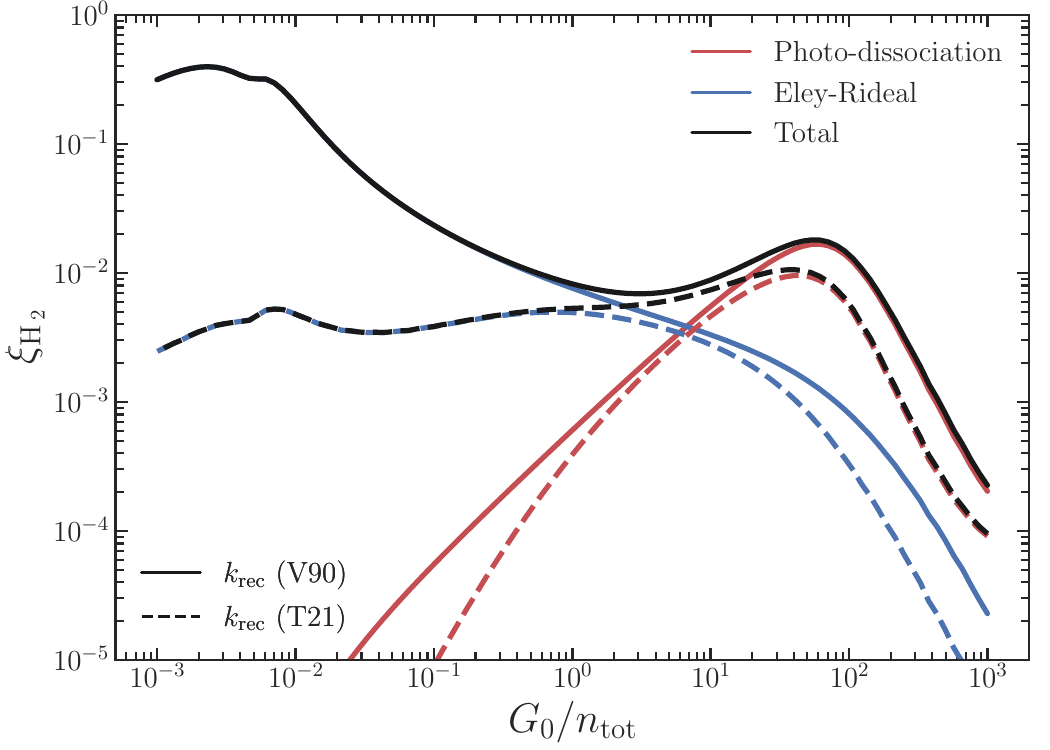}
    \caption{Molecular hydrogen formation efficiency by small PAHs (circumcoronene). We fix the radiation field intensity to the Habing parameter $G_0=100$, the free electron fraction to $X_{\rm e}=0.1$, and assume half of the H abundance is already in H$_2$. Molecular hydrogen is formed via two mechanisms: Eley-Rideal desorption (blue lines) of H$_2$ for neutral super-hydrogenated PAHs, and photo-desorption (red lines) of H$_2$ from normal and partially de-hydrogenated PAHs. The former dominates at high gas density in low radiation field, while the latter is important in a narrow regime at low density before radiation strips circumcoronene of all its H atoms. We compare the resulting efficiencies following two different models for electron recombination on PAHs by \citet{Verstraete1990IonizationGas} and \citet{Tielens2021MolecularAstrophysics}.}
    \label{fig:h2_pah_efficiency}
\end{figure}

For normally hydrogenated and partially de-hydrogenated PAHs, the absorption of a UV photon may give rise to the loss of an H atom or an H$_2$ molecule \citep{Allain1996PhotodestructionGroup}. Following the microcanonical description of bond dissociation used in Sections~\ref{subsec:pah_thermal_sputtering} and~\ref{subsec:pah_dissociation}, we compute the H$_2$ photo-desorption rate as:
\begin{align}
    k_{{\rm H}_2} = \int_{E_{\rm min}}^{13.6\rm eV}&\frac{k_{\rm diss}({\rm H}_2)}{k_{\rm diss}({\rm H}_2) + k_{\rm diss}({\rm H})+k_{\rm IR}/(n_{\rm max}+1)} \nonumber \\
    & \times\chi_{\rm pdH} \sigma_{\rm PAH}(E) \frac{F(E)}{E}\dd E,
\end{align}
where $F(E)$ is the flux density of the radiation field at energy $E$, $E_{\rm min} = n_{\rm max} \Delta \epsilon$ (see Section~\ref{subsec:pah_thermal_sputtering}), $\chi_{\rm pdH}$ is the fraction of partially de-hydrogenated PAHs, and the dissociation rates $k_{\rm diss}$ are given by Eq.~\ref{eq:pah_dissociation_rate}. We note that in this case we cannot use directly the fraction of de-hydrogenated PAHs $\chi_{\rm dH}$ to compute the rate of dissociation from normally to partially de-hydrogenated PAHs. As the PAH molecules becomes completely de-hydrogenated, relaxation upon UV photon absorption is no longer available through H or H$_2$ loss, but rather C-loss becomes important (see Section~\ref{subsec:pah_dissociation}). Hence, to maintain consistency across our model and prevent unrealistic H$_2$ production from fully de-hydrogenated PAHs, we suppress the fraction of de-hydrogenated PAHs $\chi_{\rm dH}$ when it exceeds $99$\% of the total PAH population, assuming at this point that all de-hydrogenated PAHs are bare carbon skeletons. This is done by applying a logistic equation correction to the fraction of de-hydrogenated PAHs to compute the fraction of partially de-hydrogenated PAHs:
\begin{align}
    \chi_{\rm pdH} \equiv \frac{\chi_{\rm dH}}{1+\exp[(\chi_{\rm dH}-0.99)/8\times 10^{-4}]},
\end{align}
which reproduces the width of the transition to fully de-hydrogenated PAHs in \citet{Montillaud2013EvolutionStates} and \citet{Andrews2016HydrogenationH2formation}. We note that this is only applied for small PAHs, as H$_2$ photo-desorption is not an efficient process for our large PAH clusters because UV absorption is instead followed by evaporation of the circumcoronene monomers (Eq.~\ref{eq:pah_cluster_evaporation}). Large PAHs only contribute to H$_2$ formation via the Eley-Rideal mechanisms.

In Fig.~\ref{fig:h2_pah_efficiency} we present the H$_2$ formation efficiency for small PAHs in an environment with $G_0=100$, with a fixed free electron fraction $X_{\rm e}=0.1$, and half of the H abundance already in the form of H$_2$. The H$_2$ formation efficiency is defined as \citep{Andrews2016HydrogenationH2formation,Castellanos2018PhotoinducedH2-loss}:
\begin{align}
    \xi_{{\rm H}_2}(Z) = \sum_i \frac{2k_{{\rm H}_2}(Z,i)f(Z,i)}{k^i_{\rm col}(Z)},
\end{align}
where $f(Z,i)$ is the fraction of PAHs with charge $Z$, and the sum is over all hydrogenation states for a PAH with charge $Z$ and $k_{\rm col}^i$ its hydrogen collisional rate. For neutral species we use a collisional rate with the average H addition energy barrier \citep{Rauls2008CatalyzedConditions,Boschman2015H2formationHydrogen} of $E_{\rm add} = 60$\,meV:
\begin{align}
    k_{\rm col}^0 = n_{\rm H} \sigma_{\rm geo}\sqrt{\frac{8k_{\rm B}T}{\pi m_{\rm H}}}\exp\left(-\frac{E_{\rm add}}{k_{\rm B}T}\right),
\end{align}
where $\sigma_{\rm geo}$ is the PAH geometrical cross section. For anions and (di-)cations, we follow \citet{Andrews2016HydrogenationH2formation} and consider fiducial rates of $k_{\rm col}^-=7.8\times 10^{-10}n_{\rm H}\rm cm^{3}\,s^{-1}$ and $k_{\rm col}^{+/2+}=1.4\times 10^{-10}n_{\rm H}\rm cm^3 \,s^{-1}$. The charge population of PAHs is obtained from the method described in Section~\ref{subsec:pah_optical_properties}. We have computed separately the contribution from Eley-Rideal (blue lines) and photo-dissociation (red lines) to the total molecular hydrogen formation efficiency (black lines). While keeping $G_0$ fixed, we vary the value of the total gas density ($n_{\rm tot}=n_{\rm H} + 2n_{\rm H_2}$), exploring a slice in $G_0$-$n_{\rm H}$ space. We find that at high $G_0/n_{\rm tot}$ photo-dissociation dominates H$_2$ formation for a narrow range of values, quickly dropping as the PAH population becomes dominated by totally bare carbon skeletons. H$_2$ abstraction via the Eley-Rideal mechanism is the dominant route of molecular hydrogen formation at low $G_0/n_{\rm tot}$. Eley-Rideal at high densities is highly reliant on the abundance of neutral and anionic PAH species. As discussed in Section~\ref{subsec:photoelectric_heating}, the charge state in those regimes depends on the assumptions regarding electron recombination and attachment. \citet{Tielens2021MolecularAstrophysics} predict a dominant anion population at very high gas/electron number densities, for which the collision rate is larger, resulting in a lower efficiency. This effect can be appreciated by comparing the solid curves using the Spitzer recombination rate modified by \citet{Verstraete1990IonizationGas}, and the dashed curves using the \citet{Tielens2021MolecularAstrophysics} recombination rate. Overall, we find that maximum efficiencies are on the order of $\xi_{\rm H_2}\sim 1$\%, in agreement with previous models of H$_2$ formation on PAHs \citep[e.g.][]{LePage2009MolecularMedium,Boschman2015H2formationHydrogen,Andrews2016HydrogenationH2formation,Castellanos2018PhotoinducedH2-loss}.

\section{Numerical validation}\label{sec:sims}

\subsection{Equilibrium temperature}\label{subsec:eq_tests}
\begin{figure*}
    \centering
	\includegraphics[width=\textwidth]{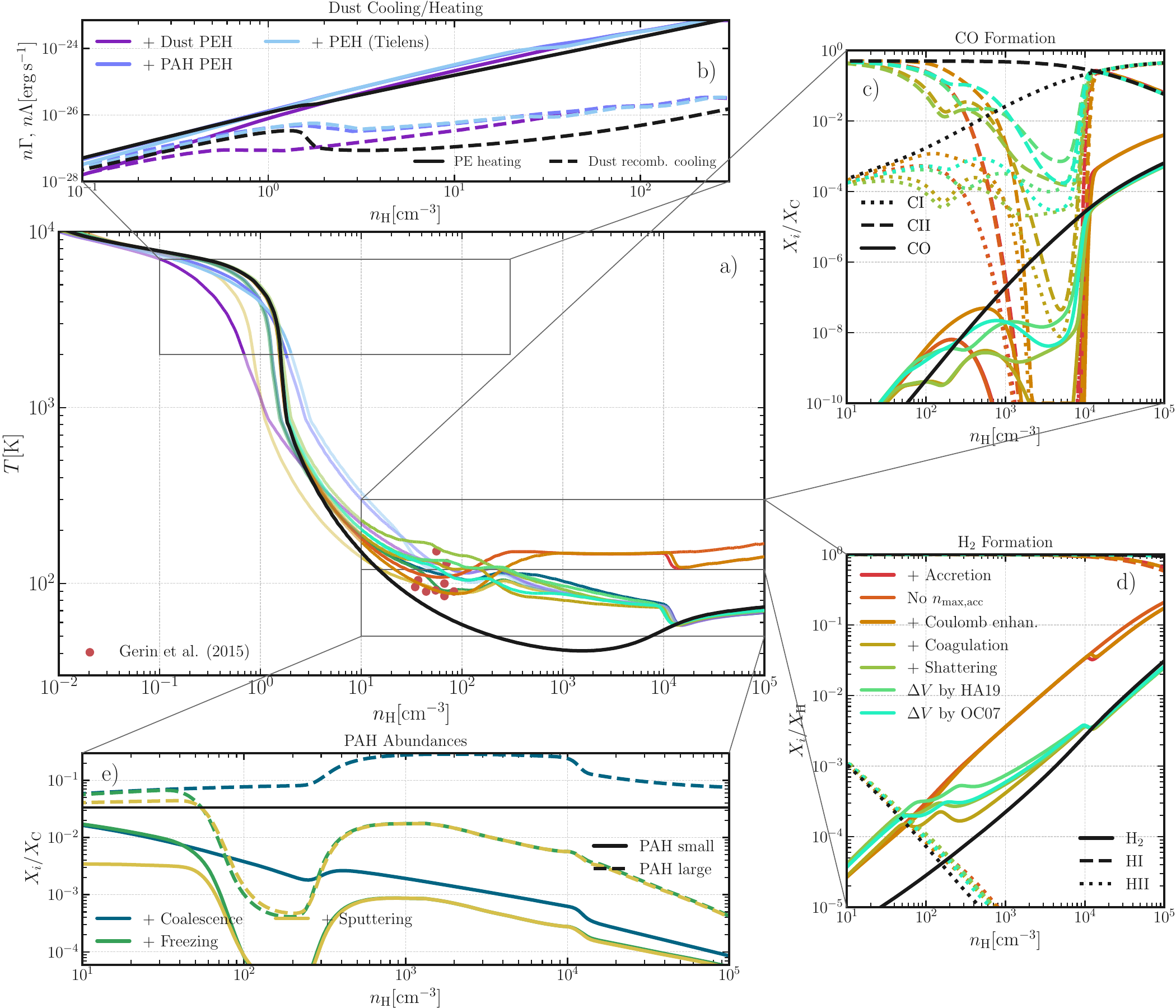}
    \caption{Temperature-density equilibrium curve comparing the fixed dust (solid black line) to the \calima~model (colour lines). Each line progressively add on top to the previous line another dust/PAH process, starting from accretion (red line) and ending in the ones including PEH from PAHs (dark and light cyan lines). From the central panel and going clockwise: a) temperature-density, b) volumetric PEH and dust recombination cooling, c) fraction of C atoms depleted on neutral C, C$^+$ and CO, d) fraction of H atoms on molecular \hmol, neutral HI, and ionised H~{\small II}, and e) fraction of C atoms depleted on small and large PAHs. Red points in panel a) are the inferred temperatures from [C~{\small II}] observations of the Galactic plane by \citet{Gerin2015CPlane}. See text for further details.}
    \label{fig:equilibrium_curve}
\end{figure*}
To begin validating the predictions from \calima, we perform a one-zone, static test of the equilibrium temperature at varying densities. We perform this by smoothly interpolating between $10^{-2}$ and $10^5\rm\,H\,cm^{-3}$ and allowing the chemistry solver to reach thermal equilibrium (i.e.~balance of heating and cooling processes), while maintaining the density fixed. All cells are initialised with $T=10^4$\,K and with all ions fully neutral. A constant $G_0=1$ and cosmic ray ionisation rate of $\eta_{\rm CR}=10^{-16}\,\rm s^{-1}\, H^{-1}$ are fixed across all densities. We assume that all cells have the same total metallicity $Z$, with solar abundances given by \citet{Asplund2009TheSun}. Depletions onto dust are taken from \citet{Zubko2004InterstellarConstraints} for oxygen, and from \citet{Dopita2000TheNebulae} for iron, magnesium, silicon and carbon. Direct sum of these depletions result in a DTM of 0.4 or a gas-to-dust (GTD) of 162. Since we force \calima~to maintain the stochiometry of the silicate grain (see Section~\ref{subsec:accretion}) we compute the true silicate dust mass using Eq.~\ref{eq:limiting_element}, which instead result in a DTM of 0.391. We then distribute dust mass in each grain size by following the resulting small-to-large (STL) ratio found in~\citet{Dubois2024GalaxiesSimulations} for the G10 simulation as in agreement with the extinction curve of the MW (i.e.~$\sim 20$\% of the dust mass in small grains). We then extend this to include $13.42$\% of the carbon dust into PAHs~\citep{Zubko2004InterstellarConstraints}, with equal mass in small and large PAHs. In panel a) of Fig.~\ref{fig:equilibrium_curve} we show (black solid line) the equilibrium curve for this elemental and dust abundances without dust evolution and using the vanilla ISM model in~\prism. In agreement with previous models \citep[e.g.][]{Wolfire2003NeutralGalaxy,Bialy2019ThermalGas,Kim2023PhotochemistrySimulations}, we find that the thermal instability is triggered for $n_{\rm H} > 1$\,cm$^{-3}$. For $n_{\rm H} \gtrsim 10^3$\,cm$^{-3}$ we find a rise in temperature, related to heating from H$_2$ formation and destruction. The red points are observational inferred kinetic temperatures from [C~{\small II}] observations of the Galactic plane by \citet{Gerin2015CPlane}. The no-evolution curve appears to deviate from these observations, predicting a steeper decrease in temperature at $10<n_{\rm H}<10^2\rm\,cm^{-3}$. This results from our choice of C depletion of 0.4991 from \citet{Dopita2000TheNebulae} instead of the larger value of 0.630 the BARE-GR-S model assumed \citep{Zubko2004InterstellarConstraints}. Together with a fixed DTM across all $n_{\rm H}$, it results in increased cooling from more available [C~{\small II}] and lower \hmol~heating due to a lower \hmol~formation for a lower dust mass compared to \calima~at high $n_{\rm H}$. This demonstrates how small variations in depletion of key elements for the cooling and heating of the ISM can have non negligible effects in the equilibrium curves, again raising a point made in \citet{Katz2022PRISM:Galaxies} that modulating abundances/depletion factors within the uncertainties of the Galaxy can return the curves into agreement with these observations. 

We now test how the growth of dust mass via gas-phase accretion can leave a signature on the final equilibrium curve. In panel (a) of Fig.~\ref{fig:equilibrium_curve} we highlight the influence of dust processes on the equilibrium curve within $10<n_{\rm H}<10^5\,\rm cm^{-3}$ in the middle inset box. Including accretion, coagulation and shattering leads to growth of the overall mass as well as transfer of mass between different grain sizes. Accretion (red and orange lines) alone is seen as a clear deviation from the fixed curve for densities above $n_{\rm H}\gtrsim 10\, \rm cm^{-3}$, leading to significantly higher gas temperatures above this density. The origin of this difference is caused by a simultaneous decrease in the O~{\small I} and C~{\small II} cooling due to accretion of these gas-phase ions, and the increased H$_2$ heating from larger dust mass/surface area. In panel d) we show the signature of the latter on the \hmol~fraction (solid lines). Dust growth is faster for smaller grains due to their larger surface-to-mass ratio, and for the same dust composition and temperature, results in a larger \hmol~formation rate~\citep{Cazaux2004HSUB2/SUBSurfaces}. 
For fixed silicate stoichiometry, Fe is fully depleted in our initial conditions. This means that while silicates cannot experience a large growth in mass from gas-phase ions, carbon is depleted to $\sim 50$\%. Therefore, carbonaceous grains compete with CO for the available free carbon. However, in our idealised test there is no inclusion of attenuation of the ISRF at high densities. This means that while CO would be expected to rapidly grow at these higher densities \citep[e.g.][]{Glover2010ModellingMedium,Glover2012IsFormation}, the continuous destruction via our un-attenuated $G_0$ releases carbon that can be accreted by grains that are definitely not destroyed at these densities and temperatures. This can also be appreciated on panel c), where gas at $10^2<n_{\rm H}<10^4\,\rm cm^{-3}$ shows large depletions of C~{\small I} and C~{\small II} compared to the fixed dust curves (black lines). This leads to complete suppression of CO at this density regime when equilibrium is reached. It is only prevented for $n_{\rm H}>10^4\,\rm cm^{-3}$ when dust grains start to become covered with ice mantles and the accretion efficiency is highly suppressed (see the curve with no $n_{\rm max,acc}$). Coulomb enhancement appears to have a non-negligible effect on these results, by preventing efficient accretion for $10^2<n_{\rm H}<10^3\,\rm cm^{-3}$ due to the prevalent positive grain charge in these neutral phases (see e.g.~Fig.~\ref{fig:dust_charge_silicate_0.005micron}).

We also explore the influence of turbulent collisions between grains in the \hmol~and CO formation. This is a static numerical experiment, so we do not have a direct measure of the gas dynamics for each density. Since this is a fundamental input for the subgrid turbulence models in \calima, we have opted by a choice of gas velocity dispersion scaling laws. We use the Larson relation \citep{Larson1981TurbulenceClouds.,Hennebelle2012TurbulentClouds} as obtained by \citet{Colman2024CloudMedium} for a large range of MHD simulations of turbulent molecular clouds:
\begin{align}\label{eq:larson_relation}
    \sigma = 5.67 \left(\frac{\langle n_{\rm H}\rangle}{10^2\rm cm^{-3}}\right)^{-0.25}\,\rm km\,s^{-1},
\end{align}
where $\langle n_{\rm H}\rangle$ is the spatial average hydrogen number density, which in our case would be the cell hydrogen number density.
Neither coagulation nor shattering alter the total dust mass, but since they can transfer mass between grains of different sizes they can affect the surface-to-volume ratio and hence the accretion and \hmol~formation rates. Adding coagulation on top of accretion leads to a rapid transfer of mass from the small to the large grains, leading to significantly lower depletions of carbon in the $10^2<n_{\rm H}<10^4\,\rm cm^{-3}$ regime, as well as lower \hmol~fractions, resembling the fixed dust fractions for $n_{\rm H}>10^4\,\rm cm^{-3}$. This is due to the fact that while accretion is halted at very high densities due to ice formation, coagulation can still take place, highly depleting the reservoirs of small grains. Adding shattering on top of coagulation allows for a return of some of the coagulated mass back into the small grain bin, particularly at $n_{\rm H}<10^3\,\rm cm^{-3}$. This can be seen as a higher \hmol~fraction in panel d). We have also compared the influence of different prescriptions for the grain relative velocity $\Delta V$. Given that the model from \citet{Hirashita2019RemodellingGalaxies} predicts slightly lower $\Delta V$ than \citet{Ormel2007AstrophysicsNote}, we find both higher coagulation and shattering rates. This results in shattering operating at higher densities (hence larger carbon depletion due to higher abundance of small grains), and coagulation at lower densities (hence lower carbon depletion due to lower abundance of small grains). We note that the inclusion of coagulation and shattering is very important in the final equilibrium temperature at these densities, displacing the gas at $10^2<n_{\rm H}<10^4\,\rm cm^{-3}$ from $\sim 150$\,K to $\sim 70$\,K.

Similarly, in panel e) we examine the influence of PAH evolution on the abundances of small and large PAHs (bottom inset box of panel a) in Fig.~\ref{fig:equilibrium_curve}). Coalescence of small PAHs to form PAH clusters is a very fast process within the high density phases, giving rise to a very steep decrease with gas density in the fraction of carbon atoms depleted onto small PAHs. Adding freezing on regular carbonaceous grains (green lines) leads to even lower abundance of small PAHs, but the more drastic difference is seen for large PAHs, which now become highly depleted at high densities. Freezing is particularly effective at $50<n_{\rm H}<300\,\rm cm^{-3}$. This is caused by the narrow region in which relative velocity between PAHs and grains is sufficiently slow to allow for van-der-Waals surface bonds, but not high enough density that depletes the majority of the grain mass into large grains with lower surface-to-volume ratio. Sputtering is also shown in panel e) (yellow lines) to illustrate the higher efficiency of destruction of PAHs in warm gas compared to regular grains.

Given our direct modelling of photo-electric heating and charging for grains and PAHs, we also compare the the results from our approach to MRN-averaged heating and recombination cooling \citep[i.e.][]{Bakes1994TheHydrocarbons,Weingartner2001PhotoelectricHeating}. In panel b) of Fig.~\ref{fig:equilibrium_curve} we show the photoelectric heating rate $n\Gamma$ (solid lines) and the recombination cooling rate $n\Lambda$ (dashed lines), comparing the result of the fixed dust model (black lines) with our live dust modelling. These models include all the processes referenced before: accretion, Coulomb enhancement, coagulation, shattering, coalescence, freezing and sputtering of PAHs, but now computing directly the efficiencies of heating and cooling from the models in Sections~\ref{subsec:dust_photoelectric_heating} and~\ref{subsec:photoelectric_heating}. Firstly, only including direct photo-electric heating from dust grains presents a significantly lower turnover in the temperature curve (see purple line in panel a)). This is in agreement with the result that large grains have a secondary contribution to photo-electric heating \citep[e.g.][]{Bakes1994TheHydrocarbons}. However, for $n_{\rm H}\gtrsim 2\,\rm cm^{-3}$ the injected energy is larger than the fixed model ratio, due to a larger dust mass caused by gas-phase accretion. Adding to this the injected energy from PAHs, this difference is even larger, which causes the turnover of these curves (violet and cyan) to be at higher densities and significantly less steep compared to the fixed dust model. However, this difference with the only-dust phot-electric heating becomes negligible for $n_{\rm H}\gtrsim 30\,\rm cm^{-3}$, as much of the PAH mass is frozen onto regular grains. We have also included a modification to the PAH charge modelling given by more efficient electron recombination from \citet{Tielens2021MolecularAstrophysics} (see also Section~\ref{ap:pah_photoelectric_heating}). We find this to have a minimal effect on the final efficiency.

We end this numerical benchmarking by noting that although the experiment was not designed to pass through the \citet{Gerin2015CPlane}, variations in dust and PAH processes encompass the scatter and slope of their result, while the fixed dust model fails to do so.

\subsection{Isolated galaxy simulations}\label{subsec:isolated_galaxy}

\subsubsection{Methods}\label{subsubsec:methods_g8}
\begin{table*}
    \centering
    \renewcommand{\arraystretch}{1.4}
    \begin{tabular}{c|c|c|p{12cm}}
         \textbf{Group Name} & $E_{\rm low}$ [eV] & $E_{\rm high}$ [eV] & \textbf{Function} \\
         \hline
         IR & 0.1 & 1.0 & Infrared radiation pressure \\
         Opt. & 1.0 & 5.6 & Radiation pressure, charging of large grains \\
         FUV & 5.6 & 11.2 & PAH cluster evaporation, photo-electric heating, Mg I, Si I, S I, Fe I ionisation \\
         LW & 11.2 & 13.6 & \hmol{} dissociation, PAH dissociation, charging of small grains and PAHs, C I ionisation \\
         EUV$_1$ & 13.6 & 15.2 & H I, N I, O I, Mg II ionisation \\
         EUV$_2$ & 15.2 & 24.59 & \hmol{}, C II, Si II, S II, Fe II, Ne I ionisation \\
         EUV$_3$ & 24.59 & 54.42 & He I, O II, C III, N II, N III, Si III, Si IV, S III, S IV, Ne II, Fe III ionisation \\
         EUV$_4$ & 54.42 & $\infty$ & He II, O III+, N IV+, C IV+, Mg III+, S V+, Si V+, Fe IV+, Ne III+ ionisation
    \end{tabular}
    \caption{Details of our choice of radiation spectrum discretisation. We indicate the photon group name, the lower and upper energy limits in eV, as well as their main roles in our thermochemistry modelling. We include ions that have their ionisation potential within each of the bins limits.}
    \label{tab:radiation_bins}
\end{table*}
We explore the results of \calima~in a dynamical system by considering isolated galaxy simulations. We make use of the G8 initial conditions presented in \citet{Rosdahl2015GalaxiesGalaxies} (see their table 1). These were also used for the original implementation of dust evolution in \ramses~\citep{Dubois2024GalaxiesSimulations} as well as in \prism~\citep{Katz2022PRISM:Galaxies}. The G8 galaxy has a halo mass of $10^{10}\,\rm M_{\odot}$ (circular velocity of $30\,\rm km\,s^{-1}$), with a galaxy mass (gas and stars) that adds up to $3.8$\% of the halo mass, hence $3.8\times 10^{8}\,\rm M_{\odot}$. The gas mass fraction is 0.171951. In order to reduce the extent of the initial starburst caused by the initial conditions of an isolated galaxy, we introduced a perturbation in the initial vertical gas velocity of $30\,\rm km\, s^{-1}$ modulated by two sinusoidal functions in the $x$ and $y$ axis of the cartesian box with wavelength of $200\,\rm pc$.
The initial metallicity $Z_{\rm init}$ of G8 is set to $0.1$, $0.3$, and $0.5\ Z_{\odot}$, using the \citet{Asplund2009TheSun} abundance ratios of elements and an initial $\rm DTM=10^{-3}$, with the depletion ratios used in Section~\ref{subsec:eq_tests} scaled to this lower DTM. Instead of using a profile of decreasing metallicity with radius \citep[e.g.][]{Katz2022PRISM:Galaxies}, we use a uniform metallicity and DTM across the disc, with gas outside of the disc at a metallicity of $10^{-4}Z_{\odot}$ and no dust. We track the abundance in O, N, C, Mg, Si, S, Fe and Ne. The ion chemical network employs eight ionisation states of O, seven for N, six for C, ten for Mg and Ne, and eleven for Si, S, and Fe. We also follow all ionisation states of H and He, in addition to \hmol~and CO. For the latter two molecular species, we also apply the correction to density coming from the unresolved turbulent density distribution (see Section~\ref{subsec:accretion}). The disc is initialised atomic and with all metals fully neutral, while the halo gas is set as fully ionised and with all metals in their highest ionisation state. We employ the `fast' solver in \citet{Katz2022RAMSES-RTZ:Hydrodynamics}. In order to provide a comparison with non-evolving dust models, we also run a suite of twin simulations with the same initial conditions and gas-phase metallicities as the \calima~simulations, but setting the dust abundances based on the scaled DTM and PAH mass fraction scaling with $12+\log(\rm O/H)$ from \citet{Remy-Ruyer2014Gas-to-dustRange,Remy-Ruyer2015LinkingPicture}. The grain size distributions are initialised to follow the equilibrium distribution obtained for the G8 simulations in \citet{Dubois2024GalaxiesSimulations}, which predicts a $\sim 0.2$ total dust mass in small grains. For these simulations, we turn off the dust evolution, only allowing for dust and PAH injection due to stellar processes (e.g. SNe and AGB winds). We also use the dust chemistry (e.g. cooling, heating, \hmol~formation) from the original version of \prism~\citep{Katz2022PRISM:Galaxies}.

Star formation is modelled using the gravo-turbulent criteria \citep{Kimm2017Feedback-regulatedReionisation} to determine local star formation efficiencies. Star formation is only allowed if the local turbulent Jeans length is unresolved and if $n_{\rm H}\geq 10\,\rm cm^{-3}$. The formed stellar particles are integer multiples of $1830\,\rm M_{\odot}$. We include feedback from SNII and SNIa using the mechanical feedback prescription of \citet{Kimm2014ESCAPESTARS}, as well as stellar winds \citep{Agertz2021VintergatanGalaxy}. Core-collapse SNe take place by stochastically sampling the SNII progenitors that leave the main sequence during each time step, using the main-sequence lifetime-mass relation by \citet{Schaerer1993Grids}. In order to be consistent with our choice of yields \citep{Limongi2018Presupernova0}, we assume that stars with $8 \leq m_*/{\rm M}_{\odot}\leq 25$ explode as SNeII with the canonical energy of $10^{51}\,\rm erg$. We also include a minimal model for the gas pre-processing caused by OB stars in the immediate gas surrounding the stellar particle, allowing for a larger amount of momentum injection \citep{Geen2015AStar}\footnote{This is included since at $\sim 4\,\rm pc$ H~{\small II} regions are mainly unresolved.}. Our choice of yields follows the implementation in the {\sc megatron} simulations \citet{Katz2024TheGalaxies}. This uses the individual element yields for SNeII from \citet{Limongi2018Presupernova0}, for SNeIa from \citet{Seitenzahl2013Three-dimensionalSupernovae}, and for AGB winds from \citet{Ritter2018NuGrid0.0001-0.02}. We direct the interested reader to \citet{Katz2024TheGalaxies} for further details. We use these yields to directly compute the seeding of dust grains and PAHs, using the prescriptions in Section~\ref{subsec:grain_seeding} and~\ref{subsec:pah_seeding}.

We make use of on-the-fly radiation transfer with the M1 moment method in \ramsesrt~\citep{Rosdahl2013Ramses-rt:Context,Rosdahl2015ARAMSES-RT}, using the GLF scheme to construct intercell fluxes. We track eight radiation bins from the IR to the EUV (see Table~\ref{tab:radiation_bins}), employing the reduced speed of light approximation setting $\tilde{c}=0.01c$. The radiation and thermo-chemistry steps are subcycled up to 500 times per each hydrodynamic step. In practice, the number of subcycles falls from 7-8 after the first star is formed to 2-3 when the first SN goes off. Radiation is emitted from stellar particles based on their mass, age, and metallicity using the BPASS v2.2.1 SED \citep{Stanway2018Re-evaluatingPopulations}, with the Chabrier initial mass function and a maximum mass of $300\,\rm M_\odot$. The mean cross-section for each radiation bin is computed based on the stellar-polling method of \ramsesrt (see Section~\ref{subsec:optical properties}). In addition, the simulation is filled with a $z=0$ UV background \citep{Haardt2012RadiativeBackground} with a self-shielding prescription that exponentially decreases the intensity of the background for $n_{\rm H} \geq 10^{-2}\,\rm cm^{-3}$. We also consider effects of radiation pressure for single scattering of UV and optical and multi-scattering in the IR \citep{Rosdahl2015ARAMSES-RT}.

Cosmic-rays are not self-consistently seeded in these simulations, so we use a fixed cosmic-ray background ionisation rate of $\eta_{\rm CR}=10^{-16}\,\rm s^{-1}\, H^{-1}$. We employ the AMR capabilities of \ramses~to reach spatial resolutions of 4.5\,pc in the highest level of refinement. We refine when a cell hosts eight times its initial mass in stars or in order to resolve the Jeans length by four cells.

We emphasise that our choice of initial conditions is not intended to resemble any particular galactic system, rather we use these G8 simulations to test \calima~in a dynamical, galactic environment.

\subsubsection{Variation in dust and PAH properties across ISM phases}\label{subsubsec:dust_variations}
\begin{figure*}
    \centering
	\includegraphics[width=\textwidth]{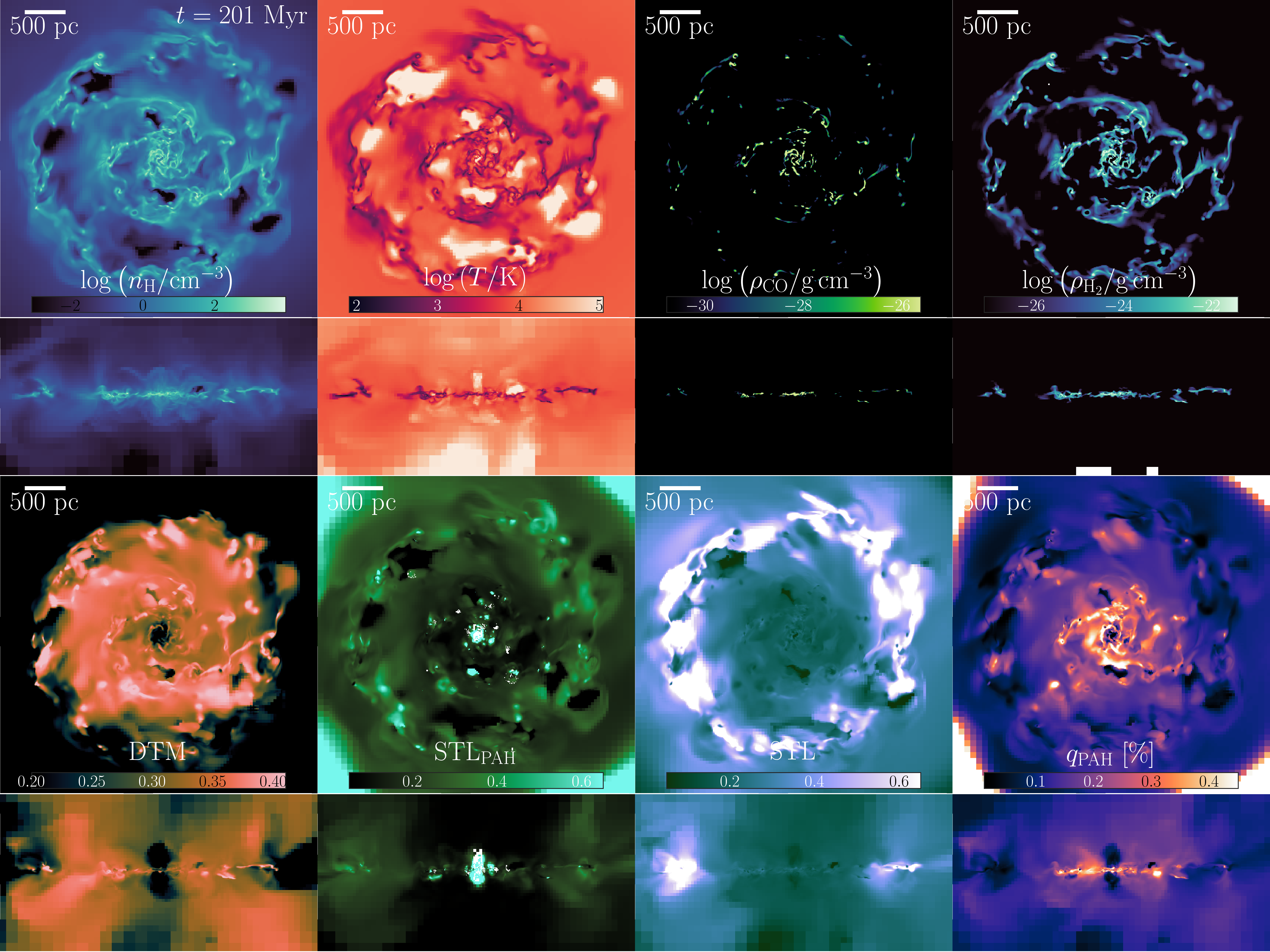}
    \caption{Density-weighted projection of the G8 galaxy with initial metallicity of $0.3Z_{\odot}$ at 200\,Myr after the beginning of the simulation. The projection has a horizontal width of 4\,kpc and a depth of 0.8\,kpc. From top-left corner in clock-wise order: (1) hydrogen number density, (2) gas temperature, (3) CO mass density, (4) \hmol~mass density, (5) $q_{\rm PAH}$, ratio of PAH mass to total dust and PAH mass, (6) smallt-to-large (STL), ratio of small dust grain mass to large grain mass, (7) STL$_{\rm PAH}$, ratio of small PAH mass to large PAH mass, and (8) DTM, fraction of metal mass depleted onto dust and PAHs. First and third rows show G8 face-on, while second and fourth show it edge-on.}
    \label{fig:G8_03_projections_grid}
\end{figure*}
We begin by examining density-weighted projections of our G8 galaxy. In Fig.~\ref{fig:G8_03_projections_grid} we show face-on (first and third row) and edge-on (second and fourth rows) projections of the $Z_{\rm init}=0.3\, Z_{\odot}$ at 200\,Myr, after the initial starburst has subsided and the SFR has stabilised ($\sim 0.02\mbox{--}0.04\,\rm M_{\odot} yr$). The galaxy exhibits a multi-phase ISM with cold, dense clumps and filaments embedded within a diffuse warm medium. Several large hot gas bubbles are visible, driven by recent star formation and SN energy injection. In the edge-on projections the disc appears razor-thin, with hot outflows piercing through the dense ISM and propelling filaments and shells outside of the plane. By looking at the projections of CO and \hmol~density it is clear that the cooler phases of the ISM are spatially constrained to the plane of the disc, with CO particularly limited to the very high density gas clumps exceeding $\gtrsim 100\,\rm cm^{-3}$. In order to quantify the relative importance of dust mass and different dust properties, we define the following useful quantities:
\begin{itemize}
    \item Dust-to-metal (DTM) mass ratio as:
    \[
    \text{DTM} = \frac{M_{\rm dust}}{M_{Z}+M_{\rm dust}},
    \]

    where $M_Z$ is the mass in gas phase metals,
    
    \item Small-to-large (STL) mass ratio for all dust grains (not including PAHs):
    \[
    \text{STL} = \frac{M_{\rm smallC}+M_{\rm smallSil}}{M_{\rm largeC}+M_{\rm largeSil}},
    \]
    as well as an equivalent for PAHs:
    \[
    \text{STL}_{\rm PAH} = \frac{M_{\rm smallPAH}}{M_{\rm largePAH}},
    \]
    
    \item $q_{\rm PAH}$ is the ratio of mass in PAHs to the total dust and PAH mass:
    \[
    q_{\rm PAH} = \frac{M_{\rm PAH}}{M_{\rm dust} + M_{\rm PAH}}.
    \]
\end{itemize}

The DTM distribution shows weak correlation with the highest-density regions, instead exhibiting variations of $\sim 2$ across the disc. Some low-DTM regions within the disc coincide with low-density, hot gas, while the inner region shows lower DTM values. This central depletion is consistent with the `DTM cavity' found in the NewCluster cosmological simulation \citep{Byun2025HowSimulation}, which they attribute to efficient dust destruction within rapidly growing bulges at high redshift. In our case, a strong outflows appears to be forming at the centre of the disc, easily seen in the edge-on projections as two bubbles with low DTM, STL and $q_{\rm PAH}$.

The STL$_{\rm PAH}$ and STL projections (lower centre panels of Fig.~\ref{fig:G8_03_projections_grid}) reveal preferential destruction and growth of small grains and PAHs. Outside the disc, dust properties reflect the initial conditions (STL$\sim 0.2$, STL$_{\rm PAH}\sim 1$), but these regions contribute negligibly to the total dust and PAH budget (see DTM and $q_{\rm PAH}$ projections) and are excluded from further analysis. Within the disc, both ratios show strong spatial variations: regions of efficient small grain destruction coincide with high gas temperatures, while zones of enhanced small grain and PAH growth align with cold, dense clumps and filaments. PAHs exhibit larger variations than silicate/carbonaceous grains due to their shorter destruction timescales, producing stronger contrasts between ISM phases. The STL$_{\rm PAH}$ projection shows a distinctive pattern around recent star formation sites: small PAH-dominated clumps (bright green) are surrounded by channels of very low PAH mass (white pixels). This spatial sequence traces PAH photo-destruction: large PAHs are first destroyed (Section~\ref{subsec:pah_evaporation}), releasing small PAHs and increasing STL$_{\rm PAH}$, followed by acetylene loss from small PAHs (Section~\ref{subsec:pah_dissociation}), which decreases STL$_{\rm PAH}$.

The $q_{\rm PAH}$ projection (lower right of Fig.~\ref{fig:G8_03_projections_grid}) shows complex behavior. While large-scale features of enhanced $q_{\rm PAH}$  correlate with molecular gas (CO and H projections) and dust-enriched regions (DTM projection), many high-$q_{\rm PAH}$  structures lack corresponding features in the dense ISM tracers. Notably, some high-$q_{\rm PAH}$ clumps exhibit low-$q_{\rm PAH}$ cores, while the corresponding CO and H distributions show no density depletion at these locations. This break in CO-PAH correlation could be related to the one observed in resolved studies of local star-forming galaxies \citep[e.g.][]{Kim2025LocalizedEmissivity}, and to the observability of PAH emission in star forming regions \citep[e.g.][]{Rodriguez2025TracingClusters}. Combined with the presence of PAHs in diffuse ISM phases, these results suggest that using PAHs as molecular gas and star formation tracers may be more complex than commonly assumed.

\begin{figure*}
    \centering
	\includegraphics[width=\textwidth]{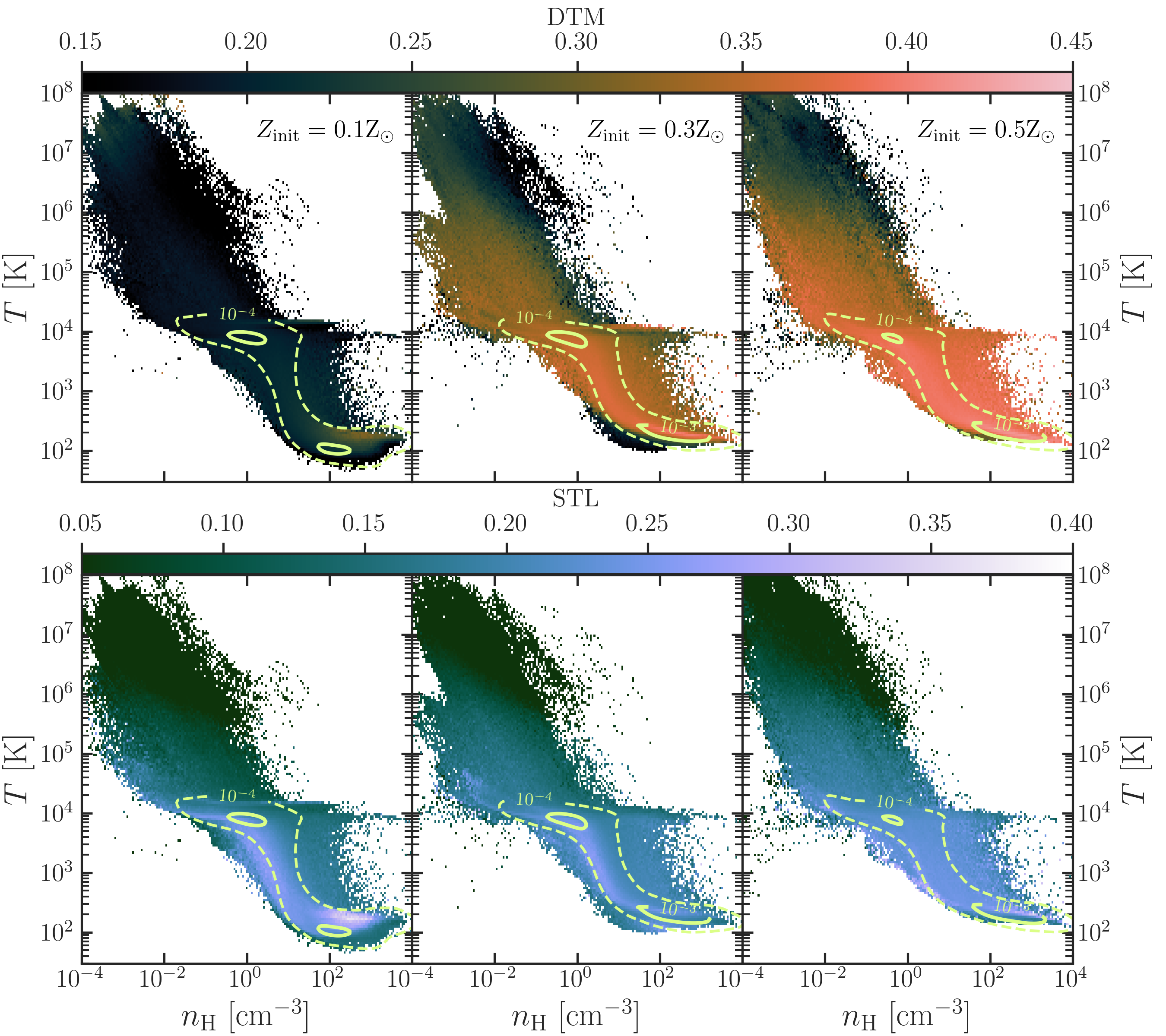}
    \caption{Mass-weighted DTM (top row) and STL (bottom row) distributions for the temperature-density phase diagrams of the G8 simulations using \calima. These are computed by stacking all outputs between 100-200\,Myr. Yellow contours indicate bins in $T$--$n_{\rm H}$ space that contribute $10^{-4}$ (dashed) and $10^{-3}$ (solid) to the total gas mass budget. Overall DTM increase is a stiff function of $Z_{\rm init}$, with $Z_{\rm init}=0.1\,Z_{\odot}$ showing a significantly lower DTM compared to the higher metallicity runs. All simulations show an overall lower DTM for $T\gtrsim 10^4\,\rm K$, with effective destruction of dust taking place for $T\gtrsim 10^6\,\rm K$. Small grains do not contribute for $T\gtrsim 10^6\,\rm K$ and appear to show a larger contribution to the dust budget in a band following the shape of the mass distribution. Maximum values of STL decrease with $Z_{\rm init}$, but the median value over the warm, diffuse ISM increases with $Z_{\rm init}$.} 
    \label{fig:DTM_STL_vs_Trho}
\end{figure*}
In Fig.~\ref{fig:DTM_STL_vs_Trho} we show temperature-density phase diagrams of the \calima~simulations with (from left to right) $Z_{\rm init}=0.1,0.3$, and $0.5\,Z_{\odot}$. These have been computed by stacking all outputs between $100\mbox{--}200$\,Myr, for a cylindrical volume centred on the G8 galaxy with radius of 1.5\,kpc and height above and below the disc of 2\,kpc. In the top row each bin in $T$--$n_{\rm H}$ space is colour-coded with the mass-weighted median DTM. Median values of DTM show a stiff dependence on $Z_{\rm init}$, with effective gas-phase accretion turning on for $Z_{\rm init}>0.1\,Z_{\odot}$, in agreement with the results from \citet{Dubois2024GalaxiesSimulations} and observations \citep[e.g.][]{Remy-Ruyer2014Gas-to-dustRange,DeVis2019ARatios,Konstantopoulou2022DustISM}. In the lowest metallicity G8 simulation we can appreciate that accretion only appears to significantly modify the median DTM for $n_{\rm H}\gtrsim 10^2\,\rm cm^{-3}$. We note that the cold phase shows a significantly broader distribution in temperature, with the lowest DTM gas capable of reaching lower gas temperatures. To aid in this analysis, we have also included contour maps showing bins in $T$-$n_{\rm H}$ space that contribute $10^{-4}$ (dashed yellow) and $10^{-3}$ (solid yellow) to the total gas mass budget. This feature of low-DTM, low-$T$ gas appears to be increasingly erased as the gas-phase metallicity is increased while maintaining the same properties of the G8 galaxy (i.e.~gas fraction, stellar mass, SFR). We will further explore in Section~\ref{subsubsec:heating_and_cooling} how the live dust modelling in \calima~directly affects the equilibrium gas distribution, particularly the cold ISM. All metallicities show an overall lower DTM for $T\gtrsim 10^4\,\rm K$, with clear signatures of thermal sputtering decreasing further the DTM for $T\gtrsim 10^6\,\rm K$. However, the low metallicity run shows a consistently higher DTM at the very tip of the hot gas phase. We argue that this is the freshly injected dust from SN. 

The growth, destruction and evolution of dust grains are highly dependent on their size, so we expect important variations in STL across different gas phases. The bottom row of Fig.~\ref{fig:DTM_STL_vs_Trho} shows the same stacked histograms but colour-coded with STL. Small grains do not survive for long for $T\gtrsim 10^6\,\rm K$, and the choice of large grain-dominated SN ejecta means that we do not see freshly injected small grains within SN-heated gas. For neutral gas, we find that the STL approaches overall values close to the ones used in the fixed dust simulations ($\sim 0.2$), but there also regions (shown as white pixels) that also indicate enhanced small grain abundance. The artificially enhanced metallicity of $0.5\, Z_{\odot}$ (compared to the mass-metallicity relation) for a galaxy as low-mass as G8 results in a warm neutral and cold ISM with particularly high STL, but in the 0.1 and $0.3\, Z_{\odot}$ simulations the high STL appears as band that follows the shape of the gas mass distribution (i.e.~dashed and solid yellow contours). The low density warm ISM shows the effects of small grain production in the turbulent shattering of large grains, with efficient gas-phase accretion on small grains taking over at higher gas densities (see the high STL cloud for $n_{\rm H}\gtrsim 10^2\,\rm cm^{-3}$ in the $Z_{\rm init}=0.1\,Z_{\odot}$). We note that the cold gas that can cool well below 100\,K shows low small mass fraction at high $n_{\rm H}$, directly linking the growth of dust mass at high density to the minimum gas temperature of the cold phase.

\begin{figure}
    \centering
	\includegraphics[width=\columnwidth]{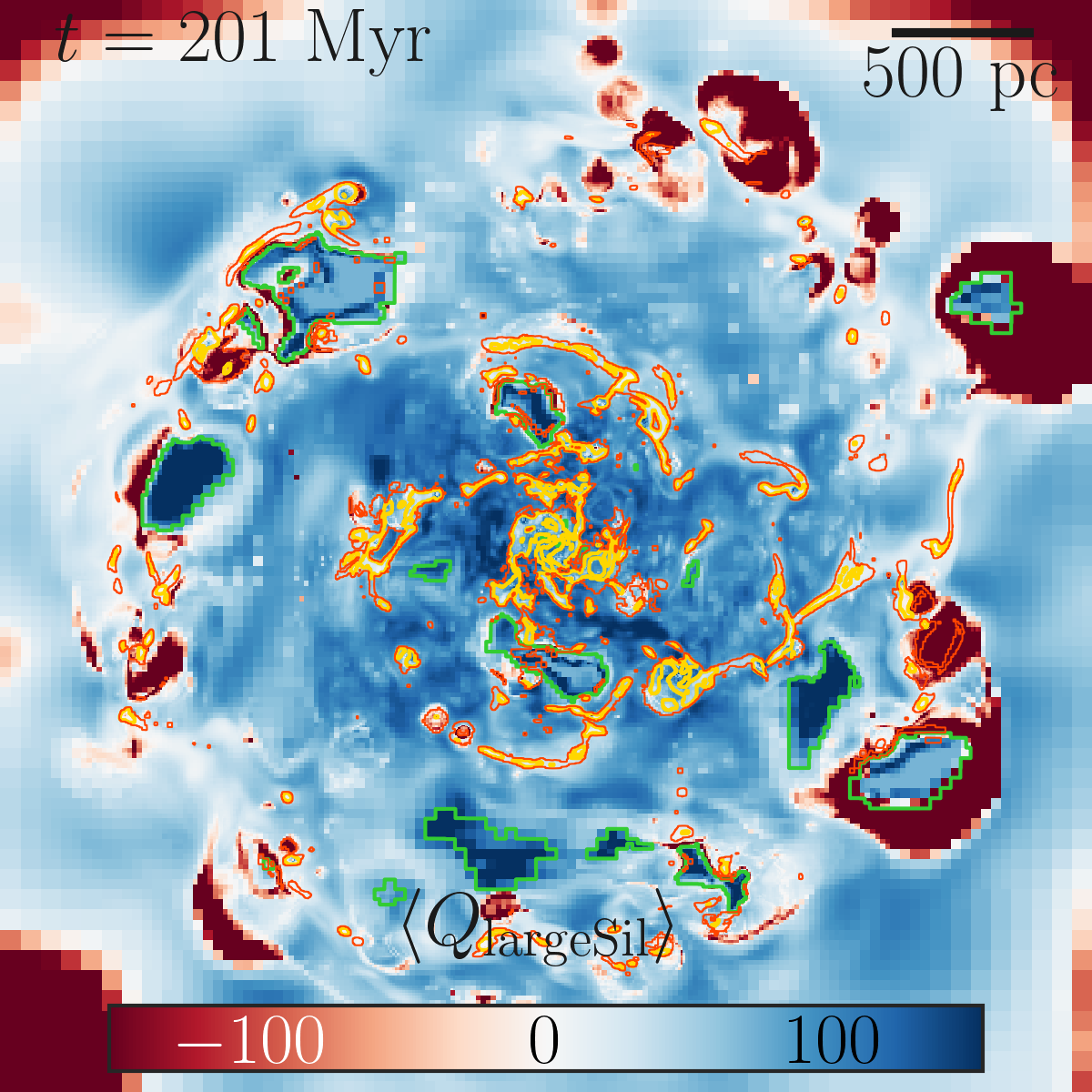}
    \caption{Same projection as Fig.~\ref{fig:G8_03_projections_grid} but showing the density-weighted mean largeSil grain charge. Green line contours indicate ionised gas ($x_{\mathrm{H}\,\scriptstyle\mathrm{II}}\geq 0.99$), yellow lines gas with $n_{\rm H}\geq 10^2\,\rm cm^{-3}$, and red lines dust collisional cooling rates exceeding $3\times 10^{-27}\,\rm erg\,s^{-1}$. The inset shows a detailed view of a young H~{\small II} region in which the transition from positively to negatively charge grains can be seen across the ionisation front, leading to enhanced dust collisional cooling.}
    \label{fig:G8_03_projection_SilCharge}
\end{figure}

A unique capability of \calima~is its self-consistent calculation of grain charging in every simulation cell. Fig.~\ref{fig:G8_03_projection_SilCharge} presents the mean charge for largeSil grains\footnote{We do not show the mean charge for smallSil and the carbonaceous grains as the qualitative behaviour is similar to largeSil with only changes in relative charging.} in the G8 simulation with $Z_{\rm init}=0.3\,Z_{\odot}$ at 200\,Myr. The grain charge distribution is a result of the balance between photo-ejection and collisions with free electrons (see Section~\ref{subsec:charging}). Across most of the ISM, grains remain positively charged, though they approach neutrality in dense, shielded regions (highlighted by yellow contours, $n_{\rm H}\geq 10^2\,\rm cm^{-3}$). Inside H~{\small II} regions (green contours indicate $x_{\rm HII}\geq 0.99$), grains become highly positively charged due to the intense radiation field (high charging parameter $\gamma$). Notably, we observe a sharp transition to negative charges at the ionization fronts (see dark red cells in the inset). This is in agreement with our predictions for high $T\gtrsim 10^4\,\rm K$ and low $\gamma$ (Fig.~\ref{fig:dust_charge_silicate_0.005micron}): immediately outside the H~{\small II} region, but lower $G_0$–reducing photoelectric emission–while the free electron density remains sufficient to drive the equilibrium distribution toward the negative limit. 
As we have discussed, grain charging has an important effect in the evolution of dust grains as well as in their interaction with the ISM. For example, by modifying the effective cross-section of the grain due to Coulomb enhancement (see e.g.~Section~\ref{subsec:accretion}) ion collision rates with grains increase. This can affect, alongside other processes, the efficiency of collisional dust cooling (Section~\ref{subsec:collisional_cooling}). In Fig.~\ref{fig:G8_03_projection_SilCharge} we have also added red contour lines representing an excess of collisional cooling rate. The majority of these dust-cooled cells are co-spatial with the previously mentioned high gas density contours (dashed yellow lines). However, some of them are co-spatial with gas cells dominated by negatively charge grains, easily distinguished in the H~{\small II} region within the inset plot. We will further discuss the role of dust collisional cooling in Section~\ref{subsubsec:heating_and_cooling}.

\begin{figure*}
    \centering
	\includegraphics[width=\textwidth]{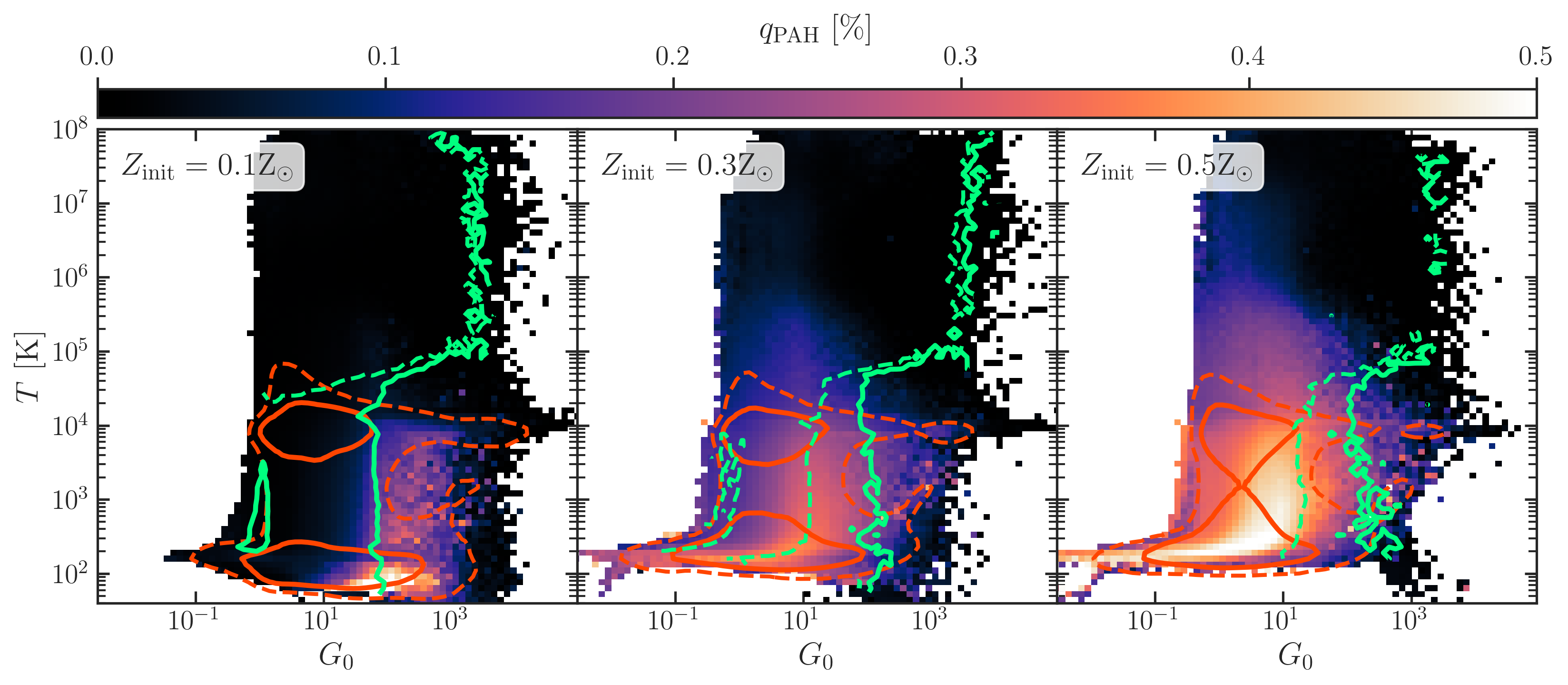}
    \caption{Mass-weighted distribution of $q_{\rm PAH}$ for different gas temperatures and Habing band strengths $G_0$. This is a result of stacking all snapshots for the G8 simulations with \calima~between $100\mbox{--}200$\,Myr, including all gas within a cylinder centred on the galaxy with radius 1.5\,kpc and height below and above the disc of 2\,kpc. Red contour lines indicate the regions where individual $T$--$G_0$ bins contribute $10^{-4}$ (dashed) and $10^{-3}$ (solid) to the total gas mass budget in the cylinder. Lime open contours show regions with $\mathrm{STL}_{\rm PAH}\geq 0.2$ (right side of dashed lines) and with $\mathrm{STL}_{\rm PAH}\geq 1$ (right side of dashed lines). PAHs are more abundant for highly shielded and cold gas, quickly becoming destroyed for $G_0\gtrsim 10^2$ and $T\gtrsim 10^4\,\rm K$. In the lowest metallicity galaxy (left panel) PAHs are predominantly present next to cool, irradiated gas, resulting from destruction processes of carbonaceous dust grains. Large PAHs dominate for $T\gtrsim 10^5\,\rm K$ for all metallicities and $G_0$, becoming important in the diffuse gas at lower $G_0$ for lower metallicities. Large PAHs dominate the cold, shielded gas. Small PAHs are more important in the low $T$, high $G_0$ gas.}
    \label{fig:TG0_qpah}
\end{figure*}
Given the contentious nature of PAH formation from the gas phase (see the discussion in Section~\ref{subsec:pah_accretion}), a careful examination of the PAH abundance with respect to other dust mass content (i.e.~$q_{\rm PAH}$) is a subject that we will explore extensively in forthcoming work. Nonetheless, we provide preliminary analysis of $q_{\rm PAH}$ for the three G8 simulations in Fig.~\ref{fig:TG0_qpah}. We have computed these by stacking the results of all snapshots between $100\mbox{--}200$\,Myr, considering all gas cells inside a cylinder co-planar with the disc of radius 1.5\,kpc and height of 2\,kpc above and below the disc plane. To guide the eye, we also include solid red contours indicating the bins in $T$--$G_0$ that contribute $10^{-4}$ and $10^{-3}$ to the total gas mass budget. For the $Z_{\rm init}=0.3,0.5Z_{\odot}$ simulations (middle and right panels) the formation of PAH via C$^+$ accretion can clearly be distinguished as the horizontal region at $T\sim 100\,\rm K$ extending towards low values of $G_0$. In the modelling used for C$^+$ accretion, this process only becomes slightly efficient for anion and neutral PAHs, which requires highly shielded gas \citep[as in e.g. TMC-1,][where $A_V\gtrsim 4$]{Chen2022ChemicalPhase}. We also find that for these simulations, the values of $q_{\rm PAH}\gtrsim 0.3\%$ are not limited to the cold ISM, but also extend to the warm neutral gas. This is a signature of efficient production of PAHs via shattering of dust grains in the turbulent warm ISM (Section~\ref{subsec:shattering}). However, there is still a clear distinction in PAH size between these two phases. We have included contours indicating the regions where the mass-weighted STL$_{\rm PAH}$ exceeds $0.2$ (dashed lime lines) and $1.0$ (solid lime lines). It shows that for these metallicities, the low $G_0$ and low $T$ phases are dominated by largePAHs (i.e.~to the left of the dashed lines), while there is a dominant contribution of smallPAHs at cold and warm phases with $G_0\gtrsim 10^2$. This overabundance of smallPAHs in this ISM phase is caused by the transfer of mass from largePAHs to smallPAHs via cluster evaporation (Section~\ref{subsec:pah_evaporation}). We note that this prediction is in direct contradiction with the suggested dominance of large PAHs close to ionising sources \citep[e.g.][]{Knight2021TracingEnvironments,Baron2025PHANGS-ML:Galaxies,Chown2024PDRs4All:Bar}. Nevertheless, we still recover the efficient destruction of PAHs in ionised gas ($T\gtrsim 10^4\,\rm K$) with $G_0\gtrsim 10\mbox{--}100$, as suggested by the decrease of PAH mid-IR emission to total IR emission within H~{\small II} regions \citep[e.g.][]{Chastenet2023PHANGS-JWSTMetallicity,Egorov2023PHANGS-JWSTMUSE,Egorov2025PolycyclicGalaxies}.

In the low-metallicity simulation ($Z_{\rm init}=0.1\,Z_{\odot}$), $q_{\rm PAH}$ growth via accretion is strongly suppressed by the reduced abundance of gas-phase C$^+$. This strong dependence on metallicity for the growth of PAHs is in agreement with observationally inferred PAH abundances in the local \citep[e.g.][]{Draine2007DustSample,Galliano2008StellarGalaxies,Remy-Ruyer2015LinkingPicture,Whitcomb2024TheSpectroscopy} and high-redshift Universe \citep[e.g.][]{Shivaei20172,Shivaei2024AMIRI}. Consequently, the majority of PAHs reside in the diffuse, irradiated ISM, identifying the shattering of carbonaceous grains as the dominant mechanism driving PAH evolution in low-metallicity environments. Given also the low abundance of smallPAHs, clustering is not sufficiently efficient, rendering the phases with larger $q_{\rm PAH}$ dominated by smallPAHs. 

\begin{figure}
    \centering
	\includegraphics[width=\columnwidth]{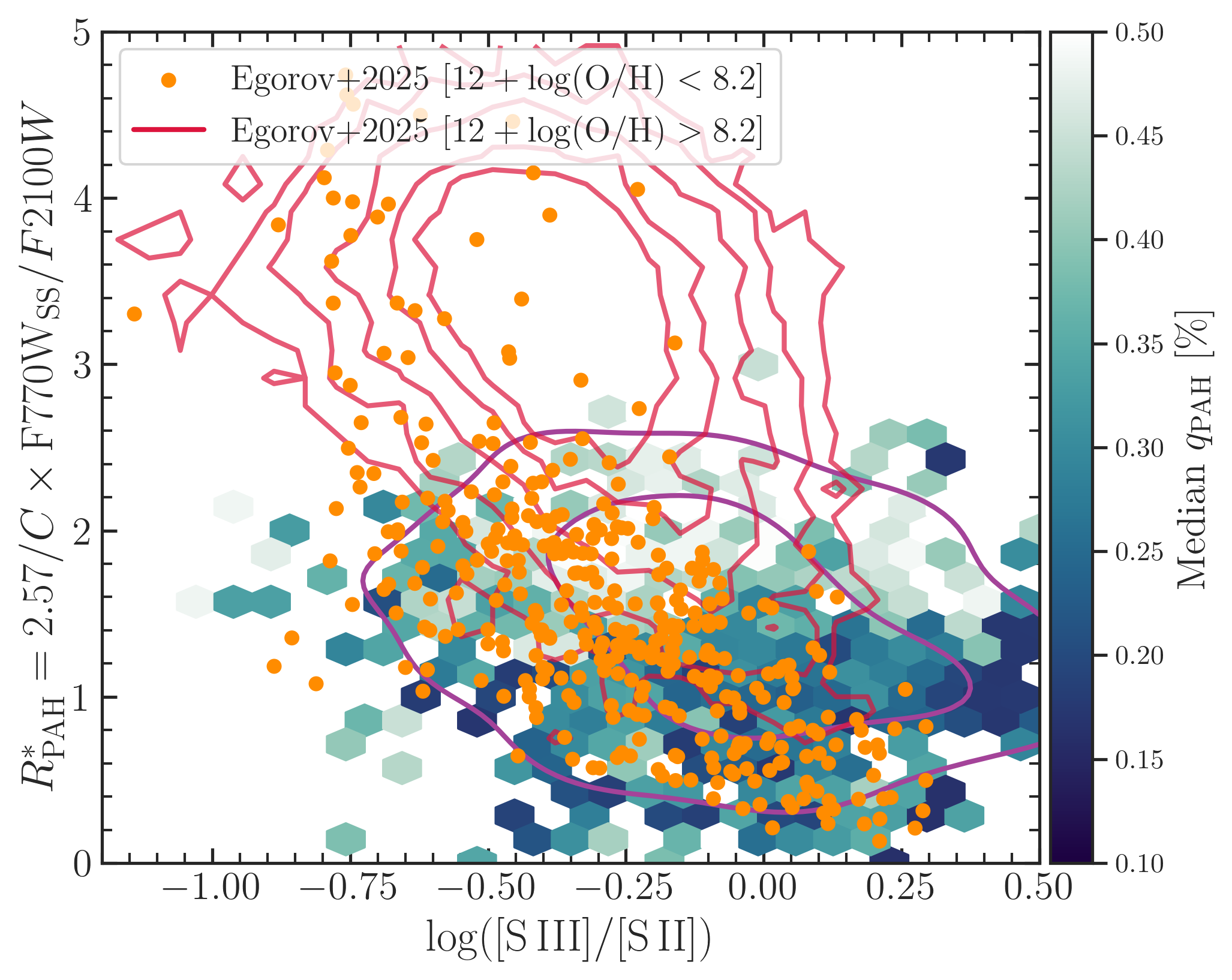}
    \caption{Histogram of an observational proxy of the PAH mass fraction, $R^*_{\rm PAH}$ \citep[][]{Egorov2025PolycyclicGalaxies}, against an observational proxy of the ionisation parameter, the $[\mathrm{S}\,\scriptstyle\mathrm{III}]\lambda9069, 9532$\AA/$[\mathrm{S}\,\scriptstyle\mathrm{II}]\lambda 6717, 6731$\AA line ratio. Hexagonal bins show the location of H~{\small II} regions in the G8 simulations using \calima, colour coded by the median $q_{\rm PAH}$ in each bin. Magenta contours indicate the 70th and 90th percentile of the pixel distribution. We include the observations of individual H~{\small II} regions and SNR from the PHANGS survey extracted in \citet{Egorov2025PolycyclicGalaxies}, separated in their low metallicity bin ($12+\log(\rm O/H)<8.2$, orange points) and high metallicity bin ($12+\log(\rm O/H)>8.2$, red contours). The distribution of H~{\small II} regions in the G8 simulations agree with the low metallicity bin, but there is no clear trend between $R^*_{\rm PAH}$ and $q_{\rm PAH}$.}
    \label{fig:Rpah_S32}
\end{figure}
We further explore the evolution of PAHs from the diffuse to the high ionisation medium using the unique capabilities of \calima. Observationally, the relative amount of PAH mass to dust mass (i.e. $q_{\rm PAH}$ is usually computed by comparing the IR luminosity in the MIR to the total IR luminosity \citep[e.g.][]{Engelbracht2005MetallicityGalaxies,Sandstrom2010TheHydrocarbons,Chastenet2023PHANGS-JWSTMetallicity,Sutter2024TheGalaxies}. In order to compute the IR luminosity of each region of our G8 simulations, we have performed mock data cubes using the Monte-Carlo radiative transfer code {\sc skirt} \citep{Camps2015SKIRT:Architecture,Baes2022MonteEmission}. These have been obtained for the three metallicity simulations at 200 Myr, obtaining face-on data cubes with a resolution of $\sim 4\,\rm pc$ per pixel and the full spectrum from 0.09 \micron~to 1000 \micron. Stellar sources for radiation are taken from the stars formed during the simulation runtime and sourcing photon packets from the BPASS v2.2.1 SED \citep{Stanway2018Re-evaluatingPopulations} using the same settings as we have used in the simulations and considering their age, metallicity and initial mass. Absorption of UV-optical light is taken into account using the predicted abundances and properties of each dust and PAH bin for each simulation cell, following the optical properties and underlying GSD for each bin used in \calima. We compute an observed proxy of $q_{\rm PAH}$ as defined by the parametrisation \citep{Egorov2025PolycyclicGalaxies}:
\begin{align}\label{eq:RPAH_egorov25}
    R^*_{\rm PAH} = \frac{2.57\times \,\rm F770W_{\rm SS}}{C \times\,\rm F2100W},
\end{align}
where F2100W is the flux density in the JWST F2100W filter, $\rm F770W_{\rm SS}=F770W-0.22\times F300M$ is the starlight-substracted correction of F770W flux density using the F300M filter \citep{Sutter2024TheGalaxies}, and $C = -8.32\times10^{-3} \log(L_\nu(\rm F770W_{\rm SS}))^2 + 0.4881\log(L_\nu(\rm F770W_{\rm SS}))-6.01$ is a correction to account for the missing information from the F1130W filter. In \citet{Egorov2025PolycyclicGalaxies} the H~{\small II} region ionisation parameter is measured using the proxy given by the $[\mathrm{S}\,\scriptstyle\mathrm{III}]\lambda9069, 9532$\AA/$[\mathrm{S}\,\scriptstyle\mathrm{II}]\lambda 6717, 6731$\AA line ratio. These can be directly obtained since \ramsesrtz~tracks the non-equilibrium abundances of individual ion populations, allowing trivial face-on projections (also at $\sim 4\,\rm pc$ resolution) of the luminosity between the $[\mathrm{S}\,\scriptstyle\mathrm{III}]$ and $[\mathrm{S}\,\scriptstyle\mathrm{II}]$ lines. In practice, the $[\mathrm{S}\,\scriptstyle\mathrm{III}]\lambda 9532$\AA line lies beyond the MUSE wavelength range used in \citet{Egorov2025PolycyclicGalaxies}, and instead they use the theoretical ratio $[\mathrm{S}\,\scriptstyle\mathrm{III}]\lambda 9532$\AA $=2.5\times [\mathrm{S}\,\scriptstyle\mathrm{III}]\lambda 9069$\AA. We apply the same strategy for these mock observational analysis. We note that our comparison is not fully on the same grounds as the original observational sample: we do not introduce instrumental noise, nor convolve our images to the spatial resolution of the observations performed with JWST and MUSE. Furthermore, we do not perform circular integration around each H~{\small II} region, and instead obtain $R^*_{\rm PAH}$ and $\log([\mathrm{S}\,\scriptstyle\mathrm{III}]/[\mathrm{S}\,\scriptstyle\mathrm{II}])$ on a per-pixel basis. However, we limit our analysis to pixels that have a sufficiently high $[\mathrm{S}\,\scriptstyle\mathrm{III}]\lambda 9069$\AA luminosity compared to H$\alpha$ (i.e. $L([\mathrm{S}\,\scriptstyle\mathrm{III}]\lambda 9069$\AA)/$L(\rm H\alpha)\geq 0.3$).

In the G8 simulations, individual pixels span a range of $\log([\mathrm{S}\,\scriptstyle\mathrm{III}]/[\mathrm{S}\,\scriptstyle\mathrm{II}])$ and $R^*_{\rm PAH}$ comparable to the low-metallicity ($12+\log(\rm O/H)<8.2$) H~{\small II} regions and SNRs studied by \citet{Egorov2025PolycyclicGalaxies}. Given the low metallicity of the simulations explored in this work, we do not reproduce the observed $R^*_{\rm PAH}$ for their high metallicity sample ($12+\log(\rm O/H)>8.2$, red contour lines), although a subset of pixels begins to populate that region. Despite the broadly consistent trend of lower $R^*_{\rm PAH}$ for higher $\log([\mathrm{S}\,\scriptstyle\mathrm{III}]/[\mathrm{S}\,\scriptstyle\mathrm{II}])$, there are notable differences from the observed data. For $\log([\mathrm{S}\,\scriptstyle\mathrm{III}]/[\mathrm{S}\,\scriptstyle\mathrm{II}])\gtrsim 0$, the simulations predict systematically higher $R^*_{\rm PAH}$ compared to the low metallicity distribution, although these have preferentially lower $q_{\rm PAH}$. These results combined into a shallower distribution, as shown by the 70th and 90th percentile contours (magenta lines). This could be another indicative of slow PAH destruction in high ionisation environments, as we have mentioned in the analysis of Fig.~\ref{fig:TG0_qpah}. We have colour coded the bins in this histogram with the median $q_{\rm PAH}$, which appear to show a preference of higher $q_{\rm PAH}$ for higher $R_{\rm PAH}$ and lower $\log([\mathrm{S}\,\scriptstyle\mathrm{III}]/[\mathrm{S}\,\scriptstyle\mathrm{II}])$. However, the limited number of H~{\small II} regions in just three simulations does not provide a reliable statistics, making a correlation between $R_{\rm PAH}$ and $q_{\rm PAH}$. In addition, it appears that while the simulations indicated $q_{\rm PAH}$ values exceeding 0.5\%, the gas cells that are sampled by these observations do not appear to be a direct measure of the median $q_{\rm PAH}$ in the G8 galaxies. We emphasise that these mock observations have not been fully tailored for a one-to-one comparison with \citet{Egorov2025PolycyclicGalaxies}, but they already suggest that the relationship between observational PAH mass parametrisations and the underlying PAH mass budget is non-trivial and warrants further investigation \citep[e.g.][]{Sandstrom2010TheHydrocarbons,Sutter2024TheGalaxies}. This comparison also illustrates the unique capability of \calima~to link theoretical models of dust and PAH evolution to direct observables.

\subsubsection{Radiation transfer with evolving dust properties}\label{subsubsec:rt_with_evol_dust}

\begin{figure*}
    \centering
	\includegraphics[width=\textwidth]{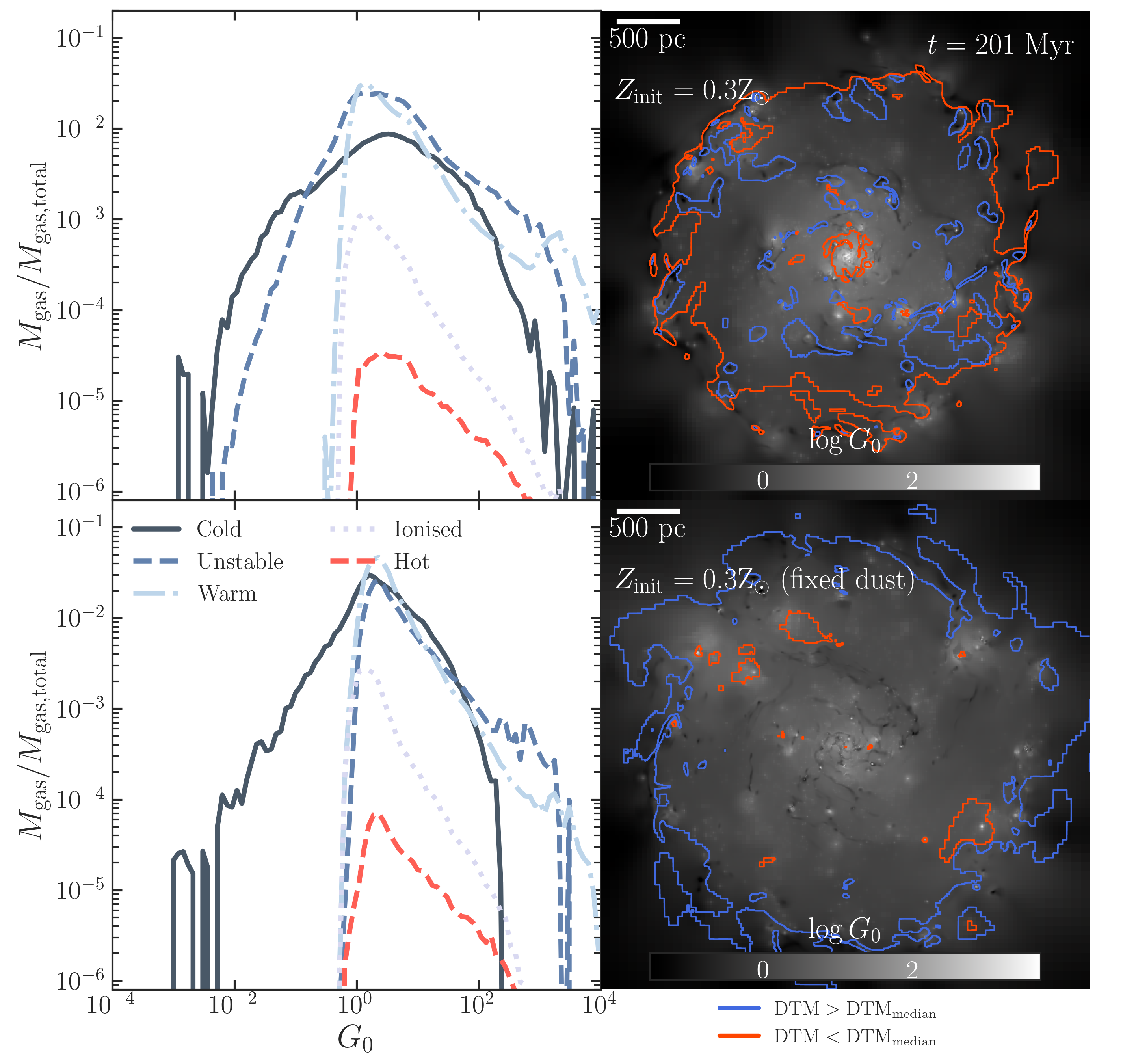}
    \caption{Comparison of $G_0$ distribution for the G8 galaxy with $Z_{\rm init}=0.3\,Z_{\odot}$ in the case of using \calima~or a fixed dust evolution. Right panels are mass-weighted projections of $G_0$ with a width of 4\,kpc and a depth of 0.8\,kpc. To the left of each projection, we include the fraction of gas mass contributed by each bin to the total gas mass budget. We have also separated the gas mass within five different phases, using the natural breaks in the gas temperature distribution. Blue (red) contour lines indicate regions in which the local DTM exceeds (lags) the mass-weighted median DTM of each simulation. While with a fixed dust the distribution of $G_0$ is highly clustered around $\sim 1$, \calima~predicts broader distributions, with shifted peaks for different phases.}
    \label{fig:G8_03_compare_G0}
\end{figure*}
As we saw in Fig.~\ref{fig:TG0_qpah}, the value of $G_0$ can vary by many orders of magnitude at a given gas temperature. In our equilibrium tests (Section~\ref{subsec:eq_tests}) we assumed a fixed value of $G_0=1$, therefore affecting the relative role of different heating mechanisms (i.e.~PEH, \hmol~pumping, etc.) and the abundance of molecules and PAHs. In our G8 simulations the value of $G_0$ is instead the result of the radiation emitted by stars formed and how it has been propagated through the ISM by \ramsesrt. Fig.~\ref{fig:G8_03_compare_G0} shows the distribution of $G_0$ in mass-weighted projections of the G8 simulations with $Z_{\rm init}=0.3\, Z_{\odot}$. We begin by examining the properties of the radiation field in the fixed dust simulations (right panels). We have calculated the mass-weighted distributions of $G_0$ for the separation of the ISM phases suggested by \citet{Kim2017Three-phaseConvergence} based on the natural breaks in the temperature distribution of the $Z_{\odot}$ ISM: 1) cold ($T< 184\,\rm K$), 2) unstable ($184\,\rm K\,<T<5050\,\rm K$), 3) warm ($5050\,\rm K\,<T<2\times 10^4\,\rm K$), 4) ionised ($2\times 10^4\,\rm K < T < 5\times 10^5\,\rm K$), and 5) hot ($T>5\times 10^5\,\rm K$). The projection shows an average $G_0\sim 1$ in the diffuse ISM, as supported by the location of the distribution peak. This projection also shows very localised variations of $G_0$, attributed to high density, shielded clumps and filaments that have not formed stars yet, and the ones that are actively star-forming. This is the source of the broadness of the cold phase distribution in $G_0$, spanning more than four orders of magnitude. The distribution of the rest of phases also show a significant broadness in $G_0$, but they quickly drop for $G_0\lesssim 1$. The similarity in the shape of these distributions for $G_0>1$ is a clear signature of the non-evolving dust abundances and properties of the fixed dust model.

The left panels of Fig.~\ref{fig:G8_03_compare_G0} compares the results of the fixed dust model to the G8 simulation using \calima. We note that both simulations have very similar star-formation rates\footnote{All G8 simulations show star-formation rates, averaged over 10\,Myr, in the range of $\sim 0.02-0.04\,\rm M_{\odot}\, yr^{-1}$.}, so the comparison in $G_0$ distributions can be considered to probe the transparency of the ISM between the \calima~simulation and the fixed dust model. Already a qualitative comparison of the two projections indicate significant differences between the two models: young star-forming regions have much larger $G_0$, while the obscuration of clumps and filaments appears to stand out even more against the diffuse background. This is clearly supported by the mass-fraction distribution of $G_0$ for the cold phase, showing a greater contribution of the values of $G_0$ away from $G_0\sim 1$. This is a direct result of the varying DTM and STL across ISM phases allowed by \calima. In both projections we have included blue (red) contours indicating regions in which the local DTM exceeds (lags) the median DTM across the disc. It is clear that large dust destructions in the inner region of the disc (recall the DTM cavity observed in Fig.~\ref{fig:G8_03_projections_grid}) gives rise to less obscure young star formation, while the growth of dust via accretion gives rise to regions of enhanced DTM and therefore lower $G_0$. This result even extends to the unstable phase, with a tail towards $G_0\sim 10^{-2}$ which was not present in the fixed dust model. For the rest of phases, the distribution still shows a hard cut-off for $G_0\sim 1$, but with broader peaks and larger contribution of cells with $G_0\gtrsim 10^2$. These results clearly demonstrates the non-trivial properties of the ISRF with on-the-fly radiative transfer and self-consistent modelling of dust evolution.

\subsubsection{Influence of dust on the molecular and neutral ISM}\label{subsubsec:molecules_with_dust}

\begin{figure}
    \centering
	\includegraphics[width=\columnwidth]{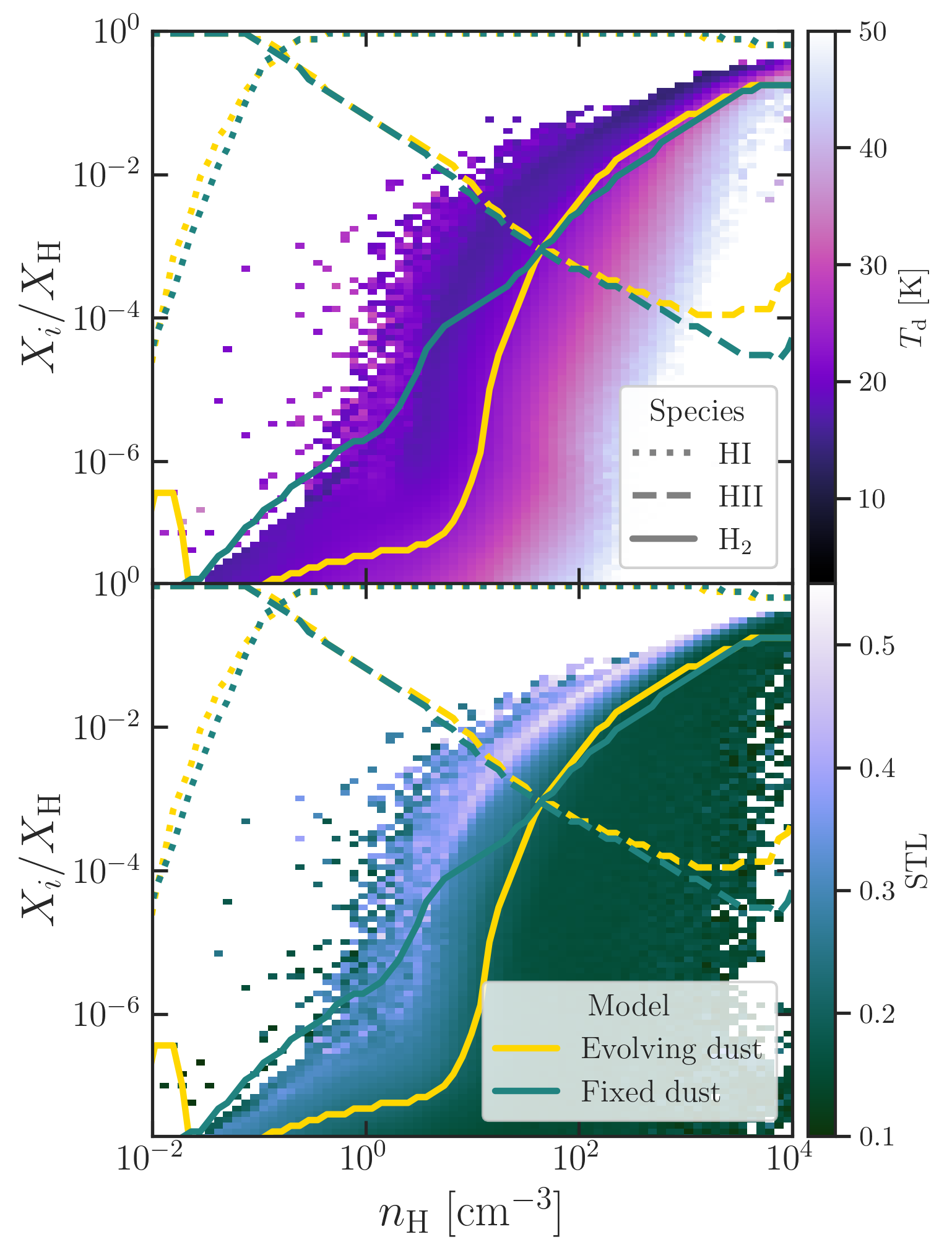}
    \caption{Mass-weighted histogram of the fraction of hydrogen in \hmol~at different $n_{\rm H}$ for all outputs between $100\mbox{--}200$\,Myr of the G8 simulation with $Z_{\rm init}=0.1\,Z_\odot$. The dashed, dotted, and solid lines show the median HI, H~{\small II} and \hmol, respectively. In the top panel we show the mass-weighted dust equilibrium temperature, and in the bottom panel the mass-weighted STL. For comparison, we also include the median HI, H~{\small II} and \hmol~for the equivalent G8 simulation with fixed DTM. \calima~predicts a lower (higher) efficiency of \hmol~formation for $n_{\rm H}\lesssim 10\,\rm cm^{-3}$ ($n_{\rm H}\gtrsim 10\,\rm cm^{-3}$) than with the fixed DTM model. Cells that at a fixed $n_{\rm H}$ have a higher \hmol~fraction are dominated by cold dust ($\lesssim 20\,\rm K$) and a larger fraction of small grains $\mathrm{STL}\gtrsim 0.3$.}
    \label{fig:xH_0.1Zsun}
\end{figure}
A direct consequence of variable dust content, grain properties, and $G_0$ is a variable efficiency of \hmol~formation in the ISM of the G8 galaxy. We explore the abundance of \hmol~with respect to total hydrogen mass at different $n_{\rm H}$ in Fig.~\ref{fig:xH_0.1Zsun} by computing 2D mass-weighted histograms for the G8 simulation with $Z_{\rm init}=0.1\,Z_{\odot}$ and using \calima. These are a stack of all simulation snapshots between $100\mbox{--}200$\,Myr, using only cells within a cylindrical volume centred in G8 with radius 1.5\,kpc and height above and below the disc of 2\,kpc. The median of the distributions for H~{\small I} (dotted), H~{\small II} (dashed) and \hmol~(solid) are given by the green lines included over the 2D histograms of \hmol. We have colour coded the \hmol~distribution with mass-weighted median dust temperature (top panel) and STL (bottom panel). We can see that the median mass distribution goes from being dominated by H~{\small II} to HI at $n_{\rm H}\gtrsim 0.1\,\rm cm^{-3}$. Molecular hydrogen is negligible for $n_{\rm H}\lesssim 10\,\rm cm^{-3}$, steeply growing in mass fraction around this density and then developing a shallower slope with $n_{\rm H}$. We compare the behaviour of the \calima~simulation to the one with fixed dust (green lines). The largest difference with the \calima~simulation is the evolution of the median \hmol~mass fraction with $n_{\rm H}$: the fixed dust model predicts much higher (up to three dex) molecular fraction for $n_{\rm H}\lesssim 10\,\rm cm^{-2}$, while above this density \calima~predicts consistently larger \hmol~fractions. In the equilibrium tests performed in Section~\ref{subsec:eq_tests} we also drew attention to this enhanced \hmol~fraction using \calima~compared to a fixed dust model. However, with higher DTM in the cold phase and in the presence of self-shielding, the allowed \hmol~fraction exceeds the predictions of the equilibrium model \citep[see also][for similar conclusions]{Katz2022PRISM:Galaxies}.

\begin{figure}
    \centering
	\includegraphics[width=\columnwidth]{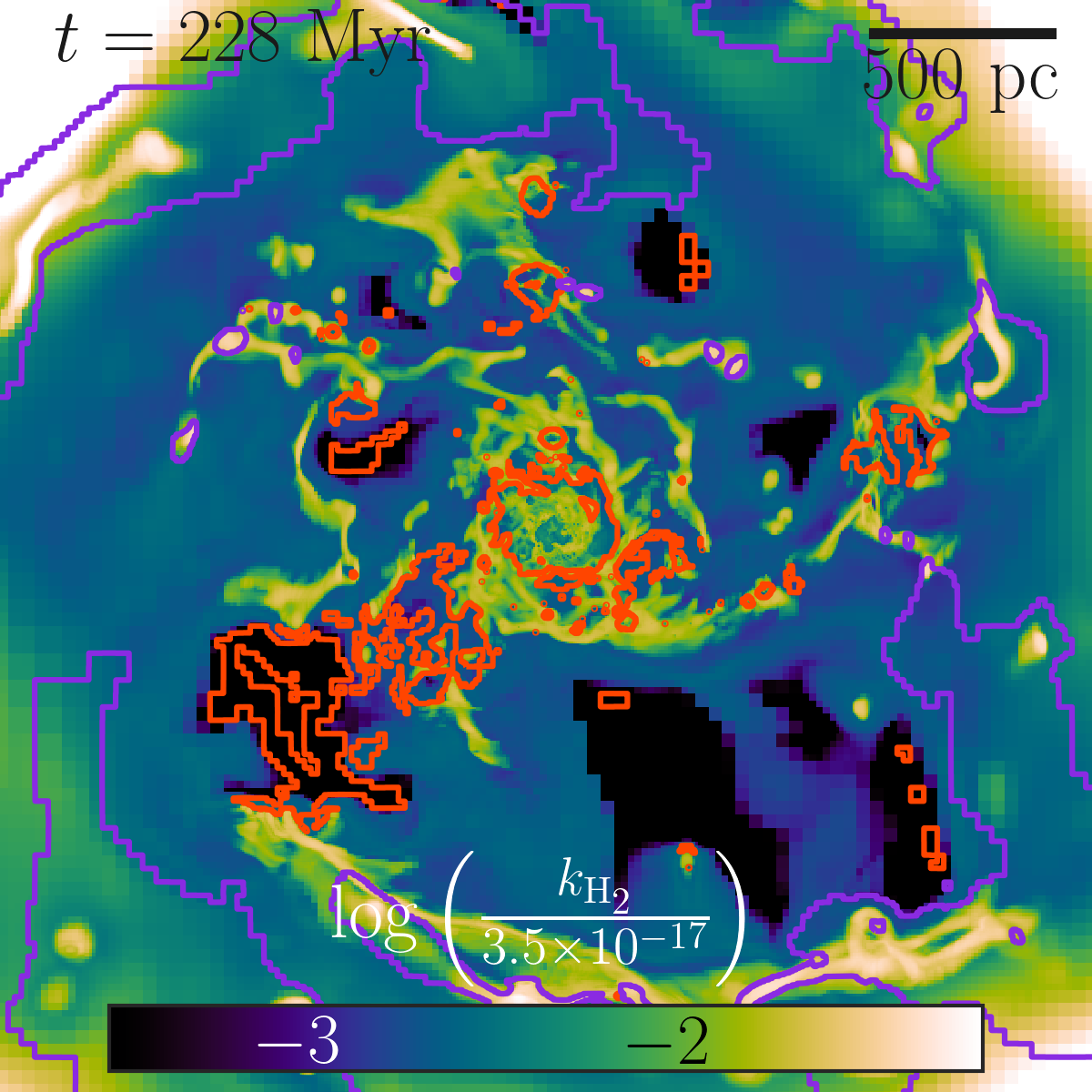}
    \caption{Mass-weighted projection of the \hmol~formation coefficient $k_{\rm H_2}$ for the G8 simulation with $Z_{\rm init}=0.1\,Z_\odot$ at $t=228\,\rm Myr$. The rates have been normalised to the canonical value of $3.5\times10^{-17}\,\rm cm^3 s^{-1}$ used in thermochemistry models including \hmol~formation on dust grains \citep[e.g.][]{Gnedin2009ModelingSimulations,Bialy2019ThermalGas}. Red contours contain gas cells with dust temperature exceeding 30\,K, while purple contours show regions where STL is larger than 0.35. Our predicted $k_{\rm H_2}$ only approaches the the canonical for the highly shielded, dense clumps and filaments (inside purple contours), with the majority of the gas showing more than 2\,dex slower \hmol~formation, particularly where hot dust lives (i.e.~inside red contours).}
    \label{fig:H2form_0.1Zsun}
\end{figure}

Molecular hydrogen formation on dust grains is considered to depend on the composition, size and temperature of the grain \citep[][]{Cazaux2002MolecularMedium,Cazaux2004HSUB2/SUBSurfaces}. By linking the dust properties to the \hmol~formation, we are capable of directly predicting how an evolving dust distribution affects the \hmol~fraction. The mass-weighted dust temperature shows that the median of the \hmol~forming gas has dust with $T_{\rm d}\sim 20\,\rm K$ at low densities, and $T_{\rm d}\lesssim 15\,\rm K$ at higher gas densities. At a fixed density, the molecular gas fraction is larger in cells that both show a lower $T_{\rm d}$ and higher STL. This effect of higher efficiency caused by larger mass fraction of small grains, is particularly clear for densities around the critical density for \hmol~formation ($n_{\rm H}\sim 1\mbox{--}10\,\rm cm^{-3}$). The reaction rate for \hmol~formation in ISM models is usually set by \citep[e.g.][]{Gnedin2009ModelingSimulations,Bialy2019ThermalGas}:
\begin{align}\label{eq:h2form_prism}
    R_d = k_{\rm H_2} f_{\rm DTG} C \sqrt{\frac{T}{100\,\rm K}} \quad[\rm cm^3s^{-1}],
\end{align}
where, $f_{\rm DTG}$ is the ratio of the local DTG to the mean value of the MW, $C$ is a clumping factor, and the coefficient $k_{\rm H_2}$ with the empirically derived value of $3.5\times 10^{-17}$ by \citet{Wolfire2008ChemicalFormation}. To further exemplify the influence of our live dust modelling on the \hmol~formation efficiency, we show in Fig.~\ref{fig:H2form_0.1Zsun} the effective local $k_{\rm H_2}$ for the \calima~G8 simulation with $Z_{\rm init}=0.1Z_\odot$ at $t=228\,\rm Myr$. We have normalised this coefficient by the canonical value of $3.5\times 10^{-17}$, such that for the same DTG, gas temperature and clumping factor, the influence of grain size and temperature drives deviations from the empirical value from \citet{Wolfire2008ChemicalFormation}. The first result that can be extracted is that across the entire G8 galaxy the value of $k_{\rm H_2}$ is much lower than $3.5\times 10^{-17}$. It only approaches this value within highly shielded and dense clumps. We also include purple contours indicating regions where the local STL exceeds 0.35, which are clearly linked with clumps of high \hmol~efficiency. The diffuse ISM shows formation coefficients up to 2 dex lower than the canonical, particularly within regions that are dominated by high dust temperature (shown as red contours). We argue that since at fixed DTM the \hmol~efficiency increases for a larger STL (i.e. larger surface-to-mass ratio) and decreases for larger dust temperature (i.e. due to lower residence time by adsorbed hydrogen), $k_{\rm H_2}$ is much lower in the ISM of a $0.1Z_\odot$ galaxy for which the diffuse ISM is more transparent and the growth of small grains is halted due to low gas-phase abundance of metals. This variability is a direct benefit from an \hmol~formation model coupled to an evolving grain size distribution and with on-the-fly radiative transfer.

\begin{figure}
    \centering
	\includegraphics[width=\columnwidth]{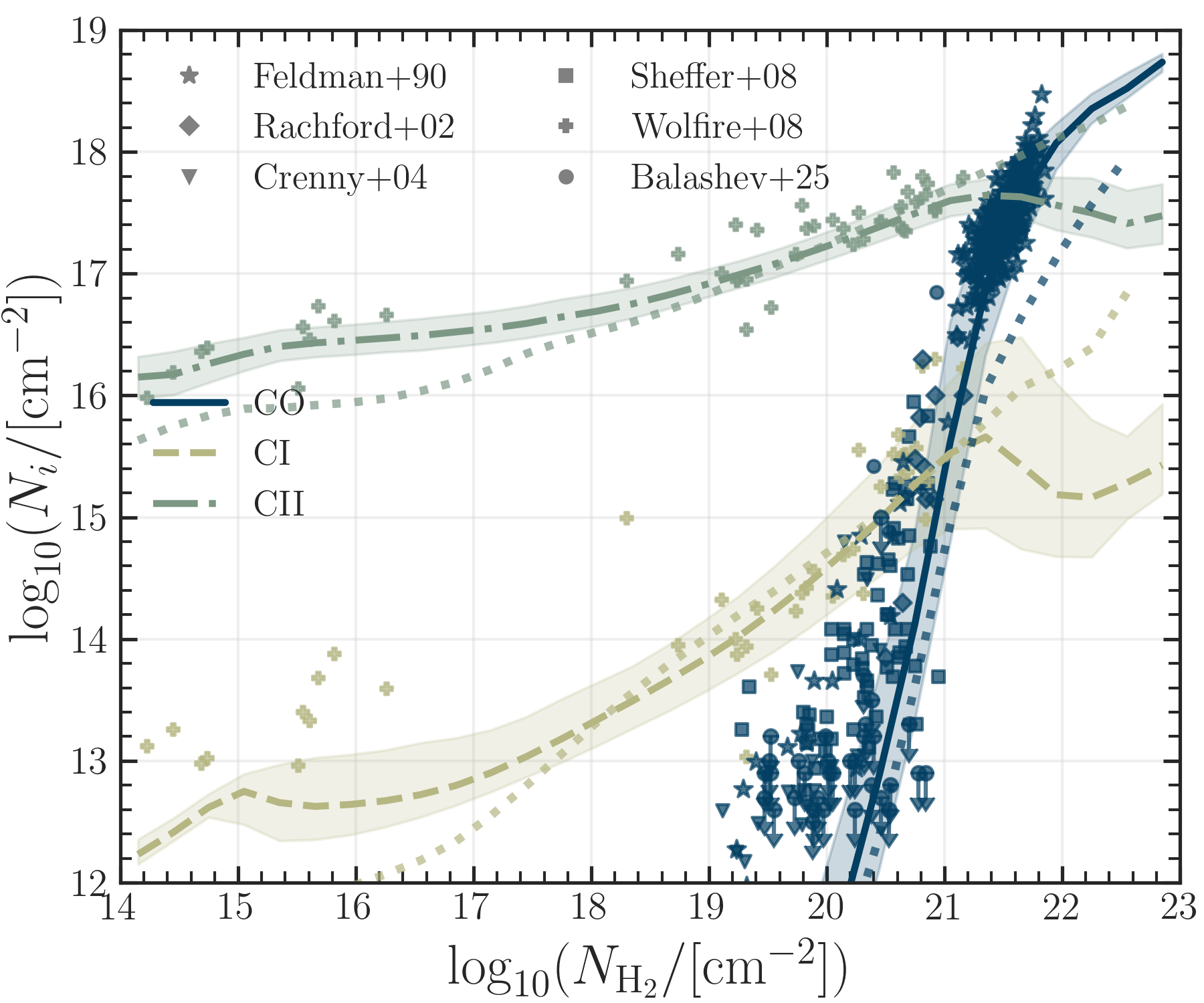}
    \caption{Median CO (solid blue), C~{\small I} (dashed olive), and C~{\small II} (dash-dotted seagreen) column densities for a stack of G8 ($Z_{\rm init}=0.1\,Z_{\odot}$) snapshots from $100\mbox{--}200$\,Myr. Dotted lines indicate the equivalent medians but for the G8 simulation with fixed DTM. For comparison, we include the observations from galactic sight lines from \citet{Federman1990ModelingChemistry}, \citet{Rachford2002AClouds}, \citet{Crenny2003ReanalysisMonoxide}, \citet{Wolfire2008ChemicalFormation}, \citet{Sheffer2008UltravioletRelationships} and for LMC and SMC sight lines (in absorption) from \citet{Balashev2025FirstRatio}. \calima~improves on the fixed dust modelling across all densities and C-bearing species (particularly for CO at high $N_{\,\rm H_2}$), although the abundance of CI at low column densities remains low compare to observations.}
    \label{fig:carbon_column_density_comparison_z0.1}
\end{figure}
As we have emphasised before, our G8 simulations do not allow a direct comparison to a system like the Small Magellanic Cloud (SMC), even less for the Large Magellanic Cloud (LMC) or the MW. Despite this, comparing column densities of the ions and molecules tracing the molecular and neutral ISM is useful to understand the details of our model \citep[e.g.][]{Gong2019AISM,Hu2021Medium,Katz2022PRISM:Galaxies,Thompson2024PredictionsChemistry}. In Fig.~\ref{fig:carbon_column_density_comparison_z0.1} we have included observations for the column densities of \hmol, CO, C~{\small I} and C~{\small II} for various sight lines in the Magellanic Clouds and in the MW \citep[][]{Federman1990ModelingChemistry,Rachford2002AClouds,Crenny2003ReanalysisMonoxide,Sheffer2008UltravioletRelationships,Wolfire2008ChemicalFormation,Balashev2025FirstRatio}. Simulation results are obtained for the \calima~simulation with $Z_{\rm init}=0.1\, Z_{\odot}$\footnote{We use the low metallicity simulation for this comparison with observations due to its more realistic gas-phase metallicity for a galaxy with a stellar mass of $10^{8}\,\rm M_{\odot}$ \citep[e.g.][]{Andrews2013TheGalaxies}.} by stacking all outputs between $100\mbox{--}200$\,Myr. Shaded regions indicate the second and third quartiles of the distribution of values at each \hmol~column density $N_{\rm H_2}$ bin. We have also included the results for the equivalent G8 simulation but with fixed dust as dotted lines. We can see that the \calima~model improves upon the predictions from the fixed dust model, by allowing for a larger CO abundance at $N_{\rm H_2}\gtrsim 10^{21}\,\rm cm^{-2}$, and larger abundances of C~{\small I} and C~{\small II} at lower densities caused by lower DTM in the diffuse neutral gas. The boost in $N_{\rm CO}$ can be explained by a larger \hmol~abundance at fixed DTM in \calima~(see Fig.~\ref{fig:xH_0.1Zsun} for $n_{\rm H}\gtrsim 10\,\rm cm^{-3}$ and the equilibrium curve in Fig.~\ref{fig:equilibrium_curve}). 

\subsubsection{Quantifying the relative role on the ISM of heating and cooling processes linked to dust and PAHs}\label{subsubsec:heating_and_cooling}

\begin{figure*}
    \centering
	\includegraphics[width=\textwidth]{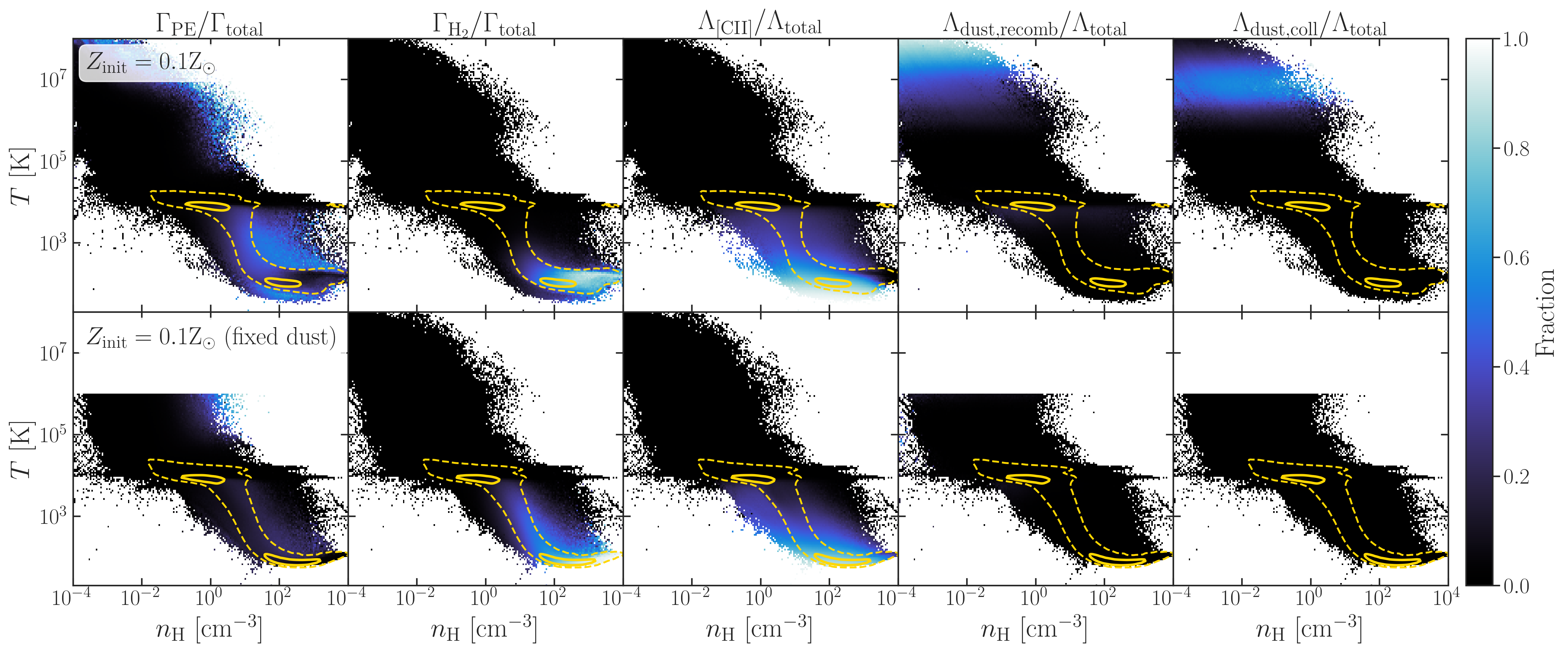}
    \caption{Temperature-density diagrams for the G8 simulation with $Z_{\rm init}=0.1\,Z_{\odot}$ showing the fraction of total contribution of heating and cooling processes–with direct link to dust and PAH evolution–to the total heating and cooling rates. These have been computed for a cylindrical region centred on the G8 galaxy with radius 1.5\,kpc and height below and above the disc of 2\,kpc. We also include yellow contours indicating bins in $T$--$n_{\rm H}$ space that contribute $10^{-4}$ (dashed lines) and $10^{-3}$ (solid lines) to the total gas mass budget. The top row is for the \calima~model while the bottom row is for the fixed dust simulation.}
    \label{fig:Trho_heatcool}
\end{figure*}
We have established that at the same initial gas-phase metallicity, the G8 simulation using \calima~results in larger \hmol~mass fractions at $n_{\rm H}\gtrsim 10\,\rm cm^{-3}$ than a fixed dust model. Since \hmol~can heat the gas by formation, dissociation, and UV pumping \citep[see section 2.2.4 in][for the modelling included in \calima]{Katz2022PRISM:Galaxies}, a larger \hmol~fraction motivates an examination of the role of \hmol~heating in the thermal state of the ISM. In order to explore the influence of an evolving dust and PAH population in the thermo-chemical properties of the ISM, we show in Fig.~\ref{fig:Trho_heatcool} the fractional contribution of (from left to right) photo-electric heating (PEH), \hmol~heating, dust and PAH recombination cooling, and dust collisional cooling to the total heating ($\Gamma_{\rm total}$) and total cooling ($\Lambda_{\rm total}$). These fractional quantities are computed on a per-cell basis, and the final 2D histogram is constructed as the median mass-weighted value at each $T$--$n_{\rm H}$ bin for the stack of simulation outputs between $100\mbox{--}200$\,Myr for the G8 simulations with $Z_{\rm init}=0.1\, Z_{\odot}$ with \calima~(top row) and fixed dust (bottom row) models. Consistent with the analysis performed in previous sections, we only consider gas within a cylinder centred on G8 with radius 1.5\,kpc and height below and above the disc of 2\,kpc. To further guide the relevance of each process to the bulk of the ISM, we have also included yellow contour lines to indicate which bins in $T$--$n_{\rm H}$ space contribute $10^{-4}$ (dashed lines) and $10^{-3}$ (solid lines) to the total gas mass budget.

As described in Section~\ref{subsubsec:methods_g8}, we initialised the G8 simulations with a fixed DTM and $q_{\rm PAH}$ based on their gas-phase metallicity and the scalings in \citet{Remy-Ruyer2014Gas-to-dustRange,Remy-Ruyer2015LinkingPicture}. Without dust and PAH evolution (only injection of dust by SN and PAHs by AGB winds) the G8 simulation with $Z_{\rm init}=0.1\,Z_{\odot}$ and fixed dust model maintains a value of $q_{
\rm PAH}\sim 1\%$, more than twice the maximum values measured for the G8 simulation using \calima~(see middle panel Fig.~\ref{fig:TG0_qpah}). Despite this difference, the fractional contribution of PEH within the diffuse unstable and cold phases is significantly increased in the \calima~simulation compared to fixed dust. In the classical implementation of PEH, the heating rate is dependent on the charging parameter $\gamma$, the dust-to-gas ratio, and the abundance of PAHs \citep[e.g.][]{Katz2022PRISM:Galaxies}, therefore becoming independent from the shape of the ISRF, the grain and PAH size distribution, the grain composition and/or the grain/PAH charge (see Sections~\ref{subsec:dust_photoelectric_heating} and \ref{subsec:photoelectric_heating}). \calima~raises important concerns about the effectiveness in capturing the variability (and full efficiency) of PEH with the classical implementations, using a single ISRF, a fixed GSD and dust/PAH properties. In future work, we will explore further the comparison between classical approaches and the self-consistent modelling of PEH for dust and PAHs in \calima~(Rodr\'iguez Montero et al., in prep.). We note that at low densities, the dominant heating mechanism is CR heating \citep[see e.g. Figure~11 in][]{Katz2022PRISM:Galaxies}.

In terms of \hmol~heating, the \calima~model indeed shows a much larger range in density in which it affects or dominates the heating of the cold ISM, as a consequence of a larger \hmol~fraction. We find that this region of enhanced \hmol~heating aligns with the contours indicating the bulk gas mass distribution, thus positioning \hmol~heating as an important driver of the dominant equilibrium distribution for the \calima~simulation. When compared to the fixed dust simulation, this histograms show that while both simulations converge to similar equilibrium temperatures for $n_{\rm H}\sim 10^4\,\rm cm^{-3}$, the \calima~simulation shows a broad distribution in temperature for $n_{\rm H}\sim 10^2\,\rm cm^{-3}$. Additionally, \hmol~appears to play a more significant role at $n_{\rm H}\lesssim 10\,\rm cm^{-3}$, as expected due to the higher \hmol~fraction at these lower gas densities in the fixed model (see Section~\ref{subsubsec:molecules_with_dust}). Overall, the equilibrium gas distribution appears to lack the minimum in gas temperature at these densities observed in the fixed dust models. This goes in agreement with the results obtained for the equilibrium tests (see Section~\ref{subsec:eq_tests}), where we argued this to be the consequence of large \hmol~heating due to more efficient formation, and a lower [C~{\small II}] cooling caused by efficient C$^{+}$ accretion by carbonaceous dust grains at $n_{\rm H}\sim 10^2\mbox{--}10^4\,\rm cm^{-3}$. To test this, the central column in Fig.~\ref{fig:Trho_heatcool} shows how the fractional contribution of [C~{\small II}] cooling differs from \calima~to the fixed dust model. While the low temperature gas at $n_{\rm H}\sim 10^2\,\rm cm^{-3}$ shows a dominant contribution of [C~{\small II}] cooling, the variable DTM (i.e.~see top left panel in Fig.~\ref{fig:DTM_STL_vs_Trho}) and the accretion of C$^+$ leads to a lower contribution of [C~{\small II}] cooling along the bulk of the gas mass distribution. This leads to a warmer unstable and cold phases of the ISM. We caution that these results are subject to our choice of initial conditions (i.e.~low gas fraction, high metallicity, low star formation). However, this may point towards the possibility of using the thermal properties of the ISM (together with observations of [C~{\small II}]) to constrain uncertain parameters in current dust accretion models.

To mimic thermal sputtering of dust and PAHs, in \citet{Katz2022PRISM:Galaxies} dust processes are turned off for $T>10^6\,\rm K$, which can be seen as the sharp cuts for dust-mediated processes above this gas temperature. While this may be a useful approximation, it appears to completely miss the role of dust recombination cooling and collisional cooling, as these appear to have an important role for gas temperatures above $10^6\,\rm K$. The updated modelling of dust thermal sputtering and collisional cooling (see Sections~\ref{subsec:sputtering} and \ref{subsec:collisional_cooling}) result in SN heated gas to retain its dust mass for sufficiently longer timescales to allow a non-negligible contribution of these processes to the cooling. In fact, for the very high temperatures ($\sim 10^7\mbox{--}10^8\,\rm K$) dust recombination can dominate the cooling of the gas. We also find that \calima~predicts a non-negligible ($\sim20\%$) contribution from recombination cooling for warm gas between $\sim 10^3\mbox{--}10^4\,\rm K$, in agreement with the original predictions from \citep[e.g.][]{Bakes1994TheHydrocarbons,Weingartner2001PhotoelectricHeating}.

\section{Discussion}\label{sec:discussion}

\subsection{Comparison to previous dust models}\label{subsec:comp_othermodels}

In Section~\ref{sec:intro}, we highlighted the recent surge of interest in dust evolution models, driven in part by the comprehensive overview of current approaches compiled by \citet{Parente2025ModelingSimulations}. This renewed focus is motivated both by the need for self-consistent forward modelling of observations from ALMA and JWST across cosmic time, and by the demand for theoretical frameworks that capture the coupled evolution of dust, PAHs, and the multi-phase ISM. Much of the modelling developed in this work builds directly on the progress made by earlier implementations in the literature. To place \calima\ in this broader context, we summarise in Table~\ref{tab:comp_to_others} the dust and PAH processes included in our framework, and indicate the extent to which each has been implemented for the first time within a galaxy formation or ISM thermochemistry model.
\begin{table*}
    \centering
    \begingroup
    \setlength{\tabcolsep}{5pt}
    \renewcommand{\arraystretch}{1.1}

    \large
    \begin{tabular}{p{4.9cm} >{\centering\arraybackslash}p{3.5cm} >{\footnotesize}p{7.3cm}}
        \toprule
        \textbf{Model feature} & \textbf{1st time?} & {\normalsize\textbf{Remarks}} \\
        \midrule
        Seeding by stellar processes & \xmark &
        Injection of dust grains by SN and winds is implemented in many live dust models for galaxy formation \citep[e.g.][]{Aoyama2017GalaxyDestruction,Hou2017EvolutionSimulation,McKinnon2018SimulatingMesh,Granato2021DustFormation,Parente2022DustVolumes,Choban2022TheFIRE,Dubois2024GalaxiesSimulations,Trayford2025ModellingMedium}.\\

        Grain charging & \cmark &
        Grain charging directly connect to the local grain and ISM properties is present in photodissociation codes \citep[e.g.][]{LePetit2006ACode,VanHoof2004GrainMedia,Ercolano20083DRegions}, but never in an ISM model nor on galaxy evolution.\\

        Dust PEH & \smark &
        PEH is usually included in ISM models using size, and ISRF-averaged efficiencies \citep[e.g.][]{Bialy2019ThermalGas,Kim2023PhotochemistrySimulations,Katz2022PRISM:Galaxies}, and with no self-consistent dependence on evolving dust properties.\\

        Accretion & \smark &
        Gas-phase accretion is a fundamental component of many dust evolution models, but it has never been connected to individual ion abundances and direct grain charging within a non-equilibrium chemistry model.\\

        Dust thermal sputtering & \smark &
        Grain-size and Coulomb effects have only been considered in post-processing tools for SN \citep{Kirchschlager2019DustDensities}, but never for dust model within a galaxy formation simulation.\\

        Dust collisional cooling & \smark &
        Implementations for low temperature cooling in clouds \citep{Hollenbach1979MOLECULEPROCESSES} and high temperature, ionised gas \citep{Dwek1981TheCondensates} are present in different forms, but never considering the finite size of grains, their chemical properties, and the evolution of their size, composition and charge.\\

        Dust non-thermal sputtering & \xmark &
        Destruction of grains for shocked gas is present in many dust evolution models.\\

        Turbulent shattering and coagulation & \smark &
        Turbulent-induced collisions of dust grains is usually included using fixed velocity dispersions \citep[e.g.][]{Dubois2024GalaxiesSimulations}, or with a scaling based on the local Jeans length \citep[e.g.][]{Li2021TheGalaxies,Narayanan2023ASimulations}, but without direct measurement of the local turbulent properties as done in this work.\\

        \hmol{} formation on grains & \smark &
        While dust live models sometimes include the local dust mass to obtain \hmol~formation rates, they do not consider variations in efficiency due to the change in grain size, composition and temperature.\\

        Dust and PAHs interaction with radiative transfer & \smark &
        Live dust models either only consider the local dust mass \citep[][]{Kannan2020SimulatingDust}, or directly do not include on-the-fly radiation transfer.\\

        Stellar PAH seeding & \cmark &
        Direct injection of PAHs either by carbonaceous grain fragmentation in SN shocks or via C-rich AGB winds has not been included in previous galaxy formation simulations with PAHs \citep{Narayanan2023ASimulations}.\\

        PAH growth & \smark &
        PAH growth has been treated as to proceed in the same way as dust grains \citep{Narayanan2023ASimulations}, without consideration of their molecular nature.\\

        PAH thermal sputtering & \cmark &
        Never included in galaxy or ISM formation models.\\

        PAH clustering & \cmark &
        Never included in galaxy or ISM formation models.\\

        PAH evaporation & \cmark &
        Never included in galaxy or ISM formation models.\\

        PAH freezing & \cmark &
        Never included in galaxy or ISM formation models.\\

        PAH photo-dissociation & \cmark &
        Never included in galaxy or ISM formation models.\\

        \hmol{} formation on PAHs & \cmark &
        Never included in galaxy or ISM formation models.\\
        \bottomrule
    \end{tabular}
    \endgroup

    \caption{Summary of the dust and PAH processes included in \calima~and their novelty in the current context of galaxy formation modelling. For each model feature, we indicate whether it is the first time is implemented in a galaxy formation/ISM model (\cmark), if it has been implemented to some extent but \calima~introduces a more self-consistent and updated modelling (\smark), and if it was implemented before in other numerical implementations of live dust modelling (\xmark).}
    \label{tab:comp_to_others}
\end{table*}

\calima~is based on the original implementation of live dust modelling within \ramses~introduced in \citet{Dubois2024GalaxiesSimulations}. The modelling here has many features inherited from \citet{Dubois2024GalaxiesSimulations}: a two-size approximation for regular dust grains, separation in carbonaceous and silicate compositions, log-normal distributions in each size bin, implementations for dust destruction in SN shocks and hot gas, shattering and coagulation, as well as a sub-grid turbulence model for gas-phase metal accretion. Despite these similarities, \calima~include fundamentally different approaches to the coupling between dust evolution and the ISM. The modelling in \citet{Dubois2024GalaxiesSimulations} did not include on-the-fly radiative transfer, cooling/heating was not treated in the fully explicit, non-equilibrium approach in \ramsesrtz, but rather by indirectly affecting the cooling due to a lowering of the effective total metallicity $Z$ used in the tabulated cooling rates in \ramses~from \citet{Sutherland1993} above $10^4\,\rm K$ and from \citet{Dalgarno1972HeatingRegions} below $10^4\,\rm K$.
In order to isolate the influence of different choice of dust modelling, we have run the same equilibrium test runs as in Section~\ref{subsec:eq_tests} but at a lower metallicity of $0.1Z_\odot$. The thermo-chemistry parameters are chosen to be the same, but we modify the choice of dust evolution modelling:
\begin{itemize}
    \item Accretion does not consider the predictions from grain charging and the gas-phase ion abundances, but instead assume that Coulomb factor $F_C=1$, except for largeC and smallSil grains in the CNM with $T < 2\times 10^4\,\rm K$ and $n_{\rm H}>10\,\rm cm^{-3}$, for which \citealt{Dubois2024GalaxiesSimulations} adopted $F_C(0.1\mu{\rm m})=0$ and $F_C(0.5{\rm nm})=10$.
    \item Coagulation follows the modelling in \citet{Aoyama2017GalaxyDestruction} which assumes that 0.5 of the gas mass-with unresolved Jeans length and with gas density above $0.1\,\rm cm^{-3}$ and temperature below $10^4\,\rm K$-is at a density of $10^3\,\rm cm^{-3}$. The small grain turbulent velocity is fixed at $0.1\,\rm km\,s^{-1}$.
    \item Shattering is based on the scaling with $n_H$ from \citet{Granato2021DustFormation} that is equivalent to having a dispersion velocity varying from $10\,\rm km\,s^{-1}$ to $1\,\rm km\,s^{-1}$ from the WIM ($1\,\rm H\,cm^{-3}$) to the molecular cloud phase ($10^3\,\rm H\,cm^{-3}$).
\end{itemize}
In \citealt{Dubois2024GalaxiesSimulations} PAHs are not modelled as a separate component, and grains are divided only into two sizes (5 nm and 0.1 \micron) and compositions (carbonaceous and silicates). We also note that in their model thermal sputtering follows the fitting curves from \citet{Hu2019ThermalMedium} to the thermal sputtering yields obtained by \citet{Nozawa2006DustUniverse}, but this is irrelevant for the temperature range of the equilibrium test.

\begin{figure}
    \centering
	\includegraphics[width=\columnwidth]{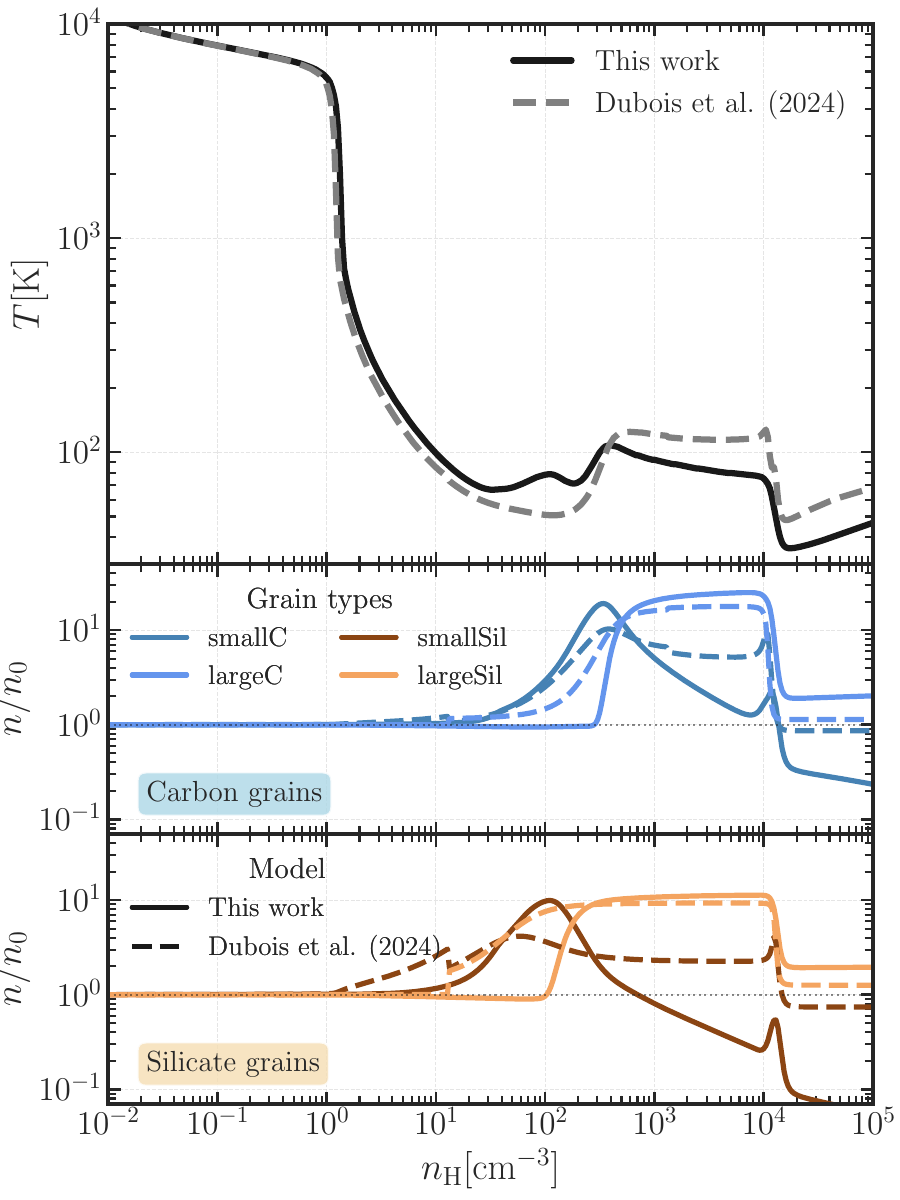}
    \caption{Equilibrium temperature test for }
    \label{fig:comp_eq_yohan}
\end{figure}
Fig.~\ref{fig:comp_eq_yohan} shows the comparison of \calima~(solid lines) and the dust parameters of \citet{Dubois2024GalaxiesSimulations} (dashed lines) for the equilibrium temperature (top panel), smallC and largeC number densities (middle panel), and smallSil and largeSil number densities (bottom panel). The number densities are normalised by their values in the initial conditions (see Section~\ref{subsec:eq_tests} for more details on the initial dust abundances), in order to better capture the variations due to dust processes at a given $n_{\rm H}$. By firstly examining the number densities of the different grain bins, it can easily be appreciated that shattering and coagulation start to have an effect on the grain size distribution at $n_{\rm H}\sim 1\,\rm cm^{-3}$ and $n_{\rm H}\sim 10\,\rm cm^{-3}$, respectively. These transitions occur at larger $n_{\rm H}$ and reaching higher values of small and large grains, in shattering- and coagulation-dominated regions, respectively. Therefore, we expect \calima~to result in larger contrasts in the STL across phases than the \citet{Dubois2024GalaxiesSimulations} predicts. We also note that due to the fixed velocity dispersion used in the coagulation model of \citet{Aoyama2017GalaxyDestruction}, we predict significantly faster depletion of small grains for $n_{\rm H}\gtrsim 10^2\,\rm cm^{-3}$. These changes in the STL across density are also imprinted in the equilibrium temperature curve (top panel of Fig.~\ref{fig:comp_eq_yohan}). Since \calima~predicts a higher (lower) abundance of small grains for $n_{\rm H}\sim 10^2\,\rm cm^{-3}$ ($n_{\rm H}\gtrsim 3\times 10^2\,\rm cm^{-3}$), the temperature is higher (lower) than using the dust parameters in \citet{Dubois2024GalaxiesSimulations}. This is caused by higher elemental addition (hence decreased metal line cooling) and higher \hmol~production (therefore more \hmol~heating) when there is a larger abundance of small grains. Overall, \citet{Dubois2024GalaxiesSimulations} tends to predict a larger fraction of small grains in the cold and molecular phases of the ISM, resulting in increased equilibrium temperatures compared to \calima.

\subsection{Limitations of our modelling}\label{subsec:limitations}
Despite the proven practical utility of the two-size approximation modelling by \citet{Hirashita2015Two-sizeGalaxies}, it possesses inherited issues of any model with a discretised GSD. While optical properties are computed as effective, locally averaged quantities obtained by weighting over the underlying GSD (see Section~\ref{subsec:optical properties}), many dust evolution processes (accretion, destruction, charging, etc.) are evaluated at the central grain size of each bin to limit the computational expense. As a result, the choice and resolution of the size discretisation can influence the inferred evolution timescales and relative importance of competing processes (e.g.~see the comparison of \calima~with \citealp{Dubois2024GalaxiesSimulations} in Section~\ref{subsec:comp_othermodels}). The immediate solution to such a problem is to increase the number of size bins \citep[e.g.][]{McKinnon2018SimulatingMesh,Aoyama2019ComparisonSurveys,Li2021TheGalaxies}, but this imposes a significant increase in the memory imprint (the GSD is evolved for every cell in the simulation) and in the computational cost of \calima~(shattering and coagulation operations increase like $N_{\rm bins}^2$ in Smoluchowski-type equations). To maintain the minimal computational expense of running simulations with \calima, we propose instead to use the moment of methods adapted to dust processing \citep[e.g.][]{Mattsson2016ModellingMoments,Mattsson2020GalacticTurbulence}.

An assumption used in many models of dust evolution for galaxy formation is the dynamical coupling of dust grains with the gas flow \citep[e.g.][]{Zhukovska2016MODELINGISM,Aoyama2017GalaxyDestruction,Gjergo2018DustSimulations,Romano2022TheGalaxy,Choban2022TheFIRE,Dubois2024GalaxiesSimulations}. Dust-gas coupling is governed by the ratio of the aerodynamic stopping time to the local gas flow timescale \citep[see Appendix~\ref{ap:dust_turbulence} and][]{Draine1979ONGAS}. While grains of all sizes remain tightly coupled to the gas in dense and turbulent phases of the interstellar medium, large grains ($a \gtrsim 0.1\,\mu{\rm m}$) are expected to become marginally coupled or decoupled in hot, low-density environments and galactic outflows, where their stopping lengths can reach tens to hundreds of parsecs \citep[e.g.][]{Hopkins2016TheClouds,Commercon2023DynamicsImplementations}. In such regimes, the assumption of perfect dust-gas coupling may lead to inaccuracies in the predicted dust distribution and its interaction with the radiation field. An additional source of uncertainty in dust dynamics arises from the influence of magnetic fields and magnetohydrodynamic (MHD) turbulence. Charged dust grains interact with magnetic fields through Lorentz forces, which can significantly modify both their acceleration and drag, particularly in diffuse and ionised phases of the ISM where grain charging is efficient (see Section~\ref{subsubsec:dust_variations}). In turbulent MHD flows, dust grains may experience gyro-motion, resonant interactions with Alfv\'enic fluctuations, and size-dependent acceleration, leading to preferential transport and spatial segregation relative to the gas \citep[e.g.,][]{Yan2004DustTurbulence,Moseley2022Acceleration,Moseley2025DustGrains}. These effects can alter effective drag timescales and enhance grain decoupling beyond what is expected from purely hydrodynamic considerations. Incorporating MHD-dependent dust forces and turbulence-driven acceleration represents an important avenue for future improvements.

A limitation of the present model is the appearance of an excess contribution of smallPAHs in regions characterised by high radiation field intensity $G_0$ and relatively low gas temperatures (i.e.~Fig.~\ref{fig:TG0_qpah} in Section~\ref{subsubsec:dust_variations}). In such environments, the balance between smallPAH destruction and formation may not be fully captured in the current treatment. While top-down formation pathways–whereby largePAHs fragment into smallPAHs under energetic processing–are expected to operate in strongly irradiated regions \citep[e.g.][]{Maragkoudakis2026PDRs4All:Bar}, observational predictions suggest that sufficiently intense radiation fields should still favour the survival and dominance of largePAHs due to their higher photostability \citep[e.g.][]{Knight2021TracingEnvironments,Chown2024PDRs4All:Bar,Baron2025PHANGS-ML:Galaxies}. Based on the modellig in Sections~\ref{subsec:pah_evaporation} and~\ref{subsec:pah_dissociation} we have compute the ratio of largePAH evaporation to smallPAH dissociation timescales. Given that smallPAHs quickly become fully dehydrogenated at high $G_0$ and that the fitting to the results of \citet{Montillaud2014AbsoluteClusters} is independent of $n_{\rm H}$, we find that evaporation becomes faster than dissociation for $G_0\gtrsim 10^2$, no matter the local gas density. Given than in these conditions PAH growth via accretion is negligible, an overabundance of smallPAHs points at the faster timescale of evaporation compared to dissociation at high $G_0$. This may suggests that either our modelling of smallPAH dissociation predicts an incorrect dependence on $G_0$ or the simple scaling of evaporation time taken from \citet{Montillaud2014AbsoluteClusters} predicts an incorrect resilience of largePAHs in intense radiation fields. Another possibility is that our description of largePAHs as PAH clusters formed from smallPAHs like circumcoronene is an incomplete approach to the evolution of PAH sizes in the ISM \citep[e.g.][]{Rapacioli2005StackedMolecules}, and instead PAHs grow by extending their planar structure instead of by stacking monomers \citep[][]{Roser2015POLYCYCLICEMISSION}. In future iterations of \calima~we will explore direct modelling of PAH evaporation, hydrogenation, etc., instead of relying in the current fitting functions.

Closely related to the problem discussed above, PAH aromatisation is an unexplored component in the current version of the model. In \calima~we track the hydrogenation sate of PAHs using parametrised prescriptions based on the results of \citet{Andrews2016HydrogenationH2formation} and \citet{Montillaud2013EvolutionStates}, allowing us to follow the number of hydrogen atoms attached to the carbon skeleton as function of local physical conditions. In this work, the hydrogenation state of PAHs is used as a \textit{proxy} for the degree of UV processing and edge saturation. While changes in hydrogenation correlate with aromatisation and aliphatisation, the latter involve structural rearrangements \citep[e.g.][]{Jones2013ThePoint,Murga2016RestructuringMedium} of the carbon skeleton that are not explicitly modelled (i.e.~dominance of sp$^3$ electron orbital hybridisation compared to sp$^2$ in aromatic carbon structures). Aromatisation is expected to occur rapidly in strong radiation fields \citep[$G_0\gtrsim 10$ for PAHs larger than $\sim 50$C,][]{Murga2019Shiva:Model}, whereas it can be significantly slower in shielded regions of the ISM. Moreover, the adpted hydrogenation modelling relies on empirical fits in $G_0$--$n_{\rm H}$ space and lacks the same level of physical self-consistency as other components of the PAH photo-chemical modelling. Planned improvements to \calima~include introducing a more complete treatment of aromatic fraction as an independent state variable, allowing for non-equilibrium evolution of both hydrogenation and aromatisation, and for these to be treated as related but distinct processes.

\section{Conclusions}\label{sec:conclusions}

\calima~provides a unified and self-consistent framework for evolving interstellar dust grains and PAHs on-the-fly in radiation-hydrodynamics simulations, embedding them as thermodynamically and radiatively active components of the multiphase ISM rather than passive tracers. It is built around a two-size (small/large), two-composition (amorphous carbonaceous/silicate) description of dust plus a dedicated PAH component (small and large clusters), so that variations in dust-to-metal ratio, small-to-large grain ratio, and PAH fraction emerge self-consistently from stellar seeding and ISM processing instead of being imposed through fixed scalings. Within this framework, the same evolving grain and PAH populations set opacities, photoelectric heating and grain-assisted recombination efficiencies, collisional gas--dust energy exchange, and \hmol~formation rates, all computed in concert with on-the-fly radiative transfer and non-equilibrium thermo-chemistry.

We have revisited and extended a broad set of microphysical processes for both dust and PAHs in a way that is tailored to galaxy-scale simulations but retains close contact with the underlying laboratory and theoretical constraints. For bulk dust, we implement stellar dust injection from SNe and AGB stars; accretion of gas-phase metals that is explicitly modulated by a turbulence-informed sub-grid density PDF and by grain charging; shattering and coagulation driven by turbulent grain relative velocities; non-thermal destruction in supernova shocks; and updated thermal sputtering that accounts for finite grain size and modern ion--solid interaction physics. For PAHs, we include formation and seeding, clustering into larger aggregates, freezing onto regular grains, photo-dissociation and evaporation of clusters, thermal and non-thermal sputtering, and a detailed treatment of charge and hydrogenation balance that feeds into both photoelectric heating and H$_2$ formation. These ingredients are consistently mapped into wavelength-dependent opacities and heating/cooling efficiencies, ensuring that radiation, thermochemistry, and grain/PAH physics evolve in synergy.

We have used equilibrium temperature--density tests to examine how dust--PAH evolution significantly modifies the canonical thermal and chemical structure of the ISM even at fixed metallicity. We highlight:
\begin{itemize}
    \item Gas-phase metal accretion, especially onto small carbonaceous grains, produces strong, density-dependent depletions in C and O that suppress C\,\textsc{ii} and O\,\textsc{i} cooling above $n_{\rm H} \sim 10~\mathrm{cm^{-3}}$, shifting equilibrium temperatures upward in the unstable and cold neutral medium compared with fixed-dust models.
    \item The competition between accretion, coagulation, and shattering reshapes the small-to-large grain ratio across density, altering the surface-to-mass ratio and therefore \hmol~formation and heating, with coagulation at high densities depleting small grains and shattering at intermediate densities partially replenishing them.
    \item PAH evolution (clustering, freezing, sputtering) rapidly depletes small PAHs in dense gas, transferring PAH carbon into clusters and grains, which in turn quenches PAH-driven photoelectric heating in the cold and molecular phases, while leaving PAHs far more abundant and radiatively active in diffuse and moderately dense gas.
    \item Direct computation of grain and PAH photoelectric efficiencies shows that MRN-averaged, fixed-efficiency recipes misestimate both the normalisation and density dependence of the heating rate, particularly once dust growth enhances small-grain mass and PAH depletion suppresses PAH abundance at higher densities.
\end{itemize}

Using isolated dwarf galaxy formations at various initial metallicities, we have explored how the same physics generates a rich, environment-dependent dust and PAH phenomenology and leaves a clear imprint on the thermodynamic and molecular structure of the ISM. The main results can be summarised as:
\begin{itemize}
    \item The dust-to-metal ratio, small-to-large grain mass ratio, PAH fraction, and local radiation field strength develop strong variations across gas phases, with enhanced small-grain and PAH abundances closely tracing the bulk of the cold and thermally unstable gas, reduced DTM and small grain mass fraction in the hot phase, and destruction of PAHs in hot and highly irradiated phases.
    \item These spatial variations lead to warmer unstable and cold phases around $n_{\rm H} \sim 10^{2}~\mathrm{cm^{-3}}$, a broader spread of equilibrium temperatures at intermediate densities, and a higher \hmol~fraction at high densities than in fixed-dust runs, even though H$_2$ is suppressed at $n_{\rm H} \lesssim 10~\mathrm{cm^{-3}}$ where dust growth and shielding are weaker. This is particularly interested in the context commonly adopted \hmol~formation efficiencies in chemistry models for the ISM.
    \item Line-of-sight column-density statistics show improved agreement with observed CO at high $N_{\rm H_2}$ and with C\,\textsc{i}/C\,\textsc{ii} in diffuse gas when dust and PAHs are evolved.
    \item The emergent, highly structured distribution of dust and PAHs produces non-trivial spatial variations in extinction, UV transparency, and photoelectric heating efficiency, indicating that simple, globally scaled dust prescriptions struggle to capture the coupling between radiation, ISM phases, and molecular chemistry in galaxy discs.
\end{itemize}

Taken together, these developments show that allowing dust and PAHs to evolve and to feed back on radiation and thermochemistry is not a minor refinement but a necessary step towards realistic ISM and galaxy formation modelling. \calima~is an major step in elevating dust and PAHs from static opacity and heating parameters to dynamically evolving agents that shape thermal stability, phase structure, and molecule formation, generating environment-dependent dust and PAH observables naturally from the underlying gas dynamics and radiation field. This establishes a pathway to interpret the rapidly growing body of dust- and PAH-sensitive observations---from C\,\textsc{ii} and CO to mid-IR PAH bands and attenuation curves---in a framework where the microphysics of grain evolution and the macroscopic evolution of galaxies and their ISM are treated as a single, self-consistent problem.

\section*{Acknowledgements}
FRM thanks Dimitra Rigopoulou, Ismael Garc\'ia-Bernete, Fergus Donnan, Nathalie Ysard, Laurent Verstraete, Bruce T. Draine, Brandon S. Hensley, Ugo Lebreuilly, Patrick Hennebelle and Gabriel Verrier for helpful discussions.
Simulation analysis made use of the Infinity Cluster hosted by Institut d'Astrophysique de Paris. We thank Stephane Rouberol for running it smoothly. The simulations included here made extensive use of the dp016 project on the DiRAC ecosystem. This work was performed using the DiRAC Data Intensive service at Leicester, operated by the University of Leicester IT Services, which forms part of the STFC DiRAC HPC Facility (www.dirac.ac.uk). The equipment was funded by BEIS capital funding via STFC capital grants ST/K000373/1 and ST/R002363/1 and STFC DiRAC Operations grant ST/R001014/1. 

This project made use of the following packages:
\textsc{cmasher} \citep{vanderVelden2020CMasher:Plots},
\textsc{matplotlib} \citep{Hunter2007Matplotlib:Environment},
\textsc{numpy} \citep{Harris2020ArrayNumPy}, and
\textsc{scipy} \citep{Virtanen2020SciPyPython}. The authors acknowledge the use of OpenAI's ChatGPT for assistance with code refactoring, and simulation analysis. The scientific analysis and results were independently verified by the authors.

\bibliographystyle{mnras}

% You should give the same name for your .bbl as your main .tex
% since it is a requirement for posting on ArXiv.
\bibliography{references_abbrev}

\begin{thebibliography}{}
\makeatletter
\relax
\def\mn@urlcharsother{\let\do\@makeother \do\$\do\&\do\#\do\^\do\_\do\%\do\~}
\def\mn@doi{\begingroup\mn@urlcharsother \@ifnextchar [ {\mn@doi@}
  {\mn@doi@[]}}
\def\mn@doi@[#1]#2{\def\@tempa{#1}\ifx\@tempa\@empty \href
  {http://dx.doi.org/#2} {doi:#2}\else \href {http://dx.doi.org/#2} {#1}\fi
  \endgroup}
\def\mn@eprint#1#2{\mn@eprint@#1:#2::\@nil}
\def\mn@eprint@arXiv#1{\href {http://arxiv.org/abs/#1} {{\tt arXiv:#1}}}
\def\mn@eprint@dblp#1{\href {http://dblp.uni-trier.de/rec/bibtex/#1.xml}
  {dblp:#1}}
\def\mn@eprint@#1:#2:#3:#4\@nil{\def\@tempa {#1}\def\@tempb {#2}\def\@tempc
  {#3}\ifx \@tempc \@empty \let \@tempc \@tempb \let \@tempb \@tempa \fi \ifx
  \@tempb \@empty \def\@tempb {arXiv}\fi \@ifundefined
  {mn@eprint@\@tempb}{\@tempb:\@tempc}{\expandafter \expandafter \csname
  mn@eprint@\@tempb\endcsname \expandafter{\@tempc}}}

\bibitem[\protect\citeauthoryear{Abel, Bryan  \& Norman}{Abel
  et~al.}{2000}]{Abel2000TheClouds}
Abel T.,  Bryan G.~L.,   Norman M.~L.,  2000, \mn@doi [ApJ]
  {10.1086/309295/FULLTEXT/}, 540, 39

\bibitem[\protect\citeauthoryear{{Abel}, {Bryan}  \& {Norman}}{{Abel}
  et~al.}{2002}]{Abel2002TheUniverse}
{Abel} T.,  {Bryan} G.~L.,   {Norman} M.~L.,  2002, \mn@doi [Sci]
  {10.1126/science.1063991}, \href
  {https://ui.adsabs.harvard.edu/abs/2002Sci...295...93A} {295, 93}

\bibitem[\protect\citeauthoryear{Abouelaziz et~al.,}{Abouelaziz
  et~al.}{1993}]{Abouelaziz1993MeasurementsCoefficients}
Abouelaziz H.,  et~al., 1993, \mn@doi [JChPh] {10.1063/1.465801}, 99, 237

\bibitem[\protect\citeauthoryear{Agertz et~al.,}{Agertz
  et~al.}{2021}]{Agertz2021VintergatanGalaxy}
Agertz O.,  et~al., 2021, \mn@doi [MNRAS] {10.1093/mnras/stab322}, 503, 5826

\bibitem[\protect\citeauthoryear{Agundez, Cabezas, Marcelino, Tercero,
  Fuentetaja, de Vicente  \& Cernicharo}{Agundez
  et~al.}{2025}]{Agundez2025A13-butadiene}
Agundez M.,  Cabezas C.,  Marcelino N.,  Tercero B.,  Fuentetaja R.,  de
  Vicente P.,   Cernicharo J.,  2025, \mn@doi [A{\&}A]
  {10.1051/0004-6361/202554343}, 697, A82

\bibitem[\protect\citeauthoryear{Allain, Leach  \& Sedlmayr}{Allain
  et~al.}{1996}]{Allain1996PhotodestructionGroup}
Allain T.,  Leach S.,   Sedlmayr E.,  1996, A{\&}A, 305, 602

\bibitem[\protect\citeauthoryear{{Allamandola}, J., {Sandford}, A., {Valero}
  \& J.}{{Allamandola} et~al.}{1988}]{Allamandola1988PhotochemicalAnalogs}
{Allamandola} J. L.,  {Sandford} A. S.,  {Valero}  J. G.,  1988, \mn@doi [Icar]
  {10.1016/0019-1035(88)90070-X}, 76, 225

\bibitem[\protect\citeauthoryear{Allamandola, Tielens, Barker, Allamandola,
  Tielens  \& Barker}{Allamandola
  et~al.}{1989}]{Allamandola1989InterstellarImplications}
Allamandola L.~J.,  Tielens A. G. G.~M.,  Barker J.~R.,  Allamandola L.~J.,
  Tielens A. G. G.~M.,   Barker J.~R.,  1989, \mn@doi [ApJS] {10.1086/191396},
  71, 733

\bibitem[\protect\citeauthoryear{Allers, Jaffe, Lacy, Draine  \&
  Richter}{Allers et~al.}{2005}]{Allers2005H_2Bar}
Allers K.~N.,  Jaffe D.~T.,  Lacy J.~H.,  Draine B.~T.,   Richter M.~J.,  2005,
  \mn@doi [ApJ] {10.1086/431919}, 630, 368

\bibitem[\protect\citeauthoryear{Altarawneh \& Ali}{Altarawneh \&
  Ali}{2024}]{Altarawneh2024FormationReview}
Altarawneh M.,  Ali L.,  2024, \mn@doi [Energy and Fuels]
  {10.1021/ACS.ENERGYFUELS.4C03513/ASSET/IMAGES/LARGE/EF4C03513{\_}0025.JPEG},
  38, 21735

\bibitem[\protect\citeauthoryear{{Andersen} \& {Bay}}{{Andersen} \&
  {Bay}}{1981}]{Andersen1981SputteringMeasurements}
{Andersen} H.~H.,  {Bay} H.~L.,  1981, in {Behrisch} R.,  ed., , Vol.~47,
  Sputtering by Particle Bombardment I: Physical Sputtering of Single-Element
  Solids.
Springer-Verlag, p.~145, \mn@doi{10.1007/3540105212_9}

\bibitem[\protect\citeauthoryear{Andrews \& Martini}{Andrews \&
  Martini}{2013}]{Andrews2013TheGalaxies}
Andrews B.~H.,  Martini P.,  2013, \mn@doi [ApJ] {10.1088/0004-637X/765/2/140},
  765, 140

\bibitem[\protect\citeauthoryear{Andrews, Candian  \& Tielens}{Andrews
  et~al.}{2016}]{Andrews2016HydrogenationH2formation}
Andrews H.,  Candian A.,   Tielens A. G. G.~M.,  2016, \mn@doi [A{\&}A]
  {10.1051/0004-6361/201628819}, 595, A23

\bibitem[\protect\citeauthoryear{Aoyama, Hou, Shimizu, Hirashita, Todoroki,
  Choi  \& Nagamine}{Aoyama et~al.}{2017}]{Aoyama2017GalaxyDestruction}
Aoyama S.,  Hou K.~C.,  Shimizu I.,  Hirashita H.,  Todoroki K.,  Choi J.~H.,
  Nagamine K.,  2017, \mn@doi [MNRAS] {10.1093/mnras/stw3061}, 466, 105

\bibitem[\protect\citeauthoryear{Aoyama, Hou, Hirashita, Nagamine  \&
  Shimizu}{Aoyama et~al.}{2018}]{Aoyama2018CosmologicalDestruction}
Aoyama S.,  Hou K.~C.,  Hirashita H.,  Nagamine K.,   Shimizu I.,  2018,
  \mn@doi [MNRAS] {10.1093/MNRAS/STY1431}, 478, 4905

\bibitem[\protect\citeauthoryear{Aoyama et~al.,}{Aoyama
  et~al.}{2019}]{Aoyama2019ComparisonSurveys}
Aoyama S.,  et~al., 2019, \mn@doi [MNRAS] {10.1093/mnras/stz021}, 484, 1852

\bibitem[\protect\citeauthoryear{Aoyama, Hirashita  \& Nagamine}{Aoyama
  et~al.}{2020}]{Aoyama2020GalaxyDistribution}
Aoyama S.,  Hirashita H.,   Nagamine K.,  2020, \mn@doi [MNRAS: Letters]
  {10.1093/mnras/stz3253}, 491, 3844

\bibitem[\protect\citeauthoryear{Asano, Takeuchi, Hirashita  \& Nozawa}{Asano
  et~al.}{2013}]{Asano2013WhatGalaxies}
Asano R.~S.,  Takeuchi T.~T.,  Hirashita H.,   Nozawa T.,  2013, \mn@doi
  [MNRAS] {10.1093/mnras/stt506}, 432, 637

\bibitem[\protect\citeauthoryear{Asano, Takeuchi, Hirashita  \& Nozawa}{Asano
  et~al.}{2014}]{Asano2014EvolutionGalaxies}
Asano R.~S.,  Takeuchi T.~T.,  Hirashita H.,   Nozawa T.,  2014, \mn@doi
  [MNRAS] {10.1093/mnras/stu208}, 440, 134

\bibitem[\protect\citeauthoryear{{Ashley} \& {Anderson}}{{Ashley} \&
  {Anderson}}{1981}]{Ashley1981EnergyDioxide}
{Ashley} J.~C.,  {Anderson} V.~E.,  1981, \mn@doi [IEEE TranNucSci]
  {10.1109/TNS.1981.4335688}, \href
  {https://ui.adsabs.harvard.edu/abs/1981ITNS...28.4131A} {28, 4131}

\bibitem[\protect\citeauthoryear{Asplund, Grevesse, Sauval  \& Scott}{Asplund
  et~al.}{2009}]{Asplund2009TheSun}
Asplund M.,  Grevesse N.,  Sauval A.~J.,   Scott P.,  2009, \mn@doi [ARA{\&}A]
  {10.1146/annurev.astro.46.060407.145222}, 47, 481

\bibitem[\protect\citeauthoryear{{Baes}, {Camps}  \& {Matsumoto}}{{Baes}
  et~al.}{2022}]{Baes2022MonteEmission}
{Baes} M.,  {Camps} P.,   {Matsumoto} K.,  2022, \mn@doi [A{\&}A]
  {10.1051/0004-6361/202244521}, \href
  {https://ui.adsabs.harvard.edu/abs/2022A&A...666A.101B} {666, A101}

\bibitem[\protect\citeauthoryear{Bakes, Tielens, Bakes  \& Tielens}{Bakes
  et~al.}{1994}]{Bakes1994TheHydrocarbons}
Bakes E. L.~O.,  Tielens A. G. G.~M.,  Bakes E. L.~O.,   Tielens A. G. G.~M.,
  1994, \mn@doi [ApJ] {10.1086/174188}, 427, 822

\bibitem[\protect\citeauthoryear{Bakes, Tielens  \& Bauschlicher}{Bakes
  et~al.}{2001}]{Bakes2001TheoreticalI.}
Bakes E. L.~O.,  Tielens A. G. G.~M.,   Bauschlicher Charles~W. J.,  2001,
  \mn@doi [ApJ] {10.1086/321501}, 556, 501

\bibitem[\protect\citeauthoryear{Balashev, Kosenko  \& Noterdaeme}{Balashev
  et~al.}{2025}]{Balashev2025FirstRatio}
Balashev S.~A.,  Kosenko D.~N.,   Noterdaeme P.,  2025, \mn@doi [A{\&}A]
  {10.1051/0004-6361/202452913}, 696, L16

\bibitem[\protect\citeauthoryear{Barlow, {Barlow}  \& J.}{Barlow
  et~al.}{1978}]{Barlow1978TheSputtering.}
Barlow M.~J.,  {Barlow}  J. M.,  1978, \mn@doi [MNRAS]
  {10.1093/MNRAS/183.3.367}, 183, 367

\bibitem[\protect\citeauthoryear{Baron et~al.,}{Baron
  et~al.}{2025}]{Baron2025PHANGS-ML:Galaxies}
Baron D.,  et~al., 2025, \mn@doi [ApJ] {10.3847/1538-4357/AD972A}, 978, 135

\bibitem[\protect\citeauthoryear{Barrera, Fuentealba, Mu{\~{n}}oz, G{\'{o}}mez
  \& C{\'{a}}rdenas}{Barrera et~al.}{2023}]{Barrera2023FormationStudy}
Barrera N.~F.,  Fuentealba P.,  Mu{\~{n}}oz F.,  G{\'{o}}mez T.,
  C{\'{a}}rdenas C.,  2023, \mn@doi [MNRAS] {10.1093/MNRAS/STAD2106}, 524, 3741

\bibitem[\protect\citeauthoryear{Bauschlicher, Bakes, Bauschlicher  \&
  Bakes}{Bauschlicher et~al.}{2001}]{Bauschlicher2001TheHydrogen}
Bauschlicher C.~W.,  Bakes E. L.~O.,  Bauschlicher C.~W.,   Bakes E. L.~O.,
  2001, \mn@doi [CP] {10.1016/S0301-0104(01)00500-6}, 274, 11

\bibitem[\protect\citeauthoryear{Bauschlicher, Peeters  \&
  Allamandola}{Bauschlicher et~al.}{2008}]{BauschlicherJr.2008ThePAHs}
Bauschlicher Jr. C.~W.,  Peeters E.,   Allamandola L.~J.,  2008, \mn@doi [ApJ]
  {10.1086/533424}, 678, 316

\bibitem[\protect\citeauthoryear{Bekki}{Bekki}{2013}]{Bekki2013CoevolutionHydrogen}
Bekki K.,  2013, \mn@doi [MNRAS] {10.1093/mnras/stt589}, 432, 2298

\bibitem[\protect\citeauthoryear{Bendo et~al.,}{Bendo
  et~al.}{2010}]{Bendo2010TheM81}
Bendo G.~J.,  et~al., 2010, \mn@doi [A{\&}A] {10.1051/0004-6361/201014568},
  518, 65

\bibitem[\protect\citeauthoryear{Bern{\'{e}} \& Tielens}{Bern{\'{e}} \&
  Tielens}{2011}]{Berne2011FormationSpace}
Bern{\'{e}} O.,  Tielens A. G. G.~M.,  2011, \mn@doi [PNAS]
  {10.1073/pnas.1114207108}, 109, 401

\bibitem[\protect\citeauthoryear{Bern{\'{e}}, Foschino, Jalabert  \&
  Joblin}{Bern{\'{e}} et~al.}{2022}]{Berne2022ContributionObservations}
Bern{\'{e}} O.,  Foschino S.,  Jalabert F.,   Joblin C.,  2022, \mn@doi
  [A{\&}A] {10.1051/0004-6361/202243171}, 667, A159

\bibitem[\protect\citeauthoryear{Bethe, {Bethe}  \& {H.}}{Bethe
  et~al.}{1930}]{Bethe1930ZurMaterie}
Bethe H.,  {Bethe}  {H.} 1930, \mn@doi [AnP] {10.1002/ANDP.19303970303}, 397,
  325

\bibitem[\protect\citeauthoryear{{Bialy} \& {Sternberg}}{{Bialy} \&
  {Sternberg}}{2019}]{Bialy2019ThermalGas}
{Bialy} S.,  {Sternberg} A.,  2019, \mn@doi [ApJ] {10.3847/1538-4357/ab2fd1},
  \href {https://ui.adsabs.harvard.edu/abs/2019ApJ...881..160B} {881, 160}

\bibitem[\protect\citeauthoryear{Bienner et~al.,}{Bienner
  et~al.}{2005}]{Bienner2005LaboratoryChemistry}
Bienner L.,  et~al., 2005, IAUS, 231, 217

\bibitem[\protect\citeauthoryear{Blair, Davidson, Fesen, Uomoto, MacAlpine  \&
  Henry}{Blair et~al.}{1997}]{Blair1997Overview}
Blair W.~P.,  Davidson K.,  Fesen R.~A.,  Uomoto A.,  MacAlpine G.~M.,   Henry
  R. B.~C.,  1997, \mn@doi [ApJ Supplement Series] {10.1086/312986}, 109, 473

\bibitem[\protect\citeauthoryear{Bloch, {Bloch}  \& {F.}}{Bloch
  et~al.}{1933}]{Bloch1933ZurMaterie}
Bloch F.,  {Bloch}  {F.} 1933, \mn@doi [AnP] {10.1002/ANDP.19334080303}, 408,
  285

\bibitem[\protect\citeauthoryear{Bocchio, Micelotta, Gautier, Jones, Bocchio,
  Micelotta, Gautier  \& Jones}{Bocchio et~al.}{2012}]{Bocchio2012SmallStudy}
Bocchio M.,  Micelotta E.~R.,  Gautier A.~L.,  Jones A.~P.,  Bocchio M.,
  Micelotta E.~R.,  Gautier A.~L.,   Jones A.~P.,  2012, \mn@doi [A{\&}A]
  {10.1051/0004-6361/201219705}, 545, A124

\bibitem[\protect\citeauthoryear{{Bocchio}, {Jones}  \& {Slavin}}{{Bocchio}
  et~al.}{2014}]{Bocchio2014AWaves}
{Bocchio} M.,  {Jones} A.~P.,   {Slavin} J.~D.,  2014, \mn@doi [A{\&}A]
  {10.1051/0004-6361/201424368}, \href
  {https://ui.adsabs.harvard.edu/abs/2014A&A...570A..32B} {570, A32}

\bibitem[\protect\citeauthoryear{{Bohdansky}}{{Bohdansky}}{1984}]{Bohdansky1984AIncidence}
{Bohdansky} J.,  1984, \mn@doi [Nuclear Instruments and Methods in Physics
  Research B] {10.1016/0168-583X(84)90271-4}, \href
  {https://ui.adsabs.harvard.edu/abs/1984NIMPB...2..587B} {2, 587}

\bibitem[\protect\citeauthoryear{Bohdansky, Roth, Bay, Bohdansky, Roth  \&
  Bay}{Bohdansky et~al.}{1980}]{Bohdansky1980AnSputtering}
Bohdansky J.,  Roth J.,  Bay H.~L.,  Bohdansky J.,  Roth J.,   Bay H.~L.,
  1980, \mn@doi [JAP] {10.1063/1.327954}, 51, 2861

\bibitem[\protect\citeauthoryear{Bolatto, Wolfire  \& Leroy}{Bolatto
  et~al.}{2013}]{Bolatto2013}
Bolatto A.~D.,  Wolfire M.,   Leroy A.~K.,  2013, \mn@doi [ARA{\&}A]
  {10.1146/annurev-astro-082812-140944}, 51, 207

\bibitem[\protect\citeauthoryear{Bolatto et~al.,}{Bolatto
  et~al.}{2024}]{Bolatto2024JWSTWind}
Bolatto A.~D.,  et~al., 2024, \mn@doi [ApJ] {10.3847/1538-4357/ad33c8}, 967, 63

\bibitem[\protect\citeauthoryear{Boschman, Reitsma, Cazaux, Schlath{\"{o}}lter,
  Hoekstra, Spaans  \& Gonz{\'{a}}lez-Maga{\~{n}}a}{Boschman
  et~al.}{2012}]{Boschman2012HydrogenationFormation}
Boschman L.,  Reitsma G.,  Cazaux S.,  Schlath{\"{o}}lter T.,  Hoekstra R.,
  Spaans M.,   Gonz{\'{a}}lez-Maga{\~{n}}a O.,  2012, \mn@doi [ApJL]
  {10.1088/2041-8205/761/2/L33}, 761, L33

\bibitem[\protect\citeauthoryear{Boschman, Cazaux, Spaans, Hoekstra  \&
  Schlath{\"{o}}lter}{Boschman et~al.}{2015}]{Boschman2015H2formationHydrogen}
Boschman L.,  Cazaux S.,  Spaans M.,  Hoekstra R.,   Schlath{\"{o}}lter T.,
  2015, \mn@doi [A{\&}A] {10.1051/0004-6361/201323165}, 579, A72

\bibitem[\protect\citeauthoryear{{Boulanger}, {F.}, {Boisssel}, {P.},
  {Cesarsky}, {D.}, {Ryter}  \& {C.}}{{Boulanger}
  et~al.}{1998}]{Boulanger1998TheSpectra}
{Boulanger} {F.} {Boisssel} {P.} {Cesarsky} {D.} {Ryter}  {C.} 1998, A{\&}A

\bibitem[\protect\citeauthoryear{Bout{\'{e}}raon, Habart, Ysard, Jones, Dartois
   \& Pino}{Bout{\'{e}}raon et~al.}{2019}]{Bouteraon2019NanoComponents}
Bout{\'{e}}raon T.,  Habart E.,  Ysard N.,  Jones A.~P.,  Dartois E.,   Pino
  T.,  2019, \mn@doi [A{\&}A] {10.1051/0004-6361/201834016}, 623, A135

\bibitem[\protect\citeauthoryear{Brandt \& Draine}{Brandt \&
  Draine}{2011}]{Brandt2011THELIGHT}
Brandt T.~D.,  Draine B.~T.,  2011, \mn@doi [ApJ]
  {10.1088/0004-637X/744/2/129}, 744, 129

\bibitem[\protect\citeauthoryear{Br{\'{e}}chignac et~al.,}{Br{\'{e}}chignac
  et~al.}{2014}]{Brechignac2014PhotoionizationCation}
Br{\'{e}}chignac P.,  et~al., 2014, \mn@doi [JChPh] {10.1063/1.4900427}, 141,
  164325

\bibitem[\protect\citeauthoryear{Bron, Le~Bourlot  \& Le~Petit}{Bron
  et~al.}{2014}]{Bron2014SurfaceFluctuations}
Bron E.,  Le~Bourlot J.,   Le~Petit F.,  2014, \mn@doi [A{\&}A]
  {10.1051/0004-6361/201322101}, 569, A100

\bibitem[\protect\citeauthoryear{Burke \& Hollenbach}{Burke \&
  Hollenbach}{1983}]{Burke1983TheTrapping}
Burke J.~R.,  Hollenbach D.~J.,  1983, \mn@doi [ApJ] {10.1086/160667}, 265, 223

\bibitem[\protect\citeauthoryear{Burke, Silk, Burke  \& Silk}{Burke
  et~al.}{1974}]{Burke1974DustPhysics}
Burke J.~R.,  Silk J.,  Burke J.~R.,   Silk J.,  1974, \mn@doi [ApJ]
  {10.1086/152840}, 190, 1

\bibitem[\protect\citeauthoryear{Burkhardt et~al.,}{Burkhardt
  et~al.}{2021}]{Burkhardt2021DiscoveryTMC-1}
Burkhardt A.~M.,  et~al., 2021, \mn@doi [ApJL] {10.3847/2041-8213/abfd3a}, 913,
  L18

\bibitem[\protect\citeauthoryear{Byun et~al.,}{Byun
  et~al.}{2025}]{Byun2025HowSimulation}
Byun G.-H.,  et~al., 2025, \mn@doi [ApJ] {10.3847/1538-4357/adfed9}, 992, 92

\bibitem[\protect\citeauthoryear{Cabezas et~al.,}{Cabezas
  et~al.}{2025}]{Cabezas2025DiscoverySpace}
Cabezas C.,  et~al., 2025, \mn@doi [A{\&}A] {10.1051/0004-6361/202556687}, 701,
  L8

\bibitem[\protect\citeauthoryear{Calura, Pipino  \& Matteucci}{Calura
  et~al.}{2007}]{Calura2007TheTypes}
Calura F.,  Pipino A.,   Matteucci F.,  2007, \mn@doi [A{\&}A]
  {10.1051/0004-6361:20078090}, 479, 669

\bibitem[\protect\citeauthoryear{Calzetti et~al.,}{Calzetti
  et~al.}{2007}]{Calzetti2007TheIndicators}
Calzetti D.,  et~al., 2007, \mn@doi [ApJ] {10.1086/520082}, 666, 870

\bibitem[\protect\citeauthoryear{Camps \& Baes}{Camps \&
  Baes}{2015}]{Camps2015SKIRT:Architecture}
Camps P.,  Baes M.,  2015, \mn@doi [Astron. Comput.]
  {10.1016/j.ascom.2014.10.004}, 9, 20

\bibitem[\protect\citeauthoryear{Canosa et~al.,}{Canosa
  et~al.}{1994}]{Canosa1994ElectronTemperature}
Canosa A.,  et~al., 1994, \mn@doi [CPL] {10.1016/0009-2614(94)00908-2}, 228, 26

\bibitem[\protect\citeauthoryear{Canosa, Laub{\'{e}}, Rebrion, Pasquerault,
  Gomet  \& Rowe}{Canosa et~al.}{1995}]{Canosa1995ReactionTemperature}
Canosa A.,  Laub{\'{e}} S.,  Rebrion C.,  Pasquerault D.,  Gomet J.~C.,   Rowe
  B.~R.,  1995, \mn@doi [Chem. Phys. Lett., Volume 245, Issue 4, p. 407-414.]
  {10.1016/0009-2614(95)01040-G}, 245, 407

\bibitem[\protect\citeauthoryear{Cardelli, Clayton, Mathis, Cardelli, Clayton
  \& Mathis}{Cardelli et~al.}{1989}]{Cardelli1989TheExtinction}
Cardelli J.~A.,  Clayton G.~C.,  Mathis J.~S.,  Cardelli J.~A.,  Clayton G.~C.,
    Mathis J.~S.,  1989, \mn@doi [ApJ] {10.1086/167900}, 345, 245

\bibitem[\protect\citeauthoryear{Carelli, Grassi  \& Gianturco}{Carelli
  et~al.}{2013}]{Carelli2013ElectronProperties}
Carelli F.,  Grassi T.,   Gianturco F.~A.,  2013, \mn@doi [A{\&}A]
  {10.1051/0004-6361/201219990}, 549, A103

\bibitem[\protect\citeauthoryear{Castellanos, Candian, Zhen, Linnartz  \&
  Tielens}{Castellanos et~al.}{2018a}]{Castellanos2018PhotoinducedH2-loss}
Castellanos P.,  Candian A.,  Zhen J.,  Linnartz H.,   Tielens A.~G.,  2018a,
  \mn@doi [A{\&}A] {10.1051/0004-6361/201833220}, 616, A166

\bibitem[\protect\citeauthoryear{Castellanos, Candian, Andrews  \&
  Tielens}{Castellanos
  et~al.}{2018b}]{Castellanos2018PhotoinducedDehydrogenation}
Castellanos P.,  Candian A.,  Andrews H.,   Tielens A. G. G.~M.,  2018b,
  \mn@doi [A{\&}A] {10.1051/0004-6361/201833221}, 616, A167

\bibitem[\protect\citeauthoryear{Cau}{Cau}{2002}]{Cau2002FormationDimers}
Cau P.,  2002, \mn@doi [A{\&}A] {10.1051/0004-6361:20020924}, 392, 203

\bibitem[\protect\citeauthoryear{Cazaux \& Tielens}{Cazaux \&
  Tielens}{2002}]{Cazaux2002MolecularMedium}
Cazaux S.,  Tielens A. G. G.~M.,  2002, \mn@doi [ApJ]
  {10.1086/342607/FULLTEXT/}, 575, L29

\bibitem[\protect\citeauthoryear{Cazaux \& Tielens}{Cazaux \&
  Tielens}{2004}]{Cazaux2004Surfaces}
Cazaux S.,  Tielens A. G. G.~M.,  2004, \mn@doi [ApJ]
  {10.1086/381775/FULLTEXT/}, 604, 222

\bibitem[\protect\citeauthoryear{Cazaux, Tielens, Cazaux  \& Tielens}{Cazaux
  et~al.}{2004}]{Cazaux2004HSUB2/SUBSurfaces}
Cazaux S.,  Tielens A. G. G.~M.,  Cazaux S.,   Tielens A. G. G.~M.,  2004,
  \mn@doi [ApJ] {10.1086/381775}, 604, 222

\bibitem[\protect\citeauthoryear{Cernicharo, Ag{\'{u}}ndez, Cabezas, Tercero,
  Marcelino, Pardo  \& De~Vicente}{Cernicharo
  et~al.}{2021a}]{Cernicharo2021PureIndene}
Cernicharo J.,  Ag{\'{u}}ndez M.,  Cabezas C.,  Tercero B.,  Marcelino N.,
  Pardo J.~R.,   De~Vicente P.,  2021a, \mn@doi [A{\&}A]
  {10.1051/0004-6361/202141156}, 649, 15

\bibitem[\protect\citeauthoryear{Cernicharo, Ag{\'{u}}ndez, Kaiser, Cabezas,
  Tercero, Marcelino, Pardo  \& De~Vicente}{Cernicharo
  et~al.}{2021b}]{Cernicharo2021DiscoveryCycles}
Cernicharo J.,  Ag{\'{u}}ndez M.,  Kaiser R.~I.,  Cabezas C.,  Tercero B.,
  Marcelino N.,  Pardo J.~R.,   De~Vicente P.,  2021b, \mn@doi [A{\&}A]
  {10.1051/0004-6361/202142226}, 655, 1

\bibitem[\protect\citeauthoryear{Cernicharo et~al.,}{Cernicharo
  et~al.}{2022}]{Cernicharo2022DiscoverySurvey}
Cernicharo J.,  et~al., 2022, \mn@doi [A{\&}A] {10.1051/0004-6361/202244399},
  663, L9

\bibitem[\protect\citeauthoryear{Chaabouni, Bergeron, Baouche, Dulieu, Matar,
  Congiu, Gavilan  \& Lemaire}{Chaabouni
  et~al.}{2012}]{Chaabouni2012StickingConditions}
Chaabouni H.,  Bergeron H.,  Baouche S.,  Dulieu F.,  Matar E.,  Congiu E.,
  Gavilan L.,   Lemaire J.~L.,  2012, \mn@doi [A{\&}A]
  {10.1051/0004-6361/201117409}, 538, A128

\bibitem[\protect\citeauthoryear{{Chastenet} et~al.,}{{Chastenet}
  et~al.}{2023}]{Chastenet2023PHANGS-JWSTMetallicity}
{Chastenet} J.,  et~al., 2023, \mn@doi [ApJL] {10.3847/2041-8213/acadd7}, \href
  {https://ui.adsabs.harvard.edu/abs/2023ApJ...944L..11C} {944, L11}

\bibitem[\protect\citeauthoryear{Chen, Li, Quan, Zhang, Chang, Li  \&
  Xiao}{Chen et~al.}{2022}]{Chen2022ChemicalPhase}
Chen L.-F.,  Li D.,  Quan D.,  Zhang X.,  Chang Q.,  Li X.,   Xiao L.,  2022,
  \mn@doi [ApJ] {10.3847/1538-4357/AC5A45}, 928, 175

\bibitem[\protect\citeauthoryear{Cherchneff, Barker, Tielens, Cherchneff,
  Barker  \& Tielens}{Cherchneff
  et~al.}{1992}]{Cherchneff1992PolycyclicEnvelopes}
Cherchneff I.,  Barker J.~R.,  Tielens A. G. G.~M.,  Cherchneff I.,  Barker
  J.~R.,   Tielens A. G. G.~M.,  1992, \mn@doi [ApJ] {10.1086/172059}, 401, 269

\bibitem[\protect\citeauthoryear{Chiar \& Tielens}{Chiar \&
  Tielens}{2005}]{Chiar2005PixieMedium}
Chiar J.~E.,  Tielens A. G. G.~M.,  2005, \mn@doi [ApJ] {10.1086/498406}, 637,
  774

\bibitem[\protect\citeauthoryear{{Choban}, {Kere{\v{s}}}, {Hopkins},
  {Sandstrom}, {Hayward}  \& {Faucher-Gigu{\`e}re}}{{Choban}
  et~al.}{2022}]{Choban2022TheFIRE}
{Choban} C.~R.,  {Kere{\v{s}}} D.,  {Hopkins} P.~F.,  {Sandstrom} K.~M.,
  {Hayward} C.~C.,   {Faucher-Gigu{\`e}re} C.-A.,  2022, \mn@doi [MNRAS]
  {10.1093/mnras/stac1542}, \href
  {https://ui.adsabs.harvard.edu/abs/2022MNRAS.514.4506C} {514, 4506}

\bibitem[\protect\citeauthoryear{Choban, Kere{\v{s}}, Sandstrom, Hopkins,
  Hayward  \& Faucher-Gigu{\`{e}}re}{Choban et~al.}{2024}]{Choban2024AGalaxies}
Choban C.~R.,  Kere{\v{s}} D.,  Sandstrom K.~M.,  Hopkins P.~F.,  Hayward
  C.~C.,   Faucher-Gigu{\`{e}}re C.~A.,  2024, \mn@doi [MNRAS]
  {10.1093/mnras/stae716}, 529, 2356

\bibitem[\protect\citeauthoryear{Chokshi, Tielens, Hollenbach, Chokshi, Tielens
   \& Hollenbach}{Chokshi et~al.}{1993}]{Chokshi1993DustCoagulation}
Chokshi A.,  Tielens A. G. G.~M.,  Hollenbach D.,  Chokshi A.,  Tielens A. G.
  G.~M.,   Hollenbach D.,  1993, \mn@doi [ApJ] {10.1086/172562}, 407, 806

\bibitem[\protect\citeauthoryear{Chown et~al.,}{Chown
  et~al.}{2024}]{Chown2024PDRs4All:Bar}
Chown R.,  et~al., 2024, \mn@doi [A{\&}A] {10.1051/0004-6361/202346662}, 685,
  A75

\bibitem[\protect\citeauthoryear{Colman et~al.,}{Colman
  et~al.}{2024}]{Colman2024CloudMedium}
Colman T.,  et~al., 2024, \mn@doi [A{\&}A] {10.1051/0004-6361/202348983}, 686,
  A155

\bibitem[\protect\citeauthoryear{Commer{\c{c}}on, Lebreuilly, Price, Lovascio,
  Laibe  \& Hennebelle}{Commer{\c{c}}on
  et~al.}{2023}]{Commercon2023DynamicsImplementations}
Commer{\c{c}}on B.,  Lebreuilly U.,  Price D.~J.,  Lovascio F.,  Laibe G.,
  Hennebelle P.,  2023, \mn@doi [A{\&}A] {10.1051/0004-6361/202245141}, 671,
  A128

\bibitem[\protect\citeauthoryear{Compi{\`{e}}gne et~al.,}{Compi{\`{e}}gne
  et~al.}{2010}]{Compiegne2010TheDustEM}
Compi{\`{e}}gne M.,  et~al., 2010, \mn@doi [A{\&}A]
  {10.1051/0004-6361/201015292}, 525, A103

\bibitem[\protect\citeauthoryear{Cosslett}{Cosslett}{1978}]{Cosslett1978RadiationReview}
Cosslett V.~E.,  1978, \mn@doi [J. Microsc.]
  {10.1111/J.1365-2818.1978.TB02454.X}, 113, 113

\bibitem[\protect\citeauthoryear{Costantini, Predehl, Costantini  \&
  Predehl}{Costantini et~al.}{2001}]{Costantini2001ChandraProperties}
Costantini E.,  Predehl P.,  Costantini E.,   Predehl P.,  2001, tysc, p.~34

\bibitem[\protect\citeauthoryear{Crenny \& Federman}{Crenny \&
  Federman}{2003}]{Crenny2003ReanalysisMonoxide}
Crenny T.,  Federman S.~R.,  2003, \mn@doi [ApJ] {10.1086/382231}, 605, 278

\bibitem[\protect\citeauthoryear{Cuppen \& Herbst}{Cuppen \&
  Herbst}{2007}]{Cuppen2007SimulationGrains}
Cuppen H.~M.,  Herbst E.,  2007, \mn@doi [ApJ] {10.1086/521014}, 668, 294

\bibitem[\protect\citeauthoryear{D'Hendecourt, Leger, D'Hendecourt  \&
  Leger}{D'Hendecourt et~al.}{1987}]{DHendecourt1987EffectGas}
D'Hendecourt L.~B.,  Leger A.,  D'Hendecourt L.~B.,   Leger A.,  1987, A{\&}A,
  180, L9

\bibitem[\protect\citeauthoryear{Dalgarno, McCray, Dalgarno  \&
  McCray}{Dalgarno et~al.}{1972}]{Dalgarno1972HeatingRegions}
Dalgarno A.,  McCray R.~A.,  Dalgarno A.,   McCray R.~A.,  1972, \mn@doi
  [ARA{\&}A] {10.1146/ANNUREV.AA.10.090172.002111}, 10, 375

\bibitem[\protect\citeauthoryear{De~Cia, Ledoux, Mattsson, Petitjean, Srianand,
  Gavignaud  \& Jenkins}{De~Cia et~al.}{2016}]{DeCia2016Dust-depletionGalaxy}
De~Cia A.,  Ledoux C.,  Mattsson L.,  Petitjean P.,  Srianand R.,  Gavignaud
  I.,   Jenkins E.~B.,  2016, \mn@doi [A{\&}A] {10.1051/0004-6361/201527895},
  596

\bibitem[\protect\citeauthoryear{De~Vis et~al.,}{De~Vis
  et~al.}{2017}]{DeVis2017Herschel-ATLAS:Relations}
De~Vis P.,  et~al., 2017, \mn@doi [MNRAS] {10.1093/mnras/stw2501}, 464, 4680

\bibitem[\protect\citeauthoryear{De~Vis et~al.,}{De~Vis
  et~al.}{2019}]{DeVis2019ARatios}
De~Vis P.,  et~al., 2019, \mn@doi [A{\&}A] {10.1051/0004-6361/201834444}, 623,
  A5

\bibitem[\protect\citeauthoryear{Dopcke, Glover, Clark  \& Klessen}{Dopcke
  et~al.}{2011}]{Dopcke2011THECLOUDS}
Dopcke G.,  Glover S.~C.,  Clark P.~C.,   Klessen R.~S.,  2011, \mn@doi [ApJL]
  {10.1088/2041-8205/729/1/L3}, 729, L3

\bibitem[\protect\citeauthoryear{Dopita, Sutherland, Dopita  \&
  Sutherland}{Dopita et~al.}{2000}]{Dopita2000TheNebulae}
Dopita M.~A.,  Sutherland R.~S.,  Dopita M.~A.,   Sutherland R.~S.,  2000,
  \mn@doi [ApJ] {10.1086/309241}, 539, 742

\bibitem[\protect\citeauthoryear{Dorschner \& Henning}{Dorschner \&
  Henning}{1995}]{Dorschner1995DustGalaxy}
Dorschner J.,  Henning T.,  1995, \mn@doi [A{\&}ARv] {10.1007/BF00873686}, 6,
  271

\bibitem[\protect\citeauthoryear{Dorschner, {Dorschner}  \& {J.}}{Dorschner
  et~al.}{1982}]{Dorschner1982InterstellarCollisions}
Dorschner J.,  {Dorschner}  {J.} 1982, \mn@doi [Ap{\&}SS] {10.1007/BF00676156},
  81, 323

\bibitem[\protect\citeauthoryear{Draine}{Draine}{1978}]{Draine1978PhotoelectricGas.}
Draine B.~T.,  1978, \mn@doi [ApJ] {10.1086/190513}, 36, 595

\bibitem[\protect\citeauthoryear{Draine}{Draine}{2003a}]{Draine2003InterstellarGrains}
Draine B.,  2003a, \mn@doi [ARA{\&}A] {10.1146/annurev.astro.41.011802.094840},
  41, 241

\bibitem[\protect\citeauthoryear{Draine}{Draine}{2003b}]{Draine2003ScatteringUltraviolet}
Draine B.~T.,  2003b, \mn@doi [ApJ] {10.1086/379118}, 598, 1017

\bibitem[\protect\citeauthoryear{Draine}{Draine}{2011}]{Draine2011PhysicsMedium}
Draine B.~T.,  2011, {Physics of the Interstellar and Intergalactic Medium}.
Princeton University Press, \mn@doi{10.2307/j.ctvcm4hzr}, \url
  {http://www.jstor.org/stable/10.2307/j.ctvcm4hzr}

\bibitem[\protect\citeauthoryear{Draine \& Li}{Draine \&
  Li}{2001}]{Draine2001InfraredGrains}
Draine B.~T.,  Li A.,  2001, \mn@doi [ApJ] {10.1086/320227}, 551, 807

\bibitem[\protect\citeauthoryear{Draine \& Li}{Draine \&
  Li}{2007}]{Draine2007Era}
Draine B.~T.,  Li A.,  2007, \mn@doi [ApJ] {10.1086/511055}, 657, 810

\bibitem[\protect\citeauthoryear{Draine \& Salpeter}{Draine \&
  Salpeter}{1979}]{Draine1979ONGAS}
Draine B.~T.,  Salpeter E.~E.,  1979, ApJ, 231, 77

\bibitem[\protect\citeauthoryear{Draine \& Sutin}{Draine \&
  Sutin}{1987}]{Draine1987CollisionalGrains}
Draine B.~T.,  Sutin B.,  1987, \mn@doi [ApJ] {10.1086/165596}, 320, 803

\bibitem[\protect\citeauthoryear{Draine \& Tan}{Draine \&
  Tan}{2003}]{Draine2003TheDust}
Draine B.~T.,  Tan J.~C.,  2003, \mn@doi [ApJ] {10.1086/376855}, 594, 347

\bibitem[\protect\citeauthoryear{Draine, Salpeter, Draine  \& Salpeter}{Draine
  et~al.}{1979}]{Draine1979DestructionDust.}
Draine B.~T.,  Salpeter E.~E.,  Draine B.~T.,   Salpeter E.~E.,  1979, \mn@doi
  [ApJ] {10.1086/157206}, 231, 438

\bibitem[\protect\citeauthoryear{Draine, Lee, Draine  \& Lee}{Draine
  et~al.}{1984}]{Draine1984OpticalGrains}
Draine B.~T.,  Lee H.~M.,  Draine B.~T.,   Lee H.~M.,  1984, \mn@doi [ApJ]
  {10.1086/162480}, 285, 89

\bibitem[\protect\citeauthoryear{Draine et~al.,}{Draine
  et~al.}{2007}]{Draine2007DustSample}
Draine B.~T.,  et~al., 2007, \mn@doi [ApJ] {10.1086/518306}, 663, 866

\bibitem[\protect\citeauthoryear{Draine, Li, Hensley, Hunt, Sandstrom  \&
  Smith}{Draine et~al.}{2021}]{Draine2021ExcitationIntensity}
Draine B.~T.,  Li A.,  Hensley B.~S.,  Hunt L.~K.,  Sandstrom K.,   Smith
  J.-D.~T.,  2021, \mn@doi [ApJ] {10.3847/1538-4357/abff51}, 917, 3

\bibitem[\protect\citeauthoryear{Dubois et~al.,}{Dubois
  et~al.}{2024}]{Dubois2024GalaxiesSimulations}
Dubois Y.,  et~al., 2024, \mn@doi [A{\&}A] {10.1051/0004-6361/202449784}, 687,
  A240

\bibitem[\protect\citeauthoryear{Dwek}{Dwek}{1998}]{Dwek1998TheGalaxy}
Dwek E.,  1998, \mn@doi [ApJ] {10.1086/305829}, 501, 643

\bibitem[\protect\citeauthoryear{{Dwek} \& {Werner}}{{Dwek} \&
  {Werner}}{1981}]{Dwek1981TheCondensates}
{Dwek} E.,  {Werner} M.~W.,  1981, \mn@doi [ApJ] {10.1086/159138}, \href
  {https://ui.adsabs.harvard.edu/abs/1981ApJ...248..138D} {248, 138}

\bibitem[\protect\citeauthoryear{{Egorov} et~al.,}{{Egorov}
  et~al.}{2023}]{Egorov2023PHANGS-JWSTMUSE}
{Egorov} O.~V.,  et~al., 2023, \mn@doi [\apjl] {10.3847/2041-8213/acac92},
  \href {https://ui.adsabs.harvard.edu/abs/2023ApJ...944L..16E} {944, L16}

\bibitem[\protect\citeauthoryear{Egorov et~al.,}{Egorov
  et~al.}{2025}]{Egorov2025PolycyclicGalaxies}
Egorov O.~V.,  et~al., 2025, \mn@doi [A{\&}A] {10.1051/0004-6361/202556427},
  703, A103

\bibitem[\protect\citeauthoryear{Elmegreen \& Scalo}{Elmegreen \&
  Scalo}{2004}]{Elmegreen2004InterstellarProcesses}
Elmegreen B.~G.,  Scalo J.,  2004, \mn@doi [ARA{\&}A]
  {10.1146/annurev.astro.41.011802.094859}, 42, 211

\bibitem[\protect\citeauthoryear{Elyajouri et~al.,}{Elyajouri
  et~al.}{2024}]{Elyajouri2024PDRs4AllBar}
Elyajouri M.,  et~al., 2024, \mn@doi [A{\&}A] {10.1051/0004-6361/202348728},
  685, A76

\bibitem[\protect\citeauthoryear{Engelbracht, Gordon, Rieke, Werner, Dale  \&
  Latter}{Engelbracht et~al.}{2005}]{Engelbracht2005MetallicityGalaxies}
Engelbracht C.~W.,  Gordon K.~D.,  Rieke G.~H.,  Werner M.~W.,  Dale D.~A.,
  Latter W.~B.,  2005, \mn@doi [ApJ] {10.1086/432613}, 628, L29

\bibitem[\protect\citeauthoryear{{Ercolano}, {Young}, {Drake}  \&
  {Raymond}}{{Ercolano} et~al.}{2008}]{Ercolano20083DRegions}
{Ercolano} B.,  {Young} P.~R.,  {Drake} J.~J.,   {Raymond} J.~C.,  2008,
  \mn@doi [ApJS] {10.1086/524378}, \href
  {https://ui.adsabs.harvard.edu/abs/2008ApJS..175..534E} {175, 534}

\bibitem[\protect\citeauthoryear{Falgarone et~al.,}{Falgarone
  et~al.}{2017}]{Falgarone2017}
Falgarone E.,  et~al., 2017, \mn@doi [Nature] {10.1038/nature23298}, 548, 430

\bibitem[\protect\citeauthoryear{Federman, Huntress, Prasad, Federman, Huntress
   \& Prasad}{Federman et~al.}{1990}]{Federman1990ModelingChemistry}
Federman S.~R.,  Huntress W.~T. J.,  Prasad S.~S.,  Federman S.~R.,  Huntress
  W.~T. J.,   Prasad S.~S.,  1990, \mn@doi [ApJ] {10.1086/168711}, 354, 504

\bibitem[\protect\citeauthoryear{Federrath \& Banerjee}{Federrath \&
  Banerjee}{2015}]{Federrath2015TheTurbulence}
Federrath C.,  Banerjee S.,  2015, \mn@doi [MNRAS] {10.1093/mnras/stv180}, 448,
  3297

\bibitem[\protect\citeauthoryear{Federrath \& Klessen}{Federrath \&
  Klessen}{2012}]{Federrath2012OnClouds}
Federrath C.,  Klessen R.~S.,  2012, \mn@doi [ApJ]
  {10.1088/0004-637X/763/1/51}, 763, 51

\bibitem[\protect\citeauthoryear{Ferland et~al.,}{Ferland
  et~al.}{2017}]{Ferland2017TheCloudy}
Ferland G.~J.,  et~al., 2017, \mn@doi [RMxAA] {10.48550/ARXIV.1705.10877}, 53,
  385

\bibitem[\protect\citeauthoryear{Ferrara}{Ferrara}{2024}]{Ferrara2024}
Ferrara A.,  2024, \mn@doi [A{\&}A] {10.1051/0004-6361/202450944}, 689, A310

\bibitem[\protect\citeauthoryear{Ferrara, Pallottini  \& Dayal}{Ferrara
  et~al.}{2022}]{Ferrara2022OnJWST}
Ferrara A.,  Pallottini A.,   Dayal P.,  2022, \mn@doi [MNRAS]
  {10.1093/mnras/stad1095}, 522, 3986

\bibitem[\protect\citeauthoryear{Fitzpatrick}{Fitzpatrick}{1999}]{Fitzpatrick1999CorrectingExtinction}
Fitzpatrick E.~L.,  1999, \mn@doi [PASP] {10.1086/316293}, 111, 63

\bibitem[\protect\citeauthoryear{Foster, Mandel, Pineda, Covey, Arce  \&
  Goodman}{Foster et~al.}{2013}]{Foster2013EvidencePerseus}
Foster J.~B.,  Mandel K.~S.,  Pineda J.~E.,  Covey K.~R.,  Arce H.~G.,
  Goodman A.~A.,  2013, \mn@doi [MNRAS] {10.1093/MNRAS/STS144}, 428, 1606

\bibitem[\protect\citeauthoryear{Fraser, McCoustra  \& Williams}{Fraser
  et~al.}{2002}]{Fraser2002TheUniverse}
Fraser H.~J.,  McCoustra M.~R.,   Williams D.~A.,  2002, \mn@doi [A&G]
  {10.1046/J.1468-4004.2002.43210.X/2/43-2-2.10-FIG011.JPEG}, 43, 10

\bibitem[\protect\citeauthoryear{Frenklach}{Frenklach}{2002}]{Frenklach2002ReactionFlames}
Frenklach M.,  2002, \mn@doi [PCCP] {10.1039/b110045a}, 4, 2028

\bibitem[\protect\citeauthoryear{{Frenklach}, {Michael}, {Feigelson}  \&
  D.}{{Frenklach} et~al.}{1989}]{Frenklach1989FormationEnvelopes}
{Frenklach} {Michael} {Feigelson}  D. E.,  1989, \mn@doi [ApJ]
  {10.1086/167501}, 341, 372

\bibitem[\protect\citeauthoryear{Fromang \& Nelson}{Fromang \&
  Nelson}{2009}]{Fromang2009GlobalSettling}
Fromang S.,  Nelson R.~P.,  2009, \mn@doi [A{\&}A]
  {10.1051/0004-6361/200811220}, 496, 597

\bibitem[\protect\citeauthoryear{Galametz et~al.,}{Galametz
  et~al.}{2016}]{Galametz2016TheN11}
Galametz M.,  et~al., 2016, \mn@doi [MNRAS] {10.1093/mnras/stv2773}, 456, 1767

\bibitem[\protect\citeauthoryear{Galliano, Dwek  \& Chanial}{Galliano
  et~al.}{2008a}]{Galliano2008StellarGalaxies}
Galliano F.,  Dwek E.,   Chanial P.,  2008a, \mn@doi [ApJ]
  {10.1086/523621/XML}, 672, 214

\bibitem[\protect\citeauthoryear{Galliano, Madden, Tielens, Peeters  \&
  Jones}{Galliano et~al.}{2008b}]{Galliano2008VariationsGalaxies}
Galliano F.,  Madden S.~C.,  Tielens A. G. G.~M.,  Peeters E.,   Jones A.~P.,
  2008b, \mn@doi [ApJ] {10.1086/587051}, 679, 310

\bibitem[\protect\citeauthoryear{Geen, Rosdahl, Blaizot, Devriendt  \&
  Slyz}{Geen et~al.}{2015}]{Geen2015AStar}
Geen S.,  Rosdahl J.,  Blaizot J.,  Devriendt J.,   Slyz A.,  2015, \mn@doi
  [MNRAS] {10.1093/mnras/stv251}, 448, 3248

\bibitem[\protect\citeauthoryear{Genzel et~al.,}{Genzel
  et~al.}{1997}]{Genzel1997WhatGalaxies}
Genzel R.,  et~al., 1997, \mn@doi [ApJ] {10.1086/305576}, 498, 579

\bibitem[\protect\citeauthoryear{Gerin et~al.,}{Gerin
  et~al.}{2015}]{Gerin2015CPlane}
Gerin M.,  et~al., 2015, \mn@doi [A{\&}A] {10.1051/0004-6361/201424349}, 573,
  A30

\bibitem[\protect\citeauthoryear{Gjergo, Granato, Murante, Ragone-Figueroa,
  Tornatore  \& Borgani}{Gjergo et~al.}{2018}]{Gjergo2018DustSimulations}
Gjergo E.,  Granato G.~L.,  Murante G.,  Ragone-Figueroa C.,  Tornatore L.,
  Borgani S.,  2018, \mn@doi [MNRAS] {10.1093/MNRAS/STY1564}, 479, 2588

\bibitem[\protect\citeauthoryear{Glatzle, Graziani  \& Ciardi}{Glatzle
  et~al.}{2022}]{Glatzle2022RadiativeRegions}
Glatzle M.,  Graziani L.,   Ciardi B.,  2022, \mn@doi [MNRAS]
  {10.1093/mnras/stab3459}, 510, 1068

\bibitem[\protect\citeauthoryear{Glover \& Clark}{Glover \&
  Clark}{2012}]{Glover2012IsFormation}
Glover S.~C.,  Clark P.~C.,  2012, \mn@doi [MNRAS]
  {10.1111/j.1365-2966.2011.19648.x}, 421, 9

\bibitem[\protect\citeauthoryear{Glover \& Jappsen}{Glover \&
  Jappsen}{2007}]{Glover2007StarDensities}
Glover S. C.~O.,  Jappsen A.,  2007, \mn@doi [ApJ] {10.1086/519445/XML}, 666, 1

\bibitem[\protect\citeauthoryear{Glover \& Mac~Low}{Glover \&
  Mac~Low}{2007}]{Glover2007SimulatingConditions}
Glover S. C.~O.,  Mac~Low M.,  2007, \mn@doi [ApJS] {10.1086/512238}, 169, 239

\bibitem[\protect\citeauthoryear{Glover, Federrath, Low  \& Klessen}{Glover
  et~al.}{2010}]{Glover2010ModellingMedium}
Glover S.~C.,  Federrath C.,  Low M.~M.,   Klessen R.~S.,  2010, \mn@doi
  [MNRAS] {10.1111/j.1365-2966.2009.15718.x}, 404, 2

\bibitem[\protect\citeauthoryear{Gnedin, Tassis  \& Kravtsov}{Gnedin
  et~al.}{2009}]{Gnedin2009ModelingSimulations}
Gnedin N.~Y.,  Tassis K.,   Kravtsov A.~V.,  2009, \mn@doi [ApJ]
  {10.1088/0004-637X/697/1/55}, 697, 55

\bibitem[\protect\citeauthoryear{Goicoechea et~al.,}{Goicoechea
  et~al.}{2025}]{Goicoechea2025PDRs4AllHydrocarbons}
Goicoechea J.~R.,  et~al., 2025, \mn@doi [A{\&}A]
  {10.1051/0004-6361/202453350}, 696, A100

\bibitem[\protect\citeauthoryear{Gomez et~al.,}{Gomez
  et~al.}{2012}]{Gomez2012DustHerschel}
Gomez H.~L.,  et~al., 2012, \mn@doi [MNRAS] {10.1111/j.1365-2966.2011.20272.x},
  420, 3557

\bibitem[\protect\citeauthoryear{Gong, Ostriker  \& Wolfire}{Gong
  et~al.}{2019}]{Gong2019AISM}
Gong M.,  Ostriker E.~C.,   Wolfire M.~G.,  2019, \mn@doi [ApJ]
  {10.3847/1538-4357/aa7561}, 843, 38

\bibitem[\protect\citeauthoryear{Goumans, {Goumans}  \& M.}{Goumans
  et~al.}{2011}]{Goumans2011HydrogenTunnelling}
Goumans T. P.~M.,  {Goumans}  M. T.~P.,  2011, \mn@doi [MNRAS]
  {10.1111/J.1365-2966.2011.18924.X}, 415, 3129

\bibitem[\protect\citeauthoryear{Graham et~al.,}{Graham
  et~al.}{1987}]{Graham1987IRASRemnants}
Graham J.~R.,  et~al., 1987, \mn@doi [ApJ] {10.1086/165438}, 319, 126

\bibitem[\protect\citeauthoryear{Granato et~al.,}{Granato
  et~al.}{2021}]{Granato2021DustFormation}
Granato G.~L.,  et~al., 2021, \mn@doi [MNRAS] {10.1093/MNRAS/STAB362}, 503, 511

\bibitem[\protect\citeauthoryear{Grassi, Bovino, Schleicher, Prieto, Seifried,
  Simoncini  \& Gianturco}{Grassi et~al.}{2014}]{Grassi2014KROME-aSimulations}
Grassi T.,  Bovino S.,  Schleicher D.~R.,  Prieto J.,  Seifried D.,  Simoncini
  E.,   Gianturco F.~A.,  2014, \mn@doi [MNRAS] {10.1093/mnras/stu114}, 439,
  2386

\bibitem[\protect\citeauthoryear{Guhathakurta, Draine, Guhathakurta  \&
  Draine}{Guhathakurta et~al.}{1989}]{Guhathakurta1989TemperatureGrains}
Guhathakurta P.,  Draine B.~T.,  Guhathakurta P.,   Draine B.~T.,  1989,
  \mn@doi [ApJ] {10.1086/167899}, 345, 230

\bibitem[\protect\citeauthoryear{Guillet, Hennebelle, Pineau~des For{\^{e}}ts,
  Marcowith, Commer{\c{c}}on  \& Marchand}{Guillet
  et~al.}{2020}]{Guillet2020DustCollapse}
Guillet V.,  Hennebelle P.,  Pineau~des For{\^{e}}ts G.,  Marcowith A.,
  Commer{\c{c}}on B.,   Marchand P.,  2020, \mn@doi [A{\&}A]
  {10.1051/0004-6361/201937387}, 643, A17

\bibitem[\protect\citeauthoryear{Haardt \& Madau}{Haardt \&
  Madau}{2012}]{Haardt2012RadiativeBackground}
Haardt F.,  Madau P.,  2012, \mn@doi [ApJ] {10.1088/0004-637X/746/2/125}, 746,
  125

\bibitem[\protect\citeauthoryear{Habart et~al.,}{Habart
  et~al.}{2003a}]{Habart2003HSUB2/SUBCloud}
Habart E.,  et~al., 2003a, \mn@doi [A{\&}A] {10.1051/0004-6361:20021489}, 397,
  623

\bibitem[\protect\citeauthoryear{Habart, Boulanger, Verstraete, Walmsley  \&
  Forets}{Habart et~al.}{2003b}]{Habart2003SomeRegions}
Habart E.,  Boulanger F.,  Verstraete L.,  Walmsley C.~M.,   Forets G. P.~d.,
  2003b, \mn@doi [A{\&}A] {10.1051/0004-6361:20031659}, 414, 531

\bibitem[\protect\citeauthoryear{Habing}{Habing}{1968}]{Habing1968TheA}
Habing H.~J.,  1968, BAIN, 19, 421

\bibitem[\protect\citeauthoryear{Hadjar, Hoekstra, Morgenstern,
  Schlath{\"{o}}lter, Hadjar, Hoekstra, Morgenstern  \&
  Schlath{\"{o}}lter}{Hadjar et~al.}{2001}]{Hadjar2001ProjectileFullerenes}
Hadjar O.,  Hoekstra R.,  Morgenstern R.,  Schlath{\"{o}}lter T.,  Hadjar O.,
  Hoekstra R.,  Morgenstern R.,   Schlath{\"{o}}lter T.,  2001, \mn@doi [PhRvA]
  {10.1103/PHYSREVA.63.033201}, 63, 033201

\bibitem[\protect\citeauthoryear{Harris et~al.,}{Harris
  et~al.}{2020}]{Harris2020ArrayNumPy}
Harris C.~R.,  et~al., 2020, \mn@doi [Nat.] {10.1038/s41586-020-2649-2}, 585,
  357

\bibitem[\protect\citeauthoryear{Hellyer}{Hellyer}{1970}]{Hellyer1970TheAsteroids}
Hellyer B.,  1970, \mn@doi [MNRAS, Vol. 148, p. 383 (1970)]
  {10.1093/MNRAS/148.4.383}, 148, 383

\bibitem[\protect\citeauthoryear{{Helou} et~al.,}{{Helou}
  et~al.}{2001}]{Helou2001EvidenceHydrocarbons}
{Helou} et~al., 2001, \mn@doi [ApJL] {10.1086/318916}, 548, L73

\bibitem[\protect\citeauthoryear{Hennebelle \& Falgarone}{Hennebelle \&
  Falgarone}{2012}]{Hennebelle2012TurbulentClouds}
Hennebelle P.,  Falgarone E.,  2012, \mn@doi [A{\&}A Review]
  {10.1007/s00159-012-0055-y}, 20, 55

\bibitem[\protect\citeauthoryear{Hennebelle, Brucy, Colman, Hennebelle, Brucy
  \& Colman}{Hennebelle et~al.}{2024}]{Hennebelle2024InefficientModel}
Hennebelle P.,  Brucy N.,  Colman T.,  Hennebelle P.,  Brucy N.,   Colman T.,
  2024, \mn@doi [arXiv] {10.48550/ARXIV.2404.17368}, p. arXiv:2404.17368

\bibitem[\protect\citeauthoryear{Henning, {Henning}  \& {Thomas}}{Henning
  et~al.}{2010}]{Henning2010CosmicSilicates}
Henning T.,  {Henning}  {Thomas} 2010, \mn@doi [ARA{\&}A]
  {10.1146/ANNUREV-ASTRO-081309-130815}, 48, 21

\bibitem[\protect\citeauthoryear{Hensley \& Draine}{Hensley \&
  Draine}{2021}]{Hensley2021ObservationalEra}
Hensley B.~S.,  Draine B.~T.,  2021, \mn@doi [ApJ] {10.3847/1538-4357/abc8f1},
  906, 73

\bibitem[\protect\citeauthoryear{Hirashita}{Hirashita}{2015}]{Hirashita2015Two-sizeGalaxies}
Hirashita H.,  2015, \mn@doi [MNRAS] {10.1093/mnras/stu2617}, 447, 2937

\bibitem[\protect\citeauthoryear{Hirashita \& Aoyama}{Hirashita \&
  Aoyama}{2019}]{Hirashita2019RemodellingGalaxies}
Hirashita H.,  Aoyama S.,  2019, \mn@doi [MNRAS] {10.1093/mnras/sty2838}, 482,
  2555

\bibitem[\protect\citeauthoryear{Hirashita \& Kuo}{Hirashita \&
  Kuo}{2011}]{Hirashita2011EffectsGrowth}
Hirashita H.,  Kuo T.~M.,  2011, \mn@doi [MNRAS]
  {10.1111/J.1365-2966.2011.19131.X}, 416, 1340

\bibitem[\protect\citeauthoryear{Hirashita \& Li}{Hirashita \&
  Li}{2013}]{Hirashita2013ConditionCores}
Hirashita H.,  Li Z.~Y.,  2013, \mn@doi [MNRAS: Letters]
  {10.1093/MNRASL/SLT081}, 434, L70

\bibitem[\protect\citeauthoryear{Hirashita \& Murga}{Hirashita \&
  Murga}{2020}]{Hirashita2020Self-consistentEvolution}
Hirashita H.,  Murga M.~S.,  2020, \mn@doi [MNRAS] {10.1093/mnras/stz3640},
  492, 3779

\bibitem[\protect\citeauthoryear{Hirashita \& Yan}{Hirashita \&
  Yan}{2009}]{Hirashita2009ShatteringTurbulence}
Hirashita H.,  Yan H.,  2009, \mn@doi [MNRAS]
  {10.1111/j.1365-2966.2009.14405.x}, 394, 1061

\bibitem[\protect\citeauthoryear{Hollenbach \& Mckee}{Hollenbach \&
  Mckee}{1979}]{Hollenbach1979MOLECULEPROCESSES}
Hollenbach D.,  Mckee C.~F.,  1979, ApJ Supplement Series, 41, 555

\bibitem[\protect\citeauthoryear{Hollenbach \& Mckee}{Hollenbach \&
  Mckee}{1989}]{Hollenbach1989MOLECULECLOUDS}
Hollenbach D.,  Mckee C.~F.,  1989, ApJ, 342, 306

\bibitem[\protect\citeauthoryear{Hollenbach, Kaufman, Bergin  \&
  Melnick}{Hollenbach et~al.}{2008}]{Hollenbach2008WaterClouds}
Hollenbach D.,  Kaufman M.~J.,  Bergin E.~A.,   Melnick G.~J.,  2008, \mn@doi
  [ApJ] {10.1088/0004-637X/690/2/1497}, 690, 1497

\bibitem[\protect\citeauthoryear{Hopkins}{Hopkins}{2013}]{Hopkins2013ATurbulence}
Hopkins P.~F.,  2013, \mn@doi [MNRAS] {10.1093/mnras/stt010}, 430, 1880

\bibitem[\protect\citeauthoryear{Hopkins \& Lee}{Hopkins \&
  Lee}{2016}]{Hopkins2016TheClouds}
Hopkins P.~F.,  Lee H.,  2016, \mn@doi [MNRAS] {10.1093/mnras/stv2745}, 456,
  4174

\bibitem[\protect\citeauthoryear{Hopkins, Kere{\v{s}}, O{\~{n}}orbe,
  Faucher-Gigu{\`{e}}re, Quataert, Murray  \& Bullock}{Hopkins
  et~al.}{2014}]{Hopkins2014}
Hopkins P.~F.,  Kere{\v{s}} D.,  O{\~{n}}orbe J.,  Faucher-Gigu{\`{e}}re C.~A.,
   Quataert E.,  Murray N.,   Bullock J.~S.,  2014, \mn@doi [MNRAS]
  {10.1093/mnras/stu1738}, 445, 581

\bibitem[\protect\citeauthoryear{Hou, Hirashita, Nagamine, Aoyama  \&
  Shimizu}{Hou et~al.}{2017}]{Hou2017EvolutionSimulation}
Hou K.~C.,  Hirashita H.,  Nagamine K.,  Aoyama S.,   Shimizu I.,  2017,
  \mn@doi [MNRAS] {10.1093/mnras/stx877}, 469, 870

\bibitem[\protect\citeauthoryear{Hou, Aoyama, Hirashita, Nagamine  \&
  Shimizu}{Hou et~al.}{2019}]{Hou2019DustSimulation}
Hou K.~C.,  Aoyama S.,  Hirashita H.,  Nagamine K.,   Shimizu I.,  2019,
  \mn@doi [MNRAS] {10.1093/MNRAS/STZ121}, 485, 1727

\bibitem[\protect\citeauthoryear{Hu, Zhukovska, Somerville  \& Naab}{Hu
  et~al.}{2019}]{Hu2019ThermalMedium}
Hu C.-Y.,  Zhukovska S.,  Somerville R.~S.,   Naab T.,  2019, \mn@doi [MNRAS]
  {10.1093/mnras/stz1481}, 487, 3252

\bibitem[\protect\citeauthoryear{Hu, Sternberg  \& van Dishoeck}{Hu
  et~al.}{2021}]{Hu2021Medium}
Hu C.-Y.,  Sternberg A.,   van Dishoeck E.~F.,  2021, \mn@doi [ApJ]
  {10.3847/1538-4357/ac0dbd}, 920, 44

\bibitem[\protect\citeauthoryear{{Hu}, {Sternberg}  \& {van Dishoeck}}{{Hu}
  et~al.}{2023}]{Hu2023Co-evolutionMedium}
{Hu} C.-Y.,  {Sternberg} A.,   {van Dishoeck} E.~F.,  2023, \mn@doi [\apj]
  {10.3847/1538-4357/acdcfa}, \href
  {https://ui.adsabs.harvard.edu/abs/2023ApJ...952..140H} {952, 140}

\bibitem[\protect\citeauthoryear{Hudgins, Allamandola, Hudgins  \&
  Allamandola}{Hudgins et~al.}{1999}]{Hudgins1999TheSize}
Hudgins D.~M.,  Allamandola L.~J.,  Hudgins D.~M.,   Allamandola L.~J.,  1999,
  \mn@doi [ApJL] {10.1086/311901}, 513, L69

\bibitem[\protect\citeauthoryear{Hunter}{Hunter}{2007}]{Hunter2007Matplotlib:Environment}
Hunter J.~D.,  2007, \mn@doi [CiSE] {10.1109/MCSE.2007.55}, 9, 90

\bibitem[\protect\citeauthoryear{Ibanez-Mejia, Walch, Ivlev, Clarke, Caselli
  \& Joshi}{Ibanez-Mejia et~al.}{2019}]{Ibanez-Mejia2019DustMedium}
Ibanez-Mejia J.~C.,  Walch S.,  Ivlev A.~V.,  Clarke S.,  Caselli P.,   Joshi
  P.~R.,  2019, \mn@doi [MNRAS] {10.1093/mnras/stz207}, 485, 1220

\bibitem[\protect\citeauthoryear{Iida et~al.,}{Iida
  et~al.}{2021}]{Iida2021IR-photonAnions}
Iida S.,  et~al., 2021, \mn@doi [PhRvA] {10.1103/PHYSREVA.104.043114}, 104,
  043114

\bibitem[\protect\citeauthoryear{Iskef, Cunningham, Watt, Iskef, Cunningham  \&
  Watt}{Iskef et~al.}{1983}]{Iskef1983ProjectedKeV}
Iskef H.,  Cunningham J.~W.,  Watt D.~E.,  Iskef H.,  Cunningham J.~W.,   Watt
  D.~E.,  1983, \mn@doi [PMB] {10.1088/0031-9155/28/5/007}, 28, 535

\bibitem[\protect\citeauthoryear{Jenkins}{Jenkins}{2009}]{Jenkins2009AMedium}
Jenkins E.~B.,  2009, \mn@doi [ApJ] {10.1088/0004-637X/700/2/1299}, 700, 1299

\bibitem[\protect\citeauthoryear{{Joblin}, {C.}, {Tielens}, M., {Allamandola},
  J., {Geballe}  \& R.}{{Joblin} et~al.}{1996}]{Joblin1996SpatialHydrocarbons}
{Joblin} {C.} {Tielens} M. A. G.~G.,  {Allamandola} J. L.,  {Geballe}  R. T.,
  1996, \mn@doi [ApJ] {10.1086/176843}, 458, 610

\bibitem[\protect\citeauthoryear{Jochims et~al.,}{Jochims
  et~al.}{1994}]{Jochims1994SizeImplications}
Jochims H.~W.,  et~al., 1994, \mn@doi [ApJ] {10.1086/173560}, 420, 307

\bibitem[\protect\citeauthoryear{{Jochims}, {Baumgaertel}  \&
  {Leach}}{{Jochims} et~al.}{1996}]{Jochims1996PhotoionizationHydrocarbons}
{Jochims} H.~W.,  {Baumgaertel} H.,   {Leach} S.,  1996, A{\&}A, \href
  {https://ui.adsabs.harvard.edu/abs/1996A&A...314.1003J} {314, 1003}

\bibitem[\protect\citeauthoryear{Jones, Tielens, Hollenbach, Jones, Tielens  \&
  Hollenbach}{Jones et~al.}{1996}]{Jones1996GrainDistribution}
Jones A.~P.,  Tielens A. G. G.~M.,  Hollenbach D.~J.,  Jones A.~P.,  Tielens A.
  G. G.~M.,   Hollenbach D.~J.,  1996, \mn@doi [ApJ] {10.1086/177823}, 469, 740

\bibitem[\protect\citeauthoryear{Jones, Fanciullo, K{\"{o}}hler, Verstraete,
  Guillet, Bocchio  \& Ysard}{Jones et~al.}{2013}]{Jones2013ThePoint}
Jones A.~P.,  Fanciullo L.,  K{\"{o}}hler M.,  Verstraete L.,  Guillet V.,
  Bocchio M.,   Ysard N.,  2013, \mn@doi [A{\&}A]
  {10.1051/0004-6361/201321686}, 558, A62

\bibitem[\protect\citeauthoryear{Jones, Ysard, K{\"{o}}hler, Fanciullo,
  Bocchio, Micelotta, Verstraete  \& Guillet}{Jones
  et~al.}{2014}]{Jones2014TheDust}
Jones A.~P.,  Ysard N.,  K{\"{o}}hler M.,  Fanciullo L.,  Bocchio M.,
  Micelotta E.,  Verstraete L.,   Guillet V.,  2014, \mn@doi [Faraday
  Discussions] {10.1039/C3FD00128H}, 168, 313

\bibitem[\protect\citeauthoryear{Jones, Koehler, Ysard, Bocchio  \&
  Verstraete}{Jones et~al.}{2017}]{Jones2017TheSolids}
Jones A.~P.,  Koehler M.,  Ysard N.,  Bocchio M.,   Verstraete L.,  2017,
  \mn@doi [A{\&}A] {10.1051/0004-6361/201630225}, 602, A46

\bibitem[\protect\citeauthoryear{Joy}{Joy}{1995}]{Joy1995AInteractions}
Joy D.~C.,  1995, \mn@doi [Scanning Microscopy]
  {https://doi.org/10.1002/sca.4950170501}, 17, 270

\bibitem[\protect\citeauthoryear{Joy, Luo, Gauvin, Hovington  \& Evans}{Joy
  et~al.}{1996}]{Joy1996ExperimentalEnergies}
Joy D.~C.,  Luo S.,  Gauvin R.,  Hovington P.,   Evans N.,  1996, Scanning
  Microscopy, 10

\bibitem[\protect\citeauthoryear{Jurac, Johnson  \& Donn}{Jurac
  et~al.}{1998}]{Jurac1998MonteGrains}
Jurac S.,  Johnson R.~E.,   Donn B.,  1998, \mn@doi [ApJ] {10.1086/305994},
  503, 247

\bibitem[\protect\citeauthoryear{Kannan, Marinacci, Vogelsberger, Sales,
  Torrey, Springel  \& Hernquist}{Kannan
  et~al.}{2020}]{Kannan2020SimulatingDust}
Kannan R.,  Marinacci F.,  Vogelsberger M.,  Sales L.~V.,  Torrey P.,  Springel
  V.,   Hernquist L.,  2020, \mn@doi [MNRAS] {10.1093/mnras/staa3249}, 499,
  5732

\bibitem[\protect\citeauthoryear{Katz}{Katz}{2022}]{Katz2022RAMSES-RTZ:Hydrodynamics}
Katz H.,  2022, \mn@doi [MNRAS] {10.1093/mnras/stac423}, 512, 348

\bibitem[\protect\citeauthoryear{Katz et~al.,}{Katz
  et~al.}{2022}]{Katz2022PRISM:Galaxies}
Katz H.,  et~al., 2022, \mn@doi [arXiv e-prints] {10.48550/arXiv.2211.04626},
  p. arXiv:2211.04626

\bibitem[\protect\citeauthoryear{{Katz} et~al.,}{{Katz}
  et~al.}{2024}]{Katz2024TheGalaxies}
{Katz} et~al., 2024, \mn@doi [arXiv] {10.48550/ARXIV.2411.07282}

\bibitem[\protect\citeauthoryear{Kawasaki, Koga  \& MacHida}{Kawasaki
  et~al.}{2022a}]{Kawasaki2022DustEffects}
Kawasaki Y.,  Koga S.,   MacHida M.~N.,  2022a, \mn@doi [MNRAS]
  {10.1093/mnras/stac1919}, 515, 2072

\bibitem[\protect\citeauthoryear{Kawasaki, Kawasaki  \& Machida}{Kawasaki
  et~al.}{2022b}]{Kawasaki2022ImplementationCore}
Kawasaki Y.,  Kawasaki Y.,   Machida M.~N.,  2022b, \mn@doi [MNRAS]
  {10.1093/MNRAS/STAC2115}, 515, 6073

\bibitem[\protect\citeauthoryear{Kelvin~Lee et~al.,}{Kelvin~Lee
  et~al.}{2021}]{KelvinLee2021TMC-1}
Kelvin~Lee K.~L.,  et~al., 2021, \mn@doi [ApJL] {10.3847/2041-8213/abe764},
  910, L2

\bibitem[\protect\citeauthoryear{Kemper, Vriend  \& Tielens}{Kemper
  et~al.}{2004}]{Kemper2004TheMedium}
Kemper F.,  Vriend W.~J.,   Tielens A. G. G.~M.,  2004, \mn@doi [ApJ]
  {10.1086/421339}, 609, 826

\bibitem[\protect\citeauthoryear{Khabazipur \& Eaves}{Khabazipur \&
  Eaves}{2023}]{Khabazipur2023DevelopmentModel}
Khabazipur A.,  Eaves N.,  2023, \mn@doi [Proc. Combust. Inst.]
  {10.1016/J.PROCI.2022.07.172}, 39, 919

\bibitem[\protect\citeauthoryear{Khan et~al.,}{Khan
  et~al.}{2025}]{Khan2025PDRs4AllJWST}
Khan B.,  et~al., 2025, \mn@doi [A{\&}A] {10.1051/0004-6361/202554096}, 699,
  A133

\bibitem[\protect\citeauthoryear{Khesali, Ghoreyshi  \& Nejad-Asghar}{Khesali
  et~al.}{2012}]{Khesali2012ThermalDiffusion}
Khesali A.~R.,  Ghoreyshi S.~M.,   Nejad-Asghar M.,  2012, \mn@doi [MNRAS]
  {10.1111/J.1365-2966.2011.20194.X/2/M{\_}MNRAS0420-2300-MU24.GIF}, 420, 2300

\bibitem[\protect\citeauthoryear{Kim \& Ostriker}{Kim \&
  Ostriker}{2017}]{Kim2017Three-phaseConvergence}
Kim C.-G.,  Ostriker E.~C.,  2017, \mn@doi [ApJ] {10.3847/1538-4357/aa8599},
  846, 133

\bibitem[\protect\citeauthoryear{Kim, Ostriker  \& Kim}{Kim
  et~al.}{2013}]{Kim2013THREE-DIMENSIONALRATES}
Kim C.~G.,  Ostriker E.~C.,   Kim W.~T.,  2013, \mn@doi [ApJ]
  {10.1088/0004-637X/776/1/1}, 776, 1

\bibitem[\protect\citeauthoryear{{Kim}, {Gong}, {Kim}  \& {Ostriker}}{{Kim}
  et~al.}{2023}]{Kim2023PhotochemistrySimulations}
{Kim} J.-G.,  {Gong} M.,  {Kim} C.-G.,   {Ostriker} E.~C.,  2023, \mn@doi
  [ApJS] {10.3847/1538-4365/ac9b1d}, \href
  {https://ui.adsabs.harvard.edu/abs/2023ApJS..264...10K} {264, 10}

\bibitem[\protect\citeauthoryear{Kim et~al.,}{Kim
  et~al.}{2025}]{Kim2025LocalizedEmissivity}
Kim J.,  et~al., 2025, \mn@doi [arXiv] {10.48550/ARXIV.2511.14833}, p.
  arXiv:2511.14833

\bibitem[\protect\citeauthoryear{Kimm \& Cen}{Kimm \&
  Cen}{2014}]{Kimm2014ESCAPESTARS}
Kimm T.,  Cen R.,  2014, \mn@doi [ApJ] {10.1088/0004-637X/788/2/121}, 788, 121

\bibitem[\protect\citeauthoryear{Kimm, Katz, Haehnelt, Rosdahl, Devriendt  \&
  Slyz}{Kimm et~al.}{2017}]{Kimm2017Feedback-regulatedReionisation}
Kimm T.,  Katz H.,  Haehnelt M.,  Rosdahl J.,  Devriendt J.,   Slyz A.,  2017,
  \mn@doi [MNRAS] {10.1093/mnras/stx052}, 466, stx052

\bibitem[\protect\citeauthoryear{Kirchschlager, Schmidt, Barlow, Fogerty, Bevan
   \& Priestley}{Kirchschlager et~al.}{2019}]{Kirchschlager2019DustDensities}
Kirchschlager F.,  Schmidt F.~D.,  Barlow M.~J.,  Fogerty E.~L.,  Bevan A.,
  Priestley F.~D.,  2019, \mn@doi [MNRAS] {10.1093/mnras/stz2399}, 489, 4465

\bibitem[\protect\citeauthoryear{Kirchschlager, Barlow  \&
  Schmidt}{Kirchschlager et~al.}{2020}]{Kirchschlager2020SilicateA}
Kirchschlager F.,  Barlow M.~J.,   Schmidt F.~D.,  2020, \mn@doi [ApJ]
  {10.3847/1538-4357/AB7DB8}, 893, 70

\bibitem[\protect\citeauthoryear{Kirchschlager, Mattsson  \&
  Gent}{Kirchschlager et~al.}{2021}]{Kirchschlager2021SupernovaTurbulence}
Kirchschlager F.,  Mattsson L.,   Gent F.~A.,  2021, \mn@doi [MNRAS]
  {10.1093/mnras/stab3059}, 509, 3218

\bibitem[\protect\citeauthoryear{Klessen, Glover, Klessen  \& Glover}{Klessen
  et~al.}{2016}]{Klessen2016PhysicalMedium}
Klessen R.~S.,  Glover S. C.~O.,  Klessen R.~S.,   Glover S. C.~O.,  2016,
  \mn@doi [SAAS] {10.1007/978-3-662-47890-5{\_}2}, 43, 85

\bibitem[\protect\citeauthoryear{Knight, Peeters, Stock, Vacca  \&
  Tielens}{Knight et~al.}{2021}]{Knight2021TracingEnvironments}
Knight C.,  Peeters E.,  Stock D.~J.,  Vacca W.~D.,   Tielens A. G. G.~M.,
  2021, \mn@doi [ApJ] {10.3847/1538-4357/ac02c6}, 918, 8

\bibitem[\protect\citeauthoryear{Kobayashi \& Tanaka}{Kobayashi \&
  Tanaka}{2010}]{Kobayashi2010FragmentationCascades}
Kobayashi H.,  Tanaka H.,  2010, \mn@doi [Icarus]
  {10.1016/J.ICARUS.2009.10.004}, 206, 735

\bibitem[\protect\citeauthoryear{Kocheril, Zagorec-Marks  \&
  Lewandowski}{Kocheril et~al.}{2025}]{Kocheril2025TerminationC6H5+}
Kocheril G.~S.,  Zagorec-Marks C.,   Lewandowski H.~J.,  2025, \mn@doi
  [Nat.Astro.] {10.1038/s41550-025-02504-y}, 9, 685

\bibitem[\protect\citeauthoryear{Konstantopoulou et~al.,}{Konstantopoulou
  et~al.}{2022}]{Konstantopoulou2022DustISM}
Konstantopoulou C.,  et~al., 2022, \mn@doi [A{\&}A]
  {10.1051/0004-6361/202243994}, 666, A12

\bibitem[\protect\citeauthoryear{{Konstantopoulou} et~al.,}{{Konstantopoulou}
  et~al.}{2024}]{Konstantopoulou2024DustComposition}
{Konstantopoulou} C.,  et~al., 2024, \mn@doi [A{\&}A]
  {10.1051/0004-6361/202347171}, \href
  {https://ui.adsabs.harvard.edu/abs/2024A&A...681A..64K} {681, A64}

\bibitem[\protect\citeauthoryear{Kr{\"u}gel}{Kr{\"u}gel}{2008}]{Krugel2008AnDust}
Kr{\"u}gel E.,  2008, An Introduction to the Physics of Interstellar Dust.
Series in Astronomy and Astrophysics, Taylor \& Francis

\bibitem[\protect\citeauthoryear{{Krumholz}}{{Krumholz}}{2014}]{Krumholz2014}
{Krumholz} M.~R.,  2014, \mn@doi [\physrep] {10.1016/j.physrep.2014.02.001},
  \href {https://ui.adsabs.harvard.edu/abs/2014PhR...539...49K} {539, 49}

\bibitem[\protect\citeauthoryear{Krumholz \& Dekel}{Krumholz \&
  Dekel}{2010}]{Krumholz2010SurvivalGalaxies}
Krumholz M.~R.,  Dekel A.,  2010, \mn@doi [MNRAS]
  {10.1111/j.1365-2966.2010.16675.x}, 406, 112

\bibitem[\protect\citeauthoryear{Krumholz \& Thompson}{Krumholz \&
  Thompson}{2013}]{Krumholz2013NumericalWinds}
Krumholz M.~R.,  Thompson T.~A.,  2013, \mn@doi [MNRAS]
  {10.1093/mnras/stt1174}, 434, 2329

\bibitem[\protect\citeauthoryear{Kunz \& Mouschovias}{Kunz \&
  Mouschovias}{2010}]{Kunz2010TheResults}
Kunz M.~W.,  Mouschovias T.~C.,  2010, \mn@doi [MNRAS]
  {10.1111/j.1365-2966.2010.17110.x}, 408, 322

\bibitem[\protect\citeauthoryear{Lamberts, de Vries  \& Cuppen}{Lamberts
  et~al.}{2014}]{Lamberts2014TheExothermicity}
Lamberts T.,  de Vries X.,   Cuppen H.~M.,  2014, \mn@doi [Faraday Discuss.]
  {10.1039/C3FD00136A}, 168, 327

\bibitem[\protect\citeauthoryear{Lange, Dominik  \& Tielens}{Lange
  et~al.}{2021}]{Lange2021StabilityDiscs}
Lange K.,  Dominik C.,   Tielens A.~G.,  2021, \mn@doi [A{\&}A]
  {10.1051/0004-6361/202140590}, 653, A21

\bibitem[\protect\citeauthoryear{Lange, Dominik  \& Tielens}{Lange
  et~al.}{2023}]{Lange2023TurbulentPAHs}
Lange K.,  Dominik C.,   Tielens A.~G.,  2023, \mn@doi [A{\&}A]
  {10.1051/0004-6361/202245108}, 674, A200

\bibitem[\protect\citeauthoryear{Langer, Glassgold, Langer  \&
  Glassgold}{Langer et~al.}{1976}]{Langer1976TimeReactions.}
Langer W.~D.,  Glassgold A.~E.,  Langer W.~D.,   Glassgold A.~E.,  1976,
  A{\&}A, 48, 395

\bibitem[\protect\citeauthoryear{Laor \& Draine}{Laor \&
  Draine}{1993}]{Laor1993SpectroscopicNuclei}
Laor A.,  Draine B.~T.,  1993, \mn@doi [ApJ] {10.1086/172149}, 402, 441

\bibitem[\protect\citeauthoryear{Larson}{Larson}{1981}]{Larson1981TurbulenceClouds.}
Larson R.~B.,  1981, \mn@doi [MNRAS] {10.1093/MNRAS/194.4.809}, 194, 809

\bibitem[\protect\citeauthoryear{Lazarian \& Yan}{Lazarian \&
  Yan}{2001}]{Lazarian2001GrainGas}
Lazarian A.,  Yan H.,  2001, \mn@doi [ApJL] {10.1086/339675}, 566, L105

\bibitem[\protect\citeauthoryear{Le~Bourlot, Le~Petit, Pinto, Roueff  \&
  Roy}{Le~Bourlot et~al.}{2012}]{LeBourlot2012SurfaceMechanisms}
Le~Bourlot J.,  Le~Petit F.,  Pinto C.,  Roueff E.,   Roy F.,  2012, \mn@doi
  [A{\&}A] {10.1051/0004-6361/201118126}, 541, A76

\bibitem[\protect\citeauthoryear{Le~Page, Snow, Bierbaum, Le~Page, Snow  \&
  Bierbaum}{Le~Page et~al.}{2001}]{LePage2001HydrogenationModel}
Le~Page V.,  Snow T.~P.,  Bierbaum V.~M.,  Le~Page V.,  Snow T.~P.,   Bierbaum
  V.~M.,  2001, \mn@doi [ApJS] {10.1086/318952}, 132, 233

\bibitem[\protect\citeauthoryear{Le~Page, Snow  \& Bierbaum}{Le~Page
  et~al.}{2009}]{LePage2009MolecularMedium}
Le~Page V.,  Snow T.~P.,   Bierbaum V.~M.,  2009, \mn@doi [ApJ]
  {10.1088/0004-637X/704/1/274}, 704, 274

\bibitem[\protect\citeauthoryear{Le~Petit, Le~Bourlot, Roueff  \&
  Nehm{\'{e}}}{Le~Petit et~al.}{2006}]{LePetit2006ACode}
Le~Petit F.,  Le~Bourlot J.,  Roueff E.,   Nehm{\'{e}} C.,  2006, \mn@doi [ApJ
  Supplement Series, Volume 164, Issue 2, pp. 506-529.] {10.1086/503252}, 164,
  506

\bibitem[\protect\citeauthoryear{Lebreuilly, Commer{\c{c}}on  \&
  Laibe}{Lebreuilly et~al.}{2019}]{Lebreuilly2019SmallMethods}
Lebreuilly U.,  Commer{\c{c}}on B.,   Laibe G.,  2019, \mn@doi [A{\&}A]
  {10.1051/0004-6361/201834147}, 626, A96

\bibitem[\protect\citeauthoryear{Lebreuilly, Vallucci-Goy, Guillet, Lombart  \&
  Marchand}{Lebreuilly et~al.}{2022}]{Lebreuilly2022ProtostellarFragmentation}
Lebreuilly U.,  Vallucci-Goy V.,  Guillet V.,  Lombart M.,   Marchand P.,
  2022, \mn@doi [MNRAS] {10.1093/MNRAS/STAC3220}, 518, 3326

\bibitem[\protect\citeauthoryear{Lee, Hopkins  \& Squire}{Lee
  et~al.}{2017}]{Lee2017TheClouds}
Lee H.,  Hopkins P.~F.,   Squire J.,  2017, \mn@doi [MNRAS]
  {10.1093/MNRAS/STX1097}, 469, 3532

\bibitem[\protect\citeauthoryear{Lewis, Ocvirk, Dubois, Aubert, Chardin, Gillet
   \& Th{\'{e}}lie}{Lewis et~al.}{2022}]{Lewis2022DUSTiERRAMSES-CUDATON}
Lewis J. S.~W.,  Ocvirk P.,  Dubois Y.,  Aubert D.,  Chardin J.,  Gillet N.,
  Th{\'{e}}lie {\'{E}}.,  2022, \mn@doi [MNRAS] {10.48550/arxiv.2204.03949},
  000, 1

\bibitem[\protect\citeauthoryear{Li \& Draine}{Li \&
  Draine}{2001}]{Li2001InfraredMedium}
Li A.,  Draine B.~T.,  2001, \mn@doi [ApJ] {10.1086/323147}, 554, 778

\bibitem[\protect\citeauthoryear{Li \& Greenberg}{Li \&
  Greenberg}{2003}]{Li2003InDust}
Li A.,  Greenberg J.~M.,  2003, in Pirronello V.,  Krelowski J.,   Manic{\`o}
  G.,  eds, Solid State Astrochemistry. Springer Netherlands, Dordrecht, pp
  37--84

\bibitem[\protect\citeauthoryear{Li, Narayanan  \& Dav{\'{e}}}{Li
  et~al.}{2019}]{Li2019The6}
Li Q.,  Narayanan D.,   Dav{\'{e}} R.,  2019, \mn@doi [MNRAS]
  {10.1093/mnras/stz2684}, 490, 1425

\bibitem[\protect\citeauthoryear{Li et~al.,}{Li
  et~al.}{2021}]{Li2021TheGalaxies}
Li Q.,  et~al., 2021, \mn@doi [MNRAS] {10.1093/MNRAS/STAB2196}, 507, 548

\bibitem[\protect\citeauthoryear{Li, Chen, Li  \& Chen}{Li
  et~al.}{2023}]{Li2023TheSurveys}
Li J.,  Chen X.,  Li J.,   Chen X.,  2023, \mn@doi [Univ]
  {10.3390/UNIVERSE9080364}, 9, 364

\bibitem[\protect\citeauthoryear{Limongi \& Chieffi}{Limongi \&
  Chieffi}{2018}]{Limongi2018Presupernova0}
Limongi M.,  Chieffi A.,  2018, \mn@doi [ApJ Supplement Series]
  {10.3847/1538-4365/aacb24}, 237, 13

\bibitem[\protect\citeauthoryear{Lindhard \& Nielsen}{Lindhard \&
  Nielsen}{1962}]{Lindhard1962NuclearDetectors}
Lindhard J.,  Nielsen V.,  1962, \mn@doi [Phys. Lett.]
  {10.1016/0031-9163(62)90229-9}, 2, 209

\bibitem[\protect\citeauthoryear{Lindhard \& Scharff}{Lindhard \&
  Scharff}{1961}]{Lindhard1961EnergyRegion}
Lindhard J.,  Scharff M.,  1961, \mn@doi [Phys. Rev.]
  {10.1103/PHYSREV.124.128}, 124, 128

\bibitem[\protect\citeauthoryear{Lindhard, {Lindhard}  \& {J.}}{Lindhard
  et~al.}{1964}]{Lindhard1964MotionCrystals}
Lindhard J.,  {Lindhard}  {J.} 1964, \mn@doi [Phys. Lett.]
  {10.1016/0031-9163(64)91133-3}, 12, 126

\bibitem[\protect\citeauthoryear{Ling, Lifshitz, Ling  \& Lifshitz}{Ling
  et~al.}{1998}]{Ling1998EnergeticsCalculations}
Ling Y.,  Lifshitz C.,  Ling Y.,   Lifshitz C.,  1998, \mn@doi [JMSp]
  {10.1002/(SICI)1096-9888(199801)33:1}, 33, 25

\bibitem[\protect\citeauthoryear{Lisenfeld \& Ferrara}{Lisenfeld \&
  Ferrara}{1998}]{Lisenfeld1998DusttoGasGalaxies}
Lisenfeld U.,  Ferrara A.,  1998, \mn@doi [ApJ] {10.1086/305354}, 496, 145

\bibitem[\protect\citeauthoryear{Liu, Zhang, Li, Bennett, Lin, Sarathy  \&
  Roberts}{Liu et~al.}{2019}]{Liu2019ComputationalAddition}
Liu P.,  Zhang Y.,  Li Z.,  Bennett A.,  Lin H.,  Sarathy S.~M.,   Roberts
  W.~L.,  2019, \mn@doi [Combust. Flame] {10.1016/J.COMBUSTFLAME.2019.01.023},
  202, 276

\bibitem[\protect\citeauthoryear{Logan, Keck, Logan  \& Keck}{Logan
  et~al.}{1968}]{Logan1968ClassicalSurfaces}
Logan R.~M.,  Keck J.~C.,  Logan R.~M.,   Keck J.~C.,  1968, \mn@doi [JChPh]
  {10.1063/1.1670153}, 49, 860

\bibitem[\protect\citeauthoryear{Loison, Rossi, Solem, Thissen, Romanzin,
  Alcaraz  \& Jacovella}{Loison et~al.}{2025}]{Loison2025EvidenceConditions}
Loison J.-C.,  Rossi C.,  Solem N.,  Thissen R.,  Romanzin C.,  Alcaraz C.,
  Jacovella U.,  2025, \mn@doi [eprint arXiv:2506.13290]
  {10.48550/ARXIV.2506.13290}, p. arXiv:2506.13290

\bibitem[\protect\citeauthoryear{Loru et~al.,}{Loru
  et~al.}{2023}]{Loru2023Detection12-diethynylbenzene}
Loru D.,  et~al., 2023, \mn@doi [A{\&}A] {10.1051/0004-6361/202347023}, 677,
  A166

\bibitem[\protect\citeauthoryear{Low \& Klessen}{Low \&
  Klessen}{2004}]{Low2004ControlTurbulence}
Low M. M.~M.,  Klessen R.~S.,  2004, \mn@doi [Rev. Mod. Phys.]
  {10.1103/RevModPhys.76.125}, 76, 125

\bibitem[\protect\citeauthoryear{Luo, Zhang  \& Joy}{Luo
  et~al.}{1991}]{Luo1991ExperimentalEnergies}
Luo S.,  Zhang X.,   Joy D.~C.,  1991, \mn@doi [Radiat. Eff. Defects Solids]
  {10.1080/10420159108220619}, 117, 235

\bibitem[\protect\citeauthoryear{Madden, Galliano, Jones  \& Sauvage}{Madden
  et~al.}{2005}]{Madden2005ISMGalaxies}
Madden S.~C.,  Galliano F.,  Jones A.~P.,   Sauvage M.,  2005, \mn@doi [A{\&}A]
  {10.1051/0004-6361:20053890}, 446, 877

\bibitem[\protect\citeauthoryear{Mallo, Ag{\'{u}}ndez, Cabezas, Roncero,
  Cernicharo  \& Molpeceres}{Mallo
  et~al.}{2025}]{Mallo2025Ion-moleculeReaction}
Mallo M.,  Ag{\'{u}}ndez M.,  Cabezas C.,  Roncero O.,  Cernicharo J.,
  Molpeceres G.,  2025, \mn@doi [A{\&}A] {10.1051/0004-6361/202557647}

\bibitem[\protect\citeauthoryear{Malloci, Mulas, Joblin, Malloci, Mulas  \&
  Joblin}{Malloci et~al.}{2004}]{Malloci2004ElectronicPhotophysics}
Malloci G.,  Mulas G.,  Joblin C.,  Malloci G.,  Mulas G.,   Joblin C.,  2004,
  \mn@doi [A{\&}A] {10.1051/0004-6361:20040541}, 426, 105

\bibitem[\protect\citeauthoryear{Maragkoudakis et~al.,}{Maragkoudakis
  et~al.}{2026}]{Maragkoudakis2026PDRs4All:Bar}
Maragkoudakis A.,  et~al., 2026, arXiv, p. arXiv:2601.23282

\bibitem[\protect\citeauthoryear{Marchand, Guillet, Lebreuilly  \&
  Mac~Low}{Marchand et~al.}{2021}]{Marchand2021FastDerivation}
Marchand P.,  Guillet V.,  Lebreuilly U.,   Mac~Low M.~M.,  2021, \mn@doi
  [A{\&}A] {10.1051/0004-6361/202040077}, 649, A50

\bibitem[\protect\citeauthoryear{Marchenko \& Moffat}{Marchenko \&
  Moffat}{2017}]{Marchenko2017SearchStars}
Marchenko S.~V.,  Moffat A.~F.,  2017, \mn@doi [MNRAS] {10.1093/MNRAS/STX563},
  468, 2416

\bibitem[\protect\citeauthoryear{Marciniak, Joblin, Mulas, Mundlapati  \&
  Bonnamy}{Marciniak et~al.}{2021}]{Marciniak2021PhotodissociationConditions}
Marciniak A.,  Joblin C.,  Mulas G.,  Mundlapati V.~R.,   Bonnamy A.,  2021,
  \mn@doi [A{\&}A] {10.1051/0004-6361/202140737}, 652, A42

\bibitem[\protect\citeauthoryear{Masson, Chabrier, Hennebelle, Vaytet  \&
  Commer{\c{c}}on}{Masson et~al.}{2015}]{Masson2015AmbipolarCase}
Masson J.,  Chabrier G.,  Hennebelle P.,  Vaytet N.,   Commer{\c{c}}on B.,
  2015, \mn@doi [A{\&}A] {10.1051/0004-6361/201526371}, 587, A32

\bibitem[\protect\citeauthoryear{Mathis, Rumpl, Nordsieck, Mathis, Rumpl  \&
  Nordsieck}{Mathis et~al.}{1977}]{Mathis1977TheGrains.}
Mathis J.~S.,  Rumpl W.,  Nordsieck K.~H.,  Mathis J.~S.,  Rumpl W.,
  Nordsieck K.~H.,  1977, \mn@doi [ApJ] {10.1086/155591}, 217, 425

\bibitem[\protect\citeauthoryear{{Mathis}, {Mezger}  \& {Panagia}}{{Mathis}
  et~al.}{1983}]{Mathis1983InterstellarClouds}
{Mathis} J.~S.,  {Mezger} P.~G.,   {Panagia} N.,  1983, \aap, \href
  {https://ui.adsabs.harvard.edu/abs/1983A&A...128..212M} {128, 212}

\bibitem[\protect\citeauthoryear{Matsumoto et~al.,}{Matsumoto
  et~al.}{2024}]{Matsumoto2024ObservationalSimulations}
Matsumoto K.,  et~al., 2024, \mn@doi [A{\&}A] {10.1051/0004-6361/202449454},
  689, A79

\bibitem[\protect\citeauthoryear{Matsunami}{Matsunami}{1980}]{Matsunami1980EnergySolids}
Matsunami N.,  1980, Technical Report IPPJ-AM-14, Energy dependence of
  sputtering yields of monatomic solids.
Institute of Plasma Physics, Nagoya University, Nagoya, Japan

\bibitem[\protect\citeauthoryear{Matsuura et~al.,}{Matsuura
  et~al.}{2014}]{Matsuura2014SpitzerMetallicities}
Matsuura M.,  et~al., 2014, \mn@doi [MNRAS] {10.1093/MNRAS/STT2495}, 439, 1472

\bibitem[\protect\citeauthoryear{Mattsson}{Mattsson}{2016}]{Mattsson2016ModellingMoments}
Mattsson L.,  2016, \mn@doi [Planetary and Space Science]
  {10.1016/J.PSS.2016.05.002}, 133, 107

\bibitem[\protect\citeauthoryear{Mattsson, {Mattsson}  \& {Lars}}{Mattsson
  et~al.}{2020}]{Mattsson2020GalacticTurbulence}
Mattsson L.,  {Mattsson}  {Lars} 2020, \mn@doi [MNRAS] {10.1093/MNRAS/STZ3359},
  491, 4334

\bibitem[\protect\citeauthoryear{McCarthy et~al.,}{McCarthy
  et~al.}{2021}]{McCarthy2021InterstellarCyanocyclopentadiene}
McCarthy M.~C.,  et~al., 2021, \mn@doi [Nature Astronomy]
  {10.1038/s41550-020-01213-y}, 5, 176

\bibitem[\protect\citeauthoryear{McCormick, Veilleux, Rupke, McCormick,
  Veilleux  \& Rupke}{McCormick et~al.}{2013}]{McCormick2013DustyGalaxiesb}
McCormick A.,  Veilleux S.,  Rupke D. S.~N.,  McCormick A.,  Veilleux S.,
  Rupke D. S.~N.,  2013, \mn@doi [ApJ] {10.1088/0004-637X/774/2/126}, 774, 126

\bibitem[\protect\citeauthoryear{McGuire, Burkhardt, Kalenskii, Shingledecker,
  Remijan, Herbst  \& McCarthy}{McGuire
  et~al.}{2018}]{McGuire2018DetectionMedium}
McGuire B.~A.,  Burkhardt A.~M.,  Kalenskii S.,  Shingledecker C.~N.,  Remijan
  A.~J.,  Herbst E.,   McCarthy M.~C.,  2018, \mn@doi [Science]
  {10.1126/science.aao4890}, 359, 202

\bibitem[\protect\citeauthoryear{McGuire et~al.,}{McGuire
  et~al.}{2021}]{McGuire2021DetectionFiltering}
McGuire B.~A.,  et~al., 2021, \mn@doi [Science] {10.1126/science.abb7535}, 371,
  1265

\bibitem[\protect\citeauthoryear{McKee \& Ostriker}{McKee \&
  Ostriker}{2007}]{McKee2007}
McKee C.~F.,  Ostriker E.~C.,  2007, \mn@doi [ARA{\&}A]
  {10.1146/annurev.astro.45.051806.110602}, 45, 565

\bibitem[\protect\citeauthoryear{McKee, {McKee}  \& {C.}}{McKee
  et~al.}{1989}]{McKee1989DustMedium}
McKee C.,  {McKee}  {C.} 1989, IAUS, 135, 431

\bibitem[\protect\citeauthoryear{McKinnon, Torrey  \& Vogelsberger}{McKinnon
  et~al.}{2016}]{McKinnon2016DustGalaxies}
McKinnon R.,  Torrey P.,   Vogelsberger M.,  2016, \mn@doi [MNRAS]
  {10.1093/mnras/stw253}, 457, 3775

\bibitem[\protect\citeauthoryear{McKinnon, Torrey, Vogelsberger, Hayward  \&
  Marinacci}{McKinnon et~al.}{2017}]{McKinnon2017SimulatingFailures}
McKinnon R.,  Torrey P.,  Vogelsberger M.,  Hayward C.~C.,   Marinacci F.,
  2017, \mn@doi [MNRAS] {10.1093/MNRAS/STX467}, 468, 1505

\bibitem[\protect\citeauthoryear{McKinnon, Vogelsberger, Torrey, Marinacci  \&
  Kannan}{McKinnon et~al.}{2018}]{McKinnon2018SimulatingMesh}
McKinnon R.,  Vogelsberger M.,  Torrey P.,  Marinacci F.,   Kannan R.,  2018,
  \mn@doi [MNRAS] {10.1093/MNRAS/STY1248}, 478, 2851

\bibitem[\protect\citeauthoryear{Mennella, Hornek{\ae}r, Thrower, Accolla,
  Mennella, Hornek{\ae}r, Thrower  \& Accolla}{Mennella
  et~al.}{2012}]{Mennella2012TheFormation}
Mennella V.,  Hornek{\ae}r L.,  Thrower J.,  Accolla M.,  Mennella V.,
  Hornek{\ae}r L.,  Thrower J.,   Accolla M.,  2012, \mn@doi [ApJL]
  {10.1088/2041-8205/745/1/L2}, 745, L2

\bibitem[\protect\citeauthoryear{Micelotta, Jones  \& Tielens}{Micelotta
  et~al.}{2010a}]{Micelotta2010PolycyclicGas}
Micelotta E.~R.,  Jones A.~P.,   Tielens A. G. G.~M.,  2010a, \mn@doi [A{\&}A]
  {10.1051/0004-6361/200911683}, 510, A37

\bibitem[\protect\citeauthoryear{Micelotta, Jones  \& Tielens}{Micelotta
  et~al.}{2010b}]{Micelotta2010PolycyclicShocks}
Micelotta E.~R.,  Jones A.~P.,   Tielens A.~G.,  2010b, \mn@doi [A{\&}A]
  {10.1051/0004-6361/200911682}, 510, A36

\bibitem[\protect\citeauthoryear{Min, Waters, De~Koter, Hovenier, Keller  \&
  Markwick-Kemper}{Min et~al.}{2007}]{Min2007TheGrains}
Min M.,  Waters L.~B.,  De~Koter A.,  Hovenier J.~W.,  Keller L.~P.,
  Markwick-Kemper F.,  2007, \mn@doi [A{\&}A] {10.1051/0004-6361:20065436},
  462, 667

\bibitem[\protect\citeauthoryear{Montet, {Montet}  \& L.}{Montet
  et~al.}{1967}]{Montet1967TheMolybdenite}
Montet G.~L.,  {Montet}  L. G.,  1967, \mn@doi [ApPhL] {10.1063/1.1755108}, 11,
  223

\bibitem[\protect\citeauthoryear{Montillaud \& Joblin}{Montillaud \&
  Joblin}{2014}]{Montillaud2014AbsoluteClusters}
Montillaud J.,  Joblin C.,  2014, \mn@doi [A{\&}A]
  {10.1051/0004-6361/201323141}, 567, A45

\bibitem[\protect\citeauthoryear{Montillaud, Joblin  \& Toublanc}{Montillaud
  et~al.}{2013}]{Montillaud2013EvolutionStates}
Montillaud J.,  Joblin C.,   Toublanc D.,  2013, \mn@doi [A{\&}A]
  {10.1051/0004-6361/201220757}, 552, A15

\bibitem[\protect\citeauthoryear{Moore, Lumsden, Ridge  \& Puxley}{Moore
  et~al.}{2005}]{Moore2005TheAV}
Moore T.~J.,  Lumsden S.~L.,  Ridge N.~A.,   Puxley P.~J.,  2005, \mn@doi
  [MNRAS] {10.1111/J.1365-2966.2005.08923.X/2/359-2-589-FIG007.GIF}, 359, 589

\bibitem[\protect\citeauthoryear{{Moseley} \& {Teyssier}}{{Moseley} \&
  {Teyssier}}{2025}]{Moseley2025DustGrains}
{Moseley} E.~R.,  {Teyssier} R.,  2025, \mn@doi [MNRAS]
  {10.1093/mnras/staf1260}, \href
  {https://ui.adsabs.harvard.edu/abs/2025MNRAS.542.1011M} {542, 1011}

\bibitem[\protect\citeauthoryear{Moseley, Teyssier  \& Draine}{Moseley
  et~al.}{2022}]{Moseley2022Acceleration}
Moseley E.~R.,  Teyssier R.,   Draine B.~T.,  2022, \mn@doi [MNRAS]
  {10.1093/mnras/stac3231}, 518, 2825

\bibitem[\protect\citeauthoryear{Mouschovias, {Mouschovias}  \&
  Ch.}{Mouschovias et~al.}{1991}]{Mouschovias1991MagneticMasses}
Mouschovias T.~C.,  {Mouschovias}  Ch. T.,  1991, \mn@doi [ApJ]
  {10.1086/170035}, 373, 169

\bibitem[\protect\citeauthoryear{Murga, Khoperskov  \& Wiebe}{Murga
  et~al.}{2016}]{Murga2016RestructuringMedium}
Murga M.~S.,  Khoperskov S.~A.,   Wiebe D.~S.,  2016, \mn@doi [Astron. Rep.]
  {10.1134/S1063772916020104}, 60, 233

\bibitem[\protect\citeauthoryear{Murga, Wiebe, Sivkova  \& Akimkin}{Murga
  et~al.}{2019}]{Murga2019Shiva:Model}
Murga M.~S.,  Wiebe D.~S.,  Sivkova E.~E.,   Akimkin V.~V.,  2019, \mn@doi
  [MNRAS] {10.1093/mnras/stz1724}, 488, 965

\bibitem[\protect\citeauthoryear{Murga, Kirsanova, Vasyunin  \&
  Pavlyuchenkov}{Murga et~al.}{2020}]{Murga2020ImpactPDRs}
Murga M.~S.,  Kirsanova M.~S.,  Vasyunin A.~I.,   Pavlyuchenkov Y.~N.,  2020,
  \mn@doi [MNRAS] {10.1093/mnras/staa2026}, 497, 2327

\bibitem[\protect\citeauthoryear{Murray, Quataert  \& Thompson}{Murray
  et~al.}{2005}]{Murray2005}
Murray N.,  Quataert E.,   Thompson T.~A.,  2005, \mn@doi [ApJ]
  {10.1086/426067}, 618, 569

\bibitem[\protect\citeauthoryear{Murray, Quataert  \& Thompson}{Murray
  et~al.}{2010}]{Murray2010TheGalaxies}
Murray N.,  Quataert E.,   Thompson T.~A.,  2010, \mn@doi [ApJ]
  {10.1088/0004-637X/709/1/191}, 709, 191

\bibitem[\protect\citeauthoryear{Nakai et~al.,}{Nakai
  et~al.}{1991}]{Nakai1991NonthermalSurfaces}
Nakai Y.,  et~al., 1991, \mn@doi [NIMPB] {10.1016/0168-583X(91)95885-H}, 58,
  452

\bibitem[\protect\citeauthoryear{Narayanan et~al.,}{Narayanan
  et~al.}{2023}]{Narayanan2023ASimulations}
Narayanan D.,  et~al., 2023, \mn@doi [ApJ] {10.3847/1538-4357/accf8d}, 951, 100

\bibitem[\protect\citeauthoryear{{Narayanan} et~al.,}{{Narayanan}
  et~al.}{2025}]{Narayanan2025TheMonsters}
{Narayanan} et~al., 2025, \mn@doi [arXiv] {10.48550/ARXIV.2509.18266}

\bibitem[\protect\citeauthoryear{Novotn{\'{y}} et~al.,}{Novotn{\'{y}}
  et~al.}{2005}]{Novotny2005RecombinationPlasma}
Novotn{\'{y}} O.,  et~al., 2005, \mn@doi [JChPh] {10.1063/1.2000927}, 123,
  104303

\bibitem[\protect\citeauthoryear{Nozawa, Kozasa  \& Habe}{Nozawa
  et~al.}{2006}]{Nozawa2006DustUniverse}
Nozawa T.,  Kozasa T.,   Habe A.,  2006, \mn@doi [ApJ] {10.1086/505639}, 648,
  435

\bibitem[\protect\citeauthoryear{Nozawa, Kozasa, Habe, Dwek, Umeda, Tominaga,
  Maeda  \& Nomoto}{Nozawa et~al.}{2007}]{Nozawa2007EvolutionMedium}
Nozawa T.,  Kozasa T.,  Habe A.,  Dwek E.,  Umeda H.,  Tominaga N.,  Maeda K.,
   Nomoto K.,  2007, \mn@doi [ApJ] {10.1086/520621}, 666, 955

\bibitem[\protect\citeauthoryear{Nozawa, Maeda, Kozasa, Tanaka, Nomoto  \&
  Umeda}{Nozawa et~al.}{2011}]{Nozawa2011FormationSupernovae}
Nozawa T.,  Maeda K.,  Kozasa T.,  Tanaka M.,  Nomoto K.,   Umeda H.,  2011,
  \mn@doi [ApJ] {10.1088/0004-637X/736/1/45}, 736, 45

\bibitem[\protect\citeauthoryear{Okada et~al.,}{Okada
  et~al.}{2013}]{Okada2013ProbingRegions}
Okada Y.,  et~al., 2013, \mn@doi [A{\&}A] {10.1051/0004-6361/201118450}, 553,
  A2

\bibitem[\protect\citeauthoryear{Omont \& Bettinger}{Omont \&
  Bettinger}{2021}]{Omont2021Intermediate-sizeHydrocarbons}
Omont A.,  Bettinger H.~F.,  2021, \mn@doi [A{\&}A]
  {10.1051/0004-6361/202140675}, 650, A193

\bibitem[\protect\citeauthoryear{Omont, Bettinger, Omont  \& Bettinger}{Omont
  et~al.}{2025}]{Omont2025CarbonHydrocarbons}
Omont A.,  Bettinger H.~F.,  Omont A.,   Bettinger H.~F.,  2025, \mn@doi
  [A{\&}A] {10.1051/0004-6361/202453102}, 696, A180

\bibitem[\protect\citeauthoryear{Omukai, Tsuribe, Schneider  \& Ferrara}{Omukai
  et~al.}{2005}]{Omukai2005ThermalEnvironments}
Omukai K.,  Tsuribe T.,  Schneider R.,   Ferrara A.,  2005, \mn@doi [ApJ]
  {10.1086/429955}, 626, 627

\bibitem[\protect\citeauthoryear{Omukai, Hosokawa  \& Yoshida}{Omukai
  et~al.}{2010}]{Omukai2010Low-metallicitySymmetry}
Omukai K.,  Hosokawa T.,   Yoshida N.,  2010, \mn@doi [ApJ]
  {10.1088/0004-637X/722/2/1793}, 722, 1793

\bibitem[\protect\citeauthoryear{Ormel \& Cuzzi}{Ormel \&
  Cuzzi}{2007}]{Ormel2007AstrophysicsNote}
Ormel C.~W.,  Cuzzi J.~N.,  2007, \mn@doi [A{\&}A]
  {10.1051/0004-6361:20066899}, 466, 413

\bibitem[\protect\citeauthoryear{Ormel, Paszun, Dominik  \& Tielens}{Ormel
  et~al.}{2009}]{Ormel2009DustDistribution}
Ormel C.~W.,  Paszun D.,  Dominik C.,   Tielens A.~G.,  2009, \mn@doi [A{\&}A]
  {10.1051/0004-6361/200811158}, 502, 845

\bibitem[\protect\citeauthoryear{Pallottini \& Ferrara}{Pallottini \&
  Ferrara}{2023}]{Pallottini2023StochasticImplications}
Pallottini A.,  Ferrara A.,  2023, \mn@doi [A{\&}A]
  {10.1051/0004-6361/202347384}, 677, 4

\bibitem[\protect\citeauthoryear{Parente, Ragone-Figueroa, Granato, Borgani,
  Murante, Valentini, Bressan  \& Lapi}{Parente
  et~al.}{2022}]{Parente2022DustVolumes}
Parente M.,  Ragone-Figueroa C.,  Granato G.~L.,  Borgani S.,  Murante G.,
  Valentini M.,  Bressan A.,   Lapi A.,  2022, \mn@doi [MNRAS]
  {10.1093/mnras/stac1913}, 515, 2053

\bibitem[\protect\citeauthoryear{Parente, {Parente}  \& {Massimiliano}}{Parente
  et~al.}{2025}]{Parente2025ModelingSimulations}
Parente M.,  {Parente}  {Massimiliano} 2025, \mn@doi [arXiv]
  {10.48550/ARXIV.2504.10585}, p. arXiv:2504.10585

\bibitem[\protect\citeauthoryear{Parker, Zhang, Kim, Kaiser, Landera, Kislov,
  Mebel  \& Tielens}{Parker et~al.}{2012}]{Parker2012LowMedium}
Parker D.~S.,  Zhang F.,  Kim Y.~S.,  Kaiser R.~I.,  Landera A.,  Kislov V.~V.,
   Mebel A.~M.,   Tielens A.~G.,  2012, \mn@doi [Proc. Natl. Acad. Sci. U. S.
  A.] {10.1073/PNAS.1113827108/SUPPL{\_}FILE/ST02.DOC}, 109, 53

\bibitem[\protect\citeauthoryear{Peeters, Hony, Van~Kerckhoven, Tielens,
  Allamandola, Hudgins  \& Bauschlicher}{Peeters
  et~al.}{2002}]{Peeters2002ThePAHs}
Peeters E.,  Hony S.,  Van~Kerckhoven C.,  Tielens A.~G.,  Allamandola L.~J.,
  Hudgins D.~M.,   Bauschlicher C.~W.,  2002, \mn@doi [A{\&}A]
  {10.1051/0004-6361:20020773}, 390, 1089

\bibitem[\protect\citeauthoryear{Peeters, Spoon  \& Tielens}{Peeters
  et~al.}{2004}]{Peeters2004PAHsFormation}
Peeters E.,  Spoon H. W.~W.,   Tielens A. G. G.~M.,  2004, \mn@doi [ApJ]
  {10.1086/423237}, 613, 986

\bibitem[\protect\citeauthoryear{Pilleri, Montillaud, Bern{\'{e}}  \&
  Joblin}{Pilleri et~al.}{2012}]{Pilleri2012EvaporatingRegions}
Pilleri P.,  Montillaud J.,  Bern{\'{e}} O.,   Joblin C.,  2012, \mn@doi
  [A{\&}A] {10.1051/0004-6361/201015915}, 542, A69

\bibitem[\protect\citeauthoryear{Pirronello, Liu, Shen  \& Vidali}{Pirronello
  et~al.}{1996}]{Pirronello1996LaboratoryInterest}
Pirronello V.,  Liu C.,  Shen L.,   Vidali G.,  1996, \mn@doi [ApJL]
  {10.1086/310464}, 475, L69

\bibitem[\protect\citeauthoryear{{Pirronello}, {Biham}, {Liu}, {Shen}  \&
  {Vidali}}{{Pirronello} et~al.}{1997}]{Pirronello1997EfficiencySilicates}
{Pirronello} V.,  {Biham} O.,  {Liu} C.,  {Shen} L.,   {Vidali} G.,  1997,
  \mn@doi [ApJL] {10.1086/310746}, \href
  {https://ui.adsabs.harvard.edu/abs/1997ApJ...483L.131P} {483, L131}

\bibitem[\protect\citeauthoryear{Pirronello, Liu, Roser, Vidali, Pirronello,
  Liu, Roser  \& Vidali}{Pirronello
  et~al.}{1999}]{Pirronello1999MeasurementsGrains}
Pirronello V.,  Liu C.,  Roser J.~E.,  Vidali G.,  Pirronello V.,  Liu C.,
  Roser J.~E.,   Vidali G.,  1999, A{\&}A, 344, 681

\bibitem[\protect\citeauthoryear{Poppe, Blum, Poppe  \& Blum}{Poppe
  et~al.}{1997}]{Poppe1997ExperimentsGrowth}
Poppe T.,  Blum J.,  Poppe T.,   Blum J.,  1997, \mn@doi [AdSpR]
  {10.1016/S0273-1177(97)00817-X}, 20, 1595

\bibitem[\protect\citeauthoryear{Popping, Somerville  \& Galametz}{Popping
  et~al.}{2017}]{Popping2017The9}
Popping G.,  Somerville R.~S.,   Galametz M.,  2017, \mn@doi [MNRAS]
  {10.1093/MNRAS/STX1545}, 471, 3152

\bibitem[\protect\citeauthoryear{{Predehl} \& {Schmitt}}{{Predehl} \&
  {Schmitt}}{1995}]{Predehl1995X-rayingHalos.}
{Predehl} P.,  {Schmitt} J.~H.~M.~M.,  1995, A{\&}A, \href
  {https://ui.adsabs.harvard.edu/abs/1995A&A...293..889P} {293, 889}

\bibitem[\protect\citeauthoryear{Purcell, {Purcell}  \& M.}{Purcell
  et~al.}{1976}]{Purcell1976TemperatureGrains.}
Purcell E.~M.,  {Purcell}  M. E.,  1976, \mn@doi [ApJ] {10.1086/154428}, 206,
  685

\bibitem[\protect\citeauthoryear{Puska, Nieminen, Puska  \& Nieminen}{Puska
  et~al.}{1983}]{Puska1983AtomsSections}
Puska M.~J.,  Nieminen R.~M.,  Puska M.~J.,   Nieminen R.~M.,  1983, \mn@doi
  [PhRvB] {10.1103/PHYSREVB.27.6121}, 27, 6121

\bibitem[\protect\citeauthoryear{Rachford et~al.,}{Rachford
  et~al.}{2002}]{Rachford2002AClouds}
Rachford B.~L.,  et~al., 2002, \mn@doi [ApJ] {10.1086/342146}, 577, 221

\bibitem[\protect\citeauthoryear{Ragone-Figueroa, Granato, Parente, Murante,
  Valentini, Borgani  \& Maio}{Ragone-Figueroa
  et~al.}{2024}]{Ragone-Figueroa2024IntertwinedSimulations}
Ragone-Figueroa C.,  Granato G.~L.,  Parente M.,  Murante G.,  Valentini M.,
  Borgani S.,   Maio U.,  2024, \mn@doi [A{\&}A] {10.1051/0004-6361/202451344},
  691, A200

\bibitem[\protect\citeauthoryear{Rapacioli, Calvo, Spiegelman, Joblin  \&
  Wales}{Rapacioli et~al.}{2005}]{Rapacioli2005StackedMolecules}
Rapacioli M.,  Calvo F.,  Spiegelman F.,  Joblin C.,   Wales D.~J.,  2005,
  \mn@doi [J. Phys. Chem. A] {10.1021/jp046745z}, 109, 2487

\bibitem[\protect\citeauthoryear{Rapacioli, Calvo, Joblin, Parneix, Toublanc
  \& Spiegelman}{Rapacioli et~al.}{2006}]{Rapacioli2006FormationMedium}
Rapacioli M.,  Calvo F.,  Joblin C.,  Parneix P.,  Toublanc D.,   Spiegelman
  F.,  2006, \mn@doi [A{\&}A] {10.1051/0004-6361:20065412}, 460, 519

\bibitem[\protect\citeauthoryear{Rau, Hirashita  \& Murga}{Rau
  et~al.}{2019}]{Rau2019ModellingGalaxies}
Rau S.~J.,  Hirashita H.,   Murga M.,  2019, MNRAS, 489, 5218

\bibitem[\protect\citeauthoryear{Rauls, Hornek{\ae}r, Rauls  \&
  Hornek{\ae}r}{Rauls et~al.}{2008}]{Rauls2008CatalyzedConditions}
Rauls E.,  Hornek{\ae}r L.,  Rauls E.,   Hornek{\ae}r L.,  2008, \mn@doi [ApJ]
  {10.1086/587614}, 679, 531

\bibitem[\protect\citeauthoryear{Reach et~al.,}{Reach
  et~al.}{2005}]{Reach2005AGalaxy}
Reach W.~T.,  et~al., 2005, \mn@doi [AJ] {10.1086/499306}, 131, 1479

\bibitem[\protect\citeauthoryear{Rebrion-Rowe et~al.,}{Rebrion-Rowe
  et~al.}{2003}]{Rebrion-Rowe2003ExperimentalFluoranthene}
Rebrion-Rowe C.,  et~al., 2003, \mn@doi [IJMSp]
  {10.1016/S1387-3806(02)00794-7}, 223-224, 237

\bibitem[\protect\citeauthoryear{Reizer, Viskolcz  \& Fiser}{Reizer
  et~al.}{2022}]{Reizer2022FormationMini-review}
Reizer E.,  Viskolcz B.,   Fiser B.,  2022, \mn@doi [Chemosphere]
  {10.1016/J.CHEMOSPHERE.2021.132793}, 291, 132793

\bibitem[\protect\citeauthoryear{Relanõ et~al.,}{Relanõ
  et~al.}{2016}]{Relano2016Dust33}
Relanõ M.,  et~al., 2016, \mn@doi [A{\&}A] {10.1051/0004-6361/201628139}, 595,
  A43

\bibitem[\protect\citeauthoryear{R{\'{e}}my-Ruyer et~al.,}{R{\'{e}}my-Ruyer
  et~al.}{2014}]{Remy-Ruyer2014Gas-to-dustRange}
R{\'{e}}my-Ruyer A.,  et~al., 2014, \mn@doi [A{\&}A]
  {10.1051/0004-6361/201322803}, 563, A31

\bibitem[\protect\citeauthoryear{R{\'{e}}my-Ruyer et~al.,}{R{\'{e}}my-Ruyer
  et~al.}{2015}]{Remy-Ruyer2015LinkingPicture}
R{\'{e}}my-Ruyer A.,  et~al., 2015, \mn@doi [A{\&}A]
  {10.1051/0004-6361/201526067}, 582, A121

\bibitem[\protect\citeauthoryear{Richter \& Howard}{Richter \&
  Howard}{2002}]{Richter2002FormationFlames}
Richter H.,  Howard J.~B.,  2002, \mn@doi [PCCP] {10.1039/B110089K}, 4, 2038

\bibitem[\protect\citeauthoryear{Rimola, Ceccarelli, Balucani  \&
  Ugliengo}{Rimola et~al.}{2021}]{Rimola2021InteractionImplications}
Rimola A.,  Ceccarelli C.,  Balucani N.,   Ugliengo P.,  2021, \mn@doi
  [Frontiers in Astronomy and Space Sciences]
  {10.3389/FSPAS.2021.655405/BIBTEX}, 8, 655405

\bibitem[\protect\citeauthoryear{Ritter, Herwig, Jones, Pignatari, Fryer  \&
  Hirschi}{Ritter et~al.}{2018}]{Ritter2018NuGrid0.0001-0.02}
Ritter C.,  Herwig F.,  Jones S.,  Pignatari M.,  Fryer C.,   Hirschi R.,
  2018, \mn@doi [MNRAS] {10.1093/mnras/sty1729}, 480, 538

\bibitem[\protect\citeauthoryear{Rodr{\'{i}}guez et~al.,}{Rodr{\'{i}}guez
  et~al.}{2025}]{Rodriguez2025TracingClusters}
Rodr{\'{i}}guez M.~J.,  et~al., 2025, \mn@doi [ApJ] {10.3847/1538-4357/ADBB69},
  983, 137

\bibitem[\protect\citeauthoryear{Romano, Nagamine  \& Hirashita}{Romano
  et~al.}{2022a}]{Romano2022DustGalaxy}
Romano L.~E.,  Nagamine K.,   Hirashita H.,  2022a, \mn@doi [MNRAS]
  {10.1093/mnras/stac1385}, 514, 1441

\bibitem[\protect\citeauthoryear{{Romano}, {Nagamine}  \& {Hirashita}}{{Romano}
  et~al.}{2022b}]{Romano2022TheGalaxy}
{Romano} L. E.~C.,  {Nagamine} K.,   {Hirashita} H.,  2022b, \mn@doi [MNRAS]
  {10.1093/mnras/stac1386}, \href
  {https://ui.adsabs.harvard.edu/abs/2022MNRAS.514.1461R} {514, 1461}

\bibitem[\protect\citeauthoryear{Rosdahl \& Teyssier}{Rosdahl \&
  Teyssier}{2015}]{Rosdahl2015ARAMSES-RT}
Rosdahl J.,  Teyssier R.,  2015, \mn@doi [MNRAS] {10.1093/mnras/stv567}, 449,
  4380

\bibitem[\protect\citeauthoryear{Rosdahl, Blaizot, Aubert, Stranex  \&
  Teyssier}{Rosdahl et~al.}{2013}]{Rosdahl2013Ramses-rt:Context}
Rosdahl J.,  Blaizot J.,  Aubert D.,  Stranex T.,   Teyssier R.,  2013, \mn@doi
  [MNRAS] {10.1093/mnras/stt1722}, 436, 2188

\bibitem[\protect\citeauthoryear{Rosdahl, Schaye, Teyssier  \& Agertz}{Rosdahl
  et~al.}{2015}]{Rosdahl2015GalaxiesGalaxies}
Rosdahl J.,  Schaye J.,  Teyssier R.,   Agertz O.,  2015, \mn@doi [MNRAS]
  {10.1093/MNRAS/STV937}, 451, 34

\bibitem[\protect\citeauthoryear{Rosdahl et~al.,}{Rosdahl
  et~al.}{2018}]{Rosdahl2018}
Rosdahl J.,  et~al., 2018, \mn@doi [MNRAS] {10.1093/mnras/sty1655}, 479, 994

\bibitem[\protect\citeauthoryear{Roser \& Ricca}{Roser \&
  Ricca}{2015}]{Roser2015POLYCYCLICEMISSION}
Roser J.~E.,  Ricca A.,  2015, \mn@doi [ApJ] {10.1088/0004-637X/801/2/108},
  801, 108

\bibitem[\protect\citeauthoryear{Roy et~al.,}{Roy
  et~al.}{2012}]{Roy2012ChangesCloud}
Roy A.,  et~al., 2012, \mn@doi [ApJ] {10.1088/0004-637X/763/1/55}, 763, 55

\bibitem[\protect\citeauthoryear{Sandstrom, Bolatto, Draine, Bot  \&
  Stanimirovi{\'{c}}}{Sandstrom et~al.}{2010}]{Sandstrom2010TheHydrocarbons}
Sandstrom K.~M.,  Bolatto A.~D.,  Draine B.~T.,  Bot C.,   Stanimirovi{\'{c}}
  S.,  2010, \mn@doi [ApJ] {10.1088/0004-637X/715/2/701}, 715, 701

\bibitem[\protect\citeauthoryear{Sandstrom et~al.,}{Sandstrom
  et~al.}{2012}]{Sandstrom2012TheEnvironment}
Sandstrom K.~M.,  et~al., 2012, \mn@doi [ApJ] {10.1088/0004-637X/744/1/20}, 744

\bibitem[\protect\citeauthoryear{{Sandstrom} et~al.,}{{Sandstrom}
  et~al.}{2023}]{Sandstrom2022PHANGS-JWSTGalaxies}
{Sandstrom} K.~M.,  et~al., 2023, \mn@doi [ApJL] {10.3847/2041-8213/aca972},
  \href {https://ui.adsabs.harvard.edu/abs/2023ApJ...944L...8S} {944, L8}

\bibitem[\protect\citeauthoryear{Schaerer, Meynet, Maeder, Schaller, Schaerer,
  Meynet, Maeder  \& Schaller}{Schaerer et~al.}{1993}]{Schaerer1993Grids}
Schaerer D.,  Meynet G.,  Maeder A.,  Schaller G.,  Schaerer D.,  Meynet G.,
  Maeder A.,   Schaller G.,  1993, A{\&}AS, 98, 523

\bibitem[\protect\citeauthoryear{Schlath{\"{o}}lter, Hadjar, Hoekstra,
  Morgenstern, Schlath{\"{o}}lter, Hadjar, Hoekstra  \&
  Morgenstern}{Schlath{\"{o}}lter
  et~al.}{1999}]{Schlatholter1999StrongFullerenes}
Schlath{\"{o}}lter T.,  Hadjar O.,  Hoekstra R.,  Morgenstern R.,
  Schlath{\"{o}}lter T.,  Hadjar O.,  Hoekstra R.,   Morgenstern R.,  1999,
  \mn@doi [PhRvL] {10.1103/PHYSREVLETT.82.73}, 82, 73

\bibitem[\protect\citeauthoryear{Schneider \& Maiolino}{Schneider \&
  Maiolino}{2024}]{Schneider2024TheSources}
Schneider R.,  Maiolino R.,  2024, \mn@doi [A{\&}A Review]
  {10.1007/s00159-024-00151-2}, 32, 2

\bibitem[\protect\citeauthoryear{Schneider \& Omukai}{Schneider \&
  Omukai}{2010}]{Schneider2010MetalsClouds}
Schneider R.,  Omukai K.,  2010, \mn@doi [MNRAS]
  {10.1111/j.1365-2966.2009.15891.x}, 402, 429

\bibitem[\protect\citeauthoryear{Schneider, Ferrara, Natarajan  \&
  Omukai}{Schneider et~al.}{2002}]{Schneider2002FirstMetals}
Schneider R.,  Ferrara A.,  Natarajan P.,   Omukai K.,  2002, \mn@doi [ApJ]
  {10.1086/339917}, 571, 30

\bibitem[\protect\citeauthoryear{Schneider, Omukai, Inoue  \&
  Ferrara}{Schneider et~al.}{2006}]{Schneider2006FragmentationDust}
Schneider R.,  Omukai K.,  Inoue A.,   Ferrara A.,  2006, \mn@doi [MNRAS]
  {10.1111/j.1365-2966.2006.10391.x}, 369, 1437

\bibitem[\protect\citeauthoryear{Schutte}{Schutte}{1995}]{Schutte1995TheIces}
Schutte W.~A.,  1995, \mn@doi [Adv. Space Res.] {10.1016/0273-1177(95)00193-I},
  16, 53

\bibitem[\protect\citeauthoryear{Seitenzahl et~al.,}{Seitenzahl
  et~al.}{2013}]{Seitenzahl2013Three-dimensionalSupernovae}
Seitenzahl I.~R.,  et~al., 2013, \mn@doi [MNRAS] {10.1093/MNRAS/STS402}, 429,
  1156

\bibitem[\protect\citeauthoryear{Seok, Hirashita  \& Asano}{Seok
  et~al.}{2014}]{Seok2014FormationGalaxies}
Seok J.~Y.,  Hirashita H.,   Asano R.~S.,  2014, \mn@doi [MNRAS]
  {10.1093/mnras/stu120}, 439, 2186

\bibitem[\protect\citeauthoryear{Seok, Koo  \& Hirashita}{Seok
  et~al.}{2015}]{Seok2015DustCloud}
Seok J.~Y.,  Koo B.-C.,   Hirashita H.,  2015, \mn@doi [ApJ]
  {10.1088/0004-637X/807/1/100}, 807, 100

\bibitem[\protect\citeauthoryear{Serra D{\'{i}}az-Cano, Jones, Serra
  D{\'{i}}az-Cano  \& Jones}{Serra D{\'{i}}az-Cano
  et~al.}{2008}]{SerraDiaz-Cano2008CarbonaceousGraphite}
Serra D{\'{i}}az-Cano L.,  Jones A.~P.,  Serra D{\'{i}}az-Cano L.,   Jones
  A.~P.,  2008, \mn@doi [A{\&}A] {10.1051/0004-6361:200810622}, 492, 127

\bibitem[\protect\citeauthoryear{Sheffer, Rogers, Federman, Abel, Gredel,
  Lambert  \& Shaw}{Sheffer et~al.}{2008}]{Sheffer2008UltravioletRelationships}
Sheffer Y.,  Rogers M.,  Federman S.~R.,  Abel N.~P.,  Gredel R.,  Lambert
  D.~L.,   Shaw G.,  2008, \mn@doi [ApJ] {10.1086/591484}, 687, 1075

\bibitem[\protect\citeauthoryear{Shivaei et~al.,}{Shivaei
  et~al.}{2017}]{Shivaei20172}
Shivaei I.,  et~al., 2017, \mn@doi [ApJ] {10.3847/1538-4357/aa619c}, 837, 157

\bibitem[\protect\citeauthoryear{{Shivaei} et~al.,}{{Shivaei}
  et~al.}{2024}]{Shivaei2024AMIRI}
{Shivaei} I.,  et~al., 2024, \mn@doi [A{\&}A] {10.1051/0004-6361/202449579},
  \href {https://ui.adsabs.harvard.edu/abs/2024A&A...690A..89S} {690, A89}

\bibitem[\protect\citeauthoryear{Shukla \& Koshi}{Shukla \&
  Koshi}{2010}]{Shukla2010AHydrocarbons}
Shukla B.,  Koshi M.,  2010, \mn@doi [PCCP] {10.1039/B919644G}, 12, 2427

\bibitem[\protect\citeauthoryear{Shukla \& Koshi}{Shukla \&
  Koshi}{2012}]{Shukla2012AMechanisms}
Shukla B.,  Koshi M.,  2012, \mn@doi [Combust. Flame]
  {10.1016/J.COMBUSTFLAME.2012.08.007}, 159, 3589

\bibitem[\protect\citeauthoryear{Shukla, Susa, Miyoshi  \& Koshi}{Shukla
  et~al.}{2008}]{Shukla2008RoleHydrocarbons}
Shukla B.,  Susa A.,  Miyoshi A.,   Koshi M.,  2008, \mn@doi [The Journal of
  Physical Chemistry A] {10.1021/jp7098398}, 112, 2362

\bibitem[\protect\citeauthoryear{{Sigmund}}{{Sigmund}}{1969}]{Sigmund1969TheoryTargets}
{Sigmund} P.,  1969, \mn@doi [PhRv] {10.1103/PhysRev.184.383}, \href
  {https://ui.adsabs.harvard.edu/abs/1969PhRv..184..383S} {184, 383}

\bibitem[\protect\citeauthoryear{{Sigmund}}{{Sigmund}}{1981}]{Sigmund1981SputteringConcepts}
{Sigmund} P.,  1981, in {Behrisch} R.,  ed., , Vol.~47, Sputtering by Particle
  Bombardment I: Physical Sputtering of Single-Element Solids.
Springer-Verlag, p.~9, \mn@doi{10.1007/3540105212_7}

\bibitem[\protect\citeauthoryear{Sigmund}{Sigmund}{1996}]{Sigmund1996Low-speedFormula}
Sigmund P.,  1996, \mn@doi [Phys. Rev. A] {10.1103/PHYSREVA.54.3113}, 54, 3113

\bibitem[\protect\citeauthoryear{Sigmund}{Sigmund}{2014}]{Sigmund2014StoppingIons}
Sigmund P.,  2014, \mn@doi [Springer Ser. Solid-State Sci.]
  {10.1007/978-3-319-05564-0{\_}8}, 179, 343

\bibitem[\protect\citeauthoryear{Silsbee, Caselli  \& Ivlev}{Silsbee
  et~al.}{2021}]{Silsbee2021IceSize}
Silsbee K.,  Caselli P.,   Ivlev A.~V.,  2021, \mn@doi [MNRAS]
  {10.1093/mnras/stab2546}, 507, 6205

\bibitem[\protect\citeauthoryear{Slavin, Dwek  \& Jones}{Slavin
  et~al.}{2015}]{Slavin2015DestructionWaves}
Slavin J.~D.,  Dwek E.,   Jones A.~P.,  2015, \mn@doi [ApJ]
  {10.1088/0004-637X/803/1/7}, 803, 7

\bibitem[\protect\citeauthoryear{Slavin, Dwek, Low  \& Hill}{Slavin
  et~al.}{2020}]{Slavin2020TheGrains}
Slavin J.~D.,  Dwek E.,  Low M.-M.~M.,   Hill A.~S.,  2020, \mn@doi [ApJ]
  {10.3847/1538-4357/ABB5A4}, 902, 135

\bibitem[\protect\citeauthoryear{Sloan et~al.,}{Sloan
  et~al.}{2007}]{Sloan2007TheAliphatics}
Sloan G.~C.,  et~al., 2007, \mn@doi [ApJ] {10.1086/519236}, 664, 1144

\bibitem[\protect\citeauthoryear{Sloan, {Sloan}  \& C.}{Sloan
  et~al.}{2017}]{Sloan2017Carbon-richNebulae}
Sloan G.~C.,  {Sloan}  C. G.,  2017, \mn@doi [P{\&}SS]
  {10.1016/J.PSS.2017.07.017}, 149, 32

\bibitem[\protect\citeauthoryear{Smolders et~al.,}{Smolders
  et~al.}{2010}]{Smolders2010WhenStars}
Smolders K.,  et~al., 2010, \mn@doi [A{\&}A] {10.1051/0004-6361/201014254},
  514, L1

\bibitem[\protect\citeauthoryear{Snow, Bierbaum, Snow  \& Bierbaum}{Snow
  et~al.}{2008}]{Snow2008IonMedium}
Snow T.~P.,  Bierbaum V.~M.,  Snow T.~P.,   Bierbaum V.~M.,  2008, \mn@doi
  [ARAC] {10.1146/ANNUREV.ANCHEM.1.031207.112907}, 1, 229

\bibitem[\protect\citeauthoryear{Sofia, Meyer, Sofia  \& Meyer}{Sofia
  et~al.}{2001}]{Sofia2001InterstellarRevisited}
Sofia U.~J.,  Meyer D.~M.,  Sofia U.~J.,   Meyer D.~M.,  2001, \mn@doi [ApJL]
  {10.1086/321715}, 554, L221

\bibitem[\protect\citeauthoryear{Spitzer, {Spitzer}, {Lyman}  \& {Jr.}}{Spitzer
  et~al.}{1948}]{Spitzer1948TheI.}
Spitzer Lyman J.,  {Spitzer} {Lyman}  {Jr.} 1948, \mn@doi [ApJ]
  {10.1086/144984}, 107, 6

\bibitem[\protect\citeauthoryear{Stanway \& Eldridge}{Stanway \&
  Eldridge}{2018}]{Stanway2018Re-evaluatingPopulations}
Stanway E.~R.,  Eldridge J.~J.,  2018, \mn@doi [MNRAS] {10.1093/mnras/sty1353},
  479, 75

\bibitem[\protect\citeauthoryear{Stepnik et~al.,}{Stepnik
  et~al.}{2003}]{Stepnik2003EvolutionFilament}
Stepnik B.,  et~al., 2003, \mn@doi [A{\&}A] {10.1051/0004-6361:20021309}, 398,
  551

\bibitem[\protect\citeauthoryear{Sutherland \& Dopita}{Sutherland \&
  Dopita}{1993}]{Sutherland1993}
Sutherland R.~S.,  Dopita M.~A.,  1993, \mn@doi [ApJS] {10.1086/191823}, 88,
  253

\bibitem[\protect\citeauthoryear{Sutter et~al.,}{Sutter
  et~al.}{2024}]{Sutter2024TheGalaxies}
Sutter J.,  et~al., 2024, \mn@doi [ApJ] {10.3847/1538-4357/AD54BD}, 971, 178

\bibitem[\protect\citeauthoryear{Taniguchi et~al.,}{Taniguchi
  et~al.}{2025}]{Taniguchi2025TheJWST}
Taniguchi K.,  et~al., 2025, \mn@doi [ApJ] {10.3847/1538-4357/AE045C}, 993, 104

\bibitem[\protect\citeauthoryear{Tappe, Rho, Reach, Tappe, Rho  \& Reach}{Tappe
  et~al.}{2006}]{Tappe2006ShockN132D}
Tappe A.,  Rho J.,  Reach W.~T.,  Tappe A.,  Rho J.,   Reach W.~T.,  2006,
  \mn@doi [ApJ] {10.1086/508741}, 653, 267

\bibitem[\protect\citeauthoryear{Temim \& Dwek}{Temim \&
  Dwek}{2013}]{Temim2013THENEBULA}
Temim T.,  Dwek E.,  2013, \mn@doi [ApJ] {10.1088/0004-637X/774/1/8}, 774, 8

\bibitem[\protect\citeauthoryear{Teyssier}{Teyssier}{2002}]{Teyssier2002}
Teyssier R.,  2002, \mn@doi [A{\&}A] {10.1051/0004-6361:20011817}, 385, 337

\bibitem[\protect\citeauthoryear{Thompson, Quataert  \& Murray}{Thompson
  et~al.}{2005}]{Thompson2005RadiationFueling}
Thompson T.~A.,  Quataert E.,   Murray N.,  2005, \mn@doi [ApJ]
  {10.1086/431923}, 630, 167

\bibitem[\protect\citeauthoryear{Thompson, Richings, Gibson,
  Faucher-Gigu{\`{e}}re, Feldmann  \& Hayward}{Thompson
  et~al.}{2024}]{Thompson2024PredictionsChemistry}
Thompson O.~A.,  Richings A.~J.,  Gibson B.~K.,  Faucher-Gigu{\`{e}}re C.~A.,
  Feldmann R.,   Hayward C.~C.,  2024, \mn@doi [MNRAS]
  {10.1093/mnras/stae1486}, 532, 1948

\bibitem[\protect\citeauthoryear{Thrower et~al.,}{Thrower
  et~al.}{2012}]{Thrower2012ExperimentalFormation}
Thrower J.~D.,  et~al., 2012, \mn@doi [ApJ] {10.1088/0004-637X/752/1/3}, 752, 3

\bibitem[\protect\citeauthoryear{Tielens}{Tielens}{2005}]{Tielens2005TheMedium}
Tielens A. G. G.~M.,  2005, {The Physics and Chemistry of the Interstellar
  Medium}.
Cambridge University Press, \mn@doi{10.1017/CBO9780511819056}, \url
  {https://www.cambridge.org/core/product/identifier/9780511819056/type/book}

\bibitem[\protect\citeauthoryear{Tielens}{Tielens}{2008}]{Tielens2008InterstellarMolecules}
Tielens A.,  2008, \mn@doi [ARA{\&}A] {10.1146/annurev.astro.46.060407.145211},
  46, 289

\bibitem[\protect\citeauthoryear{Tielens}{Tielens}{2021}]{Tielens2021MolecularAstrophysics}
Tielens A. G. G.~M.,  2021, {Molecular Astrophysics}.
Cambridge University Press, \mn@doi{10.1017/9781316718490}, \url
  {https://www.cambridge.org/core/product/identifier/9781316718490/type/book}

\bibitem[\protect\citeauthoryear{{Tielens} \& {Hagen}}{{Tielens} \&
  {Hagen}}{1982}]{Tielens1982ModelMantles}
{Tielens} A.~G.~G.~M.,  {Hagen} W.,  1982, A{\&}A, \href
  {https://ui.adsabs.harvard.edu/abs/1982A&A...114..245T} {114, 245}

\bibitem[\protect\citeauthoryear{Tielens, McKee, Seab, Hollenbach, Tielens,
  McKee, Seab  \& Hollenbach}{Tielens et~al.}{1994}]{Tielens1994TheShocks}
Tielens A. G. G.~M.,  McKee C.~F.,  Seab C.~G.,  Hollenbach D.~J.,  Tielens A.
  G. G.~M.,  McKee C.~F.,  Seab C.~G.,   Hollenbach D.~J.,  1994, \mn@doi [ApJ]
  {10.1086/174488}, 431, 321

\bibitem[\protect\citeauthoryear{Tobita et~al.,}{Tobita
  et~al.}{1992}]{Tobita1992PolycyclicImpact}
Tobita S.,  et~al., 1992, \mn@doi [CP] {10.1016/0301-0104(92)80165-R}, 161, 501

\bibitem[\protect\citeauthoryear{Totton, Misquitta  \& Kraft}{Totton
  et~al.}{2012}]{Totton2012ATemperatures}
Totton T.~S.,  Misquitta A.~J.,   Kraft M.,  2012, \mn@doi [PCCP]
  {10.1039/C2CP23008A}, 14, 4081

\bibitem[\protect\citeauthoryear{Trayford et~al.,}{Trayford
  et~al.}{2025}]{Trayford2025ModellingMedium}
Trayford J.~W.,  et~al., 2025, \mn@doi [MNRAS] {10.1093/mnras/staf2040}, 000, 1

\bibitem[\protect\citeauthoryear{Trebitsch et~al.,}{Trebitsch
  et~al.}{2021}]{Trebitsch2021TheProtoclusters}
Trebitsch M.,  et~al., 2021, \mn@doi [A{\&}A] {10.1051/0004-6361/202037698},
  653, A154

\bibitem[\protect\citeauthoryear{{Vasiliev}}{{Vasiliev}}{2025}]{Vasiliev2025DestructionSupernova}
{Vasiliev} E.~O.,  2025, \mn@doi [arXiv] {10.48550/arXiv.2512.24677}, \href
  {https://ui.adsabs.harvard.edu/abs/2025arXiv251224677V} {p. arXiv:2512.24677}

\bibitem[\protect\citeauthoryear{{Vasiliev} \& {Nath}}{{Vasiliev} \&
  {Nath}}{2025}]{Vasiliev2025DustExplosions}
{Vasiliev} E.~O.,  {Nath} B.~B.,  2025, \mn@doi [arXiv]
  {10.48550/arXiv.2512.24670}, \href
  {https://ui.adsabs.harvard.edu/abs/2025arXiv251224670V} {p. arXiv:2512.24670}

\bibitem[\protect\citeauthoryear{Verstraete}{Verstraete}{2021}]{Verstraete2021TheMedium}
Verstraete L.,  2021, \mn@doi [PAHs and the Universe]
  {10.1051/978-2-7598-2482-3-045/XML}, pp 415--426

\bibitem[\protect\citeauthoryear{{Verstraete}, {Leger}, {D'Hendecourt},
  {Defourneau}  \& {Dutuit}}{{Verstraete}
  et~al.}{1990}]{Verstraete1990IonizationGas}
{Verstraete} L.,  {Leger} A.,  {D'Hendecourt} L.,  {Defourneau} D.,   {Dutuit}
  O.,  1990, A{\&}A, \href
  {https://ui.adsabs.harvard.edu/abs/1990A&A...237..436V} {237, 436}

\bibitem[\protect\citeauthoryear{Verstraete et~al.,}{Verstraete
  et~al.}{2001}]{Verstraete2001TheModel}
Verstraete L.,  et~al., 2001, \mn@doi [A{\&}A] {10.1051/0004-6361:20010515},
  372, 981

\bibitem[\protect\citeauthoryear{Vijayan, Clay, Thomas, Yates, Wilkins  \&
  Henriques}{Vijayan et~al.}{2019}]{Vijayan2019DetailedFormation}
Vijayan A.~P.,  Clay S.~J.,  Thomas P.~A.,  Yates R.~M.,  Wilkins S.~M.,
  Henriques B.~M.,  2019, \mn@doi [MNRAS] {10.1093/mnras/stz1948}, 489, 4072

\bibitem[\protect\citeauthoryear{Virtanen et~al.,}{Virtanen
  et~al.}{2020}]{Virtanen2020SciPyPython}
Virtanen P.,  et~al., 2020, \mn@doi [Nat. Methods] {10.1038/s41592-019-0686-2},
  17, 261

\bibitem[\protect\citeauthoryear{{Voelk}, {Jones}, {Morfill}  \&
  {Roeser}}{{Voelk} et~al.}{1980}]{Voelk1980CollisionsGas}
{Voelk} H.~J.,  {Jones} F.~C.,  {Morfill} G.~E.,   {Roeser} S.,  1980, A{\&}A,
  \href {https://ui.adsabs.harvard.edu/abs/1980A&A....85..316V} {85, 316}

\bibitem[\protect\citeauthoryear{Vogelsberger, Mckinnon, O'neil, Marinacci,
  Torrey  \& Kannan}{Vogelsberger et~al.}{2019}]{Vogelsberger2019DustCooling}
Vogelsberger M.,  Mckinnon R.,  O'neil S.,  Marinacci F.,  Torrey P.,   Kannan
  R.,  2019, \mn@doi [MNRAS] {10.1093/MNRAS/STZ1644}, 487, 4870

\bibitem[\protect\citeauthoryear{Wakelam et~al.,}{Wakelam
  et~al.}{2017}]{Wakelam2017H2Observations}
Wakelam V.,  et~al., 2017, \mn@doi [Mol. Astrophys.]
  {10.1016/J.MOLAP.2017.11.001}, 9, 1

\bibitem[\protect\citeauthoryear{Weingartner \& Draine}{Weingartner \&
  Draine}{1999}]{Weingartner1999InterstellarGrains}
Weingartner J.~C.,  Draine B.~T.,  1999, \mn@doi [ApJ] {10.1086/307197}, 517,
  292

\bibitem[\protect\citeauthoryear{Weingartner \& Draine}{Weingartner \&
  Draine}{2000}]{Weingartner2000DustSMC}
Weingartner J.~C.,  Draine B.~T.,  2000, \mn@doi [ApJ] {10.1086/318651}, 548,
  296

\bibitem[\protect\citeauthoryear{Weingartner \& Draine}{Weingartner \&
  Draine}{2001a}]{Weingartner2001PhotoelectricHeating}
Weingartner J.~C.,  Draine B.~T.,  2001a, \mn@doi [ApJS]
  {10.1086/320852/FULLTEXT/}, 134, 263

\bibitem[\protect\citeauthoryear{Weingartner \& Draine}{Weingartner \&
  Draine}{2001b}]{Weingartner2001DustCloud}
Weingartner J.,  Draine B.,  2001b, \mn@doi [ApJ] {10.1086/318651/FULLTEXT/},
  548, 296

\bibitem[\protect\citeauthoryear{Weingartner \& Draine}{Weingartner \&
  Draine}{2001c}]{Weingartner2001ElectronIonHydrocarbons}
Weingartner J.~C.,  Draine B.~T.,  2001c, \mn@doi [ApJ]
  {10.1086/324035/FULLTEXT/}, 563, 842

\bibitem[\protect\citeauthoryear{Wenzel et~al.,}{Wenzel
  et~al.}{2020}]{Wenzel2020AstrochemicalCations}
Wenzel G.,  et~al., 2020, \mn@doi [A{\&}A] {10.1051/0004-6361/202038139}, 641,
  A98

\bibitem[\protect\citeauthoryear{Wenzel et~al.,}{Wenzel
  et~al.}{2024}]{Wenzel2024DetectionHydrocarbon}
Wenzel G.,  et~al., 2024, \mn@doi [Science] {10.1126/SCIENCE.ADQ6391}, 386, 810

\bibitem[\protect\citeauthoryear{Wenzel et~al.,}{Wenzel
  et~al.}{2025}]{Wenzel2025DiscoveryTMC-1}
Wenzel G.,  et~al., 2025, \mn@doi [ApJL] {10.3847/2041-8213/adc911}, 984, L36

\bibitem[\protect\citeauthoryear{Whitcomb et~al.,}{Whitcomb
  et~al.}{2024}]{Whitcomb2024TheSpectroscopy}
Whitcomb C.~M.,  et~al., 2024, \mn@doi [ApJ] {10.3847/1538-4357/ad66c8}, 974,
  20

\bibitem[\protect\citeauthoryear{Wiseman, Schady, Bolmer, Kr{\"{u}}hler, Yates,
  Greiner  \& Fynbo}{Wiseman et~al.}{2017}]{Wiseman2017EvolutionGRB-DLAs}
Wiseman P.,  Schady P.,  Bolmer J.,  Kr{\"{u}}hler T.,  Yates R.~M.,  Greiner
  J.,   Fynbo J.~P.,  2017, \mn@doi [A{\&}A] {10.1051/0004-6361/201629228},
  599, 24

\bibitem[\protect\citeauthoryear{Wolfire, McKee, Hollenbach  \&
  Tielens}{Wolfire et~al.}{2003}]{Wolfire2003NeutralGalaxy}
Wolfire M.~G.,  McKee C.~F.,  Hollenbach D.,   Tielens A. G. G.~M.,  2003,
  \mn@doi [ApJ] {10.1086/368016}, 587, 278

\bibitem[\protect\citeauthoryear{Wolfire, Tielens, Hollenbach  \&
  Kaufman}{Wolfire et~al.}{2008}]{Wolfire2008ChemicalFormation}
Wolfire M.~G.,  Tielens A. G. G.~M.,  Hollenbach D.,   Kaufman M.~J.,  2008,
  \mn@doi [ApJ] {10.1086/587688}, 680, 384

\bibitem[\protect\citeauthoryear{Yan \& Lazarian}{Yan \&
  Lazarian}{2002}]{Yan2002GrainMechanism}
Yan H.,  Lazarian A.,  2002, \mn@doi [ApJL] {10.1086/377487}, 592, L33

\bibitem[\protect\citeauthoryear{Yan, Lazarian  \& Draine}{Yan
  et~al.}{2004}]{Yan2004DustTurbulence}
Yan H.,  Lazarian A.,   Draine B.~T.,  2004, \mn@doi [ApJ] {10.1086/425111},
  616, 895

\bibitem[\protect\citeauthoryear{Zacharia, Ulbricht  \& Hertel}{Zacharia
  et~al.}{2004}]{Zacharia2004InterlayerHydrocarbons}
Zacharia R.,  Ulbricht H.,   Hertel T.,  2004, \mn@doi [Phys. Rev. B]
  {10.1103/PhysRevB.69.155406}, 69, 155406

\bibitem[\protect\citeauthoryear{Zhang et~al.,}{Zhang
  et~al.}{2010}]{Zhang2010FormationStudy}
Zhang F.,  et~al., 2010, \mn@doi [JAChS] {10.1021/JA908559V}, 132, 2672

\bibitem[\protect\citeauthoryear{Zhang, Ye, Li, Bi  \& Cong}{Zhang
  et~al.}{2025}]{Zhang2025KineticPropyne}
Zhang Z.,  Ye L.,  Li M.,  Bi Y.,   Cong H.,  2025, \mn@doi [Appl. Energy
  Combust. Sci.] {10.1016/J.JAECS.2025.100391}, 24, 100391

\bibitem[\protect\citeauthoryear{Zhao et~al.,}{Zhao
  et~al.}{2018}]{Zhao2018Low-temperatureAtmosphere}
Zhao L.,  et~al., 2018, \mn@doi [Nat. Astron.] {10.1038/s41550-018-0585-y}, 2,
  973

\bibitem[\protect\citeauthoryear{Zhao, Chen, Li, Zhao, Chen  \& Li}{Zhao
  et~al.}{2025}]{Zhao2025ObservationalPhase}
Zhao H.,  Chen B.,  Li J.,  Zhao H.,  Chen B.,   Li J.,  2025, \mn@doi [ApJL]
  {10.3847/2041-8213/AE06FE}, 991, L36

\bibitem[\protect\citeauthoryear{Zhen, Castellanos, Paardekooper, Linnartz  \&
  Tielens}{Zhen et~al.}{2014}]{Zhen2014LaboratoryChemistry}
Zhen J.,  Castellanos P.,  Paardekooper D.~M.,  Linnartz H.,   Tielens A.~G.,
  2014, \mn@doi [ApJL] {10.1088/2041-8205/797/2/L30}, 797

\bibitem[\protect\citeauthoryear{Zhen, Castellanos, Paardekooper, Ligterink,
  Linnartz, Nahon, Joblin  \& Tielens}{Zhen
  et~al.}{2015}]{Zhen2015LABORATORYFRAGMENTATION}
Zhen J.,  Castellanos P.,  Paardekooper D.~M.,  Ligterink N.,  Linnartz H.,
  Nahon L.,  Joblin C.,   Tielens A.~G.,  2015, \mn@doi [ApJL]
  {10.1088/2041-8205/804/1/L7}, 804, L7

\bibitem[\protect\citeauthoryear{Zhen et~al.,}{Zhen
  et~al.}{2016}]{Zhen2016VUVPROCESSES}
Zhen J.,  et~al., 2016, \mn@doi [ApJ] {10.3847/0004-637x/822/2/113}, 822, 113

\bibitem[\protect\citeauthoryear{Zhu, Stone  \& Bai}{Zhu
  et~al.}{2015}]{Zhu2015DUSTDIFFUSION}
Zhu Z.,  Stone J.~M.,   Bai X.~N.,  2015, \mn@doi [ApJ]
  {10.1088/0004-637X/801/2/81}, 801, 81

\bibitem[\protect\citeauthoryear{Zhukovska}{Zhukovska}{2014}]{Zhukovska2014DustSupernovae}
Zhukovska S.,  2014, \mn@doi [A{\&}A] {10.1051/0004-6361/201322989}, 562, 76

\bibitem[\protect\citeauthoryear{Zhukovska, Gail  \& Trieloff}{Zhukovska
  et~al.}{2008}]{Zhukovska2008EvolutionNeighbourhood}
Zhukovska S.,  Gail H.~P.,   Trieloff M.,  2008, \mn@doi [A{\&}A]
  {10.1051/0004-6361:20077789}, 479, 453

\bibitem[\protect\citeauthoryear{Zhukovska, Dobbs, Jenkins  \&
  Klessen}{Zhukovska et~al.}{2016}]{Zhukovska2016MODELINGISM}
Zhukovska S.,  Dobbs C.,  Jenkins E.~B.,   Klessen R.~S.,  2016, \mn@doi [ApJ]
  {10.3847/0004-637X/831/2/147}, 831, 147

\bibitem[\protect\citeauthoryear{Zhukovska, Henning  \& Dobbs}{Zhukovska
  et~al.}{2018}]{Zhukovska2018IronDepletions}
Zhukovska S.,  Henning T.,   Dobbs C.,  2018, \mn@doi [ApJ]
  {10.3847/1538-4357/aab438}, 857, 94

\bibitem[\protect\citeauthoryear{Ziegler, Biersack, Ziegler  \&
  Biersack}{Ziegler et~al.}{1985}]{Ziegler1985TheMatter}
Ziegler J.~F.,  Biersack J.~P.,  Ziegler J.~F.,   Biersack J.~P.,  1985,
  \mn@doi [this] {10.1007/978-1-4615-8103-1{\_}3}, p.~93

\bibitem[\protect\citeauthoryear{Zubko, Dwek  \& Arendt}{Zubko
  et~al.}{2004}]{Zubko2004InterstellarConstraints}
Zubko V.,  Dwek E.,   Arendt R.~G.,  2004, \mn@doi [ApJS] {10.1086/382351},
  152, 211

\bibitem[\protect\citeauthoryear{van Hoof, Weingartner, Martin, Volk  \&
  Ferland}{van Hoof et~al.}{2004}]{VanHoof2004GrainMedia}
van Hoof P. A.~M.,  Weingartner J.~C.,  Martin P.~G.,  Volk K.,   Ferland
  G.~J.,  2004, \mn@doi [MNRAS] {10.1111/j.1365-2966.2004.07734.x}, 350, 1330

\bibitem[\protect\citeauthoryear{van~der Velden}{van~der
  Velden}{2020}]{vanderVelden2020CMasher:Plots}
van~der Velden E.,  2020, \mn@doi [JOSS] {10.21105/joss.02004}, 5, 2004

\makeatother
\end{thebibliography}

\begin{appendix}

\section{Numerical implementation of equilibrium dust charging} \label{ap:dust_charging}
We compute the equilibrium probability distribution $P(Q_{\rm g})$ for a grain to
possess integer charge $Q_{
m g}$ by enforcing detailed balance among the
microphysical charging channels. The dominant processes included in the
present implementation are \citep{Weingartner2001PhotoelectricHeating}:
\begin{itemize}
  \item Photoelectric emission (rate per grain):
  \begin{equation}
    \begin{aligned}
    J_{\rm pe}(Q_{\rm g}) ={}&
    \int_{\nu_{\rm pet}}^{\nu_{\rm max}}
    \frac{c\,u_{\nu}}{h\nu}\,
    Y(Q_{\rm g},E)\,
    C_{\rm abs}(\nu,a)\, d\nu \\
    &+
    \int_{\nu_{\rm pd}}^{\nu_{\rm max}}
    \frac{c\,u_{\nu}}{h\nu}\,
    \sigma_{\rm pd}(\nu,Q_{\rm g},a)\, d\nu .
    \end{aligned}
    \end{equation}
  The first term captures the photo-emission of valence electrons, whereas the second term considers the photo-detachment of captured electrons residing in the ``lowest unoccupied molecular orbit'' when $Q_{\rm g}<0$. Here $u_{\nu}$ is the radiation energy density per unit frequency, $C_{\rm abs}(\nu,a)$ is the grain
  absorption cross section for radius $a$, $Y(Q_{\rm g},E)$ is the photoelectric yield
  as a function of emitted-electron energy $E$ and grain charge $Q_{\rm g}$, and
  $\sigma_{\rm pd}(\nu,Q_{\rm g},a)$ denotes the photo-detachment cross section. The integrals over radiation bins are limited at the top by the maximum radiation energy $\nu_{\rm max}$ (e.g. 13.6\,eV), the photo-emission threshold energy $\nu_{\rm pet}$, and the photo-detachment threshold energy $\nu_{\rm pd}$. We use dust cross sections from \cite{Draine1984OpticalGrains,Laor1993SpectroscopicNuclei,Weingartner2000DustSMC} and the dielectric tables from \cite{Draine2003ScatteringUltraviolet}.
  
  \item Electron recombination and collisional charging by ions and electrons (rate per grain, per collisional species):
  \begin{equation}
      J_{i}(Q_{\rm g}) = n_is_i(Q_{\rm g})\left(\frac{8k_{\rm B}T}{\pi m_i}\right)^{1/2}\pi a^2 \tilde{J}(Q_{\rm g},T,q_i),
  \end{equation}
  where $n_i$ is the number densities of electrons and ions with charge $q_i$ and mass $m_i$, $s_i$ their sticking coefficient, and $\tilde{J}(Q_{\rm g},T,q_i)$ is the Coulomb focusing function from \citet{Draine1987CollisionalGrains}. This expression assumes a Maxwellian distribution of collisional species.
\end{itemize}

For compact notation we denote the net upward transition rate (loss of one electron) from $Q_{\rm g}$ to $Q_{\rm g}+1$ by $R^{+}(Q_{\rm g})$ and the downward transition rate (gain of one electron) by $R^{-}(Q_{\rm g})$. The key insight from steady state is that the detailed-balance recurrence yields:
\begin{equation}
  P(Q_{\rm g}+1) = P(Q_{\rm g})\,\frac{R^{+}(Q_{\rm g})}{R^{-}(Q_{\rm g}+1)}.
\end{equation}

Normalisation is imposed by requiring $\sum_{Q_{\rm g}} P(Q_{\rm g})=1$ after truncating the charge domain to a finite bracket $Q_{\min}\ldots Q_{\max}$ chosen so that the tail probabilities at the boundaries are below a user-specified tolerance. We have found that $
\sim 10^{-8}$ is sufficient for converged distributions at all relevant ISM conditions. We also impose physical limits to the GCD, given by the most negative allowed charge $Q_{\rm min}$ before auto-ionisation occurs, and the maximum positive charge $Q_{\rm max}$ allowed by the ionisation potential (IPV) for $h\nu_{\rm max}$ \citep[see equations 24 and eq. 23 in][]{Weingartner2001PhotoelectricHeating}.

The algorithm uses the following strategy to select a reference charge $Q_{\rm peak}$ and the bounds of the GCD:
\begin{enumerate}
    \item The most probable charge $Q_{\rm peak}$ should occur approximately when $R^+(Q_{\rm g}) \simeq R^-(Q_{\rm g}+1)$. Starting at $Q_{\rm g}=0$, the code steps up or down until:
    \begin{equation}
        r(Q_{\rm g}) = \frac{R^+(Q_{\rm g})}{R^-(Q_{\rm g}+1)}
    \end{equation}
    crosses 1. The crossing point is used as $Q_{\rm peak}$.
    \item Moving up and down from $Q_{\rm peak}$, the probability distributions is computed until the lower bound $P(Q_{\rm lower})\rightarrow \epsilon$ and upper bound $P(Q_{\rm upper})\rightarrow \epsilon$ (where $\epsilon$ is the tolerance error).
\end{enumerate}
Many of the $R^{+}(Q_{\rm g})$ and $R^{-}(Q_{\rm g}+1)$ computed during the search of $Q_{\rm peak}$ can be saved and used afterwards during the computation of the $P(Q_{\rm g})$ distribution with the detail balance recurrence. This is a key optimisation feature of our method.

\begin{figure}
    \centering
	\includegraphics[width=\columnwidth]{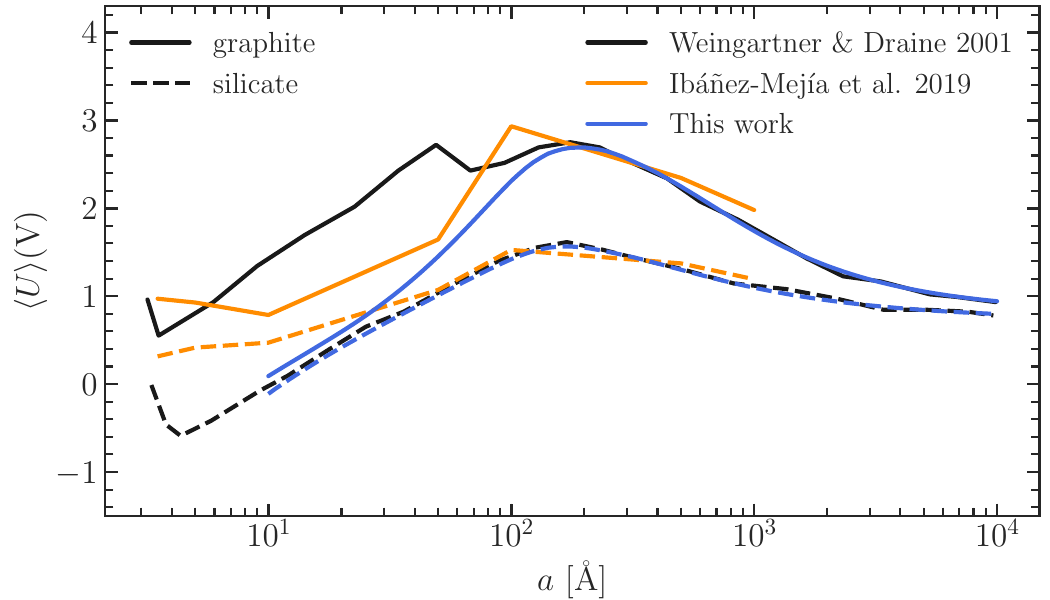}
    \caption{Average electrostatic potential of dust grains embedded in the WNM irradiated with the \citet{Mathis1983InterstellarClouds} ISRF. Solid lines are for graphite grains, while dashed lines are for silicates. We compare the results of our model (blue lines) with the results from \citet{Weingartner2001PhotoelectricHeating} (black lines) and \citet{Ibanez-Mejia2019DustMedium} (orange lines). Our results are in broad agreement with the original model by \citet{Weingartner2001PhotoelectricHeating}, although our choice of pure graphitic grains for $a\lesssim 100$\,\AA~results in lower UV absorption efficiencies compared to the PAH+graphite mixture in \citet{Weingartner2001PhotoelectricHeating}.}
    \label{fig:average_potential_comp}
\end{figure}

We test our implementation by computing the average electrostatic potential:
\begin{equation}
    U = \frac{eQ_{\rm g}}{4\pi\epsilon_0a},
\end{equation}
for grain of different sizes and compositions embedded in the \citet{Mathis1983InterstellarClouds} ISRF and within the WNM medium (i.e.~$n_{\rm e}=0.03$\,cm$^{-3}$ and $T=6000$\,K). $\epsilon_0$ is the vacuum permitivity. The results in figure 10 of \citet{Weingartner2001PhotoelectricHeating} are shown in Fig.~\ref{fig:average_potential_comp} with black solid (dashed) lines for graphite (silicate) grains. For the same compositions, our model results in the blue curves, which we have computed down to the minimum size allowed by the optical properties in \cite{Draine1984OpticalGrains}. Our results are in excellent agreement with \citet{Weingartner2001PhotoelectricHeating}, except for graphite grains of radius below $100$\,\AA. The origin of this discrepancy is the choice of optical properties for very small graphite grains in \citet{Weingartner2001PhotoelectricHeating}, which are not the pure graphite properties used in \calima, but rather with a mixture of PAH optical properties as described \citet{Li2001InfraredMedium}. Given that within this regime of very small graphite grains in \calima~we transition to the contribution of PAHs, we regard this discrepancy as a feature of our choice of dust discretisation. For comparison, we also show the prediction from fitting functions by \citet{Ibanez-Mejia2019DustMedium} as orange lines. Our results are in overall agreement with theirs in the regime of $a\gtrsim 100$\,\AA, but we find significant deviations for smaller sizes, with their prediction of small silicates exceeding both ours and \citet{Weingartner2001PhotoelectricHeating}.

\subsection{The interaction of dust grains with high energy ions}\label{ap:dust_sputtering}
\begin{table*}
    \centering
    \begin{tabular}{c|c|c|c|c|c|c|c}
         & $s$ [g\,cm$^{-3}$] & $U_0$ [eV] & $K$ &
         $\langle Z_j \rangle$ &
         $\langle m_j^{\rm atom}\rangle$ [a.m.u.] &
         $\langle N_{\rm val} \rangle$ &
         $Q_{\rm d}^*$ [dyn\,cm\,g$^{-1}$] \\
         \hline
         \textbf{Carbon}   & 2.2 & 4.0 & 0.61 & 6  & 12.011 & 4 & $8.9 \times 10^{9}$ \\
         \textbf{Silicate} & 3.3 & 5.7 & 0.10 & 12 & 24.605 & 5 & $4.3 \times 10^{10}$ \\
    \end{tabular}
    \caption{Fiducial material parameters for dust species. From left to right: bulk material density, work function, normalization factor in the photoelectric yield model, mean atomic number, mean atomic mass, average number of valence electrons, and dust specific binding energy.}
    \label{tab:dust_material_params}
\end{table*}
The thermal sputtering rate is a function of the velocity of the incident gas particles (ions, atoms or molecules) which is determined by the gas temperature \citep[e.g.][]{Barlow1978TheSputtering.,Draine1979ONGAS}:
\begin{align}\label{eq:definition_sputtering_rate}
    \frac{\dd a_i}{\dd t} = \frac{\langle m^{\rm atom}_i\rangle}{2 s_i}\sum_{k}n_k\langle Y_k v \rangle,
\end{align}
where $\langle m^{\rm atom}_i\rangle$ is the average atomic mass of the dust grain in bin $i$, $s_i$ is the material density of the dust grain, and $\langle Y_k v \rangle$ is the sputtering yield of gas species $k$ defined as the mean number of emitted atoms per incident particle averaged over the Maxwellian distribution $f_{\rm M}$:
\begin{align}
    \langle Y_k v \rangle = \int_0^{\infty}Y_k(E) v f_{\rm M}(v) \dd v,
\end{align}
where $E=m_k v^2/2+Z_keQ_{{\rm g},i}/a_i$ is the energy of the incident gas particle, $Z_k$ is the charge of the ion, $e$ is the elementary charge, and $Q_{{\rm g},i}$ and $a_i$ are the charge and radius of the dust grain bin $i$ involved in the collision. The sputtering yields due to particle $k$ impacting at normal incidence with energy $E$ is given by equation 11 in \citet{Nozawa2006DustUniverse}:
\begin{align}\label{eq:dust_sputtering_yield}
Y_k(E) &= 4.2\times 10^{14}\frac{S_k(E)}{U_0}
         \frac{\alpha_k(\mu_k)}{K\mu_k+1} \nonumber\\
       &\quad \times \left[1 - \left(\frac{E_{\rm sp}}{E}\right)^{\frac{2}{3}}\right]
         \left(1- \frac{E_{\rm sp}}{E}\right)^2 .
\end{align}
where $U_0$ is the surface binding energy in units of eV, $\mu_k=\langle m^{\rm atom}_i\rangle/m_k$ is the ratio of the average atomic mass number of the dust grain and the impinging ion, $\alpha_k$ is a function of $\mu_k$ and is discussed further below, and $K$ is a free parameter which is used to calibrate this analytic function for $Y_k(E)$ to experimental data \citep{Tielens1994TheShocks}. Incident gas particles can cause sputtering of dust grains if their energies are equal or above the threshold energy $E_{\rm sp}$ given by \citep{Bohdansky1980AnSputtering,Andersen1981SputteringMeasurements}:
\begin{align}\label{eq:sputtering_energy}
    E_{\rm sp} = \begin{cases}
        \frac{U_0}{4 \langle m^{\rm atom}_i\rangle m_k} \frac{(\langle m^{\rm atom}_i\rangle+m_k)^4}{(\langle m^{\rm atom}_i\rangle - m_k)^2} ,& \text{if } \frac{m_k}{\langle m^{\rm atom}_i\rangle} \leq 0.3 ,\\
        8 U_0 \left(\frac{m_k}{\langle m^{\rm atom}_i\rangle} \right)^{1/3} ,& \text{if } \frac{m_k}{\langle m^{\rm atom}_i\rangle} > 0.3.
    \end{cases}
\end{align}
In Eq.~\ref{eq:dust_sputtering_yield} we also introduced the nuclear stopping cross-section $S_k(E)$ in units of erg\,cm$^{2}$ given by \citep{Sigmund1981SputteringConcepts}:
\begin{align}\label{eq:nuclear_stopping_crosssection}
    S_k(E) = 4 \pi a_{\rm sc}Z_k \langle Z_{\rm i}\rangle e^2 \frac{m_k}{m_k + \langle m^{\rm atom}_i\rangle} \iota_k(\epsilon_k),
\end{align}
where $Z_k$ and $\langle Z_j \rangle$ are the atomic number for the ion and the grain average atomic number. The screening length for the interaction potential between the nuclei is given by:
\begin{align}\label{eq:nuclear_screening_length}
    a_{\rm sc} = 0.885 a_0 (Z_k^{2/3} + \langle Z_i \rangle ^{2/3})^{-1/2}, 
\end{align}
with the Bohr radius $a_0=0.529$\,\AA. The function $\iota_k(\epsilon_k)$ encapsulates the details of the screened Coulomb interaction and can be approximated by \citep{Matsunami1980EnergySolids}:
\begin{align}
    \iota_k(\epsilon_k) = \frac{3.441\sqrt{\epsilon_k}\ln (\epsilon_k+2.718)}{1+6.35\sqrt{\epsilon_k}+\epsilon_k(-1.708 + 6.882\sqrt{\epsilon_k})},
\end{align}
with the reduced energy defined as:
\begin{align}
    \epsilon_k = \frac{\langle m^{\rm atom}_i\rangle}{m_k+\langle m^{\rm atom}_i\rangle} \frac{a_{\rm sc}}{Z_k \langle Z_i \rangle e^2} E.
\end{align}
The $\alpha_k$ function encapsulates how the energy is distributed in the material upon collision of the incident particle \citep[see][for further details]{Sigmund1969TheoryTargets}. We use the updated version of the \citet{Bohdansky1984AIncidence} approximation obtained by \citet{Nozawa2006DustUniverse} based on sputtering data:
\begin{align}
    \alpha_k = \begin{cases}
        0.2 ,& \text{if } \mu_k \leq 0.5 ,\\
        0.1\mu_k^{-1}+0.25(\mu_k-0.5)^2 ,& \text{if } 0.5 < \mu_k \leq 1 ,\\
        0.3(\mu_k-0.6)^{2/3} ,& \text{otherwise}. 
    \end{cases}
\end{align}
\begin{table}
    \centering
    \begin{tabular}{c|ccccccccc}
         & $p_1$ & $p_2$ & $p_3$ & $p_4$ & $p_5$ & $p_6$ & $q_1$ & $q_2$ & $q_3$ \\
         \hline
         \textbf{C-dust} & 4.9 & 0.55 & 0.77 & 4.7 & 3.0 & 1.2 & 4.51 & 0.92 & 0.40 \\
         \textbf{Sil-dust} & 1.5 & 1.2 & 0.57 & 1.1 & 0.52 & 0.37 & 1.48 & 1.31 & 0.59 \\
    \end{tabular}
    \caption{Fitting parameters for the size-dependent correction of thermal sputtering yields. Columns list the coefficients $p_i$ and $q_i$ defining Eq.~\ref{eq:size_correction_function} for carbonaceous and silicate grains.}
    \label{tab:size_corr_fitting_params}
\end{table}

We refer to Table~\ref{tab:dust_material_params} for the relevant material parameters required for the expressions presented in this derivation. The experimental yields that were used to calibrate the analytic expression for $Y_k(E)$, and to hence derive the value of $K$, were measured for a semi-infinite target, rather than finite size dust grains \citep[e.g.][]{Tielens1994TheShocks}. The semi-infinite target approximation may break down for small grains and for sufficiently high gas temperatures. First proposed by \citet{Jurac1998MonteGrains}, a size-dependent sputtering yield arises from the consideration that if the grain size is close to the stopping length (or penetration depth, i.e.~$r_{\rm pd}$), then the energy deposition per unit mass of the grain will be maximal. We apply the results of the size-dependent sputtering yield model by \citet{Bocchio2012SmallStudy,Bocchio2014AWaves} who obtain multi-Gaussian fittings to the correction function $f(x)$ between the sputtering yield of a semi-infinite target $Y_{\infty}$ and the sputtering yield $Y_a$ of a grain of finite radius $a$, $Y_a = f(x) Y_{\infty}$. We have further modified $f(x)$ to prevent unstable behaviour in the $x \ll 1$ regime:
\begin{align}\label{eq:size_correction_function}
f(x) =
\begin{cases}
\begin{aligned}
1
&+ p_1 \exp\!\left[-\dfrac{(\ln(x/p_2))^2}{2p_3^2}\right] \\
&- p_4 \exp\!\left[-(p_5 x - p_6)^2\right],
\end{aligned}
& \text{if } x \geq 1, \\[1ex]
q_1 \exp\!\left[-\dfrac{(\ln(x/q_2))^2}{2q_3^2}\right],
& \text{otherwise}.
\end{cases}
\end{align}
where the fitting parameters are given in Table~\ref{tab:size_corr_fitting_params} and $x=a/(0.7r_{\rm pd})$. The $0.7$~factor comes from the fact that the ion usually loses the majority of its energy after it has travelled $\sim 0.7 r_{\rm pd}$ through the grain material \citep[e.g.][]{SerraDiaz-Cano2008CarbonaceousGraphite}. The penetration depth and the effective value of $Y_a$ are the result of very complex theoretical work using codes like the Stopping and Range of Ions in Matter \citep[SRIM;][]{Ziegler1985TheMatter}. For \calima~we instead estimate the value of $r_{\rm pd}$ via the integration of the total stopping power considering nuclear and electronic interactions derived in Section~\ref{subsec:collisional_cooling}. As an example, we present in Fig.~\ref{fig:sputtering_rate} the erosion rate $(\dd a/\dd t)/n_{\rm H}$ for the different dust species considering the finite size correction detailed above.

\subsection{Energy transfer in dust-gas collisions}\label{ap:dust_collisional_cooling}

The collisional model by \citet{Hollenbach1979MOLECULEPROCESSES} is based on the \textit{soft-cube} model \citep{Logan1968ClassicalSurfaces} applied to the thermal accommodation of gas particles onto dust grain surfaces \citep{Burke1983TheTrapping}. Essentially, this model assumes that a grain surface can be approximated by an array of square particles attached by a spring to an infinite mass (the lattice). The infinitesimal collisional rate onto grain surface area $\dd A$ by a gas particle with mass $m$, number density $n$, and following a Maxwell-Boltzmann velocity distribution at temperature $T$, is given by:
\begin{align}
    \dd \Gamma &= n \dd A \int_{v_z>0} \left(\frac{m}{2\pi k_{\rm B}T} \right)^{3/2}\exp\left(-\frac{mv^2}{2k_{\rm B}T}\right) v_z \dd^3 v \nonumber \\
    &= n \dd A \sqrt{\frac{k_{\rm B}T}{2\pi m}},
\end{align}
where we have assumed the $x-y$ plane is parallel to the grain surface, such that only particles with $v_z>0$ can reach the surface. Taking the integral over the full surface of the grain $4\pi a^2$ gives the total collisional rate as:
\begin{align}
    \Gamma = \pi n a^2\sqrt{\frac{8k_{\rm B}T}{\pi m}}.
\end{align}
By a similar calculation, we can derive the average kinetic energy of the colliding gas particles  $\langle E_{\rm kin}\rangle=2k_{\rm B}T$. Therefore, we can express the energy transfer rate for a gas particle with mass $m$ onto a dust grain of radius $a$ and temperature $T_{\rm d}$~\citep{Hollenbach1979MOLECULEPROCESSES}:
\begin{align}\label{eq:collisional_cooling_rate}
    H_{\rm coll}(a,T) = \sqrt{\frac{8k_{\rm B}T}{\pi m}} \pi n a^2 2k_{\rm B}(T - T_{\rm d})\mathcal{A}(T),
\end{align}
where the term in parenthesis indicates that the energy transfer goes to zero when $T=T_{\rm d}$ and allows for gas heating when $T_{\rm d}> T$. The parameter $\mathcal{A}(T)$ encapsulates the efficiency of energy transfer from the gas atom to the dust grain, termed the \textit{accommodation} factor \citep{Burke1983TheTrapping}. This is usually assumed to be of order unity, with a small dependence on the gas and dust temperatures in the range $\sim 10-1000$\,K \citep[see figure 4 in][]{Burke1983TheTrapping}. The \textit{soft-cube} model is valid as long as the surface is approximately flat from the perspective of the impinging gas atom. However, more energetic gas particles are capable of interacting with a larger region of the grain surface, potentially encountering a more uneven surface topology. This limits the validity of the model to gas temperatures $T\lesssim 10^4$\,K, which has not been enforced when implemented in many ISM or cloud collapse models~\citep[e.g.~][]{Wolfire2003NeutralGalaxy,Grassi2014KROME-aSimulations,Bialy2019ThermalGas,Kim2023PhotochemistrySimulations,Katz2022PRISM:Galaxies}. We therefore enforce a logistic curve centred on $T=10^4$\,K, allowing for a smooth transition between the regime of applicability for Eq.~\ref{eq:collisional_cooling_rate}, to the hard-sphere regime at high temperature \citep[e.g.][]{Lindhard1962NuclearDetectors,Sigmund2014StoppingIons}. We note that this will have an appreciable influence on the cooling rates at solar metallicity below $n_{\rm H}<1$\,cm$^{-3}$ \citep[e.g.~see figure 2 in][for the case of the \prism~model]{Katz2022PRISM:Galaxies}. When the low-temperature condition is satisfied, we obtain $\mathcal{A}(T)$ for atomic H and H$_2$ as \citep{Hollenbach1979MOLECULEPROCESSES}:
\begin{align}
    \mathcal{A}_{\rm H,H_2} = (1-\mathcal{A}_{\rm H,H_2}^0)\exp \left(-\sqrt{\frac{2k_{\rm B}(T+T_{\rm d})}{D_{\rm H,H_2}}} \right) + \mathcal{A}_{\rm H,H_2}^0,
\end{align}
where:
\begin{align}
    \mathcal{A}_{\rm H,H_2}^0 = \frac{2m_{\rm H,H_2}m_{\rm d}}{(m_{\rm H,H_2}+m_{\rm d})^2},
\end{align}
is the hard-sphere limit and $D_{\rm H,H_2}=500k_{\rm B}$ is the depth of the potential well (i.e.~the adsorption energy per particle).

As we have seen in Sec.~\ref{subsec:sputtering}, the ion-grain interaction in the high-temperature regime brings into play different physics. In this case, ions can get close enough to feel the grain potential and to travel through the grain material. This form of \textit{transmission} is the basis behind the thermal and non-thermal sputtering of dust grains \citep{Draine1979DestructionDust.,Nozawa2006DustUniverse}. Understanding ion transmission through the grain material requires accounting for the interaction of ions with nuclei and electrons, giving rise to an effective drag force. Unfortunately, the topic of a material's stopping power (i.e.~the average change in energy of an ion over an infinitesimal path length through the solid) at sub-relativistic velocities remains a question of great debate in the solid-state and nuclear physics community \citep[see][for a review on the different models and approximations used to account for the slowing down of ions through solids]{Sigmund2014StoppingIons}. In the high energy regime ($E\gg 1$\,MeV), the Bohr and Bethe stopping equations have been of great use in predicting the transport of heavy ions. However, these theories, when extended to the low energy regime, give negative values for the stopping power below a threshold energy. Given that an ion energy gain is not considered in these theories, the origin of these negative values is an artefact caused by the approximation limits for Bessel functions used in their derivation \citep{Sigmund1996Low-speedFormula}. Furthermore, the first Born approximation can only be applied to the most energetic particles that we will encounter in a coronal gas, so the physical meaning of these equations also becomes questionable in these regimes\footnote{We note that one of these theories for high energy stopping power, the Bethe-Bloch approximation \citep{Bethe1930ZurMaterie,Bloch1933ZurMaterie}, was used in the SN dust modelling by \citet{Kirchschlager2019DustDensities}, resulting in numerically incorrect penetration depths in the low energy regime.}. 

We instead follow the formalism developed by \citet{Sigmund1981SputteringConcepts} for the nuclear and electronic stopping by graphite and silicate materials. We recall (Section~\ref{subsec:sputtering}) that the stopping power is defined as the average energy loss per unit path length, $l$,  through the solid material \citep{Lindhard1964MotionCrystals}:
\begin{align}\label{eq:definition_stopping_power}
    \frac{\dd E}{\dd l} = -N (S_{\rm n} (E)+ S_{\rm e} (E)),
\end{align}
where $N$ is the nuclear number density, $S_{\rm n}$ is the nuclear stopping cross-section defined in Eq.~\ref{eq:nuclear_stopping_crosssection} and $S_{\rm e}$ is the electronic stopping cross-section given by the \citet{Lindhard1961EnergyRegion} equation for an electron gas:
\begin{align}\label{eq:LS61_electronic_stopping_power}
    S_{\rm e}(E) = 8 \pi \xi_{\rm e}\frac{Z_{\rm k}\langle Z_{\rm atom}\rangle}{\tilde{Z}} a_{\rm sc} e^2 \frac{v(E)}{v_0},
\end{align}
with:
\begin{align}
    \tilde{Z} = (Z_k^{2/3}+ \langle Z_i \rangle^{2/3})^{3/2},
\end{align}
for projectile $k$, and where the average grain atomic number $\langle Z_i\rangle$, and the screening length $a_{\rm sc}$ are as defined in Section~\ref{subsec:sputtering}. The factor $\xi_{\rm e}\simeq Z_k^{1/6}$ was included to improve agreement with experimental penetration depths, and $v_0$ is given by $v_0=\sqrt{2E_0/m_{\rm e}}$ where:
\begin{align}
    E_0 = \frac{\hbar^2}{2 m_{\rm e}}(3 \pi^2\langle N_{\rm val}\rangle N)^{2/3},
\end{align}
is the Fermi energy with number density provided by the material valence electrons per atom $\langle N_{\rm val}\rangle$, and $\hbar$ is the reduced Planck constant. We can now numerically integrate Eq.~\ref{eq:definition_stopping_power} to determine how an ion of initial energy $E$ slows down as it travels through the graphite and silicate materials of the dust grains modelled in \calima.

The heating rate of a dust grain of size $a$ experiencing collisions with a gas atom $k$ with mass $m_k$, number density $n_k$ and following a Maxwell-Boltzmann distribution $f(E)$ of energies $E=1/2 m_k v^2$ is given by :
\begin{align}\label{eq:collisional_heating_rate}
    H(\vec{\mathcal{P}}_i,T,n_k) = \pi a_i^2n_k F_{\rm C} \int f(E) v(E) E \zeta(\vec{\mathcal{P}}_i,E) \dd E,
\end{align}
where $\vec{\mathcal{P}}_i$ is a vector of the material properties of the grain bin $i$
\begin{align}
    \vec{\mathcal{P}}_i = (a_i,s_i,\langle Z_i \rangle, \langle m_i^{\rm atom}\rangle, \langle N_{\rm val} \rangle),
\end{align}
$F_{\rm C}$ is the Coulomb enhancement factor (see Section~\ref{subsec:accretion}), and $\zeta$ is the fraction of incident ions left in the grain, computed by integrating Eq.~\ref{eq:definition_stopping_power}. Eq.~\ref{eq:collisional_heating_rate} can easily be integrated resulting in:
\begin{align}\label{eq:collisional_heating_rate_integrated}
    H(\vec{\mathcal{P}}_i,T,n_k) = \sqrt{\frac{32}{\pi m_k}} \pi a_i^2 n_k (k_{\rm B}T)^{3/2}F_{\rm C}h(\vec{\mathcal{P}}_i,T),
\end{align}
where:
\begin{align}\label{eq:collisional_efficiency}
    h(\vec{\mathcal{P}}_i,T) = \frac{1}{2}\int u^2 \zeta(\vec{\mathcal{P}}_i,E) \exp(-u) \dd u,
\end{align}
is the distribution-averaged heating efficiency function, and $u \equiv E/(k_{\rm B}T)$. We compute $h(\vec{\mathcal{P}}_i,T)$ for the material properties and sizes of the dust grains in \calima, such that a simple linear interpolation in log-log space can be used to compute its value on-the-fly. The other quantities in Eq.~\ref{eq:collisional_heating_rate_integrated} are computed on a per-cell basis during the simulation.
\begin{figure}[t]
    \centering
    \includegraphics[width=\columnwidth]{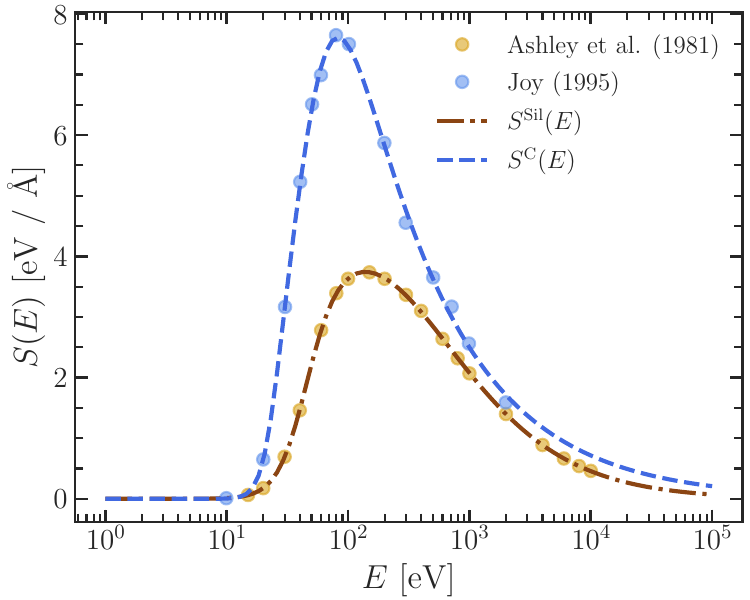}
    \caption{Experimental electron stopping power for solid carbon (blue markers) by \citet{Joy1995AInteractions} and silicon dioxide (yellow markers) by \citet{Ashley1981EnergyDioxide}. The dashed lines show the analytical fits for silicate (brown dot--dashed line) and carbonaceous grains (blue dashed line), following Eq.~\ref{eq:electron_stopping_power_solid_carbon}.}
    \label{fig:fit_electron_stopping}
\end{figure}
\begin{table}[t]
    \centering
    \begin{tabular}{c|c|c}
         & \textbf{SiO$_2$} & \textbf{C} \\
        \hline
        $a$ & $7.474$ & $-4.280 \times 10^{-4}$ \\
        $b$ & $0.892$ & $-1.221 \times 10^{-16}$ \\
        $c$ & $7.703 \times 10^{-6}$ & $-1.304 \times 10^{-7}$ \\
        $d$ & $0.997$ & $-6.502$ \\
        $e$ & $-2.918$ & $-0.985$ \\
        $f$ & $0.082$ & $-2.614 \times 10^{-4}$ \\
        $g$ & $0.223$ & $1.544$ \\
        $h$ & $0.947$ & $1.524$ \\
    \end{tabular}
    \caption{Analytical fitting parameters for Eq.~\ref{eq:electron_stopping_power_solid_carbon}, describing the electron stopping power in solid carbon (C) and silicon dioxide (SiO$_2$). Experimental data from \citet{Joy1995AInteractions} and \citet{Ashley1981EnergyDioxide}.}
    \label{tab:electron_stopping_power}
\end{table}
 In gas with temperature below $10^9$\,K, electrons are still in the sub-relativistic regime. This implies that, outside of the most energetic electrons above $\sim 10$\,keV, the first Born approximation breaks down and the Rutherford cross-section is no longer applicable. In this regime, analytic expressions (e.g.~Eq.~\ref{eq:LS61_electronic_stopping_power}) are a poor predictor of the stopping power of electrons in different solids at low energies \citep{Iskef1983ProjectedKeV,Luo1991ExperimentalEnergies,Joy1996ExperimentalEnergies}. We therefore follow the approach in \citet{Micelotta2010PolycyclicGas} (see also Section~\ref{subsec:pah_thermal_sputtering}) and instead determine the energy deposition and penetration depth by obtaining fitting functions to experimental values. For carbonaceous grains we use the experiments on carbon films by \citet{Joy1995AInteractions}, and in the absence of experimental data for astrophysically relevant silicates, we use the experimental data on SiO$_2$ stopping power by \citet{Ashley1981EnergyDioxide} (see blue and yellow markers in Fig.~\ref{fig:fit_electron_stopping}, respectively). We use the fitting equation by \citet{Micelotta2010PolycyclicGas} for the stopping power (i.e.~$\dd E/\dd l$) of the colliding electrons:
\begin{align}\label{eq:electron_stopping_power_solid_carbon}
    S(E) = \frac{h\log(1+aE_{\rm keV})}{fE_{\rm keV}^g+bE_{\rm keV}^d+cE_{\rm keV}^e}\,\rm eV\,\text{\AA}^{-1},
\end{align}
where the $E_{\rm keV}$ is the electron energy in keV and our fitting parameters are given in Table~\ref{tab:electron_stopping_power}\footnote{Note that we have re-computed the fitting values for solid carbon, which has resulted in a minor deviation from the parameters provided in \citet{Micelotta2010PolycyclicGas}. These differences are negligible.}. We present the results of this fit in Fig.~\ref{fig:fit_electron_stopping}, indicating a larger stopping power for carbon than for SiO$_2$ at the peak stopping efficiency of $\sim 10^2$\,eV.

\subsection{Subgrid modelling of dust turbulent shattering}\label{ap:dust_turbulence}

The \citeauthor{Ormel2007AstrophysicsNote} model is an updated version of the pioneering work of \citet{Voelk1980CollisionsGas}. \citeauthor{Voelk1980CollisionsGas}-like models are based on the classification of turbulent eddies according to whether their lifetime (i.e.~turnover timescale $\tau_{\rm p}$) is comparable to the stopping time $t_{\rm s}$ of the particle (Class I) or not (Class III). In our case the particle is the dust grain. This classification provides a straightforward derivation of the particle-eddy relative velocity and in turn the variance of the particle-particle relative velocity. 

We assume that dust particles and gas are coupled to each other via collisions and are in thermal equilibrium, meaning that the final relative velocity between grains is a combination of Brownian motion and the turbulent-induced velocities discussed above. The relative velocity due to thermal (Brownian) motion for two dust particles of masses $m_i$ and $m_j$ is:
\begin{align}\label{eq:brownian_relative_velocity}
    \Delta V_{ij}^{\rm B} = \sqrt{\frac{8 k_{\rm B}T(m_i+m_j)}{\pi m_im_j}}.
\end{align}
In the context of collapsing molecular clouds \citep[e.g.][]{Guillet2020DustCollapse,Kawasaki2022DustEffects} the forcing timescale for the turbulent cascade is assumed to be the free-fall time of the cloud $t_{\rm ff}$ \citep[e.g.][]{Lebreuilly2022ProtostellarFragmentation}. 
%or that the injection scale of the turbulence is the Jeans length and that the turbulent velocity is close to the isothermal sound speed $c_{\rm s}$. 
In our case, and consistent with the turbulent enhancement of the gas density PDF used for dust metal accretion (see Section~\ref{subsec:accretion}), we consider the local forcing timescale to be $\tau_{\rm L}=\Delta x_{\rm cell}/\sigma$ ($\sigma$ is the local gas velocity dispersion), and the turbulent Mach number is $\mathcal{M} = \sigma / c_{\rm s}$. Assuming a Kolmogorov turbulent spectrum, the eddy turnover time on the viscous dissipation scale is:
\begin{align}\label{eq:dissipation_timescale}
    \tau_{\eta} = \frac{\tau_{\rm L}}{\sqrt{{\rm Re}}},
\end{align}
with the Reynolds number, ${\rm Re}$, parameterised by the local thermodynamic conditions as ${\rm Re}=6.2\times 10^7 \sqrt{\frac{\rho/\mu}{10^5{\rm cm}^{-3}}} \sqrt{\frac{T_{\rm gas}}{10{\rm K}}}$ \citep{Ormel2009DustDistribution}. The motion of dust particles in a fluid is governed by the drag forces caused by gas\footnote{This modelling is pure hydrodynamical, so we leave extending this two include magnetohydrodynamical modifications for future work \citep[e.g.][]{Yan2004DustTurbulence}.}. Based on Epstein's law, the stopping time is:
\begin{align}\label{eq:Epstein_law}
    t_{\rm s} = \frac{sa}{\rho_{\rm gas}v_{\rm th}},
\end{align}
with $v_{\rm th}=\sqrt{8/\pi}c_{\rm s}$ the thermal velocity of the gas and $s$ the material density of the grain. Eq.~\ref{eq:Epstein_law} applies when the mean free path $l_{\rm mfp}$ of the gas is larger than the particle size \citep{Commercon2023DynamicsImplementations}. When this is no longer the case, one should use Stokes' law, with a transition particle size given by the condition $a_{\rm tr} \geq (9/4)l_{\rm mfp}$. Given that the largest particle size considered in this work is of~$0.1$~$\mu$m, this corresponds to $n_{\rm H} \geq 7.5\times 10^{19}$~cm$^{-3}$, well out of reach of current galaxy formation simulations. Therefore, it is safe to assume  that all dust particles modelled are well within the Epstein regime. Given these definitions of the stopping time and turbulent forcing timescale, we can define the particle Stokes' number as ${\rm St}=t_{\rm s}/\tau_{\rm L}$.

Consider Stokes' numbers for dust particles $i$ and $j$ with ${\rm St}_i\geq {\rm St}_j$ (thus $t_{\rm s}^i\geq t_{\rm s}^j$ and $m_i\geq m_j$). \citet{Ormel2007AstrophysicsNote} provide closed forms for the turbulent-induced relative velocities in three limiting regimes dictated by the viscous timescale $\tau_{\eta}$ and the turnover timescale $\tau_{\rm L}$:
\begin{align}\label{eq:OC07_relative_velocity}
\Delta V_{ij}^{\rm T} =
\begin{cases}
\begin{aligned}
\sqrt{\frac{3}{2}}\,\sigma
&\sqrt{\dfrac{{\rm St}_i-{\rm St}_j}{{\rm St}_i+{\rm St}_j}} \\
&\times
\sqrt{\dfrac{{\rm St}_i^2}{{\rm St}_i+{\rm St}_{\rm min}}
      - \dfrac{{\rm St}_j^2}{{\rm St}_j+{\rm St}_{\rm min}}},
\end{aligned}
& \text{if } \tau_{\eta} > t_{\rm s}^i, \\[1ex]
\sqrt{\frac{3}{2}}\,\sigma
\sqrt{f\!\left(\dfrac{{\rm St}_j}{{\rm St}_i}\right){\rm St}_i},
& \text{if } \tau_{\eta} \le t_{\rm s}^i < \tau_{\rm L}, \\[1ex]
\sqrt{\frac{3}{2}}\,\sigma
\sqrt{\dfrac{1}{1+{\rm St}_i} + \dfrac{1}{1+{\rm St}_j}},
& \text{if } \tau_{\rm L} \le t_{\rm s}^i .
\end{cases}
\end{align}
with the function $f$ given by
\begin{align}
    f(x) = 3.2 - (1+x) + \frac{2}{1+x}\left(\frac{1}{2.6}+\frac{x^3}{1.6+x} \right),
\end{align}
and ${\rm St}_{\rm min}\equiv \tau_{\eta}/\tau_{\rm L} = 1/\sqrt{{\rm Re}}$. The final relative velocity is obtained by combining the Brownian and turbulent contributions:
\begin{align}
    \Delta V_{ij} = \sqrt{\left(\Delta V_{ij}^{\rm B}\right)^2 + \left(\Delta V_{ij}^{\rm T}\right)^2}.
\end{align}

Recall that the collision velocity determines whether shattering, coagulation or simple bouncing\footnote{We do not consider effect of deformations on the dust grain geometrical properties.} takes place. Furthermore, in the case of shattering, the energy of the collision determines the fragment distribution that will emerge from the two-body interaction \citep{Jones1996GrainDistribution,Hirashita2009ShatteringTurbulence}, which depends on a number of material properties. \citet{Hirashita2013ConditionCores} showed that the most relevant parameter is the velocity threshold for catastrophic fragmentation (i.e.~when more than half of the target grain mass is shocked above the grain internal sound speed). Therefore, following \citet{Hirashita2013ConditionCores} we simplify the shattering model using the prescription presented by \citet{Kobayashi2010FragmentationCascades}. We assume that the disrupted mass ejected from $m_i$ is proportional to 
\begin{align}
    \phi_{ij} = \frac{E_{\rm imp}}{m_i Q_{\rm d}^*},
\end{align}
where
\begin{align}
    E_{\rm imp} = \frac{1}{2}\frac{m_im_j}{m_i+m_j}\left(\Delta V_{ij}\right)^2,
\end{align}
is the impact energy for $m_i$ colliding with $m_j$ and $Q_{\rm d}^*$ is the specific impact energy for catastrophic disruption \citep[estimated to be $\simeq 8.9\times 10^9$\,dyn\,cm\,g$^{-1}$ for carbonaceous grains and $\simeq 4.3\times 10^{10}$\,dyn\,cm\,g$^{-1}$ for silicate grains,][]{Jones1996GrainDistribution}. Using $\phi$, the ejected mass $m_{\rm ej}$ is estimated as
\begin{align}
    m_{\rm ej}(\Delta V_{ij},m_i,m_j) = \frac{\phi_{ij}}{1+\phi_{ij}}m_i.
\end{align}
The maximum $a_{\rm f,max}$ and minimum $a_{\rm f,min}$ sizes of the fragments are taken to be
\begin{align}
    m_{\rm f,max} &= 0.02 m_{\rm ej}, \\
    m_{\rm f,min} &= 10^{-6} m_{\rm f,max},
\end{align}
with the fragment mass distribution following (i.e.~given the collision of grains of masses $m_i$ and $m_j$ with relative velocity $\Delta V_{ij}$ it returns the probability of fragment of mass $m$)
\begin{align}\label{eq:fragment_distribution}
\mu_{\rm frag}(m;\Delta V_{ij},m_i,m_j) =
\begin{aligned}[t]
&\frac{(4-\alpha_{\rm f})\,m_{\rm ej}\,
       m^{(-\alpha_{\rm f}+1)/3}}
      {3\!\left[
      m_{\rm f,max}^{(4-\alpha_{\rm f})/3}
      - m_{\rm f,min}^{(4-\alpha_{\rm f})/3}
      \right]} \\
&+ (m_i - m_{\rm ej})\,
   \delta\!\left(m - m_i + m_{\rm ej}\right),
\end{aligned}
\end{align}
where we have dropped the dependence of $m_{\rm ej}(\Delta V_{ij},m_i,m_j)$ for clarity. This is a power law distribution with exponent $\alpha_{\rm f}=3.3$
in agreement with the results of~\citet{Jones1996GrainDistribution}, plus a remnant fragment of mass $m_i-m_{\rm ej}$ corresponding to last term on the RHS of the equation (where $\delta$ is the dirac delta function). 

These results need to be applied and implemented into our discretised dust distribution framework. Equation~\ref{eq:fragment_distribution} can be easily integrated over the limits corresponding to large grains, small grains, and in the case of carbonaceous grains, PAHs (see Section~\ref{sec:pahs}), in order to obtain the contribution of the shattered fragments to each bin. Integrating Eq.~\ref{eq:fragment_distribution} over the whole spectrum of fragments gives $m_i$, allowing us to define the fraction of target mass fragments that fall within  the dust grain bin centred on grain size  $a_x$ and limited by minimum and maximum masses $m_{x,{\rm min}}$ and $m_{x,{\rm max}}$:
\begin{align}\label{eq:fragment_fraction}
\chi_{\rm f}(a_x;\mathcal{M},a_i,a_j)
&\equiv
\chi_{\rm f}(m_x;\Delta V_{ij},m_i,m_j) \\
&=
\frac{1}{m_i}
\int_{m_{x,{\rm min}}}^{m_{x,{\rm max}}}
\mu_{\rm frag}(m;\Delta V_{ij},m_i,m_j)\, dm .
\end{align}
Using Eq.~\ref{eq:fragment_fraction} we define the local timescale of fragment production for a given collision model \citep{Hirashita2019RemodellingGalaxies}:
\begin{align}\label{eq:shattering_timescale}
    t_{\rm sha} (a_x;\mathcal{M},a_i,a_j) = (\sigma_{ij}\Delta V_{ij}n_{{\rm d},j} \chi_{\rm f})^{-1},
\end{align}
where we have dropped the arguments of $\chi_{\rm f}$ for clarity, and defined the collision cross section of grains of sizes $a_i$ and $a_j$ as $\sigma_{ij}=\eta_{ij} \pi (a_i+a_j)^2$, where $\eta_{ij}$ allows for corrections to the geometrical cross section. We choose the simple case $\eta_{ij}=1$. This definition of the shattering timescale allows us to obtain separate timescales for the production of shattered fragments within different grain size bins. Therefore, for carbonaceous grains we compute the timescales for the production of large grains (i.e.~inefficient shattering), small grains, PAHs and full destruction (i.e.~return to the gas phase). We do similar calculations for silicates, except that we do not follow the shattering of very small grains in this case; instead every fragment below $4.5$\,\AA~is considered to return to the gas phase. This is based on the result that smaller silicate grains are photolytically unstable in the ISM~\citep{Guhathakurta1989TemperatureGrains}. We only consider silicate-silicate, or carbon-carbon collisions, i.e.~we do not allow silicate-carbon collisions. Following \citet{Hirashita2013ConditionCores}, we posit that approaching particles have a uniform probability of hitting each other from any direction, so we obtain maximum and minimum estimates for the relative velocity given by head-on collisions and tangential collisions, respectively.

\begin{figure*}
	\includegraphics[width=\textwidth]{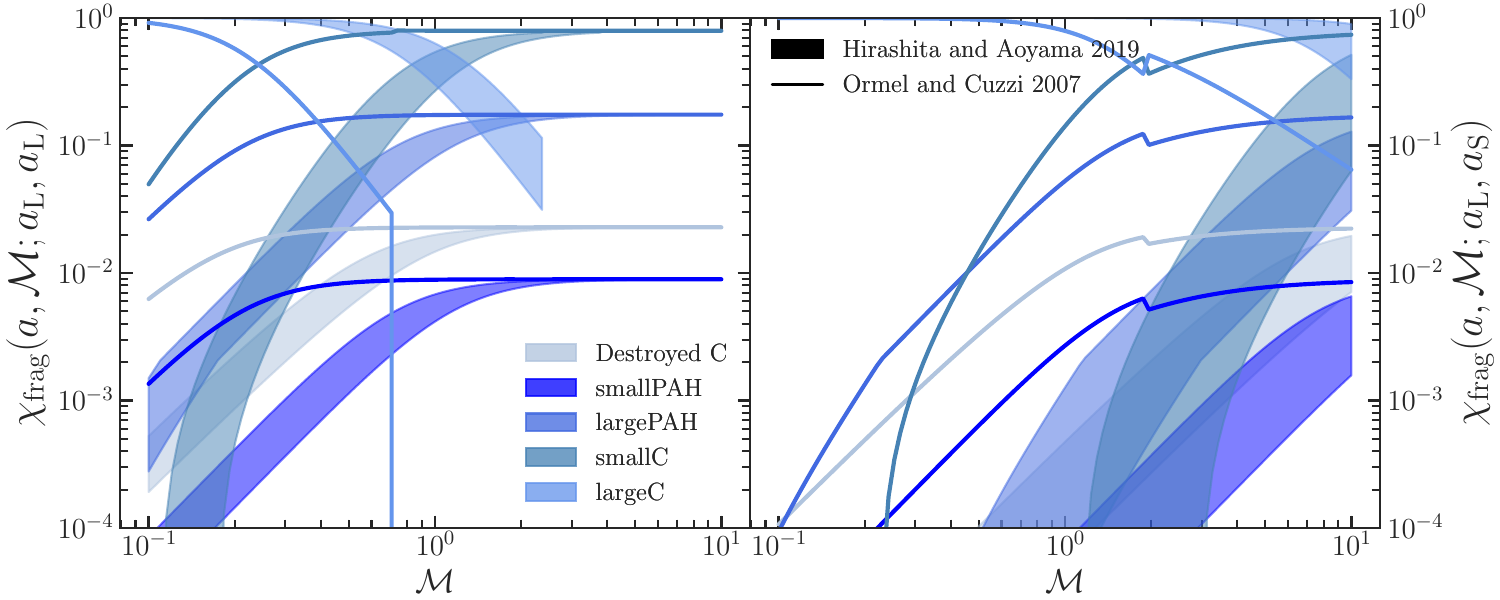}
    \caption{Carbonaceous shattering fragment mass fraction versus turbulent Mach number in the WIM ($n_{\rm H}=0.1$\,cm$^{-3}$) for the turbulent shattering models detailed in Section~\ref{subsec:shattering}. Each line and shaded area colour indicates different bins of PAHs and carbonaceous grains. Shattering is computed using either: (i) our fiducial model based on the collision velocity results of~\protect\citet{Ormel2007AstrophysicsNote} (solid lines), or (ii) our model with the basic velocity scaling from~\protect\citet{Hirashita2019RemodellingGalaxies} (shaded regions). \textit{Left panel:} Results for the models using only the collisions between two large grains with $a_{\rm L}=0.1$\,$\mu$m. \textit{Right panel:} Results for the collision between a large grain $a_{\rm L}=0.1$\,$\mu$m and a small grain $a_{\rm S}=0.01$\,$\mu$m. Collisions between large grains in the WIM allow for efficient shattering of large grains into smaller grains. The basic modelling by~\protect\citet{Hirashita2019RemodellingGalaxies} significantly underestimates the shattering efficiency at low $\mathcal{M}$. Big-small grain collisions only become relevant for supersonic turbulence.}
    \label{fig:shattering_efficiency}
\end{figure*}

Fig.~\ref{fig:shattering_efficiency} shows the results of this turbulent shattering model for carbonaceous grains in the WIM as defined in Table~1 of \citet{Yan2004DustTurbulence}. In the left panel we consider collisions of large grains with other large grains (i.e.~$m_i=m_j$), whereas in the right panel we consider the collisions of large and small grains (i.e.~$m_i>m_j$). We show the fragment mass fraction $\chi_{\rm frag}$ (see Eq.~\ref{eq:fragment_fraction}) which indicates what mass of the fragment distribution falls under the limits of each of the PAH and carbonaceous dust grain bins. In order to put our results within the context of previous treatments of the shattering timescale, we also compare the results of the \citet{Ormel2007AstrophysicsNote} model to \citet{Hirashita2019RemodellingGalaxies} (see their Appendix~C). This modelling assumes that the particle sits in the intermediate regime given in  Eq.~\ref{eq:OC07_relative_velocity}, which is a good approximation for the molecular ISM, but not for the high $\mathcal{M}$ typical of the WNM or WIM. The results using the \citet{Hirashita2019RemodellingGalaxies} model is represented in Fig.~\ref{fig:shattering_efficiency} by the shaded bands. Compared to our fiducial implementation using the \citet{Ormel2007AstrophysicsNote} model, we find that the \citet{Hirashita2019RemodellingGalaxies} model strongly underestimates the fragmentation of large grains at low to intermediate $\mathcal{M}$. This approximated form only becomes valid at very high $\mathcal{M}$. The results of the approximation are even less reliable for the case of non-equal mass collisions (see right panel). 

Independent of the velocity model assumed, shattering in large-large grain collisions in transonic turbulence results in the efficient production of small shattering fragments, allowing for $\sim 1$\% and $\sim 20$\% mass fraction in small PAHs and large PAHs. We therefore expect shattering of large grains to be an interesting process for the top-down formation of PAHs \citep{Seok2014FormationGalaxies}. Additionally, our fiducial model finds that the production of small and very small shattered fragments through the collision of large grains and small grains may becomes imporant in highly turbulent systems at high redshift~\citep[e.g.][]{Falgarone2017}.

\subsection{Revisiting PAH sputtering}\label{ap:pah_sputtering}

Ion-PAH collisions can be described by the simultaneous processes of nuclear stopping (elastic energy loss), and electronic stopping (inelastic energy loss). Nuclear stopping involves a binary collision of an ion with a single atom of a PAH molecule. If the collision energy exceeds the threshold for atom removal, the atom is ejected. For electronic stopping, the interaction is instead between the electrons of the PAH and the projectile ion. The transferred energy is then shared between the vibrational modes of the molecule, resulting in either IR emission (photo-relaxation) or in fragmentation via hydrogen or C$_2$H$_n$ loss, where $n=0,1,2$. The transfer of energy in an electronic interaction for a PAH is a topic that remains, by-and-large, theoretically unexplored. We use the theoretical modelling developed for fullerenes in \citet{Schlatholter1999StrongFullerenes} and \citet{Hadjar2001ProjectileFullerenes}. We consider that a PAH delocalised valence electrons behave like an electron gas, giving rise to an effective stopping power of the ion $S$ \citep{Sigmund1981SputteringConcepts}:
\begin{align}\label{eq:stopping power}
    S = \frac{\dd T}{\dd l} = \gamma(r_s)v,
\end{align}
where $\dd T$ is the energy loss over the path length $\dd l$ through the PAH electron gas, $\gamma$ is the friction coefficient \citep{Puska1983AtomsSections}, and $v$ is the incident ion velocity. The friction coefficient depends on the density parameter $r_s=(\frac{4}{3}\pi n_0)^{-1/3}$ which is itself a function of the valence electron density $n_0$. In the fullerene jellium model of \citet{Puska1983AtomsSections} $n_0$ assumes a spherical geometry, which requires some modifications for a disc geometry more reasonable for PAHs. The effective PAH radius $R$ is computed using Eq.~\ref{eq:pah_radius}, and the $\pi$-electron cloud thickness is fixed to $d\sim 4.31$\,\AA. Applying the jellium model to $n_0$ for PAHs \citet{Hadjar2001ProjectileFullerenes}, $n_0$ is given in terms of the coordinates $l$ and $\theta$:
\begin{align}
    n_0(l,\theta) = 0.15 \exp{-(s\cos{\theta})^2/2.7}
\end{align}
where $\theta$ is the angle between the minor axis of the PAH and the direction of the incoming ion and $s$ is the distance travelled through the PAH.
The friction parameter has been computed by \citet{Puska1983AtomsSections} for different projectile ions with the following functional form:
\begin{align}\label{eq:electron_friction_parameter}
    \gamma(r_s) = \Gamma_0 \exp \left(- \frac{r_s(s,\theta)-1.5}{R_2}\right),
\end{align}
where $\Gamma_0$ and $R_2$ are fitting parameters to their experimental results and given in Table~\ref{tab:friction_parameter_fitting} for the main ion projectiles that contribute to thermal sputtering (i.e.~H, He, C and O). The total energy transferred can then be computed by integrating Eq.~\ref{eq:stopping power} for a given incidence angle $\theta$:
\begin{align}\label{eq:ion_electronic_excitation_energy}
    T_e(\theta) = \int^{L(\theta)/2}_{-L(\theta)/2)} \gamma(r_s(l,\theta)) v \dd s,
\end{align}
where $L$ is the total path length through the PAH. Given the PAH disc geometry, $\vert \tan(\theta)\vert < \tan(\alpha)$, then $L(\theta)=d/\cos(\theta)$, and otherwise $L(\theta)=2R/\vert \sin(\theta)\vert$. This means that if the angle of incidence is $\theta=\pi/2$ the path length will be maximal, and hence so will the energy transferred.

\begin{table}
    \centering
    \begin{tabular}{c|c|c}
         Element & $\Gamma_0/a_0^2$ & $R_2$\\
         \hline
         \textbf{H} & 0.33 & 2.28\\
         \textbf{He} & 0.75 & 0.88\\
         \textbf{C} & 1.68 & 0.90\\
         \textbf{O} & 1.62 & 0.57\\
    \end{tabular}
    \caption{Fitting parameters for the electron friction (Eq.~\ref{eq:electron_friction_parameter}) taken from the theoretical predictions of \citet{Puska1983AtomsSections}. The scaling parameter $\Gamma_0$ is given in units of the the Bohr radius $a_0\simeq 5.291 \times 10^{-11}$\,m.}
    \label{tab:friction_parameter_fitting}
\end{table}

Ions colliding with a PAH may also remove a C atom from the PAH skeleton, for which we need to consider the theory of nuclear collisions by \citet{Micelotta2010PolycyclicShocks}. In this case, we only consider collisions with energies above the threshold energy, $T_0$, of C removal.  The destruction rate $R^k_{\rm N} (T,N_{\rm C})$ (the subscript N stands for `Nuclear') is given by a collisional rate weighted by the Maxwellian velocity distribution $f(v,T)$ for ion $k$ in the gas, and adding the correction due to Coulomb focusing with the $F_{\rm C}$ factor (see Section~\ref{subsec:accretion}) to the original derivation in \citet{Micelotta2010PolycyclicGas}:
\begin{align}\label{eq:nuclear_sputtering_rate}
    R^k_N(T,N_{\rm C}) = \frac{1}{2}N_{\rm C}n_k F_{\rm C} \int_{v_0}^{\infty} v \sigma_{N,k}(v,T_0)f(v,T)\dd v,
\end{align}
where $n_k$ is the local number density of ion $k$, $N_{\rm C}$ is the number of carbon atoms in the PAH, $v_0$ is the velocity corresponding to the threshold energy $T_0$, and $\sigma_{N,k}(v,T_0)$ is the nuclear interaction cross-section per C atom, taken from interpolating the results in figure 2 of \citet{Micelotta2010PolycyclicShocks}. The threshold energy $T_0$ is the minimum energy that a collision must transfer to a C atom to eject it from the PAH skeleton. However, as detailed in \citet{Micelotta2010PolycyclicShocks} the choice of $T_0$ is very uncertain, as there are no experimental nor theoretical determinations of its value for astrophysically relevant PAHs. The physical meaning of $T_0$ is also different from the threshold energy for atomic displacement used in solid state physics, which for graphite has been measured to vary from 12\,eV \citep{Nakai1991NonthermalSurfaces} up to 30\,eV \citep{Montet1967TheMolybdenite}, while for amorphous carbon the value is suggested to be lower and close to 5\,eV \citep{Cosslett1978RadiationReview}. We hence choose a value of 7.5\,eV as in \citet{Micelotta2010PolycyclicShocks} to be consistent with the  experimental data observed for pseudo-PAH materials. We found that varying the value of $T_0$ from 4.5 to 15\,eV mainly introduces a shift in the gas temperature at which sputtering starts to be relevant, from 
$\sim 5\times 10^3$\,K to $\sim 10^4$\,K.

In the coronal hot gas (up to $\sim 10^8$\,K), electrons can reach very high thermal velocities, with kinetic energies capable of significantly altering the internal energy of a PAH molecule. In these sub-relativistic regimes ($\sim 10$\,keV), elastic collisions between electrons and target nuclei are not effective, and instead inelastic interactions with target electrons dominate. Similar to the case of electronic excitations by hot ions, these inelastic interactions excite molecules, followed by either dissociation or IR photo-relaxation. However, at these low kinetic energies, the Born approximation is no longer valid (no impulsive interaction, see also Section~\ref{subsec:sputtering}), which obliges us to instead use experimentally motivated stopping powers. We use the fit to the electron stopping power $S$ of solid carbon given by Eq.~\ref{eq:electron_stopping_power_solid_carbon} with the parameters in Table~\ref{tab:electron_stopping_power}. Since $S = - \frac{\dd E}{\dd l}$, we can find the maximum pathlength $L(\theta)$ through the PAH by  numerically integrating Eq.~\ref{eq:electron_stopping_power_solid_carbon}:
\begin{align}\label{eq:stopping_power_integral}
    \int \dd l = - \int^{E_1}_{E_0} \frac{\dd E}{S(E)} = l_1 - l_0 = L(\theta),
\end{align}
where we integrate from the initial energy of the electron $E_0$ to its final energy $E_1$ upon exit from the PAH. Given that we know for a given $\theta$ the value of $L$, we can numerically solve Eq.~\ref{eq:stopping_power_integral} to determine the value of $E_1$, such that the energy transferred by the electron to the $\pi$-valence electron cloud is simply $T_{\rm elec}(\theta)=E_0-E_1$.

In the microcanonical description of a PAH \citep[e.g.][]{Tielens2005TheMedium}, and under the ergodic approximation, the microcanonical temperature $T_{\rm m}$ can be approximated by:
\begin{align}
    T_{\rm m} \simeq 2000 \left(\frac{E_{\rm in}}{1{\rm eV}}\right)^{0.4}N_{\rm C}^{-0.4}\,\rm K,
\end{align}
which is valid over the range $35-1000$\,K. This can then be related to the effective temperature of the PAH for a particular internal energy $E_{\rm in}$:
\begin{align}\label{eq:pah_effective_temperature}
    T_{\rm eff} = T_{\rm m}\left(1-0.2\frac{E_{\rm bind}}{E_{\rm in}}\right),
\end{align}
with $E_{\rm bind}$ the binding energy of the dissociated fragment (e.g.~C$_2$H$_n$). The second term in Eq.~\ref{eq:pah_effective_temperature} is a correction factor due to the finite heat bath obtained when Taylor expanding the system energy $E$ around the average energy $\langle E(T_{\rm m})\rangle$. Therefore, the effective PAH temperature for a given excitation energy $T_{\rm e}$ (in eV) is approximately \citep{Draine2001InfraredGrains,Tielens2005TheMedium}:
\begin{align}
    T_{\rm eff} \simeq 2000 \left(\frac{T_{\rm e}}{N_{\rm C}} \right)^{0.4} \left(1-0.2\frac{E_{\rm bind}}{T_{\rm e}} \right)\, \rm K.
\end{align}

In the Gibbs microcanonical distribution, the bond dissociation rate $k_{\rm diss}$ is estimated via the Arrhenius law \citep{Tielens2005TheMedium}:
\begin{align}\label{eq:pah_dissociation_rate}
    k_{\rm diss} = k_0(T_{\rm eff}) \exp \left(- \frac{E_{\rm bind}}{k_{\rm B}T_{\rm eff}}\right),
\end{align}
where $k_0$ is a coefficient estimated by:
\begin{align}
    k_0(T_{\rm eff}) = \frac{k_{\rm B}T_{\rm eff}}{h}\exp \left(1 + \frac{\Delta S}{\mathcal{R}}\right),
\end{align}
where $h$ is Planck's constant, $\Delta S$ is the change in entropy due to dissociation, and $\mathcal{R}$ is the ideal gas constant. We use the dissociation parameters for the acetylene fragment (C$_2$H$_2$) given in \citet{Micelotta2010PolycyclicGas} and \citet{Ling1998EnergeticsCalculations}.
We now need to consider the competition between this dissociation pathway and IR photo-relaxation. Computing the probability for dissociation requires evaluating the photon emission and photo-dissociation rates over the energy cascade resulting from each IR photon emission. This quickly becomes a difficult numerical task, but because of the steep dependence of the dissociation rate $k_{\rm diss}$ on the internal energy (Eq.~\ref{eq:pah_dissociation_rate}), one can simplify this computation by evaluating the internal energy before IR cooling sets in \citep{Tielens2021MolecularAstrophysics}. We follow an approach to this method presented in \citet{Micelotta2010PolycyclicGas}, in which we assume all photons have the same energy $\Delta \epsilon =0.16$\,eV, corresponding to a typical C-C vibrational mode. In this case, the normalised dissociation probability\footnote{Note that in \citet{Micelotta2010PolycyclicGas} $P(n_{\rm max})$ is defined as the un-normalised probability, i.e.~is missing the second term in the denominator.} is given by:
\begin{align}\label{eq:pah_dissociatio_probability}
    P(n_{\rm max}) = \frac{k_0 \exp [-E_{\rm bind}/(k_{\rm B}T_{\rm av})]}{k_{\rm IR}/(n_{\rm max}+1)+k_0 \exp [-E_{\rm bind}/(k_{\rm B}T_{\rm av})]},
\end{align}
where $T_{\rm av}$ is the geometric mean of the PAH effective temperature between the initial excitation energy and the final equilibrium internal energy after $n_{\rm max}$ IR photon emissions. Naively, $n_{\rm max}$ could be estimated as $n_{\rm max}=(T_{\rm eff}-E_{\rm bind})/\Delta \epsilon$, assuming that the probability for dissociation after each IR photon emission is the same. However, an accurate computation of the probability indicates that these probabilities decrease. In \citet{Micelotta2010PolycyclicGas} they instead determine the value of $n_{\rm max}$ as the step at which the probability has decreased by a factor of 10 with respect to the previous step. We have obtained a simple scaling from their $n_{\rm max}$ for different $N_{\rm C}$ given by $n_{\rm max}= N_{\rm C}/5$. Once the dissociation probability $P(v,\theta,N_{\rm C})$ is computed, we can obtain the mean dissociation rate for a given particle velocity $v_k$:
\begin{align}
R_{\rm e}^k(v,N_{\rm C}) &=
v_k n_k F_{\rm C} \nonumber \\
&\quad \times
\int_0^{\pi/2}
\sigma_{\rm PAH,geo}(N_{\rm C},\theta)\,
P(v,\theta,N_{\rm C})\,
\sin\theta\, \dd\theta .
\end{align}
where here $k$ goes over ions and electrons as well, and the PAH geometrical cross-section seen by a particle colliding with a PAH depends on the angle $\theta$ as follows:
\begin{align}
    \sigma_{\rm PAH,geo}(N_{\rm C},\theta) = \pi R(N_{\rm C})^2 \cos \theta + 2R(N_{\rm C})d\sin \theta.
\end{align}
This can be integrated over the full Maxwellian distribution to give the total collision rate constant at a given $T$:
\begin{align}\label{eq:electronic_sputtering_rate}
    R_{\rm e}^k(T,N_{\rm C}) = \int_{v_{0}}^{\infty} R_{\rm e}^k(v,N_{\rm C})f(v,T)\dd v,
\end{align}
where $v_0$ is the impinging electron or ion velocity corresponding to the binding energy of the acetylene group $E_{\rm bind}$.

\begin{table}
    \centering
    \begin{tabular}{c|c|c}
        Fragment & $E_0$ [eV] & $\Delta S$ [cal K$^{-1}$ mol$^{-1}$]\\
        \hline
        \textbf{H} & 4.3 & 11.8\\
        \textbf{H$_2$} & 3.52 & -12.69\\
        \textbf{C$_2$H$_2$} & 4.6 & 10.0\\
    \end{tabular}
    \caption{Dissociation parameters used for our PAHs with normal hydrogen coverage.}
    \label{tab:my_label}
\end{table}
\begin{figure*}
     \centering
     \begin{subfigure}[b]{0.48\textwidth}
         \centering
         \includegraphics[width=\textwidth]{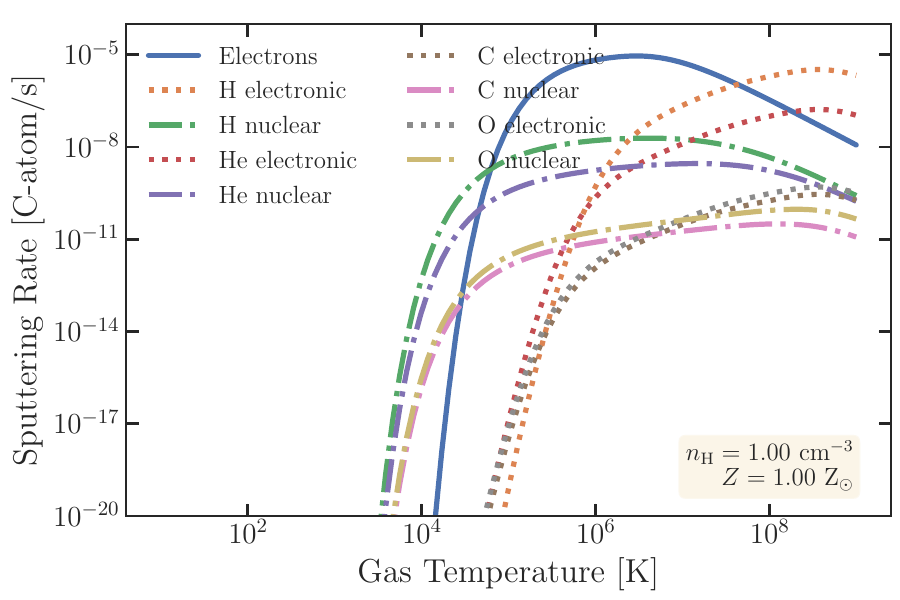}
         \caption{Results for small PAHs ($N_{\rm C}=54$). Nuclear collisions are only important at the onset of sputtering at $\sim 10^4$\,K, quickly overcome by the free electron destruction rate, peaking at $\sim 10^6$\,K. Ion electronic excitations play an important role at very high gas temperatures.}
         \label{fig:smallpah_thermal_sputtering}
     \end{subfigure}
     \hfill
     \begin{subfigure}[b]{0.48\textwidth}
         \centering
         \includegraphics[width=\textwidth]{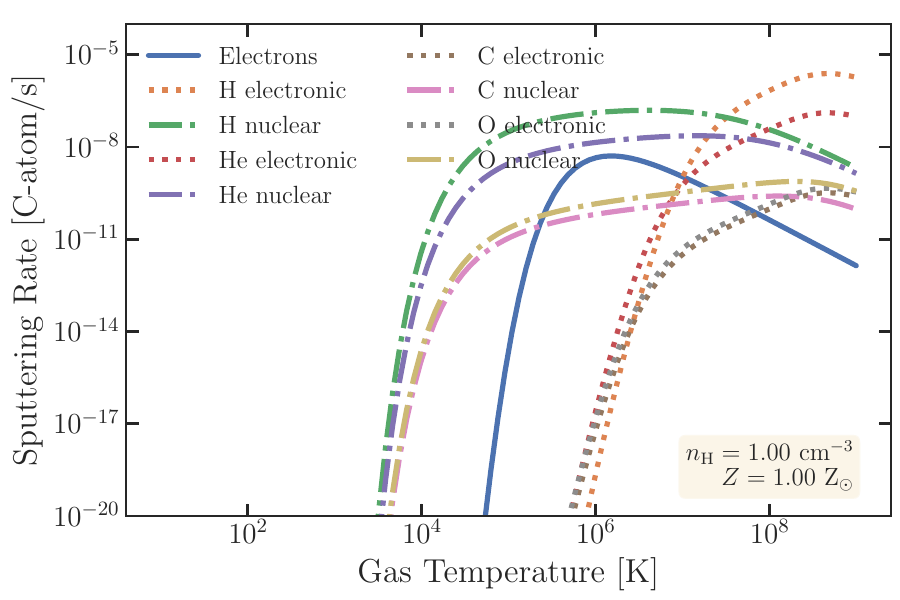}
         \caption{Results for large PAHs ($N_{\rm C}=418$). In this case, nuclear collisions dominate across all temperatures, with direct C collision important up to $\sim 10^7\,\rm K$, and nuclear electronic excitations at higher temperatures. Collisions with free electrons are less relevant.}
         \label{fig:largepah_thermal_sputtering}
     \end{subfigure}
        \caption{Sputtering rates for different PAH sizes due to free electron collisions (solid blue line), ion electronic excitations (dashed lines), and nuclear collisions (dot-dashed lines), separated by different ions: H, He, C and O. These rates have been computed for a gas density of $1\,\rm cm^{-3}$ and assuming solar metal abundances \citep{Asplund2009TheSun}.}
        \label{fig:pah_thermal_sputtering}
\end{figure*}
In Fig.~\ref{fig:smallpah_thermal_sputtering} we present the results of our PAH sputtering model for small PAHs ($N_{\rm C}=54$) embedded in ionised gas with density of $1\, \rm cm^{-3}$ and metal solar abundance following \citet{Asplund2009TheSun}. We drop the dependence on the factor $F_{\rm C}$ as this depends on the ion charge. Therefore, these rate constants are in units of 
C-atom/s (such that the sputtering mass-loss timescale can be computed with Eq.~\ref{eq:pah_sputtering_timescale}), for electronic ion collisions (Eq.~\ref{eq:electronic_sputtering_rate} with the probability obtained for the excitation energy given by Eq.~\ref{eq:ion_electronic_excitation_energy}), nuclear ion collisions (Eq.~\ref{eq:nuclear_sputtering_rate}), and free electron collisions (Eq.~\ref{eq:electronic_sputtering_rate} with the deposited energy given by Eq.~\ref{eq:stopping_power_integral}). For the ions, we separate the contributions into electronic and nuclear rates coming from the dominant ions: H, He, C and O. Thermal sputtering becomes a relevant process for PAHs at $\sim 10^4$\,K, i.e.~approximately 1\,dex lower in temperature than thermal sputtering for regular dust grains (see Section~\ref{subsec:sputtering}). At this temperature, the nuclear collisions dominate the destruction rate, with H and He dominating due to their larger abundance with respect to C and O. Above $\sim 10^4$\,K the destruction rate by free electrons grows sharply becoming the dominant sputtering agent in warm ionised gas (blue solid line). The electron sputtering peaks at $\sim 10^6$\,K, decreasing at higher temperatures due to decreasing stopping power (Eq.~\ref{eq:stopping power}) at these thermal energies. Ion electronic collisions start to dominate the PAH sputtering rate only when gas temperatures reach $\sim 10^8$\,K, again, mainly driven by H and He collisions. 

We have also computed these rates for our large PAHs ($N_{\rm C}=418$) in Fig.~\ref{fig:largepah_thermal_sputtering}. Due to direct dependence of nuclear sputtering rate on $N_{\rm C}$, large PAHs are dominated by C direct collisions up to $\gtrsim 10^7\,\rm K$. Also, the $N_{\rm C}$-dependence is implicit in the electron excitations, which means that electron and ion electronic collisions differ significantly from their smaller PAH counterparts. We find that for large PAHs, free electron collisions are subdominant at all temperatures.

\subsection{Equilibrium charge distribution of PAHs and their PEH and recombination cooling}\label{ap:pah_photoelectric_heating}
To be consistent with the rest of the photo-physical processes considered for PAHs, we use the distribution-averaged cross-sections in Section~\ref{subsec:pah_optical_properties} which have been computed for the small and large PAH bins based on the theoretical curves by \citet{Li2001InfraredMedium}. In this model we assume that with both grain sizes we can explore the charge parameter space $Q \in {-1,0,1,2}$. Larger grains are capable of reaching higher ionisation states than smaller grains. While this could be the case for PAH clusters ($\sim 10\rm\,\text{\AA}$), we have opted to maintain the di-cation as the maximum ionisation state for all PAHs \citep[see also][]{Andrews2016HydrogenationH2formation,Ibanez-Mejia2019DustMedium}. Experimental data for the ionisation potential ($IP$) of the di-cation is limited \citep{Zhen2016VUVPROCESSES}, but it seems that it is higher than 13.6\,eV for small PAHs, and theoretical modelling argue for a negligible variation at least up to circumcircumcoronene \citep[$N_{\rm C}=96$,][]{Andrews2016HydrogenationH2formation}. For the rest of the charge states, we use the $IP$ fitting function from \citet{Jochims1996PhotoionizationHydrocarbons,Verstraete1990IonizationGas} and updated by \citet{Wenzel2020AstrochemicalCations}. We assume that $IP^0=0$ for $Q=-1$. Based on the experimental studies of neutral \citep{Jochims1996PhotoionizationHydrocarbons,Verstraete1990IonizationGas} and cationic \citep{Zhen2016VUVPROCESSES} PAHs \citep{Wenzel2020AstrochemicalCations}, empirical functions can be derived for the electron yield of different PAHs. For the neutral species, we use the yields from \citet{Jochims1996PhotoionizationHydrocarbons}\footnote{Energy values are expressed in eV here and in the following.}:
\begin{align}\label{eq:ionisiation_yield_pah_neutral}
	Y(E,Q) = \begin{cases}
      \frac{E - IP^+}{9.2} & , E\leq IP^+ + 9.2\\
      1 & , E> IP^+ + 9.2
    \end{cases},
\end{align}
with $IP^+$ referring to the ionisation potential from $Q=0$ to $Q=1$. In the case of the first cation, we use the yields from \citet{Wenzel2020AstrochemicalCations}:
\begin{align}\label{eq:ionisiation_yield_pah_cation}
	Y(E,Q)= \begin{cases}
		0 &, E < IP^{2+}\\
		\frac{\alpha}{11.3-IP^{2+}} (E - IP^{2+}) &, IP^{2+} \leq E < 11.3\\
		\alpha &, 11.3 \leq E < 12.9\\
		\frac{\beta(N_{\rm C})-\alpha}{2.1}(E-12.9) &, 12.9 \leq E < 15.0\\
		\beta(N_{\rm C}) &, E \geq 15.0
	\end{cases},
\end{align}
where $\alpha =0.3$ and the correction factor $\beta$ is given by:
\begin{align}\label{eq:beta_correction}
	\beta(N_{\rm C}) = \begin{cases}
		0.59+8.1\times 10^{-3} N_{\rm C} &, 32\leq N_{\rm C} < 50\\
		1 &, N_{\rm C} \geq 50
	\end{cases}.
\end{align}
We set the ionisation yield of the anion equal to 1. This is an approximation, as recent experimental data \citep{Iida2021IR-photonAnions} has shown that electron detachment does not occur for the energy expected from electron affinity for the pentacene anion (C$_{22}$H$_{14}$), but instead experiments show efficient detachment for photon energies $E$ beyond 2.8\,eV. It remains unclear how this extrapolates to the larger PAH molecules and PAH clusters that we consider in \calima, but given the little effect on the PEH efficiency of having $IP^{0}> 0$ \citep{Berne2022ContributionObservations}, we have maintained this approximation across our PEH model.

The ionisation cross-section is defined as:
\begin{align}\label{eq:ionisation_cross_section}
	\sigma_{\rm ion}(N_{\rm C},E,Q) = Y(E,Q)\langle C_{\rm abs}(N_{\rm C},E,Q)\rangle,
\end{align}
which we assume to be in units of m$^2$ in this section and $C_{\rm abs}$ is the absorption cross-section in Section~\ref{subsec:pah_optical_properties}.
From this definition we can compute the ionisation rate for a PAH molecule of charge $Q$ and number of carbons $N_{\rm C}$:
\begin{align}\label{eq:pah_photo_ionisation_rate}
	k_{\rm pe} (N_{\rm C},Q) = \int_{IP^{Q+1}}^{13.6} \frac{\sigma_{\rm ion}(N_{\rm C},E,Q) F(E)}{E}\dd E,
\end{align}
where $F(E)$ is the flux density of the radiation field. Photo-ionisation is only allowed for photons with energy $E$ above the ionisation potential and below the Lyman limit. The recombination rate is calculated using Spitzer's method adapted by \citet{Verstraete1990IonizationGas} for cations:
\begin{align}
	k_{\rm rec} = 4.27\times 10^{-13} N_{\rm C}(1+\Phi) \left(\frac{T}{300{\rm K}}\right)^{1/2} \,{\rm cm^3\,s^{-1}},
\end{align}
where we have defined the dimensionless quantity:
\begin{align}
    \Phi = \frac{eU}{k_{\rm B} T} = N_{\rm C}^{-1/2}\left(\frac{T}{1.85 \times 10^5 {\rm K}}\right)^{-1} 
\end{align}
where $U$ is the mean electrostatic potential evaluated at the radius of the PAH. This equation can be extended to allow all cations ($Q> 0$):
\begin{align}\label{eq:pah_recombination_rate}
    k_{\rm rec} = 4.27\times 10^{-13} N_{\rm C}(1 + \Phi (1+Q))\left(\frac{T}{300{\rm K}}\right)^{1/2}\,\rm cm^3\,s^{-1}.
\end{align}
The electron attachment rate comes from empirical rates fitted to the quantum mechanical computations of \citet{Carelli2013ElectronProperties}:
\begin{align}
    k_{\rm att} = 2.74\times 10^{-9} \left(\frac{T}{300\,\rm K}\right)^{0.11} \exp \left(\frac{1.11\,\rm K}{T}\right)\,\rm cm^3\,s^{-1}.
\end{align}
For the fitting parameters $a$, $b$ and $c$ we use the results for coronene, the largest molecule in their study. Both electron recombination and attachment rates are very uncertain aspects of our PEH modelling. 

Another option available in \calima~for the electron recombination rate (see Section~\ref{subsec:eq_tests}) is the one given by \citet{Tielens2005TheMedium}:
\begin{align}\label{eq:pah_recombination_rate_Tielens}
    k_{\rm rec} \simeq 1.3\times 10^{-6} \sqrt{N_{\rm C}\frac{300\, \rm K}{T}}\,\rm cm^3\,s^{-1},
\end{align}
which is independent of $Q$. Experimentally measured rates for ionised small PAHs \citep{Abouelaziz1993MeasurementsCoefficients,Rebrion-Rowe2003ExperimentalFluoranthene,Novotny2005RecombinationPlasma,Bienner2005LaboratoryChemistry} fall well below the rate given by Eq.~\ref{eq:pah_recombination_rate_Tielens}. One explanation for this unexpected low sticking efficiency in small PAHs could be that dissociative recombination becomes important. When PAH cations recombine with electrons their internal energy is increased by the value of $IP$. This excited state needs to reach a new stable state, which could trigger molecular dissociation. However, for the $IP$ of circumcoronene cations (5.9 and 8.8\,eV for the first and second ionisation states) the IR relaxation rate is about 2 orders of magnitude larger than the dissociation rate \citep[e.g.][]{Montillaud2013EvolutionStates,Andrews2016HydrogenationH2formation}, and for larger PAHs their $IP$ is even smaller. This means that, for the PAHs of interest in this model, recombination is expected to be non-dissociative. Therefore, we assume that this extrapolated stability to ISM-relevant PAHs results in a roughly constant $k_{\rm rec}$ value of $10^{-5}$\,cm$^3$\,s$^{-1}$, close to the prediction of Eq.~\ref{eq:pah_recombination_rate}. Similarly, for electron attachment to neutral PAHs, the electron affinity increases with PAH size and for $N_{\rm c}>30$ it exceeds 1\,eV and auto-ionisation is very unlikely. Therefore, from the PAH polarisability, we obtain \citep{Tielens2021MolecularAstrophysics}:
\begin{align}\label{eq:electron_attachment_rate}
    k_{\rm att} \simeq 1.3 \times 10^{-7} N_{\rm C}\,\rm cm^3\,s^{-1}.
\end{align}
Again, this rate falls well above the experimental results of \citet{Tobita1992PolycyclicImpact} and \citet{Canosa1994ElectronTemperature}, which reported a rate close to $10^{-9}$\,cm$^3$\,s$^{-1}$. Given the lack of experimental studies on astrophysically relevant PAHs, we have opted to use the more conservative rates from \citet{Verstraete1990IonizationGas} and \citet{Carelli2013ElectronProperties}. These rates are highly uncertain, so both implementations of recombination and attachment are implemented in \calima, allowing future work to further constraint what prescription better agrees with PAH charging observations. In the future, these rates will need to be revised once further progress has been made on the experimental measurements of large PAHs.

By assuming equilibrium between the different processes driving the evolution of the different population charges, we obtain \citep{Berne2022ContributionObservations}:
\begin{align}\label{eq:pah_charge_fractions}
    f^- &= \left(1  + \frac{k_{\rm det}}{k_{\rm att}n_{\rm e}} + \frac{k_{\rm det}k_{\rm pe}^0}{k_{\rm att}k_{\rm rec}^+ n_{\rm e}^2} + \frac{k_{\rm det}k_{\rm pe}^0k_{\rm pe}^+}{k_{\rm att}k_{\rm rec}^+k_{\rm rec}^{2+}n_{\rm e}^3} \right)^{-1},\\
    f^0 &= \left(1 + \frac{k_{\rm att}n_{\rm e}}{k_{\rm det}} + \frac{k_{\rm pe}^0}{k_{\rm rec}^+ n_{\rm e}} + \frac{k_{\rm pe}^0 k_{\rm pe}^+}{k_{\rm rec}^+k_{\rm rec}^{2+}n_{\rm e}^2} \right)^{-1},\\
    f^+ &= \left(1 + \frac{k_{\rm rec}^+n_{\rm e}}{k_{\rm pe}^0} + \frac{k_{\rm pe}^+}{k_{\rm rec}^{2+}} + \frac{k_{\rm att}k_{\rm rec}^+ n_{\rm e}^2}{k_{\rm det}k_{\rm pe}^0} \right)^{-1},\\
    f^{2+} &= \left(1 + \frac{k_{\rm rec}^{2+}n_{\rm e}}{k_{\rm pe}} \right)^{-1},
\end{align}
where $f^-$, $f^0$, $f^+$ and $f^{2+}$ are the fractions of PAH molecules in the anion, neutral, mono-cation, and di-cation states, respectively, and $k_{\rm det}=k_{\rm pe}^-$. Based on this, we can compute the power injected by the photo-electrons from the first three charge states above:
\begin{align}
    P_e^{\rm tot} = f^-P_e^- + f^0P_e^0 + f^+ P_e^+,
\end{align}
where the power injected in the gas by the charge state $Q$ is defined as:
\begin{align}
    P_e ^Q = \int_{IP^{Q+1}}^{13.6}\gamma_e(E)(E - IP^{Q+1}) \sigma_{\rm ion}(Nc,E,Q)\frac{F(E)}{E}\dd E,
\end{align}
with $\gamma_e(E)$ encapsulating the fraction of the energy which goes into the kinetic energy $E_{\rm K}$ of the photo-electron. Energy conservation requires $E-IP=\Delta E^*(E)+E_{\rm K}(E)+E_{\rm K}^{\rm PAH}(E)$, where $\Delta E^*$ is the change in internal energy of the PAH upon excitation by ionisation and $E_{\rm K}^{\rm PAH}$ is its kinetic energy. PAH molecule has much larger mass than the electron's and due to conservation of momentum, the kinetic energy of the PAH molecule is negligible compared to that of the electron. Therefore, assuming that the photon energy is primarily used for the ionisation reaction, the internal energy of the PAH is barely modified, allowing to define the energy averaged partition coefficient \citep{DHendecourt1987EffectGas}:
\begin{align}
    \langle \gamma_e (E)\rangle = \biggl \langle \frac{E_{\rm K}(E)}{E-IP} \biggr \rangle .
\end{align}
We use the estimated value of $\langle \gamma_e (E)\rangle = 0.46 \pm 0.06$ obtained by \citet{Berne2022ContributionObservations} from the spectroscopic measurements performed for the coronene molecule by \citet{Brechignac2014PhotoionizationCation}. The final outcome of any PEH model is the total heating efficiency, $\epsilon_{\rm PAH}$, defined as the ratio of the power injected by the photo-electrons and the power absorbed by the PAHs:
\begin{align}
\epsilon_{\rm PAH}(N_{\rm C})
&= \frac{P_e^{\rm tot}}{P_{\rm abs}^{\rm tot}} \\
&= \frac{P_e^{\rm tot}}
       {f^- P_{\rm abs}^-
        + f^0 P_{\rm abs}^0
        + f^+ P_{\rm abs}^+
        + f^{2+} P_{\rm abs}^{2+}} .
\end{align}
where the power absorbed by each PAH charge state is:
\begin{align}
    P_{\rm abs}^Q = \int_{0}^{13.6\, \rm eV} \langle \sigma_{\rm abs}(N_{\rm C},E,Q)\rangle F(E) \dd E.
\end{align}
This definition of the PEH efficiency is similar to that used by \citet{Weingartner2001PhotoelectricHeating} but differs from \citet{Bakes1994TheHydrocarbons} as they only consider photons in the range $6-13.6$\,eV. From this, the local PEH rate for each PAH size bin writes:
\begin{align}
    \Gamma_{\rm PAH}(N_{\rm C}) = \epsilon_{\rm PAH} P_{\rm abs}^{\rm tot} n_{\rm PAH}(N_{\rm C}),
\end{align}
with $n_{\rm PAH}(N_{\rm C})$ being the local number density of PAH molecules with carbon number $N_{\rm C}$.

\begin{figure*}
    \centering
	\includegraphics[width=\textwidth]{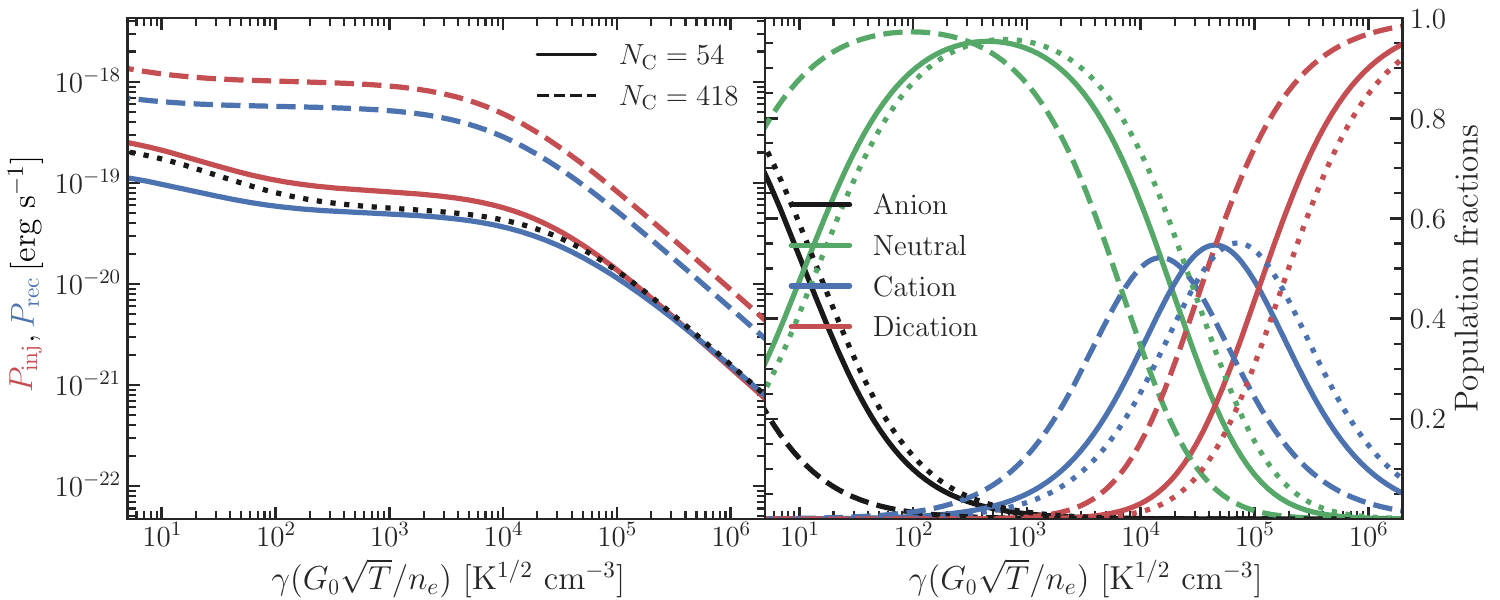}
    \caption{Photo-electric heating, recombination cooling and PAH charge distributions embedded in the \citet{Draine1978PhotoelectricGas.} at constant $T=10^4$\,K and varying charging parameter $\gamma$. We separate the contribution from small (solid lines) and large (dashed lines) PAHs, comparing also to the results from \citet{Berne2022ContributionObservations} for $N_{\rm C}=54$ (dotted lines). \textit{Left panel}: per PAH injected (red) and recombined (blue) power evolution with $\gamma$. \textit{Right panel}: PAH charge fraction for anions (black), neutral (green), cation (blue), and dication (red). Given the $\lesssim 50$\% lower cross-section for anion and neutral PAHs used in \citet{Berne2022ContributionObservations} compared to ours, we find $\sim 50$\% higher injected power at $\gamma < 10^4$.}
    \label{fig:peh_heating_charge}
\end{figure*}
In Fig.~\ref{fig:peh_heating_charge} we show the results of our modelling of the PAH charge distribution and PEH for the small (solid lines) and large (dashed lines) PAHs. In order to allow benchmarking against the modelling in \citet{Berne2022ContributionObservations} (dotted lines) for $N_{\rm C}=54$, we use the \citet{Draine1978PhotoelectricGas.} ISRF, fixing the gas temperature to $10^4$\,K and varying the value of $n_{\rm e}$ to span the range of charging parameter $\gamma$ of interest for PEH. Since the anion and neutral PAH cross-sections from \citet{Malloci2004ElectronicPhotophysics} are on average $\sim 30$\% smaller than ours, we find on a difference of $\sim 20$\% in the computed injected power (solid red line). This difference can also be appreciated in the population fractions of each charge state in the right panel of Fig.~\ref{fig:peh_heating_charge}. The most effective PAH charges for photo-electron ejection are the anion and the neutral, which dominate for $\gamma \lesssim 10^4$\,K$^{1/2}$\,cm$^{-3}$, explaining the plateau in $\epsilon_{\rm PAH}$ observed below this ionisation parameter value. We do not include it in this plot, but using the recombination and attachment rates from \citet{Tielens2021MolecularAstrophysics} as described above has a critical effect on the distribution of PAH charges, allowing for the dominance of anions ($Z=-1$) at low $\gamma$, and therefore an efficiency 4 times larger at $10^2<\gamma <10^4\,\rm K^{1/2}\,cm^{-3}$.

\end{appendix}

\vspace{1cm}

\end{document}